\documentclass[nopreprintline,11pt]{elsarticle} 

\usepackage[T1]{fontenc}
\usepackage[utf8]{inputenc}

\usepackage{lmodern}
\usepackage{textcomp}
\usepackage{microtype}

\usepackage[english]{babel}
\usepackage{csquotes}

\usepackage{amsmath}
\usepackage{amsfonts}
\usepackage{amssymb}
\usepackage{amsthm}
\usepackage{latexsym}
\usepackage{mathtools}

\usepackage{geometry} 
\usepackage[toc,page]{appendix}
\usepackage{lineno} 

\usepackage{xcolor}

\usepackage{float} 
\usepackage{multirow}
\usepackage{longtable} 

\usepackage{placeins} 

\usepackage{csquotes}
\usepackage{natbib}
\setcitestyle{authoryear,open={(},close={)}}
\newcommand{\textcite}{\cite}

\usepackage{graphicx}
\usepackage{setspace}

\usepackage{thmtools}

\declaretheorem[
	name=Theorem,
	numberwithin=section
	]{thm}

\declaretheorem[
	name=Proposition,
	sibling=thm,
	]{prop}
\declaretheorem[
	name=Corollary,
	sibling=thm,
	]{cor}
\declaretheorem[
	name=Remark,
	sibling=thm,
	style=remark
	]{rem}
\declaretheorem[
	name=Result,
	sibling=thm,
	]{result}
\declaretheorem[
	name=Assumption,
	style=remark
	]{asm}

\newcommand\Plost{P_{\rm{ext}}}
\newcommand\Pfix{P_{\rm{fix}}}
\newcommand\Psur{P_{\rm{sur}}}
\newcommand\Pexinf{P_{\rm{ext}}^\infty}
\newcommand{\sA}{s \alpha}
\def\A{\alpha}
\def\barA{\bar\A}
\def\Th{\Theta}
\def\sN{\nu}
\newcommand{\be}{\beta}
\newcommand{\ep}{\epsilon}

\newcommand{\ga}{\gamma}

\newcommand{\la}{\lambda}
\newcommand{\si}{\sigma}
\newcommand{\Ga}{\Gamma}
\newcommand{\De}{\Delta}
\newcommand{\bartfix}{\bar{t}_{\rm{fix}}}
\newcommand{\bartloss}{\bar{t}_{\rm{loss}}}
\newcommand\tfix{t_{\rm{fix}}}
\newcommand\tloss{t_{\rm{loss}}}
\newcommand\stilde{\tilde{s}(\A)}
\newcommand\Pexinftyfl{P_{\rm{ext}}^\infty}
\newcommand\Pexnfl{P_{\rm{ext}}^{(n)}}
\newcommand\pfl{p^{(\rm{fl})}}

\newcommand{\textblue}[1]{\textcolor{black}{#1}}
\newcommand{\EV}[1]{\operatorname{E}#1}
\newcommand{\Var}[1]{\operatorname{Var}#1}

\newcommand{\Poi}[1]{\operatorname{Poi}#1}
\newcommand{\Prob}[1]{\operatorname{Prob}#1}
\newcommand{\Exp}[1]{\operatorname{Exp}#1}
\newcommand{\Ei}[1]{\operatorname{Ei}#1}
\newcommand{\ProdLog}[1]{\operatorname{ProductLog}#1}
\defcitealias{Haldane1927}{Haldane's (1927)}
\defcitealias{Hill1982a}{Hill's (1982)}
\defcitealias{Hoellinger2019}{H\"ollinger's (2019)}
\defcitealias{Lande1976}{Lande's (1976) equation}
\defcitealias{Mathematica}{{\it Mathematica}}

\usepackage[pdftex,pdfpagelabels]{hyperref}
\hypersetup{pdftitle={manuscript},
			pdfauthor={Hannah Götsch}}

\numberwithin{equation}{section}
\numberwithin{figure}{section}
\begin{document}
\normalbaselineskip18pt
\baselineskip18pt
\allowdisplaybreaks[1]
\setcounter{page}{1}

\journal{TPB}

\begin{frontmatter}
\title{Polygenic dynamics underlying the response\\
 of quantitative traits to directional selection}

\author[address1,address2]{Hannah G\"otsch\corref{cor1}}
\cortext[cor1]{Corresponding author}
\ead{hannah.goetsch@univie.ac.at}
\author[address1]{Reinhard B\"urger}

\address[address1]{Faculty of Mathematics, University of Vienna, 1090 Vienna, Austria}
\address[address2]{Vienna Graduate School of Population Genetics, Austria}

\begin{abstract}

We study the response of a quantitative trait to exponential directional selection in a finite haploid population, both at the genetic and the phenotypic level. We assume an infinite sites model, in which the number of new mutations per generation in the population follows a Poisson distribution (with mean $\Th$) and each mutation occurs at a new, previously monomorphic site. Mutation effects are beneficial and drawn from a distribution. Sites are unlinked and contribute additively to the trait. Assuming that selection is stronger than random genetic drift, we model the initial phase of the dynamics by a supercritical Galton-Watson process. This enables us to obtain time-dependent results. We show that the copy-number distribution of the mutant in generation $n$, conditioned on non-extinction until $n$, is described accurately by the deterministic increase from an initial distribution with mean 1. This distribution is related to the absolutely continuous part $W^+$ of the random variable, typically denoted $W$, that characterizes the stochasticity accumulating during the mutant’s sweep. A suitable transformation yields the approximate dynamics of the mutant frequency distribution in a Wright-Fisher population of size $N$. Our expression provides a very accurate approximation except when mutant frequencies are close to 1. On this basis, we derive explicitly the (approximate) time dependence of the expected mean and variance of the trait and of the expected number of segregating sites. Unexpectedly, we obtain highly accurate approximations for all times, even for the quasi-stationary phase when the expected per-generation response and the trait variance have equilibrated. The latter refine classical results. In addition, we find that $\Th$ is the main determinant of the pattern of adaptation at the genetic level, i.e., whether the initial allele-frequency dynamics are best described by sweep-like patterns at few loci or small allele-frequency shifts at many. The number of segregating sites is an appropriate indicator for these patterns. The selection strength determines primarily the rate of adaptation. The accuracy of our results is tested by comprehensive simulations in a Wright-Fisher framework. We argue that our results apply to more complex forms of directional selection.

\end{abstract}

\end{frontmatter}

\textbf{Keywords:} polygenic adaptation, mutation, infinite-sites model, Galton-Watson branching process, Wright-Fisher model

\hypersetup{pageanchor=true}

\nolinenumbers
\modulolinenumbers[1]
\renewcommand\linenumberfont{\tiny}

\section{Introduction}

Phenotypic adaptation is ubiquitous and can occur rapidly as a response to artificial selection or to a gradual or sudden change in the environment. On the basis of accessible phenotypic measurements, the response of the mean phenotype can often be predicted or described by simple equations, such as the breeder's equation or its evolutionary counterpart, \citetalias{Lande1976} \citep{FalconerMackay1996,WalshLynch2018}. During this process genotype frequencies change, and new genotypes and new alleles can be generated by recombination and mutation. Even if the response of the mean phenotype is simple to describe, it is a highly complex, often insurmountable, task to infer the response on the genetic level.

There are multiple reasons for this problem. The response of a trait may be due to small frequency changes at a large number of loci of minor effect \citep{HillKirkpatrick2010, Barghi_etal2019, Boyle_etal2017}, as originally proposed by \cite{Fisher1918} and idealized and formalized in the infinitesimal model \citep{Bulmer1980, BartonEtheridgeVeber2017}. The selection response may also be due to selective sweeps at a few loci of large effect \citep{Kaplan_etal1989, Hermisson2005, MesserPetrov2013, Stephan2019selective}. Finally, a mixture of both is a third realistic scenario \citep{Lande1983, Chevin2008, Chevin2019}. In principle, different genetic architectures in combination with different initial conditions can lead to similar phenotypic responses, at least as far as the mean phenotype is concerned. Even if loci contribute additively to the trait and their number and effects are known, it is notoriously difficult to derive the allele-frequency dynamics at individual loci and the resulting dynamics of the trait's mean and variance from the selection pressure acting on the trait. Additional complications arise if linkage disequilibrium, epistasis, or pleiotropy have to be taken into account \citep{Turelli1990, Buerger2000, BartonKeightley2002, ZhangHill2005review, WalshLynch2018}.  

For quite some time the focus in studying adaptation at the genetic and molecular level has been on selective sweeps, first hard sweeps  and more recently soft sweeps which can occur from standing variation or immigration and are not restricted to mutation-limited evolutionary scenarios \citep{HermissonPennings2017review}. Although footprints of hard and soft selective sweeps are detectable in genetic data by various methods \citep{Nielsen2005, Huber2016, Setter2020volcanofinder}, they seem to be sparse in the data, especially hard sweeps in humans \citep{Pritchard2010, Hernandez_etal2011}. The detection of polygenic responses by small allele-frequency changes at loci of minor effect is much more demanding. It became feasible at a larger scale only when sequencing became cheaper \citep{Berg2014}. Genome-wide association studies (GWASs) became an important tool in identifying associations between genetic variants, typically single-nucleotide polymorphisms (SNPs), and phenotypic traits. Although GWASs are highly valuable in elucidating the genetic architecture underlying phenotypic traits,  they provide no direct information on the contribution of the causal variants to adaptation of the trait, which can be deleterious, neutral or advantageous. More refined approaches, such as experimental evolution methods that yield time-series data, have been pursued successfully to demonstrate truly polygenic adaptation \citep[reviewed in][]{Sella2019,Barghi2020}.

Theoreticians only recently started to explore when adaptation of polygenic traits is likely to be based on selective sweeps at a few loci vs.\ on small polygenic allele-frequency changes. Essential for this line of research is a good understanding of which genetic architectures and parameters (number and effects of loci, distribution of mutation effects, linkage, epistasis, pleiotropy) as well as population sizes and forms and strengths of selection can cause which adaptive patterns. By this we mean sweep patterns (hard or soft), polygenic patterns, or a mixture of both. 

Some studies focused on polygenic traits under stabilizing selection, where the trait mean is close to the optimum or at most a few phenotypic standard deviations away and the selective response is due to standing genetic variation established previously by mutation-selection-(drift) balance. The findings are qualitatively similar, whether the population size is so large that random genetic drift can be ignored \citep{deVladar2014, Jain2017} or moderately large so that it plays a role but is weaker than selection \citep{JohnStephan2020, Hayward2021}. Roughly, if most standing variation is due to many polymorphic loci of small effect, then the response of the mean is mainly caused by small frequency changes. Only under exceptional circumstances do selective sweeps occur and drive the response. This result is not unexpected because (i) loci of small effect are more like to carry alleles of intermediate frequency and (ii) close to the phenotypic optimum fine tuning at the contributing loci is important. 

Other studies focused on polygenic adaptation under directional selection \citep{Hoellinger2019}, for instance caused by a sudden big environmental shift \citep{Thornton2019, Hoellinger2023}. Although the assumptions on the genetic architecture and on the form of directional selection differ among these studies, they show that selective sweeps at several loci (parallel or successive) occur more often than in the above discussed scenario. The main determinant of the adaptive pattern at the genomic level is the so-called population-wide background mutation rate, but not the strength of selection or the population size \citep{Hoellinger2019}.

In this study, we also explore how various evolutionary and genetic factors determine the rate and pattern of polygenic adaptation, both at the phenotypic and the genetic level. However, our assumptions depart in several respects from those imposed in previous studies. First, we consider a trait subject to exponential directional selection, which is nonepistatic, does not induce linkage disequilibrium, and leads to a constant selection strength $s$. Second, in contrast to the investigations cited above, we focus on the response due to \emph{de novo} mutations, i.e., we ignore standing variation. Third, but similar to \cite{Thornton2019}, we assume an infinite sites model. The number of mutations per generation follows a Poisson distribution and each mutation occurs on a new, previously monomorphic locus. Loci are unlinked and contribute additively to the trait. Mutation effects are beneficial and follow a general distribution. This is justified by the crucial assumption that in our population of size $N$ selection is stronger than random genetic drift ($Ns>1$), so that prolonged segregation or fixation of non-advantageous mutations can be ignored. These assumptions enable us to derive analytical expressions for the time dependence of the mutant frequency distribution at each locus, of the expected number of segregating sites, and of the expected mean and variance of the trait. 

\textblue{To what extent are these assumptions biologically plausible? That selection can be (much) stronger than random genetic drift is supported by numerous studies of rapid adaptation in response to environmental change. These include the evolution of industrial melanism \citep{CookSaccheri2013}, pesticide resistance \citep{Hawkins_etal2019}, 
coat color in mice \citep{Linnen_etal2013}, and life-history traits in guppies \citep{Reznick_etal1997}. Available estimates of selection coefficients in some of these cases reach 0.4, with some published values of $Ns$ exceeding 1000 \citep[e.g.][]{Linnen_etal2013}.}

\textblue{Recent genetic analyses shed considerable light on the genetic basis of the response of several rapidly adapting traits. The response can be due to a single \citep[][peppered moth]{vantHof_etal2016} or several (\citealt[][fungicide resistance]{Hawkins_etal2019}; \citealt[][deer mice]{Linnen_etal2013}) \emph{de-novo} mutations; it can be highly polygenic from standing variation \citep[][herbicide resistance]{Hawkins_etal2019}; and it can arise from a combination of standing polygenic variation and large-effect genes or new mutations (\citealt[][insecticide resistance]{Hawkins_etal2019}; \citealt[][guppies]{Whiting_etal2022}). Obviously, the response in artificial selection experiments started from isogenic lines is always caused by \emph{de-novo} mutations. 
}

\textblue{Although our basic model is set up such that the initial response is solely due to new mutations, variation builds up quickly and the results in Section \ref{sec:stationary} show that correct predictions emerge even for the quasi-stationary phase when a balance between new mutation, directional selection, and random drift has been reached. During this phase the response of the trait is caused by a combination of standing variation and new mutation.
On the basis of numerous theoretical studies, we argue in Section \ref{Disc:Outlook} that our neglect of epistasis and linkage disequilibrium imposes only weak limitations on our results unless selection has a significant non-directional component  (e.g., stabilizing, as in late phases of adaptation) or usable genetic variation is concentrated in a small genomic region.}

The paper is structured as follows. In Section \ref{sec:Model}, we introduce our basic model and give an overview of the analytical concepts regarding the one-locus model and its extension to infinitely many sites. Also our simulation setting is described. 
In Section \ref{sec:dyn_allele_freq}, we derive an explicit, approximate expression for the mutant frequency distribution in a finite Wright-Fisher population as a function of time. This becomes feasible by using a supercritical Galton-Watson process with a quite general offspring distribution to describe the spreading of a beneficial mutant. 
Comparison with simulation results of a corresponding Wright-Fisher model demonstrates that our approximation is highly accurate for a wide range of parameters and a large time range. 

Key results are explicit and computationally efficient time-dependent expressions for the expected mean and the expected variance of an additive quantitative trait under exponential selection. They are presented in Section \ref{sec:MeanVar}, seem to be entirely new, and provide highly accurate approximations to corresponding Wright-Fisher simulations. Interestingly, they even allow the derivation of expressions for the long-term, quasi-stationary response of the trait's mean and variance. They are not only derived from first principles by assuming the infinite sites model but also recover and refine classical results. Proofs are given in Appendix \ref{Proofs_MeanVar}. In Section \ref{sec:segsites} (and Appendix \ref{app:segsites}), we derive explicit, approximate expressions for the evolution of the expected number of segregating sites. They are based on a combination of branching process methods with (in part new) approximations for the expected time to loss or to fixation of a new beneficial mutant. The latter are deduced and numerically tested in Appendix \ref{meanfixtime}. In Section \ref{sec:indicator-sweep-shift}, we use the approximation for the number of segregating sites to characterize the numerically determined initial response patterns. This allows us to examine the genomic patterns associated with the early phase of phenotypic adaptation.

Because this paper is rather technical, we summarize, explain, and discuss our findings comprehensively in Section \ref{sec:discussion}. A central quantity in our theory is the expected number $\Th$ of new beneficial mutations occurring per generation in the total population. We provide quantitative support to previous findings in related but different contexts that sweep-like patterns will occur if $\Th$ is sufficiently much smaller than 1, whereas polygenic shifts will drive the initial selection response if $\Th$ is much greater than 1. Other model parameters have only a minor influence on the initial adaptive pattern, but may have a major influence on the rate of response of the trait. We propose to use the expected number of segregating sites as an indicator for distinguishing between sweep-like vs.\ polygenic shift-like patterns in the early phases of adaptation. In Section~\ref{Disc:Outlook}, we discuss the applicability of our approach to more general selection scenarios and to more general genetic architectures as well as the limitations inherent in our model assumptions.
\section{The model}\label{sec:Model}

We assume a haploid, panmictic population which has discrete generations and a constant population size of $N$. Our main goal is to explore the dependence of the pattern and rate of allele-frequency change at the loci driving the adaptation of a quantitative trait on the model's parameters. To achieve this aim, we derive the dynamics from first principles. First we set up the basic model for the quantitative trait, then the underlying one-locus model, the multilocus model, which we assume to be an infinite-sites model, and the simulation model. A glossary of symbols is provided in Table \ref{table_notation}.

\subsection{The quantitative-genetic model}\label{sec:QGmodel}

We consider a quantitative trait subject to nonepistatic directional selection and investigate the evolution of the distribution of allele frequencies at the loci contributing to this trait during adaptation.
The fitness of an individual with phenotypic value $G$ is given by $\exp(sG)$, where the selection coefficient $s>0$ is assumed to be constant in time. The trait is determined additively by a (potentially infinite) number of diallelic loci at which mutation occurs. We ignore environmental contributions to the phenotype which, in the absence of genotype-environment interaction and of epistasis, would only weaken selection on the underlying loci. Therefore, genotypic and phenotypic values are equal.  

Each locus $i$ can be in the ancestral state $a_i$ or in the derived state $A_i$. The effect on the phenotype of a locus in the ancestral state is zero, whereas a mutation at locus $i$ contributes a value $\A_i>0$ to the phenotype $G$. Thus, the phenotype of an individual is given by $G = \sum_i \A_i \delta_{A_i}$, where $\delta_{A_i}=1$ if locus $i$ is in the derived state, and $\delta_{A_i}=0$ otherwise. 
The fitness of the ancestral allele is normalized to $1$. Therefore, the fitness of the ancestral genotype, which has value $G=0$, is also one. Because of our additivity assumption and the shape of the fitness function, a mutation at locus $i$ has a fitness of $\sigma_i = e^{\sA_i}>1$, where $s>0$. Hence, all mutations are advantageous. Because we mostly assume $Ns>1$, deleterious mutations are very unlikely to reach high frequency or become fixed, and thus will not contribute to a permanent selection response of the mean phenotype.

Unless stated otherwise, we assume that initially all loci are in the ancestral state and adaptation occurs from \emph{de novo} mutations. New mutations occur from the ancestral state to the derived state and back-mutations are ignored. We define the population-wide \emph{advantageous} mutation rate by $\Th = N U$, where $U$ is the expected number of mutations per individual that increase the trait value. Thus, $\Th$ is the expected number of new mutations per generation that affect the trait. We assume an \emph{infinite sites model}, i.e., each mutation hits a new locus. This assumption is appropriate if per-locus mutation rates are small enough such that the next mutation occurs only after the previous one becomes lost or fixed. 
 
In the infinite sites model we investigate two different scenarios. In the first, all mutation effects are equal. In the second, mutation effects $\A_i$ at any locus $i$ are drawn from a distribution defined on $[0,\infty)$ that has a density $f$. Its mean is denoted by $\bar\A$ and all its higher moments are assumed to exist. Thus, the loci are exchangeable and all mutations are beneficial. For specification of analytical results and their comparison with simulation results, we use an exponential distribution \citep[for empirical support, see][]{Eyre-Walker&Keightley2007, Tataru2017} and a truncated normal distribution (see Remark \ref{rem:truncatedGaussian}).

\subsection{The one-locus model}\label{sec:1Lmodel}

We consider a single diallelic locus in isolation and assume that allele frequencies evolve according to a haploid Wright-Fisher model with selection. The (relative) fitness of the ancestral allele $a$ and of the derived allele $A$ (a new mutant) are set to $1$ and $\sigma= e^s$, $s>0$. Therefore, the selective advantage of the mutant relative to the wildtype is $e^s-1\approx s$. We assume that in generation $n=0$ the new beneficial mutant $A$ occurs as a single copy and no other mutation occurs while this allele is segregating. 

Because it seems unfeasible to derive analytic results for the time dependence of the allele frequency for either the Wright-Fisher model or its diffusion approximation, we approximate the stochastic dynamics in the initial phase, where stochasticity is most important, by a branching process \citep[e.g.][]{Athreya1972, Allen2003, Haccou2005}. During this initial phase, interactions between mutant lineages can be ignored if $N$ is large enough. The first step is to approximate the evolution of the frequency distribution of the mutant by a Galton-Watson process. In Section \ref{sec:dyn_allele_freq}, we will approximate this discrete process by a continuous one and couple it with the deterministic allele-frequency dynamics to obtain an analytically accessible model for the long-term evolution of the mutant in a population of size $N$.

To be specific, we consider a Galton-Watson process $\{Z_n\}$, where $Z_n=\sum_{j=1}^{Z_{n-1}}\xi_j$, $Z_0=1$, and $\xi_j$ denotes the (random) number of offspring of individual $j$ in generation $n-1$. Thus, $Z_n$ counts the number of descendants of the mutant that emerged in generation 0. We assume that the $\xi_i$ are independent, identically distributed random variables having at least three finite moments, where 
\begin{linenomath}\begin{equation}\label{def_momentsxi}
	\EV [\xi_j] =\si>1 \;\text{ and }\; \Var[\xi_j] = v \,.
\end{equation}\end{linenomath}
Therefore, the process $\{Z_n\}$ is supercritical. The $\xi_i$ are also independent of $n$.  We denote the generating function of $\xi_j$ (hence of $Z_1$) by $\varphi$. 

Whenever we consider extinction or survival probabilities as functions of $\si=e^{s\A}$, we assume that $v=2\si/\rho$  holds for the underlying family of offspring distributions, where $\rho>0$ is a constant (cf.~Remark \ref{Psur=csalpha}). In this case, we also assume that the probability of ultimate extinction is continuous in $\si=e^{s\A}$ and decreases with increasing $\si$.
When we compare our analytical results with simulation results from a Wright-Fisher model, we assume that the offspring distribution is Poisson, because this provides a close approximation to the binomial distribution if the population size $N$ is sufficiently large, especially if $Ns\gg1$ \citep[cf.][]{Lambert2006}.

\subsection{The infinite-sites model}\label{sec:ISmodel}

We assume that the trait is determined by a potentially infinite number of loci at which new, beneficial mutants can occur. We assume an infinite-sites model, so that every mutation occurs at a new, previously monomorphic locus. 

The initial population starts evolving in generation $\tau=0$ in which the first mutations may occur.
Prior to generation 0, the ancestral allele is fixed at every site potentially contributing to the trait. We assume that in each generation $\tau\ge0$, the total number of new mutants emerging in the population follows a Poisson distribution with mean $\Th$. Each mutant occurs at a new site (locus). Because selection is multiplicative in our model and random genetic drift is a weak force relative to selection (because $N$ is assumed to be large), we ignore linkage disequilibrium among loci. Hence, we assume that the allele-frequency distributions at the different loci evolve independently.

According to the mutation model, the $i$th mutation will occur at a random time $\tau_i$. The corresponding locus will be labeled locus $i$, and the mean offspring number of the mutant - when the mutant is still rare in the population - is denoted by $\si_i>1$. For mathematical convenience, we approximate the discrete distribution of the waiting time $\tau_i$ by the Erlang distribution with density 
\begin{linenomath}\begin{align}\label{Erlang}
	m_{i,\Th} (t) & = \frac{\Th ^i t ^{i-1}}{(i-1)!} \exp (-\Th  t )\,, \;\text{ if } t>0
\end{align}\end{linenomath}
(and $m_{i,\Th} (t)=0$ if $t\le0$). Throughout, we use $t$ as an integration variable and emphasize that generations $\tau$ are discrete. The mean waiting time until mutation $i$ is $i/\Th$. We note that if mutations occurred according to a Poisson point process with rate $\Th$, the Erlang distribution would be the exact waiting-time distribution between $i$ events. Here, we use it as an approximation.
The total number of mutation events $k$ until generation $\tau$ is Poisson distributed with  mean $\Th \tau$
\begin{linenomath}\begin{align}\label{Poi(Thetatau)}
	\Poi_{\Th\tau} (k) & = \frac{(\Th\tau )^k}{k!} \exp (-\Th\tau )\,.
\end{align}\end{linenomath}

\subsection{The simulation model}\label{SIMmodel}

The simulation algorithm written in \textit{C++} is based on previous versions by \cite{ThesisIlse} and \cite{ThesisBen}. 
With the above assumptions, we generate Wright-Fisher simulations forward in discrete time to examine the adaptation of a trait under selection, mutation and random genetic drift in the case of an originally monomorphic ancestral population. The first mutation events can occur in generation $0$. The algorithm acts on each locus because we assume that loci evolve independently. The basic idea for the algorithm is that each generation is represented by a vector, whose length is determined by the number of mutated loci. A mutation event occurs according to the mutation rate, adds a new locus and thus increases the length of the vector by one. The entries of the vector contain the number of mutants at each locus. Random genetic drift is included via binomial sampling with sampling weights due to selection \citep[][p.\ 24]{Ewens2004}.
We use the \textit{GNU Scientific Library} \citep{GNU} and the \textit{MT19937} \citep{Matsumoto} random number generator.
Unless stated otherwise, a simulation for a set of parameters consists of $20\,000$ replicate runs. In almost all cases, $20\,000$ replicates yielded standard error bars smaller than the size of the symbols in the figures; in a few cases, more runs were performed. Therefore, error bars are not shown. Runs are stopped at a given number $n$ (or $\tau$) of generations. From these replicate runs, means (and variances) of the quantities of interest are computed, or histograms if distributions are the objects of interest.
\section{Dynamics of allele-frequency distributions}\label{sec:dyn_allele_freq}

It is well known that, conditioned on fixation, the frequency of a beneficial mutant grows faster than predicted by the corresponding deterministic model \citep[e.g.][]{MaynardSmithHaigh1974, OttoBarton1997}. For models related to ours, it has been shown that conditioned on fixation and once a certain threshold has been exceeded, the evolution of the frequency of mutants can be approximated by a deterministic increase from a random variable, often labeled $W$, that describes the stochasticity that accumulates during the spread of the mutant \citep{Desai2007, Uecker2011}. The latter authors called this the \emph{effective initial mutant population size}. \citet{Martin2015} employed a variant of this approach and approximated the initial and the final phase of evolution by a Feller process conditioned on fixation. Based on this, they derived a semi-deterministic approximation for the distribution of the (beneficial) allele frequency at any given time. 

Below, we develop a refined variant that provides a highly accurate, analytically tractable, and computationally efficient approximation for the frequency distribution of a beneficial mutant in any generation $n$ in a diallelic haploid Wright-Fisher population of size $N$ (Result \ref{thm:g(p)}). Its extension in eqs.~\eqref{tilde{h}^{(i)}} and \eqref{tildeX^{(i)}} to the $i$th mutant occurring in the infinite sites model paves the way for an analytical treatment of our applications in Sections \ref{sec:MeanVar} and \ref{sec:sweepshift}.

\subsection{Approximating the random variable $W$}\label{sec:general_single_locus}

Our starting point is the supercritical Galton-Watson process $\{Z_n\}$ defined in Section \ref{sec:1Lmodel}. The probability of ultimate extinction, denoted by $\Pexinf(\si)$, is the unique fixed point of the offspring generating function $\varphi$ in $(0,1)$, i.e., it satisfies $\varphi(\Pexinf)=\Pexinf$ and $0<\Pexinf<1$. When no confusion can occur, we suppress the dependence of $\Pexinf$ on the mean offspring number $\si>1$.

We define $W_n=Z_n/\si^n$. According to a classical result \citep[][p.\ 154]{KestenStigum1966, Athreya1972, Haccou2005} there exists a random variable $W$ such that
\begin{linenomath}
\begin{equation}\label{defW}
	\lim_{n\to\infty} W_n = W
\end{equation}
\end{linenomath}
with probability 1. The limiting random variable $W$ has expectation $\EV[W]=1$, satisfies $W=0$ precisely if the process dies out, and $W$ is positive and absolutely continuous\footnote{Absolute continuity of a function $f$ is equivalent to having a derivative $f'$ almost everywhere and $f'$ being Lebesgue integrable.} otherwise. We define the random variable $W^+$ corresponding to the absolutely continuous part of $W$ by
\begin{linenomath}
\begin{equation}\label{defW+}
	\Prob[W\le x] = \Pexinf + (1-\Pexinf)\Prob[W^+\le x]\,, \quad \text{where $x  \geq 0$}\,,
\end{equation}
\end{linenomath}
so that $W^+$ has a proper distribution with density $w^+$. Because $\EV[W]=1$, eq.~\eqref{defW+} informs us that $\EV[W^+]=(1-\Pexinf)^{-1}$.

In general, not much is known about the shape of the distribution $W^+$. However, for a geometric offspring distribution with mean $\si$, $W^+$ is exponential with parameter $1-1/\si$ \citep[for a proof, see e.g.][pp.\ 155--156]{Haccou2005}. As pointed out by a reviewer, $W^+$ is exponential if and only if the offspring distribution is a fractional linear, or modified geometric, distribution (which has a fractional linear generating function). We treat the properties of the associated Galton-Watson processes that are relevant in our context in Appendix \ref{app_frac_linear}. We will use them when extending results in Sections \ref{sec:MeanVar} and \ref{sec:sweepshift} to cases when the effective population size $N_e$ differs from the actual population size $N$.

We denote the Laplace transform of $W$ by $\Phi(u) = \EV [e^{-uW}]$, where $u \geq 0$. It is well known \citep[p.\ 10]{Athreya1972} that $\Phi$ is uniquely determined as the solution of Poincar\'e's (or Abel's) functional equation
\begin{linenomath}
\begin{equation}\label{Phi_varphi}
	\Phi(u)=\varphi(\Phi(u/\si))\,.
\end{equation}
\end{linenomath} 
In general, this cannot be solved explicitly. However, if the offspring distribution is geometric, then \eqref{Phi_varphi} becomes
\begin{linenomath}
\begin{equation}\label{Phi_varphi_geo}
	\Phi(u) = \frac{1}{1+\si(1-\Phi(u/\si))}\,.
\end{equation}
\end{linenomath}
This is (uniquely) solved by the Laplace transform of $W$ with (cumulative) distribution
\begin{linenomath}
\begin{equation}\label{distW_geo}
	\Prob [W \leq x] = \Pexinf + (1-\Pexinf)\left( 1- \exp [-(1-\Pexinf)x] \right)\,.
\end{equation}
\end{linenomath}
Hence, $W^+$ is exponentially distributed with parameter $1-\Pexinf = 1-\si^{-1}$ \citep[p.\ 155]{Haccou2005}. 

For a Poisson offspring distribution, which has the generating function $\varphi(x) = e^{-\si(1-x)}$, we use the approximation $\varphi(x)\approx(1+\si(1-x))^{-1}$ which has the same error as the first-order Taylor approximation near $x=1$, but is much more accurate on the whole interval $[0,1]$. With this approximation, we obtain from \eqref{Phi_varphi} that $\Phi(u) \approx \Bigl(1+\si(1-\Phi(u/\si))\Bigr)^{-1}$.
Therefore, for a Poisson offspring distribution, we approximate $W^+$ by an exponential distribution with parameter $1-\Pexinf$, where $\Pexinf$ is the corresponding extinction probability (see Sect.~\ref{sec:Plost(n)}).

\begin{rem} \label{rem:W+}
A straightforward calculation shows that
\begin{linenomath}\begin{equation}\label{VarW+}
	\Var[W^+] = \frac{1}{1-\Pexinf}\left(\Var[W] +1 -  \frac{1}{1-\Pexinf} \right)\,,
\end{equation}\end{linenomath}
where $\Var[W]=v/(\si(\si-1))$ and $v$ is the variance of the offspring distribution \citep[cf.][p.~4]{Athreya1972}.
If the offspring distribution is geometric with mean $\si>1$, then $\Pexinf=1/\si$, $\Var[W^+]=\si^2/(\si-1)^2=\EV[W^+]^2$, as is required for an exponential distribution. For a Poisson offspring distribution with mean $\si>1$, it follows that the coefficient of variation of $W^+$ has the series expansion $1-\tfrac13(\si-1)+O((\si-1)^2)$. It can also be shown that the skew has the expansion $2-(\si-1) + O((\si-1)^2)$. This suggests that as $\si$ decreases to 1, the accuracy of the exponential approximation increases. Therefore, our approximations perform best in the slightly supercritical case, i.e., if $\si$ is only slightly larger than 1. It is straightforward to infer that for a Poisson offspring distribution, $W^+$ is not gamma distributed because the skew is too small.
\end{rem}

\begin{rem}
As mentioned above and shown in Appendix \ref{app:fraclin_W_n_W} for self-containedness, the distribution of $W^+$ is exponential if and only if the offspring distribution is fractional linear. Our argument above shows that $W^+$ is approximately exponential with parameter $1-\Pexinf$ for a Poisson offspring distribution, and second and third moments converge to that of an exponential as $\si\to1$. We expect that this is the case for a large class of offspring distributions, i.e., when for $\si\to 1$ the generating function is asymptotically equivalent to that of a fractional linear offspring distribution. Indeed,  
\cite{Uecker2011} derived an exponential distribution for $W^+$ for a birth-death branching processes with time-dependent mean offspring number (as they consider a model with temporal variation in its population size and the selection pressure). \cite{Martin2015} show that $W^+$ is approximately exponential with parameter $1-\Pexinf$ for Feller diffusions. They start the deterministic phase in their Wright-Fisher model with selfing from a gamma distribution because they admit one or more initially segregating mutants.
\end{rem}

\subsection{Extinction probability by generation $n$ for a Poisson offspring distribution}\label{sec:Plost(n)}

Because we use a Poisson offspring distribution for comparisons of our analytic results with simulation results from the corresponding Wright-Fisher model, we summarize the properties of the probabilities of extinction of the mutant until generation $n$. 
We assume that the mutant starts as a single copy in generation $n = 0$, and the offspring distribution is Poisson with mean $\si=e^s$, $s>0$. Then the probability $\Pexinf(\si)$ of ultimate loss, or extinction, of the mutant is the unique solution of $\varphi(x) = e^{-\si(1-x)} =x$ in $(0,1)$. It can be expressed in terms of the (principal branch of the) product logarithm, or the Lambert function:
\begin{linenomath}\begin{equation}\label{plost*}
	\Pexinf(\si) = -\frac{\ProdLog[-\si\exp(-\si)]}{\si}\,.
\end{equation}\end{linenomath}
With $\si=e^s$, this has the series expansion
\begin{linenomath}\begin{equation}\label{Plost*app}
	\Pexinf(\si) = 1 - 2s + \frac{5s^2}{3}  + O(s^3)\,.
\end{equation}\end{linenomath}
The extinction probability $\Pexinf(e^s)$ decreases from $1$ to $0$ as $s$ increases from $0$ to $\infty$. We write $\Psur = 1- \Pexinf$ for the (long-term) survival probability. It follows directly from \eqref{Plost*app} that \citetalias{Haldane1927} approximation, $2s$, serves as an upper bound for $\Psur$ and approaches $\Psur$ as $s\to0$ (indeed, Haldane derived an equation equivalent to $e^{-\si(1-x)} =x$).
 
\begin{figure}[t]
\centering
\includegraphics[width=0.7\textwidth]{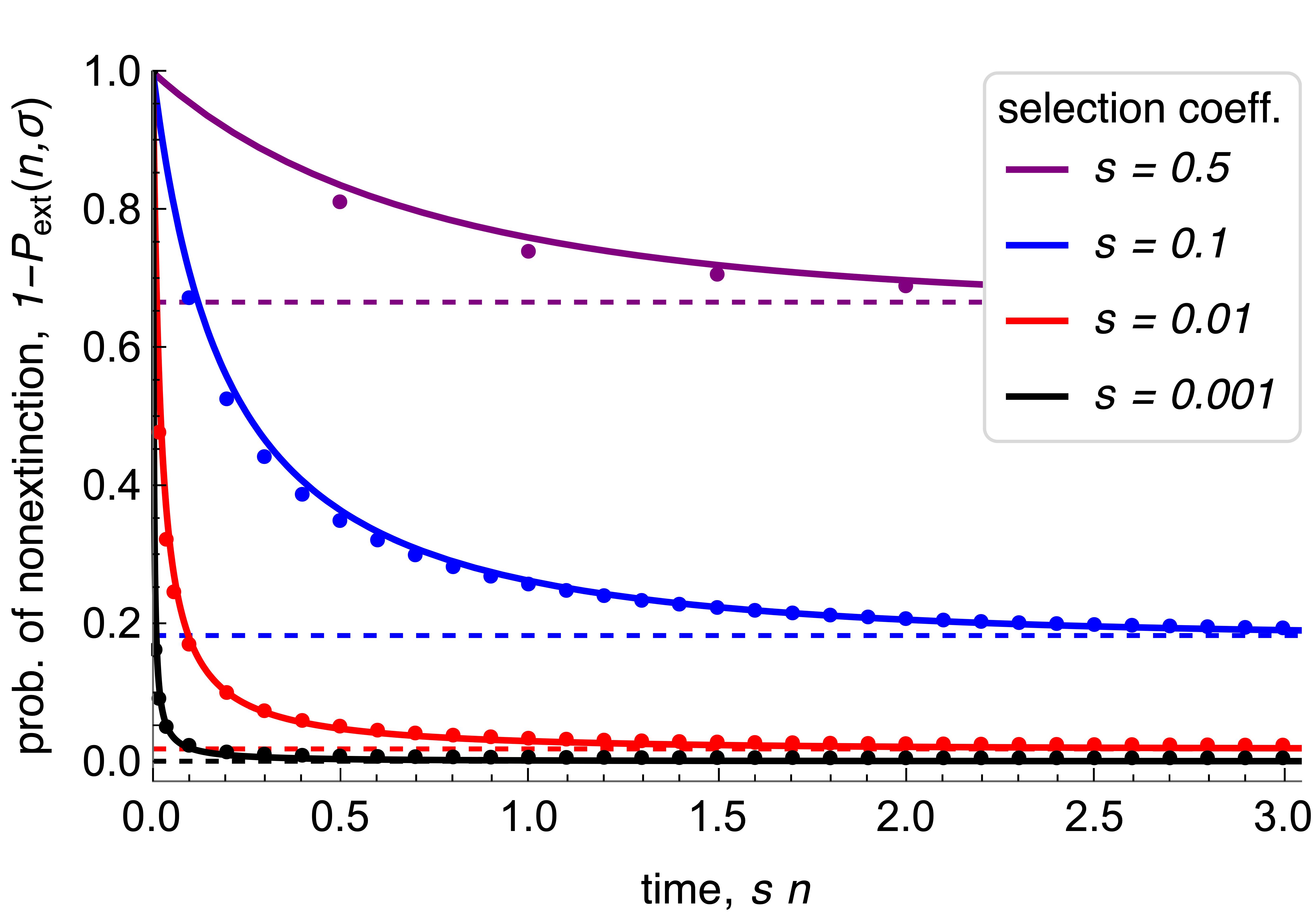}
\caption[$\Plost(n,\si)$]{Probability of non-extinction of a mutant by generation $n$, $1-\Plost(n,\si)$, for various values of the mean $\si=e^s$ of a Poisson offspring distribution. Time is scaled to facilitate comparison for different selection coefficients $s$. 
The dots show $1-\Plost(n,\si)$ as obtained from numerical iteration of \eqref{plost(tau)}. The solid curves show the respective approximations (and upper bounds) obtained from \eqref{Psurvn/Psurinfty} with $\Pexinf(\si)$ from \eqref{plost*}. The dashed lines show the survival probabilities $\Psur(\si) = 1-\Pexinf(\si)$, as calculated from \eqref{plost*}.}
\label{figplosttau}
\end{figure}

We define $\Plost(n,\si)$ as the probability that the mutation gets lost before generation $n$, i.e., $1-\Plost(n,\si)$ is the probability that the mutation is still segregating in generation $n$. If no confusion can occur, we write $\Plost(n)$ instead of $\Plost(n,\si)$. Then $\Plost(n)$ satisfies the recursion
\begin{subequations}\label{plost(tau)}
\begin{linenomath}\begin{align}
	\Plost(0) & = 0,  \\
	\Plost(n) & = \exp [ - \si (1-\Plost(n - 1))]. \label{plost(tau)b}
\end{align}\end{linenomath}
\end{subequations}
because $\Plost(n)=\varphi_{(n)}(0)=\varphi\bigl(\varphi_{(n-1)}\bigr)(0)$, the $n$th iteration of the generating function $\varphi$ of the Poisson distribution. By monotonicity of $\Plost(n)$ and continuity of $\varphi$ it follows that $\Plost(n,\si)\to\Pexinf(\si)$ for $n \to \infty$ and for every $\si$. Indeed, it is well known that $\Plost(n,\si)\to\Pexinf(\si)$ holds for very general offspring distributions.

An explicit formula for $\Plost(n)$ seems to be unavailable. However, the recursion \eqref{plost(tau)} is readily evaluated numerically and convergence of $1-\Plost(n,\si)$ to $\Psur(\si)=1-\Pexinf(\si)$ is fast (Fig.~\ref{figplosttau}). As is documented by Figure \ref{figplosttau} and supported by extensive numerical evaluation, the inequality
\begin{linenomath}\begin{equation}
	\frac{1-\Plost(n,\si)}{1-\Pexinf(\si)} \le 1 + \frac{\Pexinf(\si)}{\si^n-\Pexinf(\si)}  \label{Psurvn/Psurinfty}
\end{equation}\end{linenomath}
apparently holds always, and the upper bound serves as a highly accurate approximation unless $\si$ is large and $n$ is very small.\textblue{\footnote{\textblue{Indeed, this inequality can be proved for Poisson distributions (and some other families), even in the slightly stronger form when $\si^n$  is replaced by $(1/\varphi'(\Pexinf(\si)))^n$. Also counter examples exist (RB, unpublished results).}}} For fractional linear offspring distributions, equality holds in \eqref{Psurvn/Psurinfty} (Appendix \ref{app_fl_basics}). Indeed, convergence of $1-\Plost(n,\si)$ to $1-\Pexinf(\si)$ at the geometric rate $1/\si$ follows immediately from Corollary 1 in Sect.~1.11 of \citet{Athreya1972}. The time $T_\ep$ needed for the probability of non-extinction by generation $n$ to decline to $(1+\ep)(1-\Pexinf(\si))$ is explicitly given in \eqref{Tep_FL_general} for the fractional linear case, and it provides an accurate approximation and upper bound for the Poisson case (see Appendix~\ref{app_frac_lin_compare}).

\begin{rem}\label{Psur=csalpha}
For a large class of offspring distributions having bounded third moments, $\Psur(\si)$ has an approximation of the form $\frac{2(\si-1)}{v}+O((\si-1)/v)^2)$ as $\si$ decreases to 1 \citep{Athreya1992}. If $\si=e^{s\A}$ and $s\A$ is small, this yields $\Psur(e^{s\A})\approx \rho s\A$, where $\rho=1$ for a geometric offspring distribution, and $\rho=2$ for a Poisson distribution or any other distribution with $v=\si$; cf.~\eqref{Plost*app} and \eqref{Psur_fracLin}. By our monotonicity and continuity assumption on the extinction probability (Sect.~\ref{sec:1Lmodel}), $\Psur$ is monotone increasing in $\si$, $\Psur(e^{s\A})\le\min\{\rho s\A,1\}$ for every $s\A$, and $\Psur(e^{s\A})\to\rho s\A$ as $s\A\to0$.
\end{rem}

\subsection{Evolution of the distribution of mutant frequencies at a single locus}\label{sec:single_locus_evolve}

\begin{figure}[t]
\centering
\includegraphics[width=0.7\textwidth]{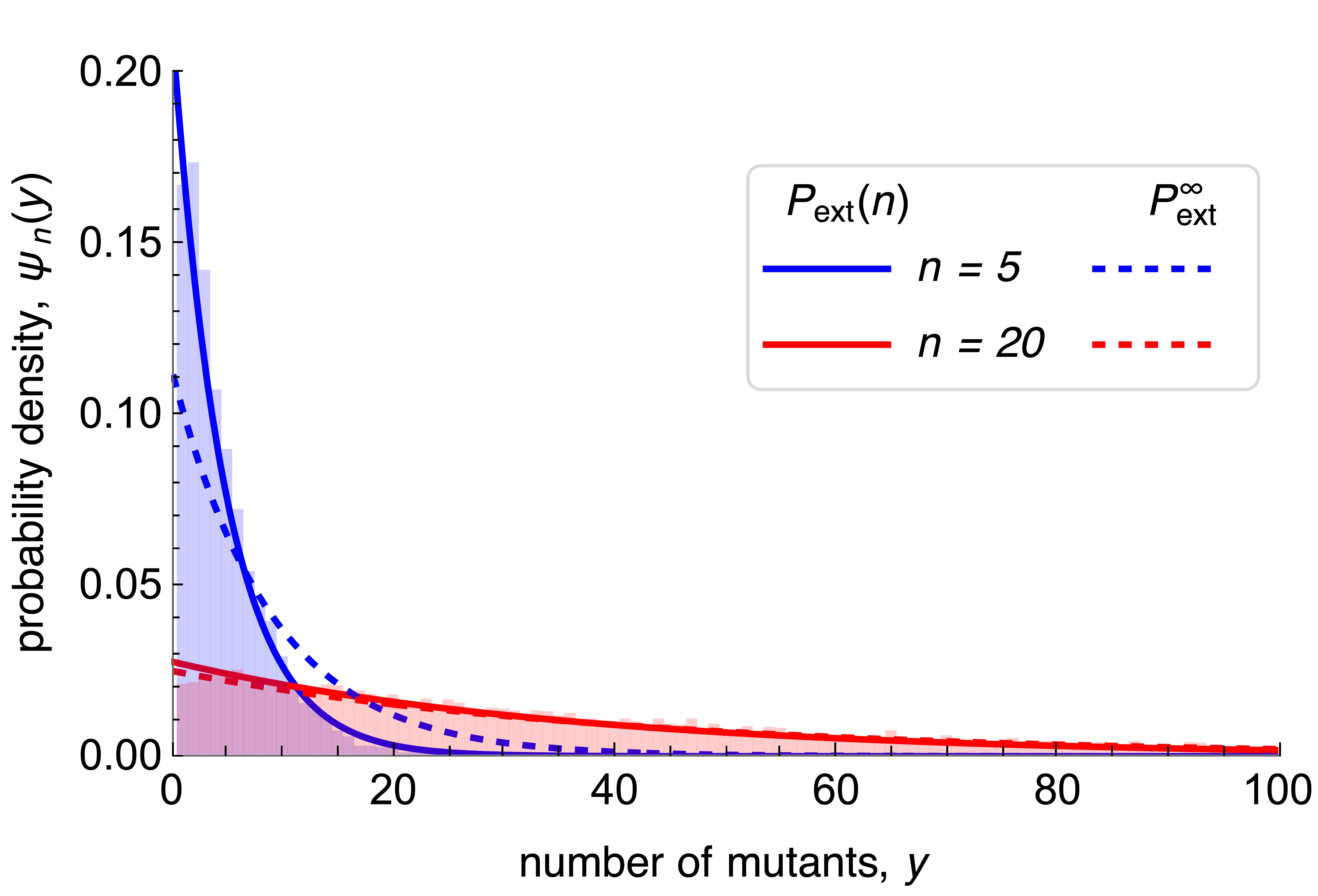}
\caption[$\psi_n(y)$]{The probability density $\psi_n(y)$ of the number of descendants of a single mutation in generation $n=5$ (blue) and $n=20$ (red). Solid curves show the density $\psi_n(y)$ of \eqref{Psin_exp}. Dashed curves show its simplification obtained by substituting $\Pexinf$ for $\Plost(n)$ in \eqref{Psin_exp}, which is equivalent to conditioning on long-term survival. The mean of the Poisson offspring distribution is $\si=e^s$ with $s=0.1$. The histograms are from Wright-Fisher simulations with a population size of $N=10^4$, conditioned on non-extinction until generation $n$.}
\label{figpsi}
\end{figure}

Now we start the investigation of the evolution of the frequency distribution of a mutation that occurred as a single copy in generation $n=0$. Our first goal is to approximate the distribution of $Z_n$ conditioned on $Z_n>0$ by a simple continuous distribution. Therefore, we define the distribution of the (discrete) random variable $W_n^+$
\begin{linenomath}
\begin{equation}\label{defWn+}
	\Prob[W_n\le x] = \Plost(n) + (1-\Plost(n))\Prob[W_n^+\le x]\,, \quad \text{where $x  \geq 0$}\,.
\end{equation}
\end{linenomath}
Then $\Prob[Z_n\le y| Z_n>0] = \Prob[\si^n W_n^+\le y]$. We approximate this distribution by the exponential distribution
\begin{equation}\label{Psin_exp}
	\Psi_n(y) = 1 - e^{-\la_n y}\,, \;\text{where}\; y\ge0 \;\text{and}\; \la_n = \frac{1-\Plost(n)}{\si^n} \,,
\end{equation}
and denote the corresponding random variable by $Y_n$ and its density by $\psi_n$. For fractional linear offspring distributions, this approximation is best possible in a rigorous sense (see Appendix \ref{app_frac_linear}). Because of the convergence of $W_n^+$ to $W^+$, the exponential distribution $\Psi_n$ will provide a close approximation to the true distribution of $W^+_n$ (with the possible exception of very small values of $n$) if $W^+$ is approximately exponential. Figure \ref{figpsi} shows that for a Poisson offspring distribution, the exponential density $\psi_n$ provides an excellent approximation for the number of mutants in generation $n$, conditioned on non-extinction until $n$, in a Wright-Fisher model. As $\si$ decreases to 1, the accuracy of the approximation will increase.   

The above approximation, $\Prob[Z_n\le y| Z_n>0] \approx \Psi_n(y)$, informs us that in an infinite population the distribution of the number of mutants after $n$ generations, conditioned on non-extinction until generation $n$ and originating from a single mutation, can be described by a deterministic exponential increase that starts with an exponentially distributed initial mutant population (according to $\psi_0$) of mean size 1.
Then the mean of this population in generation $n$ is $\si^n/(1-\Plost(n))$, which approaches $\si^n/\Psur$ as $n$ increases, and the distribution remains exponential. Thus, according to this model a beneficial mutant destined for survival does indeed grow faster than expected from the corresponding deterministic model (which has a growth rate of $\si$), in particular in the early generations.

As already noted, \cite{Desai2007}, \cite{Uecker2011}, and \cite{Martin2015} used similar approaches, but conditioned on fixation, instead of non-extinction until the given generation. Therefore, they took the random variable $W^+$, or rather its exponentially distributed approximation with the fixation probability as parameter, as their \emph{effective initial mutant population size} (which on average is larger than one). For large $n$, the distribution of $\si^{n} W^+$ has the same asymptotic behavior as $\Psi_n$ in \eqref{Psin_exp}, but our distribution $\Psi_n$ provides a more accurate approximation for the early phase of the spread of the mutant (Figure~\ref{figpsi}) because in each generation it has the correct mean. The distributions shown by dashed curves in Figure \ref{figpsi} are obtained by conditioning on long-term survival in the branching process.

Now we apply the above results to derive an approximation for the frequency of a mutant in a finite population of large and constant size $N$, conditioned on non-extinction until generation $n$. Our starting point is the approximate exponential growth of the number of mutants, as given by the distributions $\Psi_n$ in \eqref{Psin_exp} of the random variables $Y_n$. Because in a finite population exponential growth is impossible, we follow the population genetics tradition and use relative frequencies (probabilities). We assume that the number of resident types remains at its initial frequency of $N-1\approx N$ (because $N$ is large and residents produce one offspring). 
We define the random variables $X_n$, measuring the relative frequency of the mutants in the total population in generation $n$, by
\begin{linenomath}\begin{equation}\label{def_X_n}
	X_n = \frac{Y_n}{Y_n+N}\,.
\end{equation}\end{linenomath}
We note that in the absence of stochastic variation ($\Var[Y_n]=0$), we have $Y_n=\si^n Y_0$ and therefore $p(n)=X_n = Y_n/(Y_n+N-1)$ solves the corresponding deterministic diallelic selection equation, \eqref{eqdet}, if $p(0)=Y_0/N=1/N$. Because \eqref{def_X_n} is equivalent to $Y_n=NX_n/(1-X_n)$, the density of $X_n$ is approximated by 
\begin{linenomath}\begin{equation}
	g_{a(n)} (x)= \psi_n\left(N \frac{x}{1- x}\right) \frac{d}{d x}\left(\frac{Nx}{1-x}\right)\,,
\end{equation}\end{linenomath}
where $\psi_n$ is the density of $\Psi_n$ in \eqref{Psin_exp}. A simple calculation yields \eqref{g(p)} in our key result of this section:

\begin{result}\label{thm:g(p)}
Conditioned on non-extinction until generation $n$ and for given $N$,  the discrete distribution of the mutant frequency originating from a single mutation in generation $0$ with fitness $\si$ can be approximated by the distribution of the positive continuous random variable $X_n$ with density
\begin{subequations}\label{g(p)}
\begin{linenomath}\begin{equation} 
	g_a (x) = \frac{a}{(1 - x)^2} \cdot  \exp \left[ - a \frac{x}{1-x}\right],
\end{equation}\end{linenomath}
where
\begin{linenomath}\begin{equation}\label{a(t)}
	a = a(n, \si, N)  = N (1-\Plost(n,\si)) \si^{-n}. 
\end{equation}\end{linenomath}
\end{subequations}
\end{result}

This result does not provide a controlled approximation of the Wright-Fisher or the associated diffusion dynamics, simply because the time dependence of allele frequencies is prohibitively difficult to deal with analytically in these stochastic dynamical systems. However, its utility and accuracy will be demonstrated below.

\begin{rem}\label{MartinLambert}
Our expression \eqref{g(p)} for the density $g_a(x)$ has the same structure as that given by \cite{Martin2015} in their equation (4). The difference is that their $\be_t$ decays exponentially with $t$ with a constant parameter, whereas our $a=a(n,\si,N)$ decays exponentially with a parameter depending on $n$. This difference originates from the fact that we condition on non-extinction until $n$, whereas \cite{Martin2015} condition on fixation.
\end{rem}

\begin{figure}[t]
\centering
\includegraphics[width=0.8\textwidth]{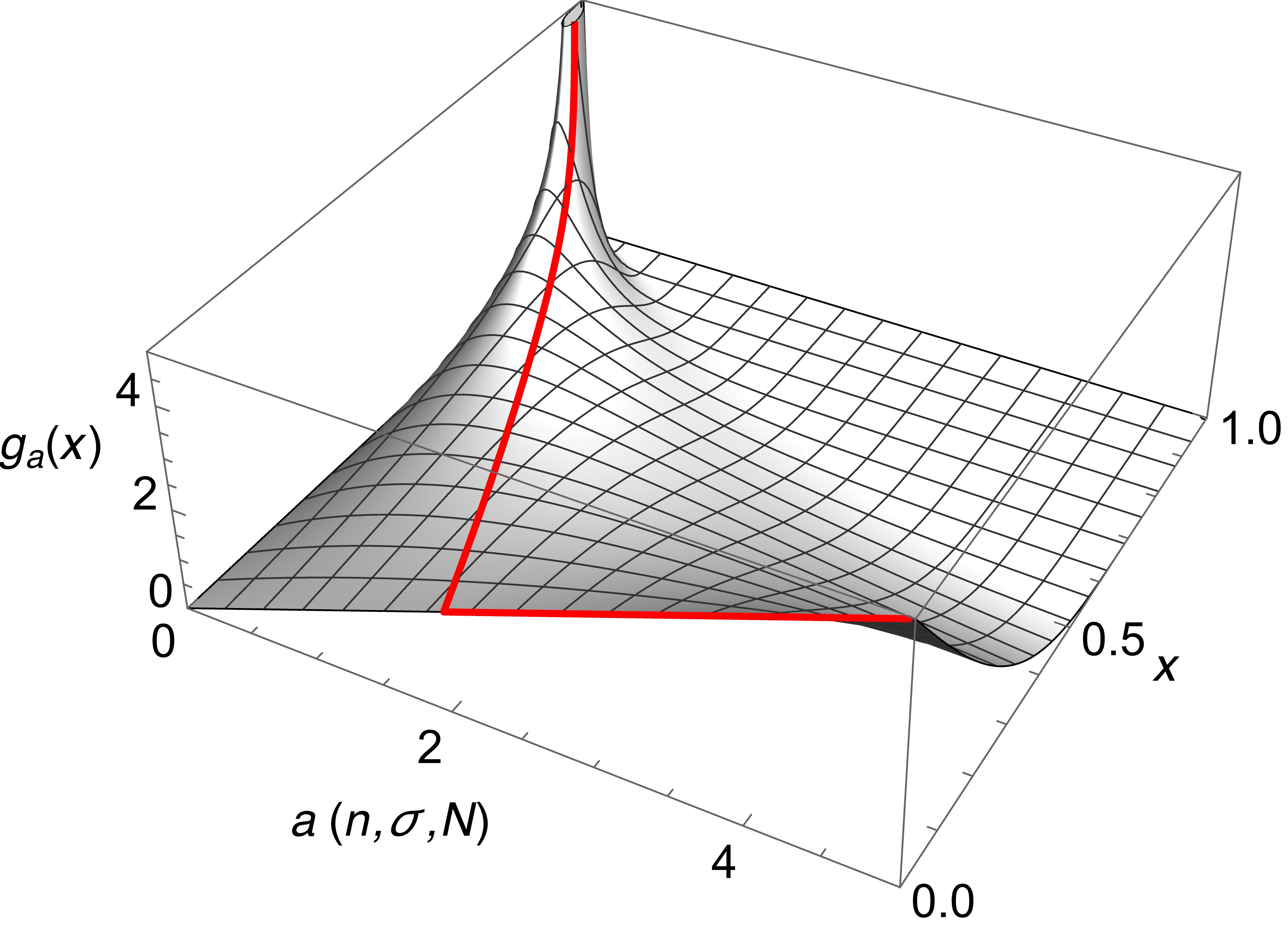}
\caption[$g_a (x)$ - 3D plot]{The probability density $g_a (x)$ in \eqref{g(p)} is shown as a function of $x$ and $a$. The red curve represents the unique maximum $g_a(\hat{x}(a))$ at $\hat{x}(a)=1-\frac{a}{2}$ for $a<2$ and the unique maximum at zero for $a>2$.}
\label{fig_g3D}
\end{figure}

To interpret and understand the analytic results derived below, we first need to study the properties of the density $g_a (x)$ in \eqref{g(p)} and of its constituent terms. The only parameter on which the mutant frequency distribution $g_a (x)$ in \eqref{g(p)} depends is the compound parameter $a=a(n,\si,N)$. This dependence is displayed in Fig.~\ref{fig_g3D}. Obviously, $a(n,\si,N)$ is proportional to $N$. As a function of $n$, $a(n,\si,N)$ decreases approximately exponentially from its maximum value $N$, assumed at $n=0$, to 0 as $n\to\infty$. As a function of $\si$ (or $s$), $a(n,\si,N)$ decreases as well (see Fig.~\ref{fig_a}).

We observe that $g_{a(n, \si, N)}(0)=a(n, \si, N)$. If $a>2$, then $g_a (x)$ attains its maximum at $x=0$ (Fig.~\ref{fig_g3D}). In this case $g_a (x)$ decays with increasing $x$ (and approximately exponentially for small $x$). This will always be the case in the initial phase of the mutant's invasion.
If $a<2$, which will occur for sufficiently large $n$, then $g_a (x)$ has an unique maximum $g_a (\hat{x}) = \frac{4}{a} \exp(a-2)$ at $\hat{x} = 1- \frac{a}{2} \in (0,1]$, which shifts from $x=0$ to $x=1$ as $a$ decreases (see the red curve in Fig.~\ref{fig_g3D}). 

A crucial quantity in $a$ is also the probability of nonextinction by generation $n$, $1-\Plost(n,\si)$. Figure \ref{figplosttau} documents the approach of $1-\Plost(n,\si)$ to $1-\Pexinf(\si)$ as $n$ increases and compares it with the approximation \eqref{Psurvn/Psurinfty}. The convergence is quick if $s\gtrapprox0.1$ (Section \ref{sec:Plost(n)}). For small $s$, however, $1-\Plost(n)$ is much larger than $1-\Pexinf$ for many more generations. Consequently, simplifying $g_a$ by substituting $\Pexinf$ for $\Plost(n)$ will considerably decrease the accuracy of the approximation during this initial phase (cf.~Fig.~\ref{figpsi}, which shows this effect for $\psi_n$).

\begin{figure}[t]
\centering
\includegraphics[width=0.7\textwidth]{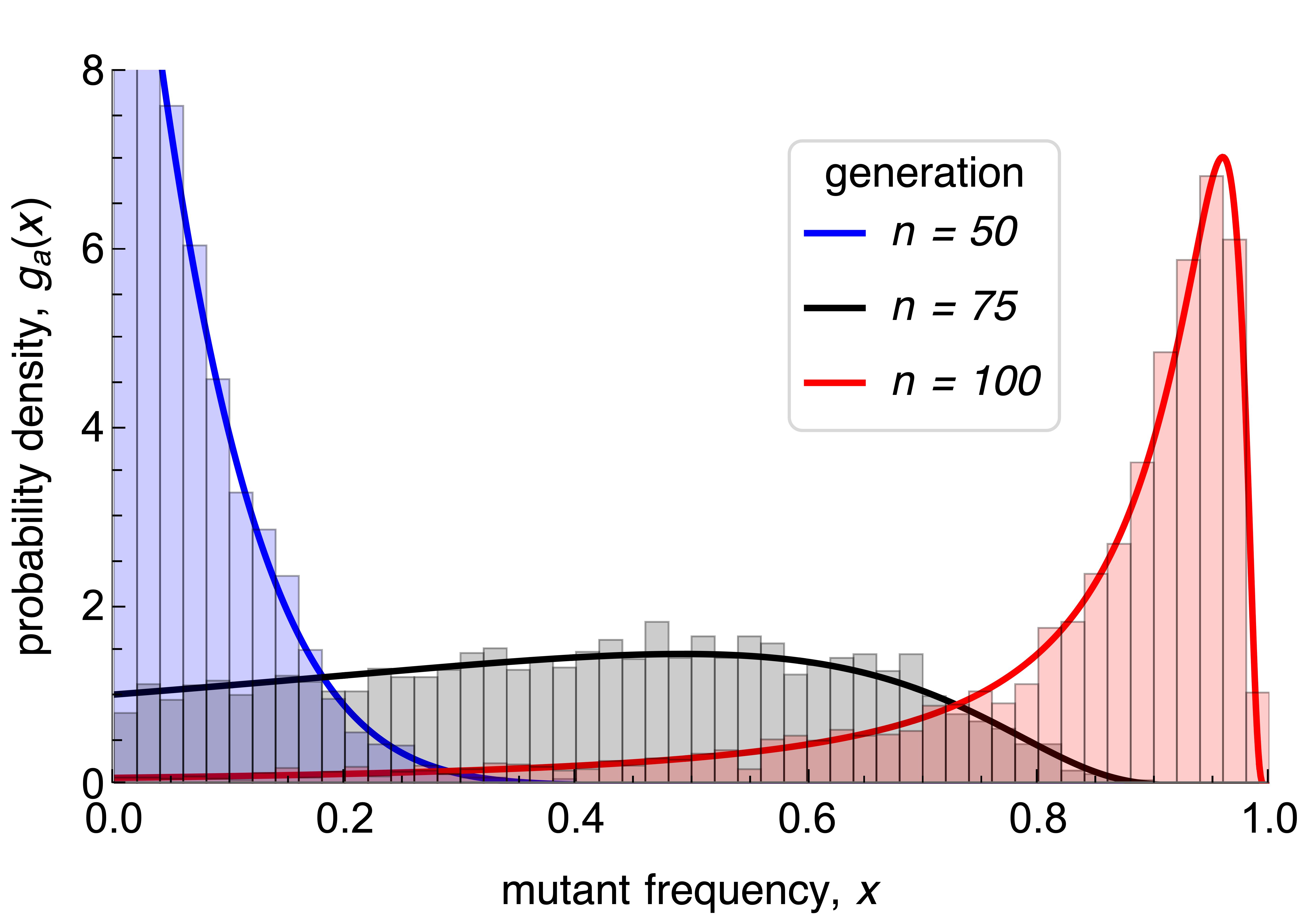}
\caption[$g_a (x)$]{The curves show the probability density $g_{a(n,\si,N)} (x)$ in \eqref{g(p)} of the mutant frequency in different generations $n$, as indicated in the legend. The histograms are obtained from simulations of the corresponding Wright-Fisher model. The population size is $N=10^4$ and the mean of the Poisson offspring distribution is $\si=e^s$ with $s=0.1$.}
\label{fig_g(p,a)}
\end{figure}

Figures \ref{fig_g(p,a)} and \ref{fig_g(p,a)2} compare the analytic approximation \eqref{g(p)} for the distribution of mutant frequencies at various generations with the corresponding histograms obtained from  Wright-Fisher simulations. They show that the approximation is very accurate in the initial phase whenever $Ns\ge1$. If $Ns\ge100$, it remains accurate almost until fixation of the mutant. If $Ns$ is of order 10, the approximation remains accurate for a surprisingly long period, essentially until the (mean) mutant frequency exceeds $\tfrac12$, i.e., until it has become the dominating variant. That the branching-process approximation underestimates the effects of random genetic drift when the mutant reaches high frequency is an obvious consequence of its assumptions. In Section~\ref{sec:single-locus_mean_var} we will see that the mean of $g_a$ provides a highly accurate approximation to the true mean even if the mutation is close to fixation and $Ns$ is much less than 100.

We note that Wright-Fisher simulations show that for $s>0.5$ (i.e., very strong selection), the distribution of the mutant frequencies cannot be approximated by the density $g_a$.
The reason is that for large $\si$ and a Poisson offspring distribution, the density $\psi_n$ of $Y_n=N \frac{X_n}{1-X_n}$ deviates too much from an exponential distribution. That for large $\si$, the variance becomes much smaller than the mean, can be inferred directly from the coefficient of variation given in Remark \ref{rem:W+}. In the following we exclude very strong selection and assume $s<0.5$. In fact, we focus on parameters satisfying $s\leq 0.1$ and $Ns \ge 10$.

\subsection{Evolution of the distribution of mutant frequencies in the infinite-sites model}\label{sec:evo_mutdist_ISM}

We use the model outlined in Section \ref{sec:ISmodel}. In particular, $\tau$ refers to the generation since the initial population started to evolve. Mutations $i=1,2,3,\ldots$ occur at generations $\tau_i$, as outlined above. We note that $\tau$ differs from $n$, as used in the above subsections on a single locus, because for locus $i$ the generation number in the Galton-Watson process is $n=0$ at time $\tau_i$. 
From  \eqref{Erlang} we recall the Erlang distribution and define
\begin{linenomath}\begin{equation}\label{Mi}
	M_{i,\Th} (\tau) = \int_0^{\tau} m_{i,\Th} (t) d t = 1-\frac{\Gamma (i, \Th \tau)}{\Gamma (i)}\,,
\end{equation}\end{linenomath}
which approximates the probability that $i$ mutations have occurred by generation $\tau$. Here, $\Ga(n,z)=\int_z^\infty x^{n-1}e^{-x}\,dx$ denotes the incomplete Gamma function \citep[][Chap.~6.5]{Abramowitz1964}. We recall  that $\lim_{z\to0}\Ga(n,z)=\Ga(n) =(n-1)!$ (if $n$ is a positive integer), and $\Gamma(0,z)=E_1(z)$, where $E_1(z)= \int_z^{\infty} x^{-1} e^{-x}\, dx$ is the exponential integral.

\begin{figure}[t]
\centering
\begin{tabular}{ll}
A & B \\
\includegraphics[width=0.45\textwidth]{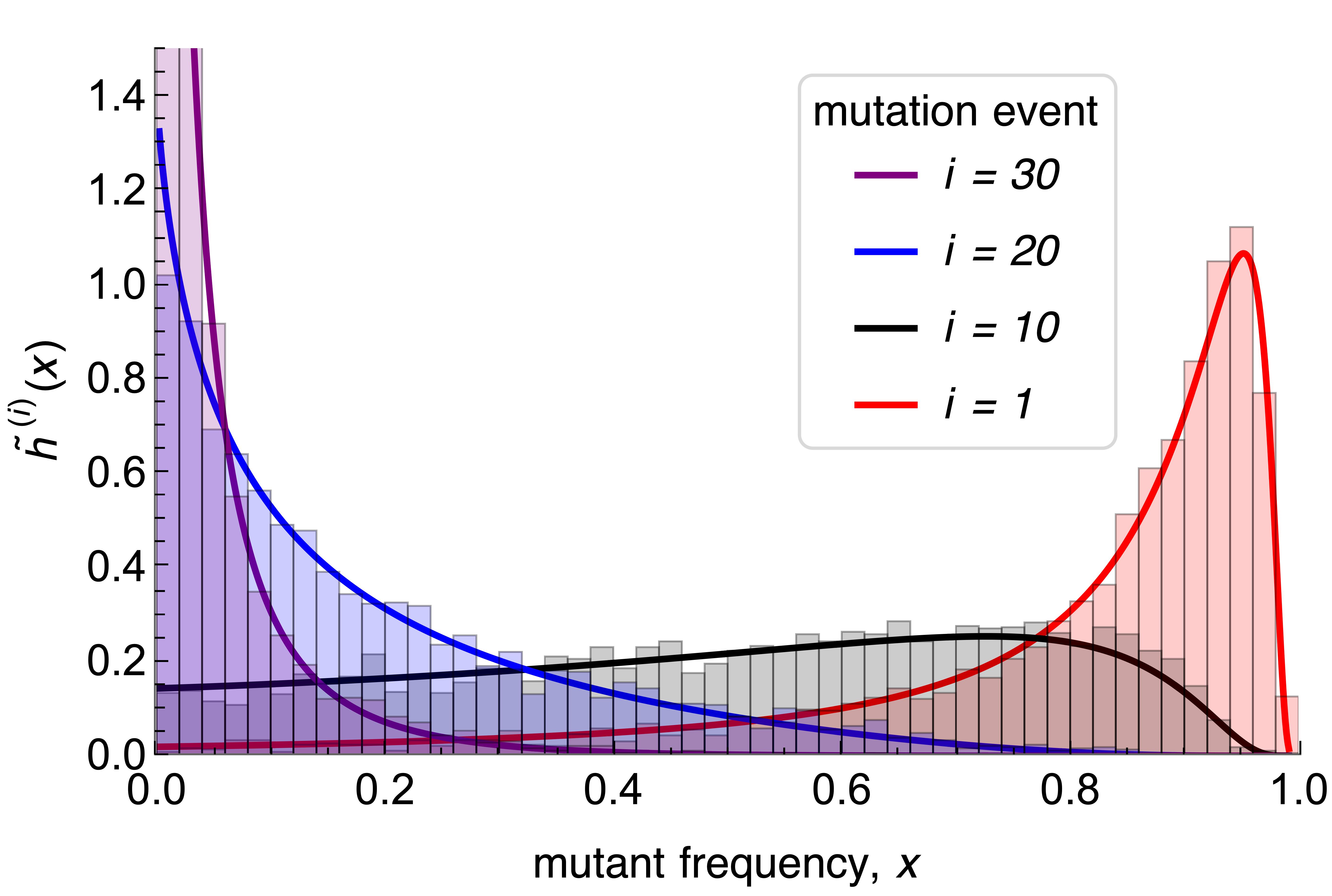} &\;
\includegraphics[width=0.45\textwidth]{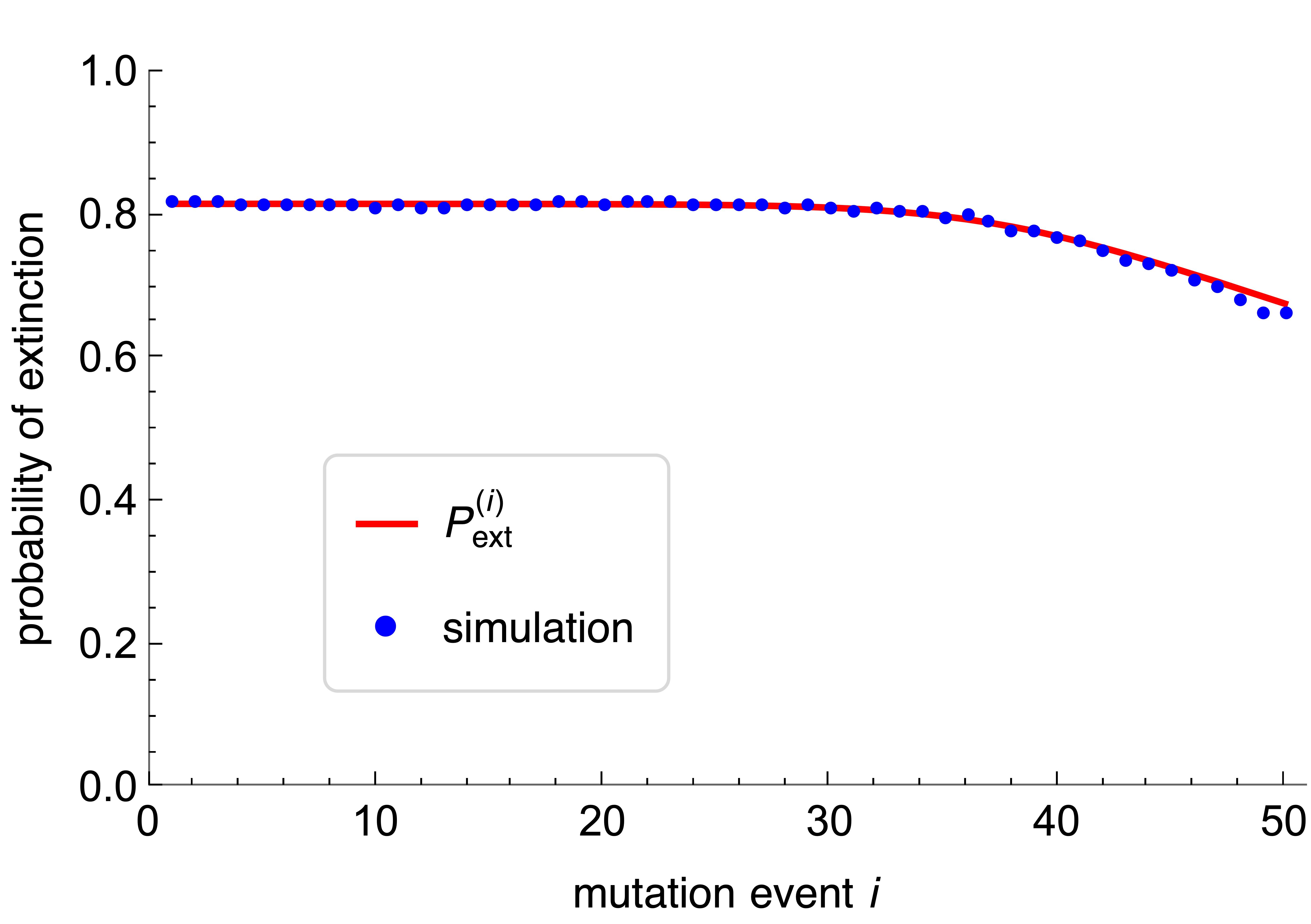}
\end{tabular}
\caption[$h^{(i)}$ and $\Plost^{(i)}$]{The curves in panel A show the absolutely continuous part $\tilde h_\tau^{(i)} (x)$ in \eqref{tilde{h}^{(i)}} of the mutant frequency $\tilde X^{(i)}_\tau$ at the $i$th locus (indicated in the legend) in generation $\tau=100$. The histograms are obtained from simulations of the corresponding Wright-Fisher model. The population-wide mutation rate is $\Th=0.5$, the population size is $N=10^4$ and the mean of the Poisson offspring distribution is $\si=e^s$ with $s=0.1$. The red curve in panel B displays the probability $\Plost^{(i)}$ in \eqref{plosti(tau)} that the mutation at locus $i$ has been lost by generation $\tau=100$. The blue symbols show the simulation results from the corresponding Wright-Fisher model. For larger $i$, the simulated values would lie below the theoretical curve because in some replicate runs considerably less than 50 mutations occur until generation 100. We note that the shape of $h_\tau^{(i)}$ is essentially identical to that of $\tilde h_\tau^{(i)}$, except that the vertical range has to be rescaled. For smaller $\tau$ or smaller $s$ slight differences become visible (results not shown).}
\label{fig_hISM}
\end{figure}

In the following, we write $[t]$ for the nearest integer to $t$, and we define
\begin{linenomath}\begin{equation}\label{a(t)_continuous}
	a(t,\si_i,N) = N (1-\Plost([t],\si_i)) \si_i^{-t} \quad\text{for every (real) } t\ge0\,,
\end{equation}\end{linenomath}
where $\si_i$ is the fitness of the mutant at locus $i$.
From Result \ref{thm:g(p)} we infer that, conditioned on non-extinction until generation $\tau$, the discrete distribution of the frequency of mutations at the locus at which the $i$th mutation event occurred, locus $i$ for short, can be approximated by the distribution of the continuous random variable with density
\begin{linenomath}\begin{equation}\label{hi(p)} 
	h^{(i)} (x) = h^{(i)}_{\tau,\si_i,N,\Th}(x) = \frac{1}{M_{i,\Th} (\tau)} \int_0^{\tau} g_{a(\tau -t,\si_i,N)} (x)\,  m_{i,\Th} (t)\, d t\,.
\end{equation}\end{linenomath}
Here, we used that the $i$th mutant has occurred approximately $\tau-[t]$ generations in the past, where $t$ is Erlang distributed with parameters $i$ and $\Th$. 

The probability that the mutation at locus $i$ has been lost by generation $\tau$, i.e., that the ancestral type is again fixed at this locus in generation $\tau$, can be approximated  by
\begin{linenomath}\begin{equation}\label{plosti(tau)}
	\Plost^{(i)}(\tau, \si_i,\Th) 
	= \frac{1}{M_{i,\Th} (\tau)} \int_0^{\tau} \Plost(\tau -[t],\si_i )\, m_{i,\Th} (t)\, d t
\end{equation}\end{linenomath}
(Fig.~\ref{fig_hISM}B).
Below, we shall need unconditioned distributions of the mutant frequencies.
Therefore, we define
\begin{linenomath}\begin{equation}\label{tilde{h}^{(i)}}
	\tilde{h}^{(i)} (x) =\tilde{h}^{(i)}_{\tau,\sigma_i,N,\Th}(x) = \frac{1}{M_{i,\Th}(\tau)}\int_0^{\tau} \bigl(1-\Plost([\tau-t],\sigma_i)\bigr) \, g_{a(\tau-t, \sigma_i, N)}(x)\,  m_{i,\Th}(t)\, dt
\end{equation}\end{linenomath}
(see Fig.~\ref{fig_hISM}A). Then $\int_0^1 \tilde{h}^{(i)}_{\tau,\sigma_i,N,\Th}(x) dx = 1-\Plost^{(i)}(\tau,\sigma_i,\Th)$ is the probability of nonextinction of the $i$th mutant until generation $\tau$. Therefore, in the absence of conditioning on non-extinction, we obtain for the probability that in generation $\tau$ the frequency $\tilde X^{(i)}_\tau$ of the $i$-th mutant is less or equal $p\in [0, 1]$:
\begin{linenomath}
\begin{equation}\label{tildeX^{(i)}}
	\Prob [\tilde X^{(i)}_\tau \leq p] = \Plost^{(i)}(\tau,\sigma_i,\Th) + \int_0^p \tilde{h}^{(i)}_\tau(x)\, dx\,.
\end{equation}
\end{linenomath}

\FloatBarrier
\section{Evolution of the phenotypic mean and variance}\label{sec:MeanVar}

Based on the results obtained above, we are now in the position to derive approximations for the expected response of the phenotype distribution from first principles. As usual in quantitative genetics, we concentrate on the evolution of the phenotypic mean and the phenotypic variance, i.e., the within-population variance of the trait. For this purpose, 
we will need the first and second moments pertaining to the density $g_a(x)$. They are given by 
\begin{subequations}\label{gammas}
\begin{linenomath}\begin{equation}\label{mean_g}
	\ga_1(a) = \int_0^1 x g_a (x)\, dx = 1 - a e^a E_1(a) 
\end{equation}\end{linenomath}
and
\begin{linenomath}\begin{equation}\label{moment2_g}
	\ga_2(a)  = \int_0^1 x^2 g_a (x)\, dx = 1 + a - a (2+a) e^a E_1(a)\,,
\end{equation}\end{linenomath}
respectively, where $E_1(a)= \int_a^{\infty} x^{-1} e^{-x}\, dx$ is the exponential integral. In addition, we will need the within-population variance of the mutant's allele frequency,
\begin{linenomath}\begin{equation}\label{var_g}
	\ga(a)  = \int_0^1 x(1-x) g_a (x)\, dx = a(1+a) e^{a} E_1(a) -a\,.
\end{equation}\end{linenomath}
\end{subequations}
By the properties of the exponential integral, the mean $\ga_1(a)$ decays from one to zero as $a$ increases from 0 to $\infty$ \citep[][Chap.~5.1]{Abramowitz1964}. Together with the properties of $a=a(t,\si,N)$, this implies that $\ga_1(a(t))$ increases from approximately $1/N$ to 1 as $t$ increases from 0 to $\infty$ (see Fig.~\ref{fig_ContrMeanVar}A). For the variance $\ga(a)$, we find that $\ga(a)\to0$ as either $a\to0$ or $a\to\infty$, and $\ga$ assumes a unique maximum at $a\approx0.719$, where $\ga(0.719)\approx 0.196$. As function of $t$, we have $\ga(a(0))\approx 1/N$ and $\ga(a(t))\to0$ as $t\to\infty$ (see Fig.~\ref{fig_ContrMeanVar}B). 

The value $t_M$ of $t$ at which the variance is maximized can be computed from \eqref{a(t)_continuous} with $a\approx0.719$. Because $\Plost([t],\si)$ converges to $\Pexinf(\si)$ quickly (Sects.~\ref{sec:Plost(n)} and \ref{app_frac_lin_compare}), we obtain $t_M \approx \ln(2.8Ns/v)/s$, where $v$ is the variance of the offspring distribution. For a Poisson offspring distribution, $N=10^4$ and $s=0.001$ or $s=0.1$, we obtain $t_M\approx3.33$ or $7.84$, which is in excellent agreement with Fig.~\ref{fig_ContrMeanVar}B.

\begin{rem}
For large values of $a$ (such as $a>80$), we encountered numerical problems when using \citetalias{Mathematica} to evaluate $\ga_1(a)$ or $\ga(a)$. The reason is that it multiplies the then huge value $e^a$ with the tiny value $E_1(a)$, which is set to $0$ if it falls below a prescribed value. However, the right-hand side of
\begin{equation}
	e^a E_1(a) = \int_1^\infty \frac{1}{x} e^{a(1-x)}\,dx
\end{equation}
can be integrated numerically without problems. An efficient, alternative possibility is to use the approximation $e^a E_1(a) \approx \frac{1}{a}(1-\frac{1}{a})$ (e.g., if $a>50$), which follows from the asymptotic expansion \eqref{e^x E_1(x)}.
\end{rem}

Beginning with Section \ref{sec:inf_many_sites}, we define the expected phenotypic mean, $\bar G(\tau)$, and the expected phenotypic variance, $V_G(\tau)$, of the trait in generation $\tau$ in terms of the distributions of the random variables $\tilde X^{(i)}_{\tau}$ defined in \eqref{tildeX^{(i)}} and \eqref{tilde{h}^{(i)}}. The definitions of $\bar G(\tau)$ and $V_G(\tau)$ are given in eqs.~\eqref{def_barGtau} and \eqref{def_VGtau}. (In the introductory Section \ref{sec:single-locus_mean_var} a simpler random variable can be used.) Therefore, our results below are precise mathematical statements about these quantities, and not statements about the mean and variance of the trait derived from the corresponding Wright-Fisher model. The figures illustrate the accuracy at which Wright-Fisher simulation results can be approximated by our theory, which we base on the \emph{assumption} that allele-frequency distributions evolve as stated in Result \ref{thm:g(p)}.  A Poisson offspring distribution is used only for comparison with the simulation results or when stated explicitly.

\subsection{A single locus}\label{sec:single-locus_mean_var}

We start by investigating the evolutionary dynamics of the trait's mean $\bar{G}_i$ and variance $V_{G_i}$ caused by the contribution of a single locus $i$. Let $\A_i$ be the phenotypic effect of the mutant and $\si_i=e^{s\A_i}$ its fitness effect. Further, let $X^{(i)}_\tau$ be the random variable describing the mutant's frequency $\tau$ generations after its occurrence, conditioned on non-extinction until generation $\tau$. Thus, $X^{(i)}_\tau$ corresponds to $X_n$ defined in \eqref{def_X_n} with $\tau=n$, $\si_i=\si$, and $i$ indicating the locus. The density of $X^{(i)}_\tau$ is $g_{a(N,\si_i,\tau)}$, as given in \eqref{g(p)}. The mutant's expected contribution to the phenotypic mean is $\A_i \EV[X^{(i)}_\tau] $ and to the phenotypic variance it is $\A_i^2 \EV[X^{(i)}_\tau(1-X^{(i)}_\tau)]$. By employing equations \eqref{mean_g} and \eqref{var_g}, we can define the contributions of locus $i$ to the expected mean and the expected variance of the trait by   
\begin{linenomath}\begin{equation} 
	\bar{G}_i (\tau)  = \A_i \int_0^1 x g_{a(\tau,\si_i)}(x)\, dx =\A_i \ga_1(a(\tau,\si_i))  \label{Gi}
\end{equation}\end{linenomath}
and
\begin{linenomath}\begin{equation}
	V_{G_i} (\tau)  = \A_i^2\int_0^1 x(1-x) g_{a(\tau,\si_i)}(x)\, dx = \A_i^2  \ga(a(\tau,\si_i))\,, \label{VGi} 
\end{equation}\end{linenomath}
respectively.

\begin{figure}[t]
\centering
\begin{tabular}{ll}
A & B \\
\includegraphics[width=0.45\textwidth]{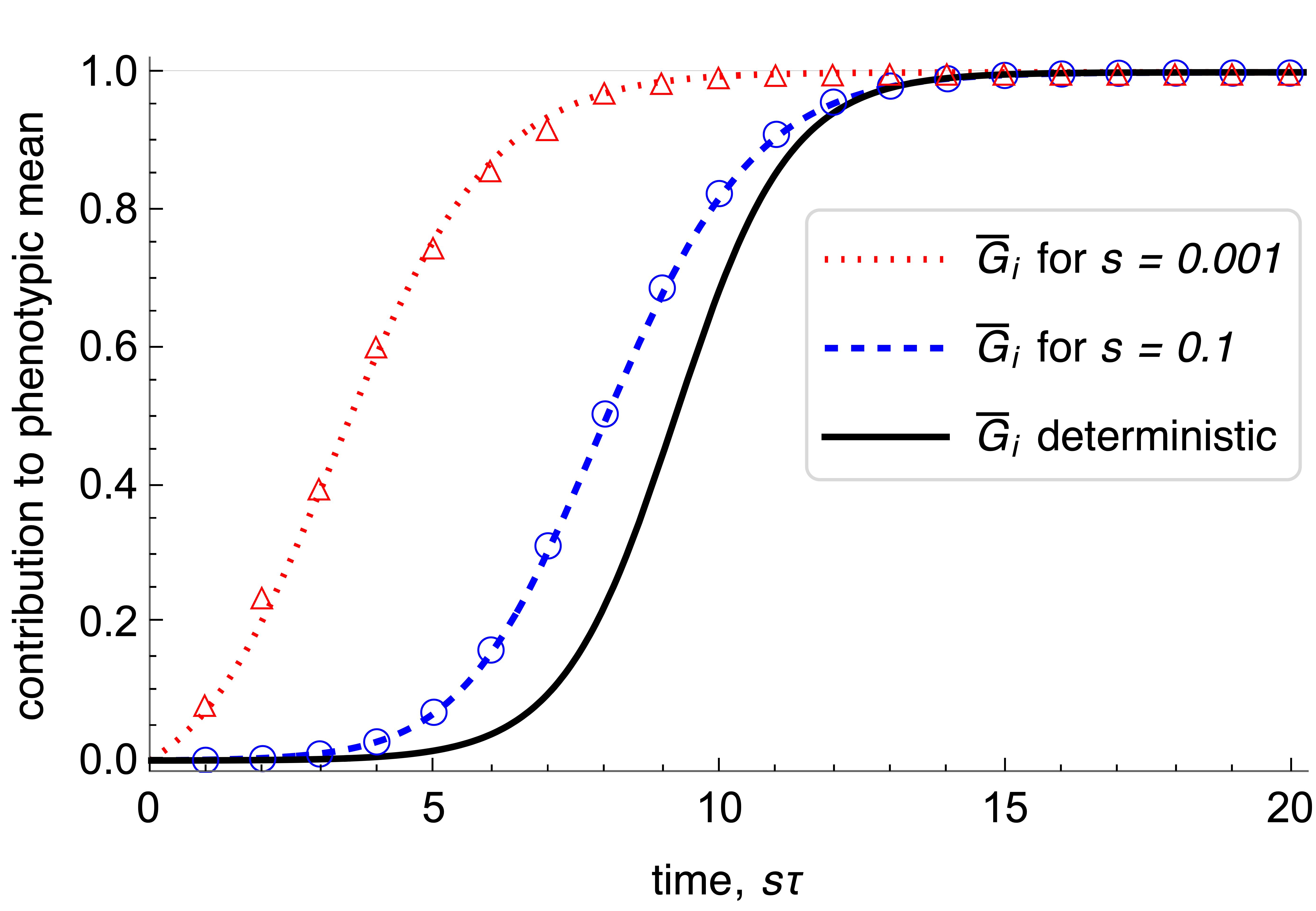} &
\includegraphics[width=0.45\textwidth]{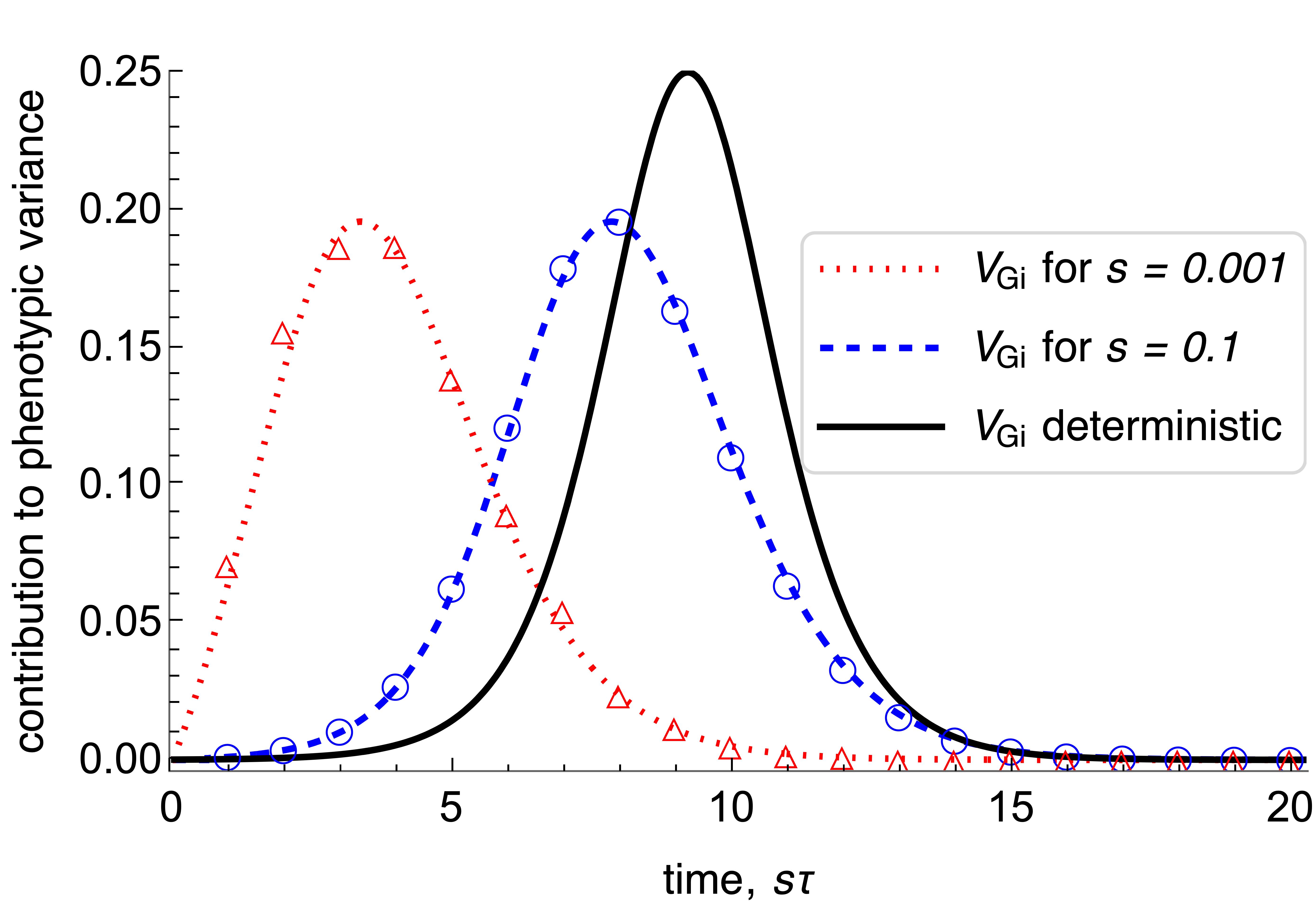}
\end{tabular}
\caption[$\bar{G}_i$ and $V_{G_i}$]{Contribution of a single locus to the expected phenotypic mean $\bar G_i(\tau)$ (panel A) and to the expected phenotypic variance $V_{G_i}(\tau)$ (panel B), $\tau$ generations after the mutation has occurred. The dashed and the dotted curves show $\bar G_i(\tau)$ in \eqref{Gi} and $V_{G_i}(\tau)$ in \eqref{VGi} for $s=0.1$ and $s=0.001$, respectively. The symbols show the simulation results from the corresponding Wright-Fisher models. The solid curve shows mean and variance computed from the solution \eqref{eqdet} of the deterministic model with $p(0)=1/N$. In this case, the selection coefficient affects the evolution of the mutant frequency, hence of the mean and the variance, only through a proportional change in time scale. Rescaling time by the selection coefficient $s$ makes the deterministic solution independent of $s$. In a finite population, weaker selection leads to a disproportionally faster spreading of the mutant, which is conditioned on non-extinction up to generation $\tau$. The population size is $N=10^4$, and the mutation effect is normalized to $\A_i = 1$. Because we have $\A_i=1$, $\bar G_i(\tau)$ and $V_{G_i}(\tau)$ coincide with the expected mean \eqref{mean_g} and variance \eqref{var_g}, respectively, of the mutant frequency distribution $g_a(x)$ defined in \eqref{g(p)}. In this and subsequent figures, analytical expressions are displayed as continuous curves.}
\label{fig_ContrMeanVar}
\end{figure}

In Figure \ref{fig_ContrMeanVar} the branching process approximations $\bar{G}_i (\tau)$ and $V_{G_i} (\tau)$ (dashed curves) are compared with simulation results from the diallelic Wright-Fisher model (symbols) and with $\A_i p(\tau)$ and $\A_i^2p(\tau)(1-p(\tau))$ (solid curves), where 
\begin{linenomath}\begin{equation}
	p(\tau) = \frac{\si^\tau p(0) }{1+(\si^\tau-1)p(0)} \label{eqdet}
\end{equation}\end{linenomath}
is the solution of the discrete, deterministic selection equation $p(\tau+1) = \si p(\tau)/\bar w(\tau)$, and $\bar w(\tau) = 1+(\si-1) p(\tau)$ is the mean fitness \citep[e.g.,][p.~30]{Buerger2000}. Figure \ref{fig_ContrMeanVar} shows excellent agreement between the branching process approximations and the Wright-Fisher simulations. It also demonstrates that in a finite population and conditioned on non-extinction until generation $\tau$, the mutant spreads faster than in a comparable infinite population. The smaller the selective advantage is, the faster the (relative) initial increase of the mutant frequency, and hence of the mean and the variance --- on the time scale $s\tau$. This is related to the fact that, to leading order in $Ns$ (for $Ns\gg1)$, the scaled fixation time of a mutant of effect $s$ in a Wright-Fisher model is 
\begin{linenomath}\begin{equation}\label{tfixsimp}
	\tfix\approx \frac{2\ln(2Ns)}{s}\,,
\end{equation}\end{linenomath}
which is a simplified version of eq.~\eqref{tfixHP}. For $N=10^4$ and $s=0.1$, $s=0.001$, we obtain from \eqref{tfixHP} $s\tfix\approx 16.4$, $7.0$, respectively, which provide good estimates for the time to completion of the sweep (Fig.~\ref{fig_ContrMeanVar}). Additional numerical support for the expected duration time of $\tfix$ is provided by Fig.~\ref{fig_MeanVarEqu}. Indeed, the expected duration time \eqref{tfixsimp} of a selective sweep has been derived by \citet[][Lemma 3.1]{Etheridge_etal2006}.

The phenomenon described above is well-known \citep[see also][]{DurrettSchweinsberg2004,Uecker2011}. The intuitive explanation is that mutants that initially spread more rapidly than corresponds to their (small) selective advantage have a much higher probability of becoming fixed than mutants that remain at low frequency for an extended time period. The smaller the selection coefficient $s$, the more important a quick early frequency increase is.

\subsection{Diffusion approximations and length of the phases of adaptation}\label{sec:length_of_phases}

Although our results and approximations are derived on the basis of a branching process approximation, the beginnings and lengths of the phases when they are most accurate are best described using diffusion approximations for the (expected) fixation times. This is motivated and supported by the results of \citet{Etheridge_etal2006} and our simulation results. 
Because in diffusion theory quantities are typically given as functions of the selection coefficient (and the population size), we denote the selection coefficient of a mutant of effect $\A$ by
\begin{linenomath}\begin{equation}\label{stilde}
	\stilde = e^{s\A}-1\,.
\end{equation}\end{linenomath}

For given $N$ and $s$, we will need the expected mean fixation time of a single mutant (destined for fixation). 
The expectation is taken with respect to the distribution $f$ of the mutation effects $\A$ which has mean $\bar\A$:
\begin{linenomath}\begin{equation}\label{bartfix}
	\bartfix(s,f, N) = \frac{\int_0^{\infty}\tfix(\stilde, N) \Pfix(\stilde, N) f(\A)\, d\A}{\int_0^{\infty}\Pfix(\stilde, N) f(\A)\, d\A}  \,.
\end{equation}\end{linenomath}
Here, $\tfix(\stilde, N)$ denotes the mean time to fixation of a single mutant with effect $\A$ and $\Pfix(\stilde, N)$ its probability of fixation, both in a population of size $N$.  For comparison with simulation results, we use the well-known diffusion approximation \eqref{PfixDif} for $\Pfix$ (where \eqref{PfixDif_Ne} yields \eqref{PfixDif} if $N_e=N$) and the simplified version \eqref{tfix} of the diffusion approximation for $\tfix$. Details and an efficient numerical procedure to compute $\bartfix(s,\bar\A, N)$ are described in Appendix~\ref{meanfixtime_f}.
If $f$ is exponential with mean 1, an accurate approximation of $\bartfix(\stilde,N)$ is 
\begin{linenomath}\begin{equation}
 	 \bartfix(s,\bar\A,N) \approx \frac{2}{s\bar\A}\ln(2Ns\bar\A)  \label{bartfix_app3}
\end{equation}\end{linenomath}
if $2/N\le s\bar\A \le 0.2$ (see Appendix~\ref{meanfixtime_f}). For $\bar\A=1$, the approximation \eqref{bartfix_app3} corresponds to the simplified approximation \eqref{tfixsimp} for equal mutation effects.

We note from \eqref{PfixDif} and \eqref{Plost*app} that $\Pfix(\stilde, N) > 1 - e^{2\stilde} > \Psur(e^{s\A})$ holds for every $N$, where $\Psur$ is based on a Poisson offspring distribution. We emphasize that we distinguish between the fixation probability $\Pfix$ in the diffusion approximation of the Wright-Fisher model and the survival probability $\Psur$ in the Galton-Watson process, although they are quantitatively very similar if $Ns$ is sufficiently large and $s$ small. We use the former only to compute fixation times, whereas the latter is used in our results and their derivation.

\begin{rem}
Our approach can be extended to cases when the effective population size $N_e$ differs from the actual population size $N$. This occurs if the variance of the offspring distribution differs from its mean. We follow \citet[][p.~120]{Ewens2004} and define the variance-effective population size by $N_e=N/(v/\si)$ (because we assume large $N$, we simplified his $N-\tfrac12$ to $N$). This yields the diffusion approximation
\begin{linenomath}\begin{equation}\label{PfixDif_Ne}
	\Pfix(s,N,N_e) = \frac{1-e^{-2sN_e/N}}{1-e^{-2N_es}}
\end{equation}\end{linenomath}
instead of \eqref{PfixDif}. Numerical comparison of this fixation probability with the survival probability $\Psur$ computed for fractional linear offspring distributions from \eqref{Pexinftyfl} shows that these values are almost identical if $Ns>1$ and $v\ge \si/2$. For instance, if $s=0.1$ (i.e., $\si=1.1$), $N=1000$, $v=\si/2, \si, 5\si$, we obtain
$\Pfix(s,N,N_e) \approx 0.3297, 0.1813, 0.0392$, respectively, and $\Psur^{({\rm FL})}(1+s)\approx 0.3333, 0.1818, 0.0392$. For the corresponding Poisson offspring distribution, $\Psur^{({\rm Poi})}\approx0.1761$ is obtained. Essentially identical numbers are obtained if $N=100$. Interestingly, the value $0.1761$ coincides almost exactly with the true fixation probability of $\approx 0.1761$ in the Wright-Fisher model with $N=1000$. Indeed, it has been proved that the diffusion approximation always overestimates the true fixation probability in the Wright-Fisher model \citep{BuergerEwens1995}. These considerations are an additional reason why we distinguish between $\Psur$ and $\Pfix$.
\end{rem}

\subsection{Infinitely many sites}\label{sec:inf_many_sites}

Now we proceed with our infinite-sites model.
Let $\A_i$ denote the effect of the $i$th mutation (at locus $i$) on the trait. Then its fitness effect is $\si_i=e^{s\A_i}$.
The (unconditioned) distribution of the frequency $\tilde X^{(i)}_\tau$ of the $i$th mutant in generation $\tau$ was defined in \eqref{tildeX^{(i)}} and its absolutely continuous part $\tilde{h}^{(i)}_\tau$ in \eqref{tilde{h}^{(i)}}.
Below we shall need the $k$-th moment of $\tilde X^{(i)}_\tau$: 
\begin{linenomath}\begin{equation}
	\EV \left[  (\tilde X^{(i)}_\tau)^k \right] = \int_0^1 x^k \tilde{h}^{(i)}_\tau(x)\, dx \,. \label{Ek}
\end{equation}\end{linenomath}

Now we are in the position to introduce general expressions for the expectations of the mean phenotype $\bar{G}(\tau) = \bar{G}(\tau,f,s,N,\Th)$ and of the phenotypic variance $V_G(\tau) = V_G(\tau,f,s,N,\Th)$ in any generation $\tau$, where we recall that the monomorphic population starts to evolve at $\tau=0$ and mutations occur when $\tau>0$. The mutation effects $\A_i$ are drawn independently from the distribution $f$. Because we assume linkage equilibrium in every generation $\tau$, the random variables $\tilde X^{(i)}_{\tau}$ $(i=1,2,3,\ldots)$ are mutually independent. 
For a given sequence of mutation events and given mutant frequency $x^{(i)}_\tau$ at locus $i$ in generation $\tau$, the phenotypic mean is $\sum_i \A_i x^{(i)}_\tau$ and the phenotypic variance is $\sum_i \A_i^2 x^{(i)}_\tau(1-x^{(i)}_\tau)$. Therefore, by taking expectations with respect to the Poisson distributed mutation events and the evolutionary trajectories of allele frequencies, we define the expected phenotypic mean $\bar{G}$ and variance $V_G$ as follows:
\begin{linenomath}\begin{equation}\label{def_barGtau}
	\bar{G}(\tau) = \sum_{n=1}^{\infty} \Poi_{\Th \tau}(n) \left( \sum_{i=1}^{n} \int_0^\infty \A_i \EV \left[ \tilde X^{(i)}_{\tau,e^{\sA_i},\Th} \right] f(\A_i)\, d\A_i \right) 
\end{equation}\end{linenomath} 
and
\begin{linenomath}\begin{equation}\label{def_VGtau}
	V_G(\tau)  =  
		\sum_{n=1}^{\infty} \Poi_{\Th \tau}(n) \left( \sum_{i=1}^{n} \int_0^\infty \A_i^2 \EV \left[ \tilde X^{(i)}_{\tau,e^{\sA_i},\Th}(1-\tilde X^{(i)}_{\tau,e^{\sA_i},\Th} ) \right] f(\A_i)\, d\A_i \right) \,. 
\end{equation}\end{linenomath}
The reader may note that the assumption of linkage equilibrium is not needed for the mean phenotype, and that assuming vanishing pairwise linkage disequilibria would be sufficient for the variance.

\begin{prop}\label{prop:general_barG_VG(tau)}  
The expected mean and variance of the trait, as defined in \eqref{def_barGtau} and \eqref{def_VGtau}, respectively, can be expressed as
\begin{linenomath}\begin{equation}
	\bar{G}(\tau) = \Th \int_0^\infty \A f(\A) \int_0^\tau \bigl(1-\Plost([t],e^{\sA}) \bigr) \ga_1(a(t, e^{\sA}))\, dt \, d\A 	\label{barGtau}
\end{equation}\end{linenomath}
and
\begin{linenomath}\begin{equation}
	V_G(\tau)  = \Th \int_0^\infty \A^2 f(\A) \int_0^\tau \bigl(1-\Plost([t],e^{\sA}) \bigr) \ga(a(t, e^{\sA}))\, dt \, d\A\,.	\label{VGtau}
\end{equation}\end{linenomath}
In particular, the population-wide mutation rate $\Th$ affects both quantities only as a multiplicative factor.
\end{prop}

The proof is relegated to Appendix~\ref{Proofs_MeanVar}.

\begin{figure}[t!]
\centering
\begin{tabular}{ll}
A & B \\
\includegraphics[width=0.45\textwidth]{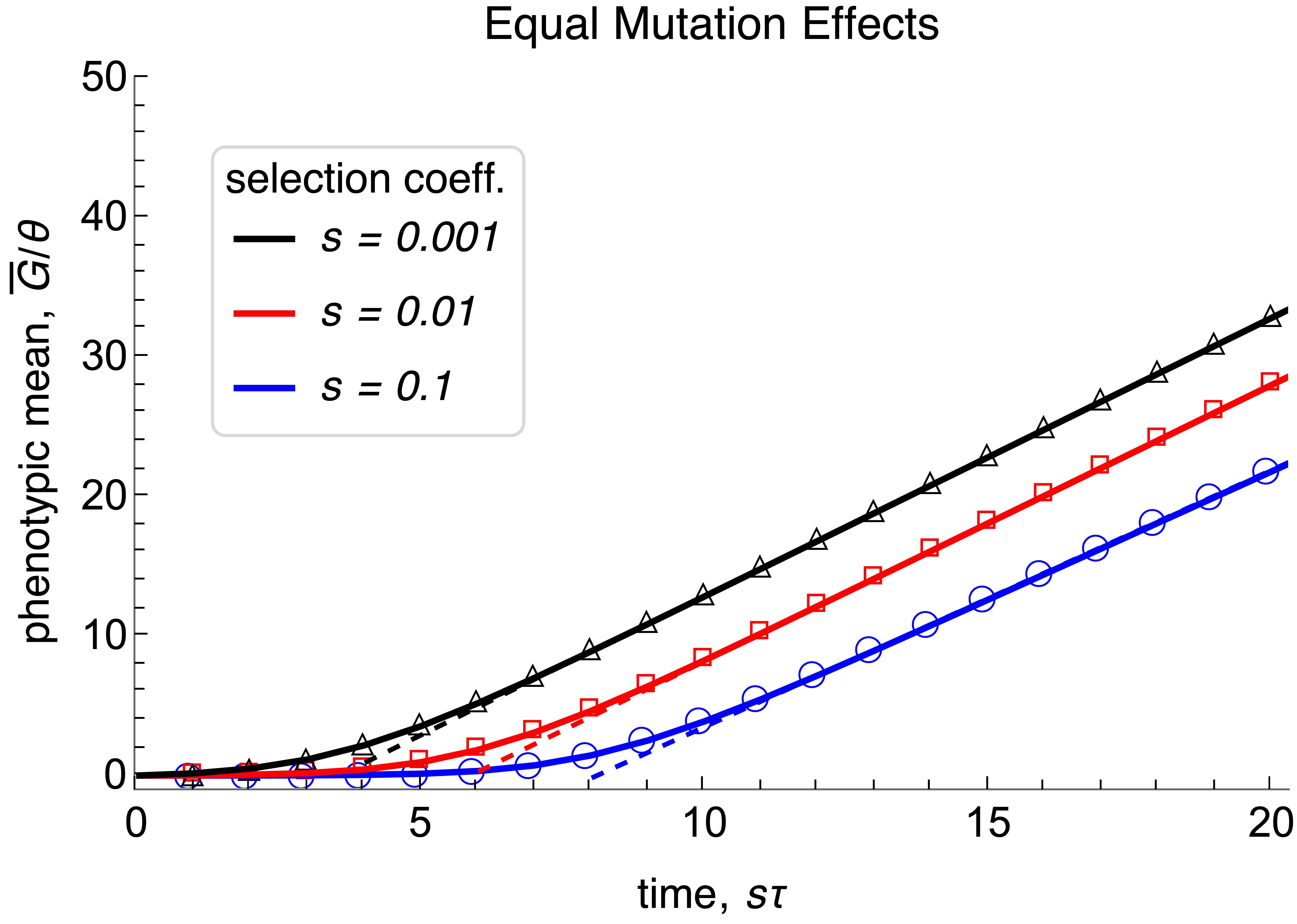} &
\includegraphics[width=0.45\textwidth]{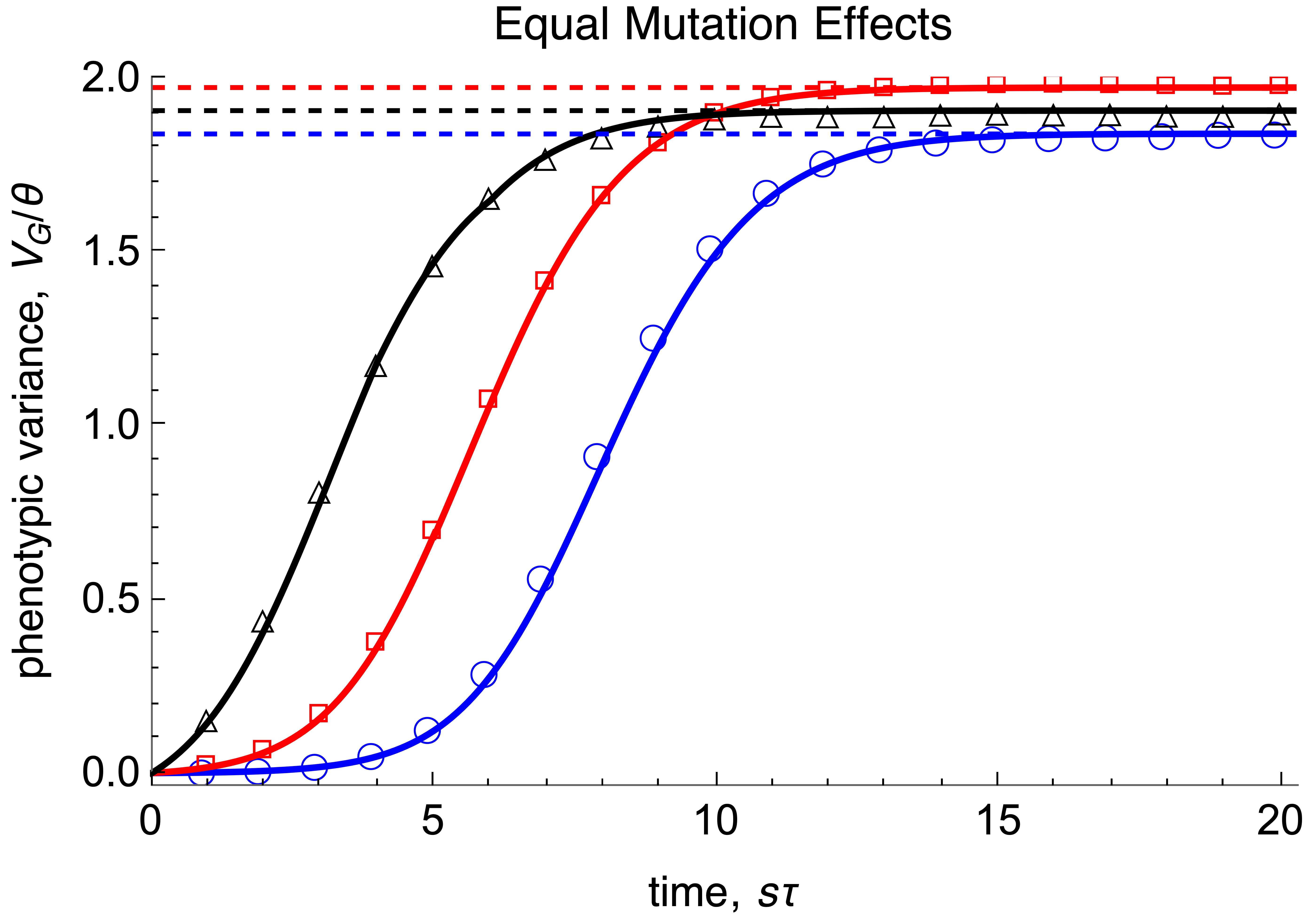} \\
C & D \\
\includegraphics[width=0.45\textwidth]{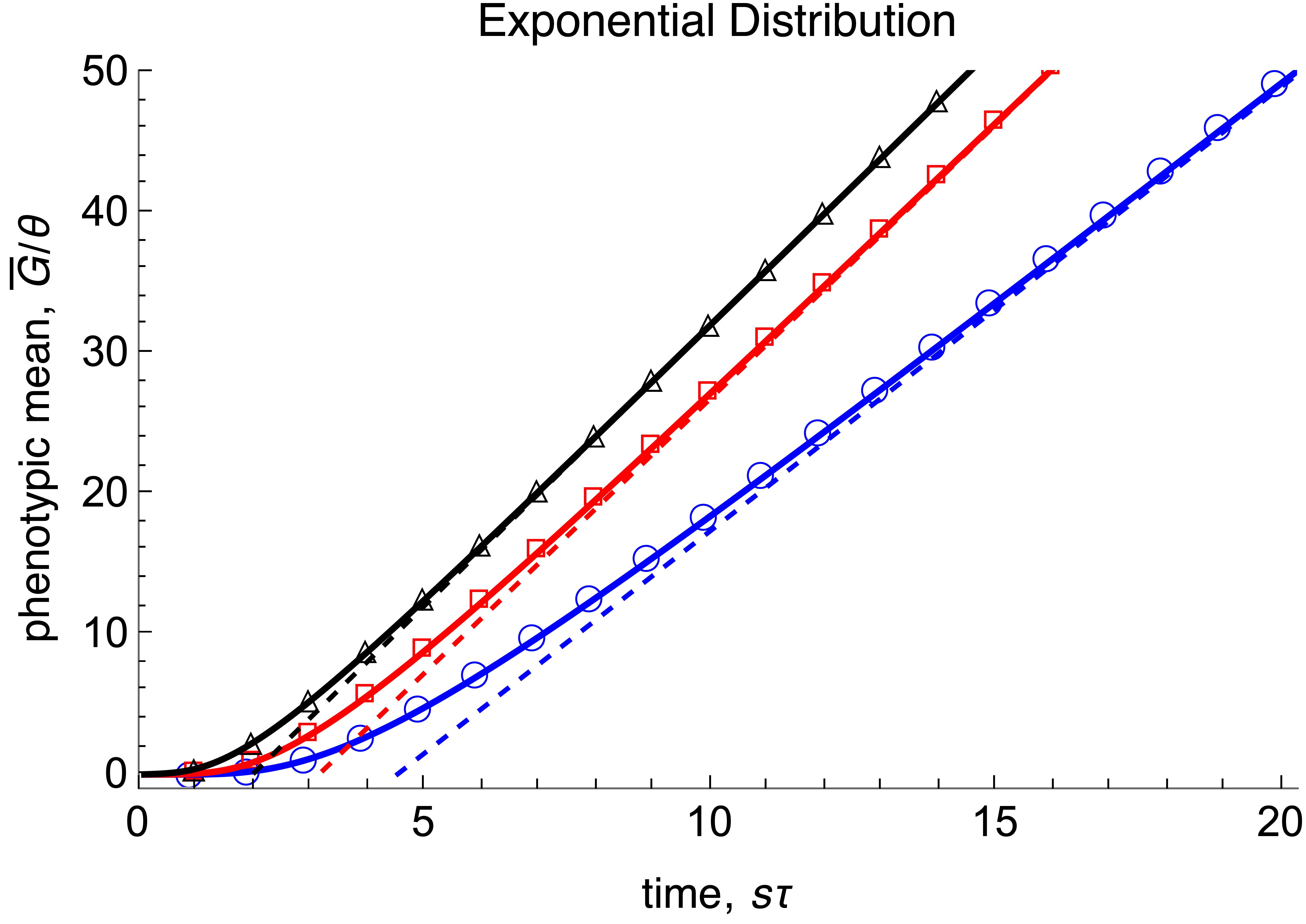} &
\includegraphics[width=0.45\textwidth]{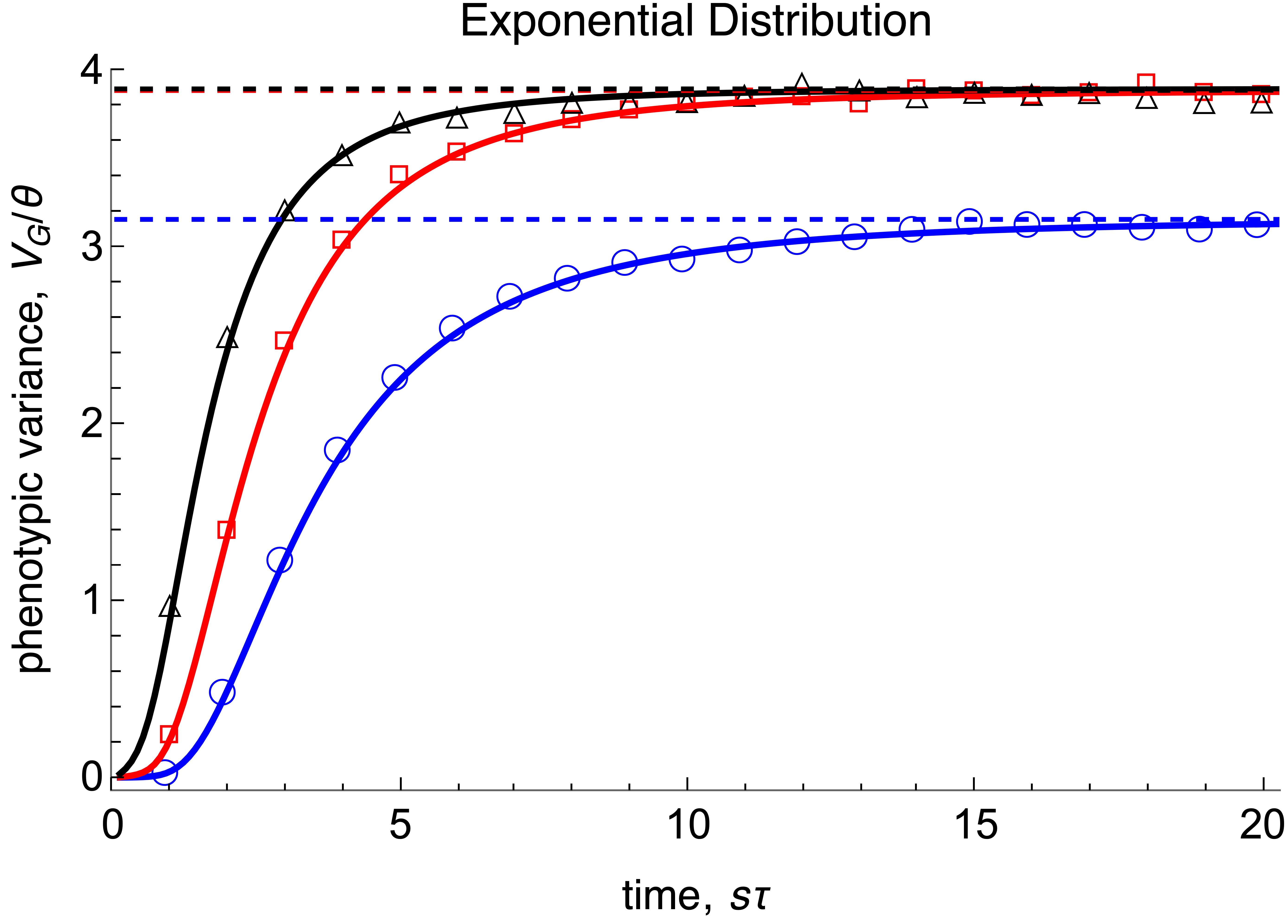} 
\end{tabular}
\caption[$\bar{G}$ and $V_{G}$ - Stationary phase]
{The scaled expected phenotypic mean $\bar G(s\tau)/\Th$ (panels A and C) and the scaled expected phenotypic variance $V_G(s\tau)/\Th$ (panels B and D) are shown for the selection coefficients $s$ given in the legend.  Generations $\tau$ are scaled by $s$ to facilitate comparison of curves and data with differing $s$. In A and B all mutation effects are $\A_i=1$, and in C and D the distribution $f$ of mutation effects is exponential with mean $\bar\A=1$.  The solid curves show $\bar G(s\tau)/\Th$ and $V_G(s\tau)/\Th$ as computed for equal effects from \eqref{Efinal} and  \eqref{Varfinal}, respectively, and for exponentially distributed effects from \eqref{barGtau} and \eqref{VGtau}.
The dashed curves in A and C show the approximation \eqref{barG_new} for the mean close to and at stationarity. The dashed lines in B and D show the stationary value of the variance, i.e., \eqref{VG*_prop}. The values of the mean fixation times multiplied by $s \,(=0.1, 0.01, 0.001)$ are $s\tfix\approx15.6$, $11.7$, $7.0$ for equal effects $\A=1$ in A and B, and $s\bartfix\approx15.8$, $10.6$, $5.9$ for exponentially distributed effects in C and D, where $\tfix$ is calculated from \eqref{tfixHP} and $\bartfix$ from \eqref{bartfix}. The symbols are from Wright-Fisher simulations with $\Th=\tfrac12$. The population size is $N=10^4$. Analogous results for a truncated normal distribution ($\A\ge0$ and $\bar\A=1$) are shown in Figure \ref{fig_nor}. }
\label{fig_MeanVarEqu}
\end{figure}

Figure \ref{fig_MeanVarEqu} displays evolutionary trajectories of the expected mean and variance given in Proposition~\ref{prop:general_barG_VG(tau)} and compares them with simulation results from corresponding Wright-Fisher models. On this scale of resolution the agreement is excellent, both for equal and exponentially distributed effects. After an initial phase the trajectories reach a quasi-stationary phase, in which the phenotypic variance has stabilized and the phenotypic mean increases  approximately linearly in time. Similar to the one-locus case treated above, the expected mean fixation times given in the figure caption provide decent approximations for the average time required to enter the quasi-stationary phase. For refined comparisons with more explicit approximations for the quasi-stationary phase and for the initial phase we refer to Fig.~\ref{fig_VarSEqu} and to Fig.~\ref{fig_MeanVarInitial1},  respectively. 

Notable in all panels of Fig.~\ref{fig_MeanVarEqu} is the faster increase of the mean and the variance for smaller selection coefficients on the time scale $s\tau$ (see the discussion in the connection with Fig.~\ref{fig_ContrMeanVar}). In addition, for the same selection coefficient the evolutionary response is, especially initially, much faster for exponentially distributed mutation effects (panels C and D) than for equal mutation effects (panels A and B), as shown by different steepnesses (A vs.\ C) and shapes (B vs.\ D) of the trajectories. The reason is that mutations of larger effect tend to fix earlier and more frequently than mutations of smaller effect. Therefore, large mutation effects speed up the response in the initial phase, whereas small mutation effects delay the entry into the quasi-stationary phase, with both types occurring in an exponential distribution. As shown by the numerical values of $s\tfix$ and $s\bartfix$ in the caption of Fig.~\ref{fig_MeanVarEqu}, this results in similar mean fixation times for equal and for exponentially distributed mutation effects, and thus in similar average entry times into the quasi-stationary phase. Finally, with an exponential distribution the equilibrium variance is about twice as large as for equal effects (see also Corollary~\ref{cor:equilibrium_simp}).

Whereas the simulation results for $2 \Th \geq 1$ are accurately described by the analytic approximations for the phenotypic mean and the phenotypic variance (Figure \ref{fig_MeanVarEqu}), for $2 \Th \ll 1$ much more fluctuation is observed in the simulations (Fig.~\ref{fig_MeanVar}). The reason is that the time between successful mutation events is much larger for smaller $\Th$, and this results in more pronounced stochastic effects. For small $\Th$ one sweep after another occurs, whereas for large $\Th$ parallel sweeps are common. This will be quantified in Section \ref{sec:segsites} using the number of segregating sites in the population.

\begin{rem}\label{rem: mutation_discrete_time}
In general, the integrals in \eqref{barGtau} and \eqref{VGtau} cannot be simplified. Of course, for equal effects $\A$, $\int_0^\infty \A^k f(\A) \int_0^\tau \ldots dt \, d\A$ simplifies to $\A^k\int_0^\tau \ldots dt$. An alternative approach is to assume that the $i$th mutation occurs at its mean waiting time $i/\Th$ and the number of mutation events until time $\tau$ is (approximately) $[\Th\tau]$.
For given mutation effects $\A_i$, we obtain instead of eqs.~\eqref{barGtau} and \eqref{VGtau} the simple approximations
\begin{linenomath}\begin{equation}  
	\bar{G}(\tau) \approx \sum_{i=0}^{[\Th \tau ]-1}  \bigl( 1-\Plost([i/\Th],  \sigma_i) \bigr) \bar{G}_i \left(i/\Th\right) 	\label{Efinal}
\end{equation}\end{linenomath}
and
\begin{linenomath}\begin{equation} 
	V_G(\tau)\approx \sum_{i=0}^{[ \Th \tau ]-1}  \bigl( 1-\Plost([i/\Th],  \sigma_i) \bigr) V_{G_i} \left(i/\Th\right) \,,  \label{Varfinal}
\end{equation}\end{linenomath}
respectively, where we have used the one-locus results \eqref{Gi} and \eqref{VGi}. The derivation is given in Appendix \ref{app:mutation_discrete}. A meaningful application of these formulas requires $[\Th \tau]>1$. For values of $\Th$ smaller than 1, we use linear interpolation between $\Th=0$ and $\Th=1$ (Appendix \ref{app:mutation_discrete}). The approximations \eqref{Efinal} and \eqref{Varfinal} are evaluated much faster than the integrals in Proposition~\ref{prop:general_barG_VG(tau)} and are applied in Fig.~\ref{fig_MeanVarEqu}A,B and Fig.~\ref{fig_MeanVar} to the case of equal mutation effects. These figures demonstrate their high accuracy.
\end{rem}

In the following we derive simpler and more explicit expressions for the expected phenotypic mean and variance of the trait for different phases of adaptation. First, we characterize their long-term behavior, then we provide simpler approximations of the exact expressions in Proposition \ref{prop:general_barG_VG(tau)}, and finally we derive very simple approximations for their initial increase.

\subsection{Approximations for the quasi-stationary phase}\label{sec:stationary}

Although our approach was designed to study the early phase of adaptation, we show here that it yields accurate and simple approximations for the quasi-stationary phase, when the equilibrium variance has stabilized and the response of the expected mean is approximately linear in time. This phase is reached when the flux of incoming and fixing mutations balances. This is the case after about $\tau_c$ generations (see Fig.~\ref{fig_MeanVarEqu}), where
\begin{linenomath}\begin{equation}\label{tau_c}
	\tau_c = 1/\Th + \bartfix \,,
\end{equation}\end{linenomath}
is the expected time by which the first mutation, which appeared at (about) $\tau = 1/\Th$,
becomes fixed. If $\tau \ge \tau_c$, then from the $\Th \tau$ mutations, which are expected to arise until generation $\tau$, $\Th(\tau-\tau_c)+1 = \Th (\tau - \bartfix)$ mutations are expected to have had enough time to go to fixation. Fixed mutations contribute $\A$ to the $\bar{G}$ and 0 to the $V_{G}$. The mutants from the remaining $\Th \tau_c- 1 = \Th\bartfix$ mutation events either segregate in the population or have been lost before generation $\tau$.

In the results derived below, we will need the following assumptions.

\begin{asm} \label{A1}
Let $N\to\infty$ and assume 
\begin{linenomath}\begin{equation}\label{scaling_N_s}
	Ns^K = C^K 
\end{equation}\end{linenomath}
where $C>0$ is an arbitrary constant and the constant $K$ satisfies $K\ge2$.
\end{asm}

\begin{asm} \label{A2}
Offspring distributions of mutants of effect $\A$ satisfy $v=2\si/\rho$, where $\rho$ is a positive constant, and have uniformly bounded third moments as $\si$ decreases to 1 \citep{Athreya1992}. In addition, the inequality \eqref{Psurvn/Psurinfty} holds.
\end{asm}

We note that the scaling Assumption \ref{A1} for $s$ and $N$ reflects the fact that, compared to diffusion approximations when $K=1$, in our model selection is stronger than random genetic drift. In \citet{Boenkost_etal2021}, Assumption \ref{A1} with $K>2$ is called moderately strong selection and used to prove that the fixation probability of a single mutant in Cannings models is asymptotically equivalent to $2(\si-1)/v$ (in our notation) as $N\to\infty$. 
The first requirement in Assumption \ref{A2} implies that $\lim\limits_{s\A\to\infty}\Psur(e^{s\A})/(\rho s \A)=1$ (see Remark \ref{Psur=csalpha}). It is satisfied by fractional linear and by Poisson offspring distributions. As discussed in Sect.~\ref{sec:Plost(n)} and shown in Appendix \ref{app_frac_linear}, equality holds in \eqref{Psurvn/Psurinfty} for fractional linear offspring distributions and \eqref{Psurvn/Psurinfty} is also fulfilled for Poisson offspring distributions (see also Fig.~\ref{figplosttau}).

We denote the values of $\bar G(\tau)$ and $V_G(\tau)$ in the quasi-stationary phase by $\bar{G}^*(\tau)$ and $V_{G}^*$, respectively. 
For notational simplicity, we set
\begin{linenomath}\begin{equation}\label{nu}
	\sN = \sN(s,\A,N) = N\Psur(e^{\sA})\,.
\end{equation}\end{linenomath}
We recall from Section \ref{sec:QGmodel} that the mutation distribution $f$ has bounded moments. Here is our first main result. 

\begin{prop}\label{prop:equilibrium}
(1) At stationarity, the expected per-generation response  of the phenotypic mean, $\De\bar{G}^* = \lim\limits_{\tau\to\infty}(\bar{G}(\tau+1)-\bar{G}(\tau))$, is given by
\begin{linenomath}\begin{equation} \label{barG*_prop}
	\De\bar{G}^* = \Th\int_0^\infty \A \Psur(e^{s\A}) f(\A)\, d\A \,, 
\end{equation}\end{linenomath}
where $\De\bar{G}^*$ depends on $\Th$, $s$, and $f$. 

(2) Assuming \ref{A1} and \ref{A2}, the expected phenotypic variance at stationarity, $V_G^*=\lim\limits_{\tau\to\infty}V_G(\tau)$, has the approximation
\begin{linenomath}\begin{equation} \label{VG*_prop}
	V_G^* = \Th \int_0^{\infty}  \frac{\A }{s} \Psur(e^{\sA}) \sN e^\sN E_1(\sN) f(\A)\, d\A\ + O\Bigl(N^{-K_1/K}\Bigr) \,, 
\end{equation}\end{linenomath} 
for every choice of $K_1>1$ and $K>K_1+1$, whence $K>2$ needs to hold.
\end{prop}

The proof of this Proposition is relegated to Appendix~\ref{Proofs_MeanVar}. We discuss the choice of $K_1$ and $K$ in the error term in Remark \ref{rem:cor}.

\begin{rem}
For mutation effects equal to $\A$, \eqref{barG*_prop} simplifies to
\begin{linenomath}\begin{equation} 
	\De\bar{G}^* = \Th \A \Psur(e^{s\A})\,, \label{DebarG*_equal}
\end{equation}\end{linenomath}
and \eqref{VG*_prop} simplifies to
\begin{linenomath}\begin{equation} 
	V_G^*(\A, s, N, \Th) \approx \Th \frac{\A }{s} \Psur(e^{\sA}) \sN e^\sN E_1(\sN)\,. \label{VG*_equal}
\end{equation}\end{linenomath}
\end{rem}

\begin{rem}\label{rem_prop_equilibrium}
(1) The expression \eqref{barG*_prop} for the expected per-generation response of the mean phenotype reflects the (permanent) response due to mutations that are going to fixation. Therefore, during the quasi-stationary phase, the phenotypic mean increases linearly with time as mutants reach fixation. This response is independent of the population size (in constrast to the early response; see below). In Appendix \ref{Proofs_MeanVar} (Remark \ref{Rem:barG*}), we provide the approximation \eqref{barG_new} for $\bar{G}(\tau, f, s, N, \Th)$ when $\tau$ is sufficiently large. It is shown in Fig.~\ref{fig_MeanVarEqu}.

(2) Because only segregating mutations contribute to the phenotypic variance, the number of segregating sites and the phenotypic variance remain constant over time once the quasi-stationary phase has been reached. The term $\frac{\A }{s}\sN e^\sN E_1(\sN)$ in \eqref{VG*_prop} and \eqref{VG*_equal} is precisely the total variance accumulated by a mutant with effect $\A$ during its sweep to fixation. This follows immediately from \eqref{VGi} and \eqref{approx_ga_infty1}.
\end{rem}

The following result provides much simpler approximations and not only recovers but refines well known results from evolutionary quantitative genetics. For $n\ge2$ we write $\A_n(f)=\int_0^\infty \A^n f(\A) d\A$ for the $n$th moment about zero of the mutation distribution $f$. 

\begin{cor}\label{cor:equilibrium_simp}
We assume a Poisson offspring distribution and the assumptions in Proposition \ref{prop:equilibrium}.
Then the following simple approximations hold in the quasi-stationary phase.

(1) The per-generation change of the mean phenotype at stationarity can be approximated as follows:
\begin{linenomath}\begin{equation}
	\De\bar G^*\approx 2\Th s \left( \A_2(f) - \frac{5s}{6}\A_3(f) \right) \,. \label{DebarG*}
\end{equation}\end{linenomath}

(2) For mutation effects equal to $\A$, the change in the mean is
\begin{linenomath}\begin{equation}
	\De\bar G^* \approx 2\Th s{\A}^2 (1-\tfrac{5}{6}s\A)\,. \label{DebarG*equal}
\end{equation}\end{linenomath}

(3) For an exponential distribution $f$ of mutation effects with expectation $\bar\A$, the change in the mean is
\begin{linenomath}\begin{equation}
	\De\bar G^* \approx 4\Th s{\bar\A}^2 (1-\tfrac{5}{2}s\bar\A)\,. \label{DebarG*exp}
\end{equation}\end{linenomath}

(4) The stationary variance has the approximation 
\begin{linenomath}\begin{equation}
	V_G^*(f, s, N, \Th) \approx \frac{1}{s}\De\bar G^* - \frac{\Th\bar\A}{Ns}\,. \label{VG*app}
\end{equation}\end{linenomath}

(5) If mutation effects are equal to $\A$, then \eqref{VG*app} and \eqref{DebarG*equal} yield
\begin{linenomath}\begin{equation}\label{VG*Nsconst}
	V_G^* \approx 2\Th\A^2 \left (1 - \frac{5}{6}s\A - \frac{1}{2Ns\A} \right)\,.
\end{equation}\end{linenomath}

(6) If $f$ is exponential (with mean $\bar\A$), then \eqref{VG*app} and \eqref{DebarG*exp} yield
\begin{linenomath}\begin{equation}\label{VG*Ns}
	V_G^* \approx 4\Th{\bar\A}^2\left(1-\frac{5}{2}s\bar\A-\frac{1}{4Ns\bar\A} \right)\,.
\end{equation}\end{linenomath}
\end{cor}

The proof is given in Appendix \ref{proof_of_Corollary}. It is based on series expansions of the expressions given in Proposition \ref{prop:equilibrium}. 

\begin{rem}\label{rem:cor}
(1) Assumption \ref{A1} yields $s=O(N^{-1/K})$ and $1/(Ns) = O(N^{-1+1/K})$. Thus, $s$ is of lower (or equal) order than $1/(Ns)$ as $N\to\infty$. The error in \eqref{VG*_prop} is $O(N^{-K_1/K})$, where $K_1>1$ and $K>K_1+1$. By the latter assumption, we have $N^{-K_1/K}< N^{-1/K}$, but $N^{-K_1/K} > N^{-1+1/K}$. Thus, the term of order $1/(4Ns\barA)$ in \eqref{VG*Ns} is not secured because the error term is of larger order. With more accurate estimates in our proof of \eqref{VG*_prop} a smaller error term might be obtainable. 
For illustration, we choose $K_1=3/2$ and $K=3$. Then we obtain $s=O(N^{-1/3})$, $1/(Ns)= O(N^{-2/3})$, and the error term is of order $O(N^{-1/2})$.

(2) Approximations analogous to \eqref{DebarG*} can be obtained from \eqref{barG*_prop} for other offspring distributions by series expansion of the survival probability $\Psur(e^{s\A})$, or simply from $\Psur(e^{s\A})\approx\rho s\A$ (cf.~Remark \ref{Psur=csalpha}).  
\end{rem}

\begin{figure}[t!]
\centering
\begin{tabular}{ll}
A & B \\
\includegraphics[width=0.45\textwidth]{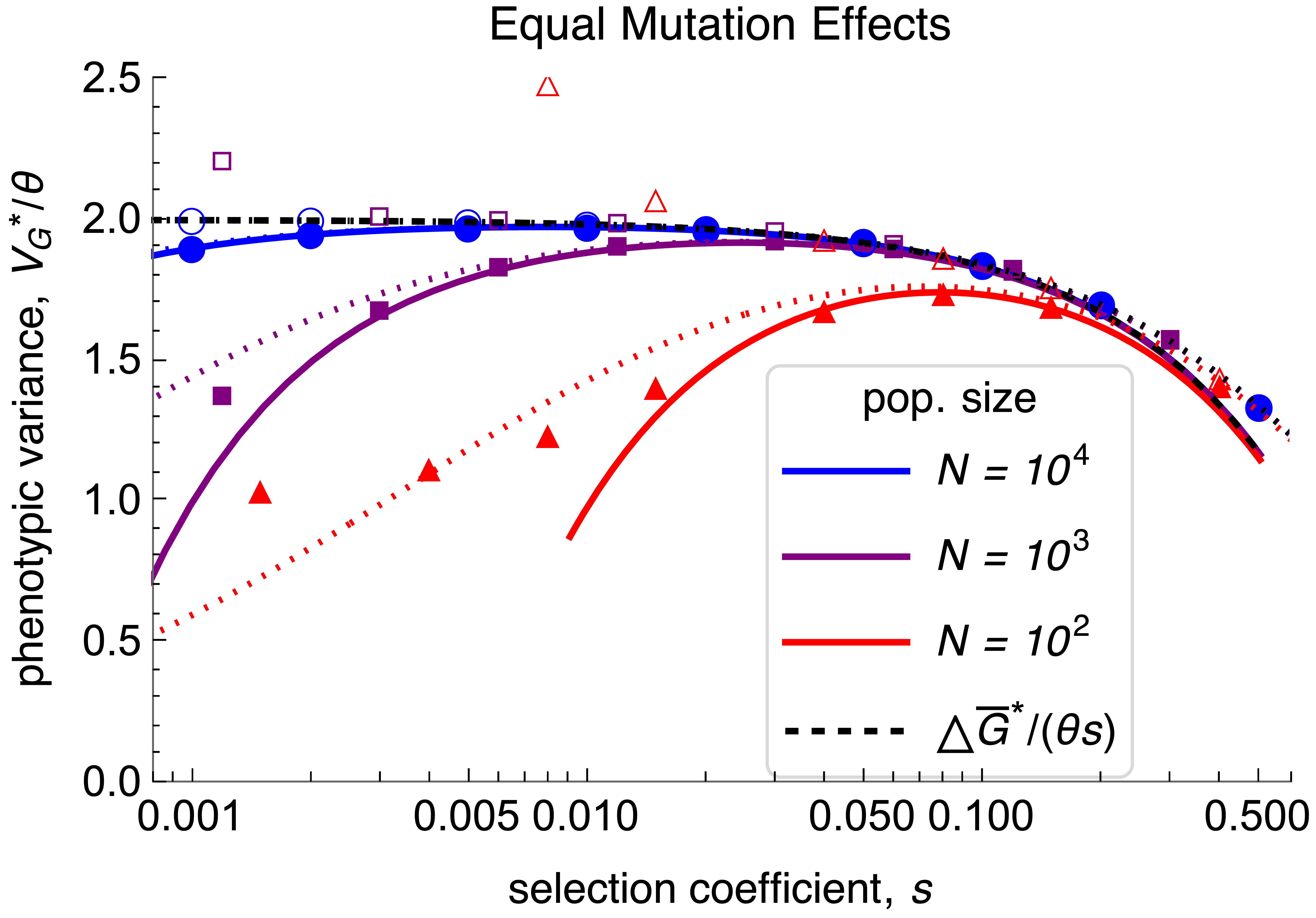} & 
\includegraphics[width=0.45\textwidth]{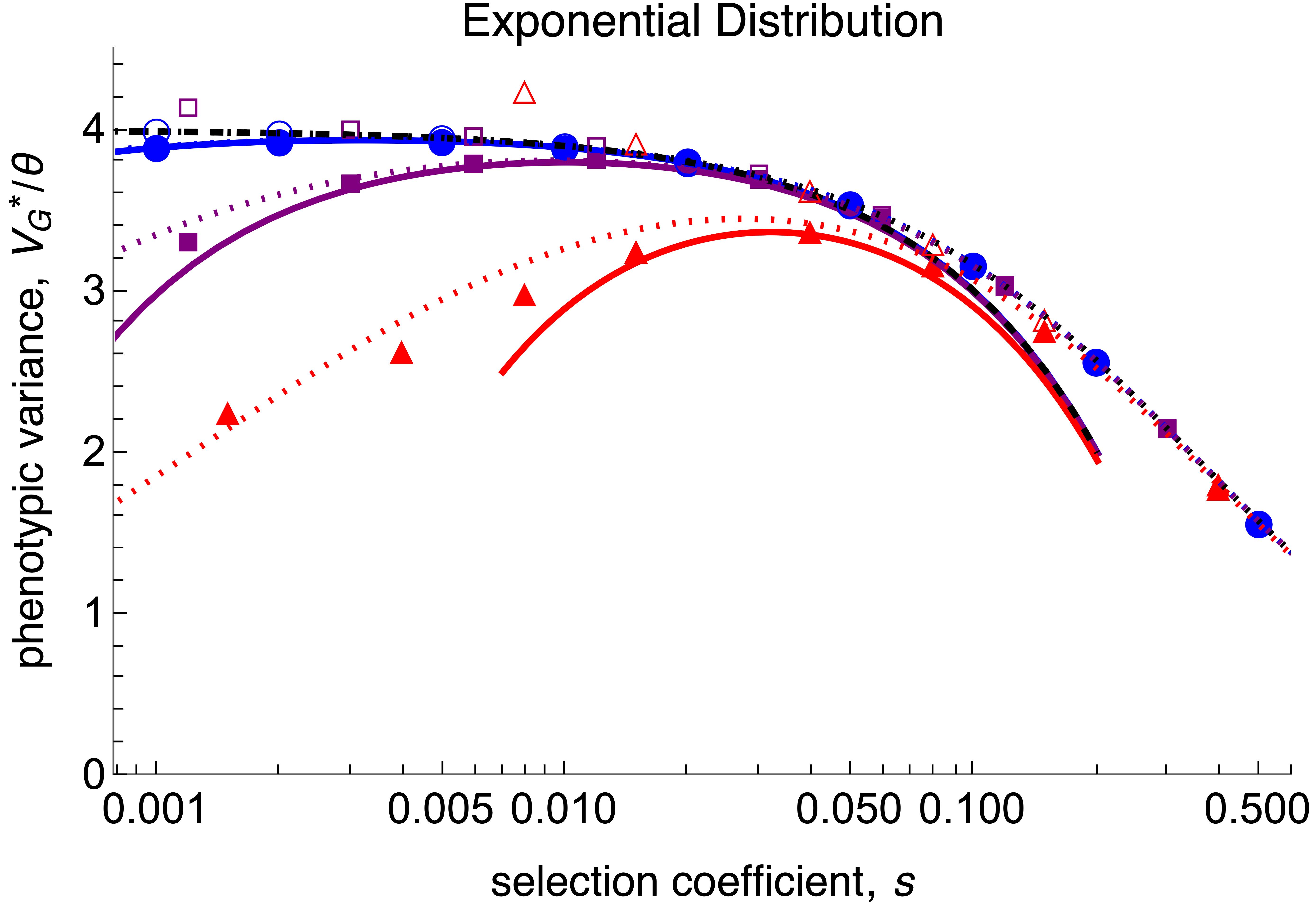}
\end{tabular}
\caption[$V_{G}^* (s)$]{The scaled expected phenotypic variance $V_G^*/\Th$ in the quasi-stationary phase for mutation effects equal 1 (panel A) and for exponentially distributed effects with mean 1 (panel B). The solid curves show the simple approximations \eqref{VG*Nsconst} (in A) and \eqref{VG*Ns} (in B) as functions of the selection coefficient $s$ for three population sizes $N$ (see legend). The most accurate available approximations are \eqref{VG*_equal} and \eqref{VG*_prop}; they are shown as dotted curves in A and B, respectively. Simulation results from the Wright-Fisher model with $\Th=\tfrac12$ are shown as symbols (circles for $N=10^4$,  squares for $N=10^3$,  triangles for $N=100$). In addition,  the scaled expected per-generation change in the phenotypic mean, $\Delta \bar{G}^*/(\Th s)$, is displayed (black dashed curves); it as calculated from \eqref{DebarG*equal} and \eqref{DebarG*exp} in (A) and (B), respectively.
For the Wright-Fisher simulations, filled symbols are used for the scaled phenotypic variance and open symbols for the change in the scaled phenotypic mean. Note that in a Wright-Fisher model, $\Delta \bar{G}^*/(\Th s)$ diverges to infinity if $s\to0$ because neutral mutations can become fixed. 
Analogous results for a truncated normal distribution ($\A\ge0$) with mean $\bar\A=1$ are shown in Figure \ref{fig_nor}.}
\label{fig_VarSEqu}
\end{figure}

In Figure~\ref{fig_VarSEqu} the approximations derived in Proposition~\ref{prop:equilibrium} and Corollary~\ref{cor:equilibrium_simp} (shown as curves) are compared with results from Wright-Fisher simulations (shown as symbols). The figure shows that the approximation \eqref{VG*_prop} for the stationary variance (with its simplification \eqref{VG*_equal} for equal mutation effects) is highly accurate in a parameter range containing $5 \le Ns\bar\A \le 0.5N$ (dotted curves in A and B), although it has been derived under the assumption $Ns\bar\A\gg1$. The corresponding simplified approximations \eqref{VG*Nsconst} and \eqref{VG*Ns} (solid curves), derived under the additional assumption $s\ll1$, are accurate in a range containing $5 \le Ns\bar\A \le 0.1N$. Whereas these approximations correctly predict a decrease in $V_G^*/\Th$ as $N$ decreases, the analytically derived dependence of $V_G^*/\Th$ on $s$ exaggerates the decrease of variance observed  for small $s$ in the Wright-Fisher simulations. Indeed, in the limit of $s\to0$ (neutrality), the true scaled variance $V_G^*/\Th$ should converge to $\A^2$ if mutation effects are equal to $\A$, and to $2\bar\A^2$ if they are exponentially distributed with mean $\bar\A$. 
This is indicated by the Wright-Fisher simulations and follows from the classical result \citep[e.g.,][]{ClaytonRobertson1955, LynchHill1986} that in the absence of selection, the diploid stationary variance is to a close approximation $2\Th \int_{-\infty}^\infty \A^2 f(\A) d\A$, where $f$ is the density of a mutation distribution with arbitrary mean. The missing factors of 2 in comparison to the classical neutral result are due to the assumption of haploidy. The reason why for small $s$ our approximation \eqref{VG*_prop} (dotted curves) decays below the neutral prediction is that for weak selection ($Ns\bar\A\le1$), the allele-frequency dynamics are increasingly driven by random genetic drift which is not appropriately reflected by the branching-process approximation. 

The approximations \eqref{DebarG*equal} and \eqref{DebarG*exp} for $\Delta \bar{G}^*$ (black dashed curves in panels A and B of Fig.~\ref{fig_VarSEqu}) accurately match the response calculated from Wright-Fisher simulations (open symbols) in a range containing $2/N \le s\bar\A \le 0.1$. If higher-order terms in $s$ had been included, then these approximations would be accurate for $s\bar\A$ up to $0.5$. (The dotted curves for the variance are accurate up to this value because they are computed from \eqref{VG*_equal} and \eqref{VG*_prop}, which do not use series expansion in $s$.) Comparison of the terms of order $s^2$ in the approximations \eqref{DebarG*equal} and \eqref{DebarG*exp} with the term $\frac{\Th\A}{Ns}$ in \eqref{VG*app} shows that the latter can be neglected in \eqref{VG*app} if $cs^2\bar\A^2\gg 1/N$, where $c=\tfrac53$ for equal mutation effects and $c=10$ for exponentially distributed effect. The other way around, random genetic drift will distort the classical relation $\Delta \bar{G}^* \approx s V_G^*$ between the response of the mean and the genetic variance if (approximately) $s\bar\A<\sqrt{10/(cN)}$. For equal mutation effects ($c=\tfrac53$), this becomes $s\bar\A<2.4/\sqrt{N}$; for exponentially distribution effects ($c=10$), it becomes $s\bar\A<1/\sqrt{N}$.  Both relations confirm the observations in Fig.~\ref{fig_VarSEqu}, where $\bar\A=1$.

\begin{rem}\label{rem:truncatedGaussian}
In quantitative genetic models often a normal distribution with mean 0 is assumed for the mutation effects. To apply our analysis, which ignores mutations with a negative effect, to this setting, we have to adapt this distribution and use a truncated normal distribution instead. If the original normal distribution has variance $v_N^2$, then the truncated normal (restricted to positive effects) is defined by $f(\A) = \frac{2}{\sqrt{2\pi v_N^2}}\exp(-\A^2/(2v_N^2))$ if $\A\ge0$ (and 0 otherwise). Also our mutation rate $\Th$ will correspond to $\Th/2$ in a quantitative genetic model in which mutations have (symmetric) positive and negative effects.  This truncated normal distribution has mean $\bar\A=\sqrt{2v_N^2/\pi}$. Taking this as the parameter, we obtain $\A_2(f) = v_N^2 = \tfrac12{\bar\A}^2\pi$ and $\A_3(f) = {\bar\A}^3\pi$. Therefore, \eqref{DebarG*} yields
\begin{linenomath}\begin{equation}
	\De\bar G^* \approx \pi\Th s {\bar\A}^2(1-\tfrac{5}{3}s\bar\A)\,, \label{DebarG*tn}
\end{equation}\end{linenomath}
instead of \eqref{DebarG*exp}, and the stationary variance is approximately 
\begin{linenomath}\begin{equation}
	V_G^*  \approx \pi \Th \bar{\A}^2 \left( 1 - \frac{5}{3} s \bar{\A} - \frac{1}{\pi N s \bar{\A}} \right)\,. \label{VG*tn}
\end{equation}\end{linenomath}
\end{rem}

After this excursion to the long-term evolution of the trait, we first provide approximations for $\bar G(\tau)$ and $V_G(\tau)$ in Proposition \ref{prop:general_barG_VG(tau)} that can be evaluated expeditiously. Then we turn to the initial phase of adaptation.

\subsection{Approximations for the dynamics of $\bar G(\tau)$ and $V_G(\tau)$} \label{sec:Dynamics_approx}

\begin{prop}\label{prop:Dynamics}
We posit Assumptions \ref{A1} and \ref{A2}. 

(1) The phenotypic mean has the approximation
\begin{linenomath}\begin{align}
	\bar{G}(\tau, f, s, N, \Th) 
		 &= \Th\, \int_0^{\infty} \frac{1}{s}\Psur(e^{s\A}) \Bigl( \exp(\sN e^{-\sA\tau}) E_1(\sN e^{-\sA\tau}) - e^\sN E_1(\sN) \Bigr) f(\A)\, d\A \notag\\
			&\qquad + O\Bigl( N^{-1+2/K} \Bigr) \;\text{ as } N\to\infty \,, \label{barG_init}
\end{align}
where $K>3$ is required. 

(2) The phenotypic variance has the approximation  
\begin{align}
	V_G(\tau, f, s, N, \Th) 
		 &=  \Th\, \int_0^{\infty} \frac{\A}{s} \Psur(e^{s\A}) \Bigl( \sN e^\sN E_1(\sN) - \sN e^{-\sA\tau} \exp (\sN e^{-\sA\tau}) E_1(\sN e^{-\sA\tau}) \Bigr) f(\A)\, d\A  \notag\\
		&\qquad + O\Bigl( N^{-K_1/K} \Bigr)  \;\text{ as } N\to\infty \,, \label{VG_init}
\end{align}\end{linenomath}
where $\sN=N\Psur(e^{\sA})$, $K_1>1$ and $K>K_1+1$ (whence $K>2$ is necessary).
\end{prop}

The proof is provided in Appendix \ref{App:proof of prop:Dynamics}.
For many offspring distributions, including the Poisson, an insightful simplification is obtained by applying $\Psur(e^{\sA})\approx \rho\sA$ for an appropriate constant $\rho$ (cf.~Remark~\ref{Psur=csalpha}). 

We note that the error term is not satisfactory for the very early phase of evolution when $\tau$ is still very small. For this very early phase, we derive simple explicit approximations with time-dependent error terms in Proposition \ref{prop:Initial2}. For large $\tau$, $V_G(\tau, f, s, N, \Th)$ converges to $V_G^*$ because the time-dependent term in \eqref{VG_init} vanishes. Therefore, this approximation will be very accurate for large $\tau$. Analogously, $\bar{G}(\tau, f, s, N, \Th)$ provides an accurate approximation for large $\tau$ as shown in Remark \ref{Rem:barG*} and illustrated in Figure \ref{fig_MeanVarEqu}.C.

\begin{rem}\label{rem:Dynamics}
For equal mutation effects $\A$, the approximations \eqref{barG_init} and \eqref{VG_init} simplify to
\begin{linenomath}\begin{equation}
	\bar{G}(\tau, f, s, N, \Th) 
		 \approx \frac{\Th}{s}\Psur(e^{s\A}) \Bigl( \exp(\sN e^{-\sA\tau}) E_1(\sN e^{-\sA\tau}) - e^\sN E_1(\sN) \Bigr) \label{barG_init_equal}
\end{equation}
and 
\begin{equation}
	V_G(\tau, f, s, N, \Th) 
		 \approx  \frac{\Th\A}{s} \Psur(e^{s\A}) \Bigl( \sN e^\sN E_1(\sN) - \sN e^{-\sA\tau} \exp (\sN e^{-\sA\tau}) E_1(\sN e^{-\sA\tau}) \Bigr) \,, \label{VG_init_equal}
\end{equation}\end{linenomath}
respectively.  Again, it is insightful to keep in mind the approximation $\Psur(e^{\sA})\approx \rho\sA$.
\end{rem}

\subsection{Approximations for the initial phase} \label{sec:initial}
For the very early phase of adaptation simple explicit approximations can be derived for $\bar{G}(\tau)$ and $V_G(\tau)$.

\begin{prop}\label{prop:Initial2}
We require Assumptions \ref{A1} and \ref{A2}. In addition, we assume $0\le\tau\le1/(4\barA s)$ and that the mutation distribution $f$ is exponential with mean $\barA$.

(1) The phenotypic mean and the phenotypic variance have the approximations
\begin{linenomath}\begin{equation}
	\bar{G}(\tau) = \frac{\Th\tau}{N}\biggl[\frac{\barA}{1-\barA s \tau} + O\Bigl(N^{-1+1/K}\Bigr)\biggr]\,, \label{barG_smalltau_a}
\end{equation}\end{linenomath}
and
\begin{linenomath}\begin{equation}\label{VG_smalltau_a}
	V_G(\tau) =  \frac{2\Th\barA^2\tau}{N}\biggl[\frac{1-s\barA\tau/2}{(1-s\barA\tau)^2} + O\Bigl(N^{-1+1/K}\Bigr)\biggr]\,,
\end{equation}\end{linenomath}
respectively (and the error terms are independent of $\tau$).

(2) For sufficiently small $s$ the leading order terms have the series expansions
\begin{linenomath}\begin{equation}
	\bar{G}(\tau) \approx \frac{\Th\bar\A\tau}{N}\Bigl(1+s\bar\A\tau+(s\bar\A\tau)^2+(s\bar\A\tau)^3 \Bigr) \label{barG_smalltau_series}
\end{equation}\end{linenomath}
and
\begin{linenomath}\begin{equation}
	V_G(\tau) \approx \frac{2\Th{\bar\A}^2\tau}{N} \Bigl(1 + \tfrac32s\bar\A\tau  + 2(s\bar\A\tau)^2 + \tfrac52 (s\bar\A\tau)^3 \Bigr)\,. 
	 \label{VG_smalltau_series}
\end{equation}\end{linenomath}
\end{prop}

The proof is given in Appendix \ref{app:proof_prop:Initial2}.

\begin{rem}\label{Gbar_smalltau_equal}
If the mutation effects are equal to $\A$, then the approximations
\begin{linenomath}\begin{equation} 
	\bar{G}(\tau) \approx  \frac{1}{\A} V_G(\tau) \approx \frac{\Th}{N s} \left(e^{s\A\tau} -1 \right)	 
	\label{VG_smalltau_const}
\end{equation}\end{linenomath}
are obtained. The errors are of order $\tau\,O\Bigr((Ns)^{-1}\Bigr)=\tau\,O\Bigr(N^{-1+1/K}\Bigr)$. The validity of this approximation, in particular of the error term, requires only $e^{2s\A\tau }=O(1)$, thus effectively $s\A\tau=O(1)$ (see Appendix \ref{app:proof_prop:Initial2}).
\end{rem}

\begin{figure}[t]
\centering
\begin{tabular}{ll}
A & B \\
\includegraphics[width=0.45\textwidth]{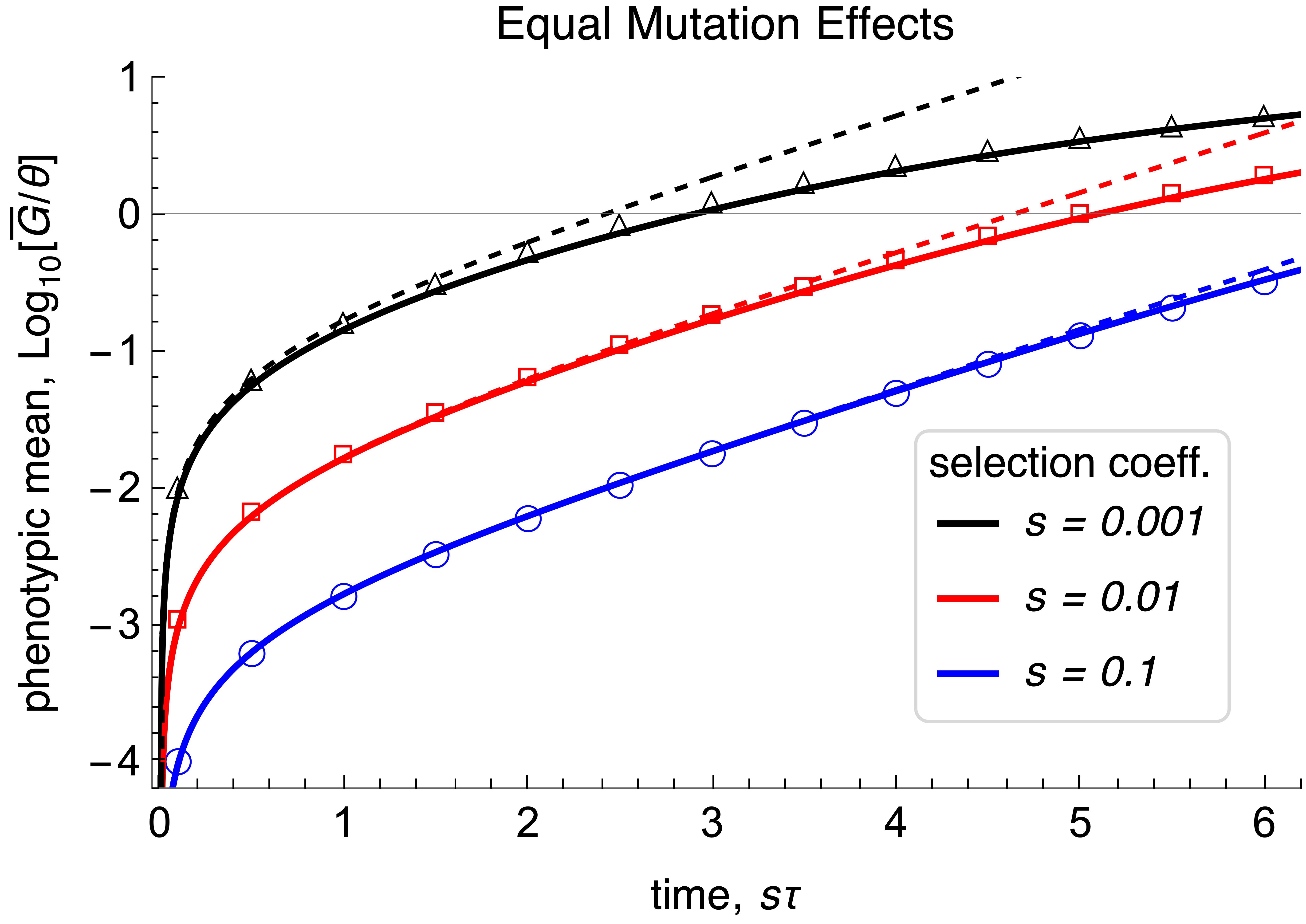} &
\includegraphics[width=0.45\textwidth]{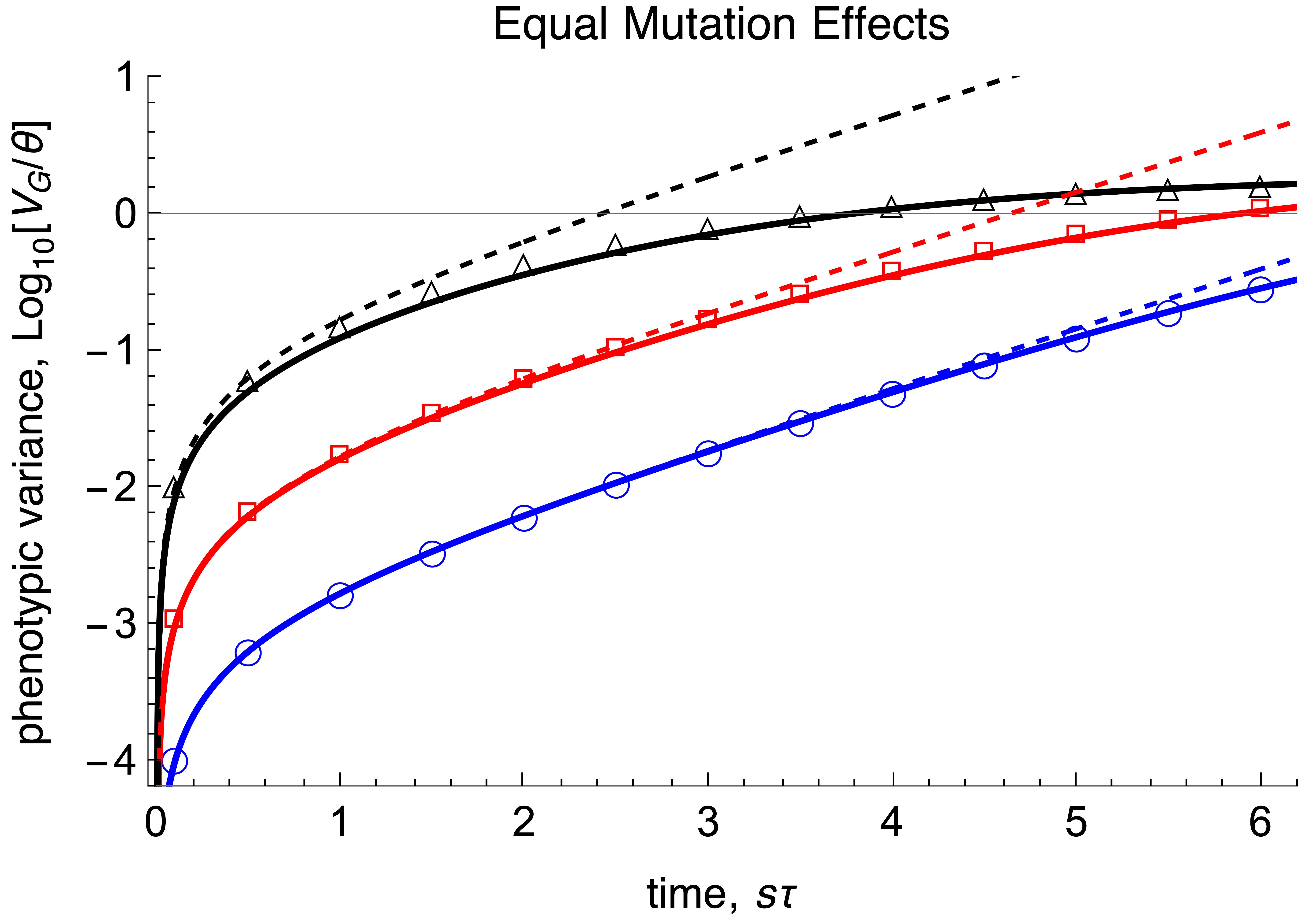}\\
C & D \\
\includegraphics[width=0.45\textwidth]{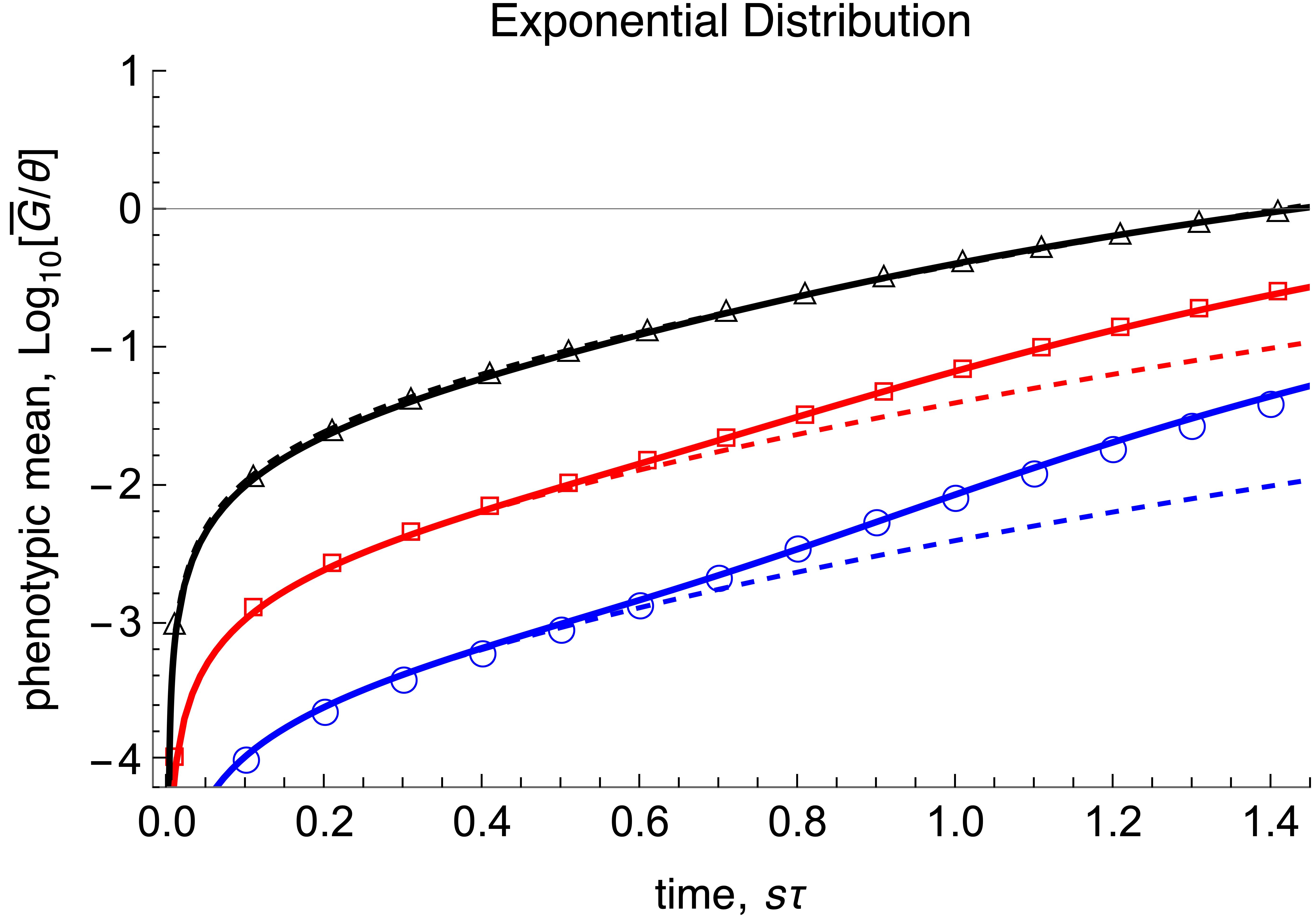} &
\includegraphics[width=0.45\textwidth]{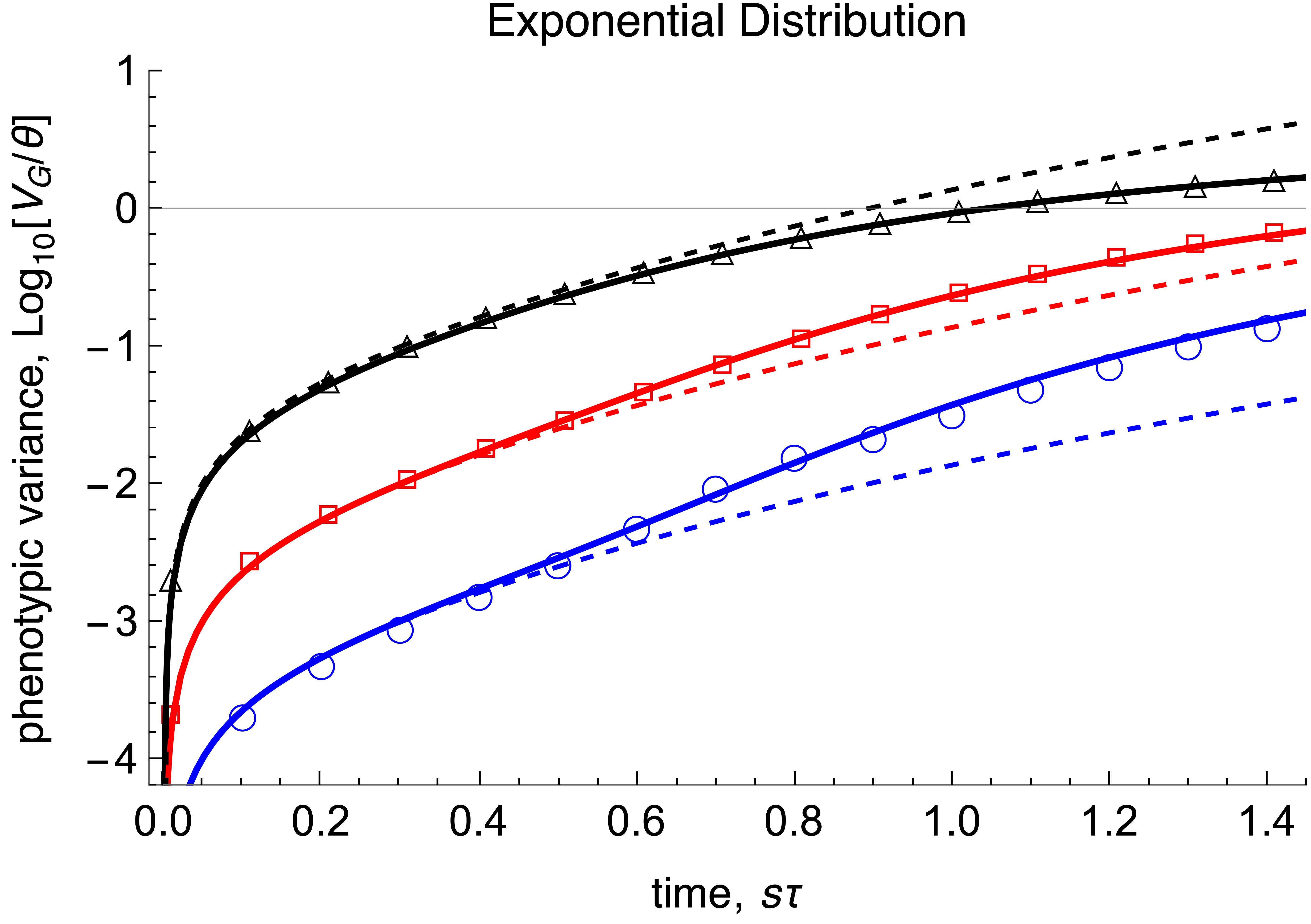}
\end{tabular}
\caption[$\bar{G}$ and $V_{G}$ - Initial phase]{The scaled phenotypic mean $\log_{10}(\bar G(s\tau)/\Th)$ (panels A, C)  and the scaled phenotypic variance $\log_{10}(V_G(s\tau)/\Th)$ (panels B, D) in the initial phase for selection coefficients $s$ as given in the legend. For equal mutation effects $\A=1$, the solid curves in A and B show the approximations \eqref{barG_init_equal} and \eqref{VG_init_equal}, respectively. The dashed curves show the simplified approximation in \eqref{VG_smalltau_const}. For exponentially distributed effects with $\bar\A=1$ the solid curves in C and D show the approximations \eqref{barG_init} and \eqref{VG_init}, respectively. The dashed curves show the simplified approximations \eqref{barG_smalltau_series} and \eqref{VG_smalltau_series}. In all cases, symbols present results from Wright-Fisher simulations with $\Th=\tfrac12$, the population size is $N=10^4$, and the offspring distribution is Poisson.  }
\label{fig_MeanVarInitial1}
\end{figure}

In Figure \ref{fig_MeanVarInitial1} the approximations for the expected phenotypic mean and variance obtained in Proposition \ref{prop:Dynamics} and Proposition \ref{prop:Initial2} are compared with Wright-Fisher simulations for various selection coefficients. For equal mutation effects (panels A and B) the approximations are accurate in a much wider range of values $s\tau$ than for exponentially distributed effects (panels C and D). In both cases, the more complicated approximations, \eqref{barG_init_equal} and \eqref{VG_init_equal} (solid curves in A and B) and \eqref{barG_init} and \eqref{VG_init} (solid curves in C and D), are accurate for a much longer time span (in fact for $\tau\to\infty$ as discussed above) than the corresponding simple approximations, \eqref{VG_smalltau_const} (dashed curves in A, B) and \eqref{barG_smalltau_series}, \eqref{VG_smalltau_series} (dashed curves in C, D).  For an exponential mutation distribution, the simple approximations are accurate if, approximately, $\tau<1/(2s\bar\A)$. For weak selection, this can still be quite a long time span. The approximations \eqref{barG_smalltau_a} and \eqref{VG_smalltau_a} are not shown because on this scale of resolution they are almost indistinguishable from the corresponding simple series expansions if $\tau\lessapprox 1/(2s\bar\A)$, but then diverge as $\tau s\barA\to1$.
For large $s\tau$, the approximations in Proposition \ref{prop:Dynamics} and Remark \ref{rem:Dynamics} for $\bar G(\tau)$ and $V_G(\tau)$ are almost identical to the exact expressions in Proposition \ref{prop:general_barG_VG(tau)} (results not shown). Visible differences on the scale of resolution in our figures occur only for small or moderately large $s\tau$ if $s=0.001$ and $N=10^4$, i.e., $Ns=10$ (see Figure \ref{fig_VarInitial_comparison}).

The approximations \eqref{barG_smalltau_a} and \eqref{VG_smalltau_a} inform us that initially, as long as $s\barA\tau\ll1$, $\bar{G}(\tau)$ and $V_G(\tau)$ increase nearly linearly. Comparison with \eqref{VG_smalltau_const} shows that for an exponential mutation distribution with mean $\barA$ this early increase is considerably larger than for mutation effects equal to $\barA$. These approximations also show that initially the relation $\De\bar G \approx s V_G$, typically expected if the trait variance is high, fails; cf.~\eqref{VG*app}. We note that \eqref{barG_smalltau_a} and \eqref{VG_smalltau_a} are always overestimates because their error terms are negative (see Appendix \ref{app:proof_prop:Initial2}).

\FloatBarrier
\section{Sweep-like vs.\ polygenic shift-like patterns}\label{sec:sweepshift}

An important role in examining whether sweep-like patterns or polygenic shift-like patterns characterize the early phase of adaptation is played by the number of segregating sites. If sweeps occur successively, at most one segregating site is expected for most of the time. As the number of segregating sites increases, parallel sweeps occur and the sweep pattern will be transformed to a shift-like pattern. If adaptation is dominated by subtle allele frequency shifts, the number of segregating sites is expected to be large. Of course, there will be intermediate patterns, where no clear distinction between a sweep and a shift pattern can be made. Roughly, we call a pattern of adaptation `sweep-like' if a mutant has already risen to high frequency, so that its loss is very unlikely, when the next mutant starts its rise to (likely) fixation. We call a pattern shift-like if several or many mutants increase in frequency simultaneously (see Fig.~\ref{fig:sweep_shift_new}).

We emphasize that we focus on the patterns occurring during the early phases of adaptation because by our assumption that mutations are beneficial and random genetic drift is weak ($s>0$ and $Ns\gg1)$, all mutants that become `established' will sweep to fixation, independently of other parameters such as $\Th$. In addition, although exponential selection may provide a good approximation for selection occurring in a moving optimum model or by continued truncation selection of constant intensity, it can approximate selection caused by a sudden shift in the phenotypic optimum only during the early phase when the population mean is still sufficiently far away from the new optimum (see Discussion).

\subsection{Number of segregating sites}\label{sec:segsites}

To characterize the patterns of adaptation, an approximation for the expected number of segregating sites, $\EV[S]$, is needed. This expectation is taken not only with respect to the distribution of the number of segregating sites, but also with respect to the mutation distribution $f$, and it assumes a Poisson offspring distribution. For a precise definition see eq.~\eqref{EV_S_def} in Appendix \ref{app:segsites}. 

We assume the infinite sites model, as used in Sects.~\ref{sec:inf_many_sites} and onwards. However, we complement it by \emph{diffusion approximations} for the fixation and the loss probabilities as well as for the corresponding expected times to fixation and loss (see Appendix \ref{meanfixtime}). To derive a simple approximation $\bar S$ for $\EV[S]$, we distinguish between mutations that eventually become fixed, which occurs with probability $\Pfix(\stilde,N)$ (see \eqref{stilde} for $\stilde$), and those that get lost, which occurs with probability $1-\Pfix(\stilde,N)$. In addition, we assume that mutations that are destined for fixation and occurred $t \ge \tfix(\stilde,N)$ generations in the past are already fixed. Thus, in a given generation $\tau$, mutations destined for fixation contribute to $\bar S$ if and only if they occurred less than or exactly $\min\{\tfix(\stilde,N)-1,\tau \}$ generations in the past (including 0 generations in the past, i.e., now, in generation $\tau$). Therefore, the number of generations in which mutants occur that contribute to the current variation is $\min\{\tfix(\stilde,N),\tau+1\}$. This is consistent with the fact that mutations can already occur in generation $\tau=0$. Analogously, mutations destined for loss contribute to $\bar S$ at time $\tau$ only if their time $t$ of occurrence (prior to $\tau$) satisfies $0 \le t\le\min\{\tloss(\stilde,N)-1,\tau \} $, where $\tloss(s,N)$ denotes the expected time to loss of a single mutant with selective advantage $s$ in population of size $N$; see \eqref{tloss}. 
In summary, we define the approximation $\bar S$ as follows:
\begin{linenomath}\begin{align}\label{Seg}
	\bar S (\tau,f,s,N,\Th) & = \Th  \int_0^{\infty}  f(\A) \Pfix(\stilde,N) \min\{\tfix(\stilde,N),\tau+1 \} \, d\A \notag \\
			&\quad+ \Th  \int_0^{\infty}  f(\A) (1-\Pfix(\stilde,N)) \min\{\tloss(\stilde,N),\tau+1 \} \, d\A \,.
\end{align}\end{linenomath}

\begin{figure}[ht!]
\centering
\begin{tabular}{ll}
A & B \\
\includegraphics[width=0.44\textwidth]{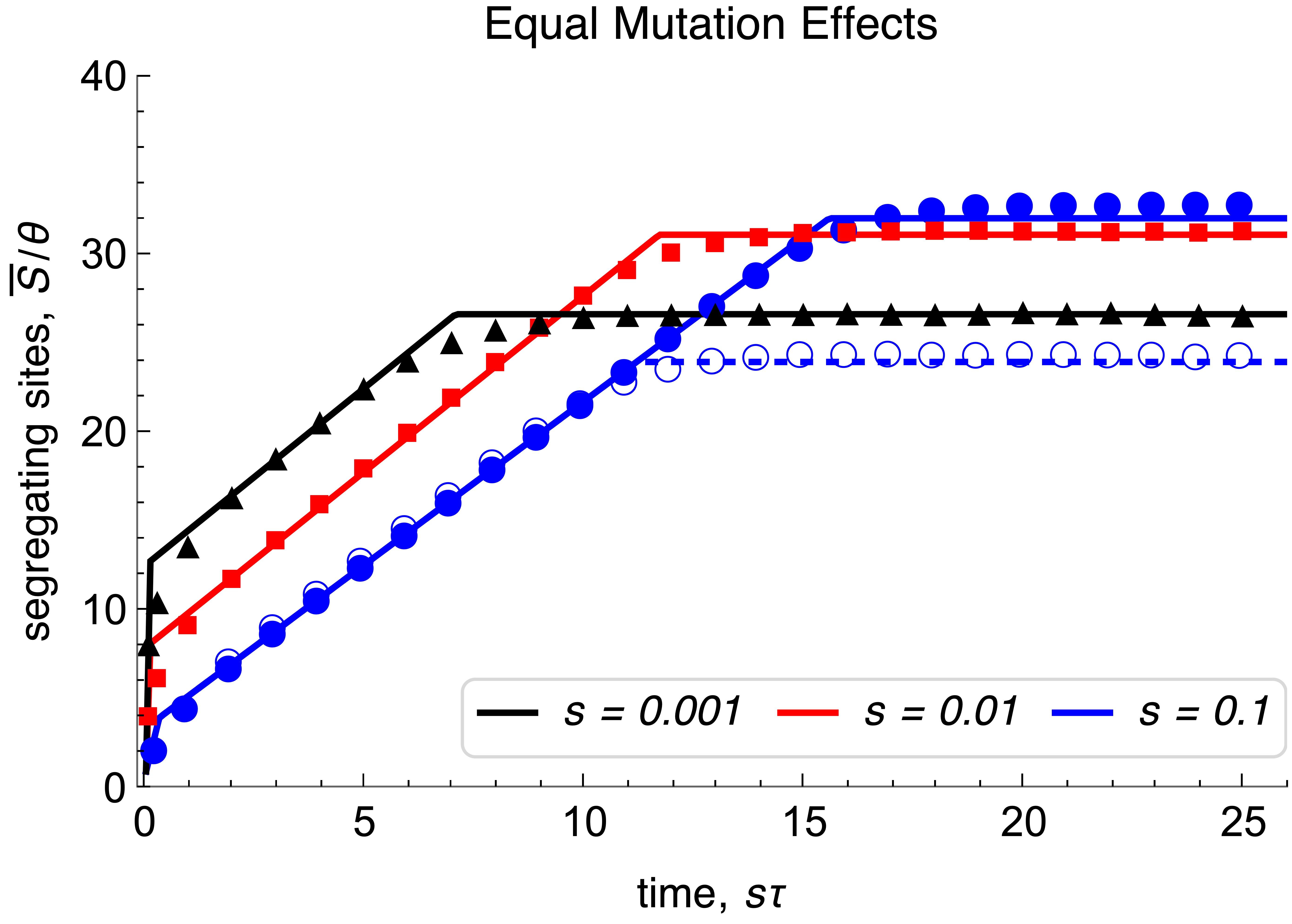} &
\includegraphics[width=0.44\textwidth]{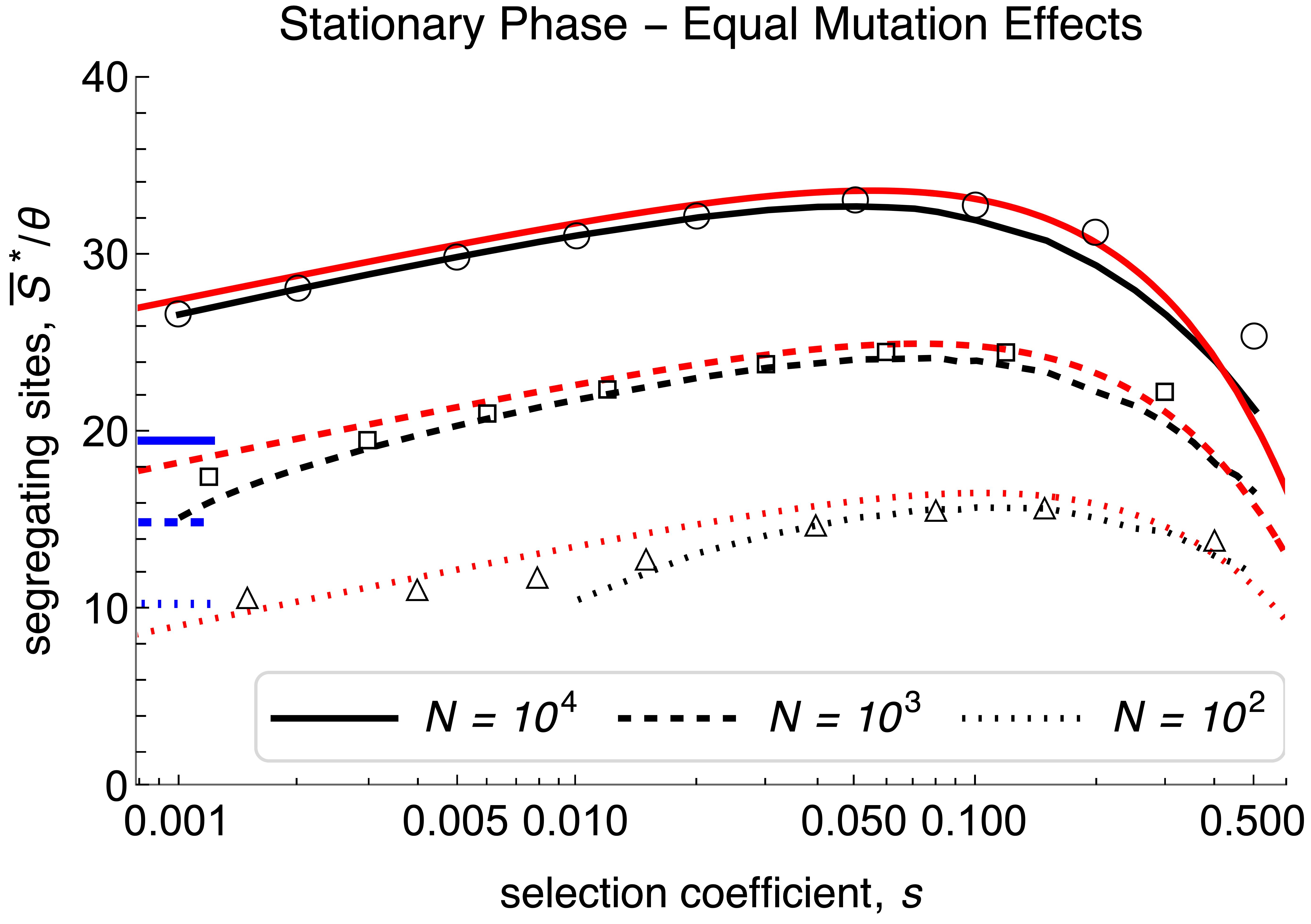} \\
C & D \\
\includegraphics[width=0.44\textwidth]{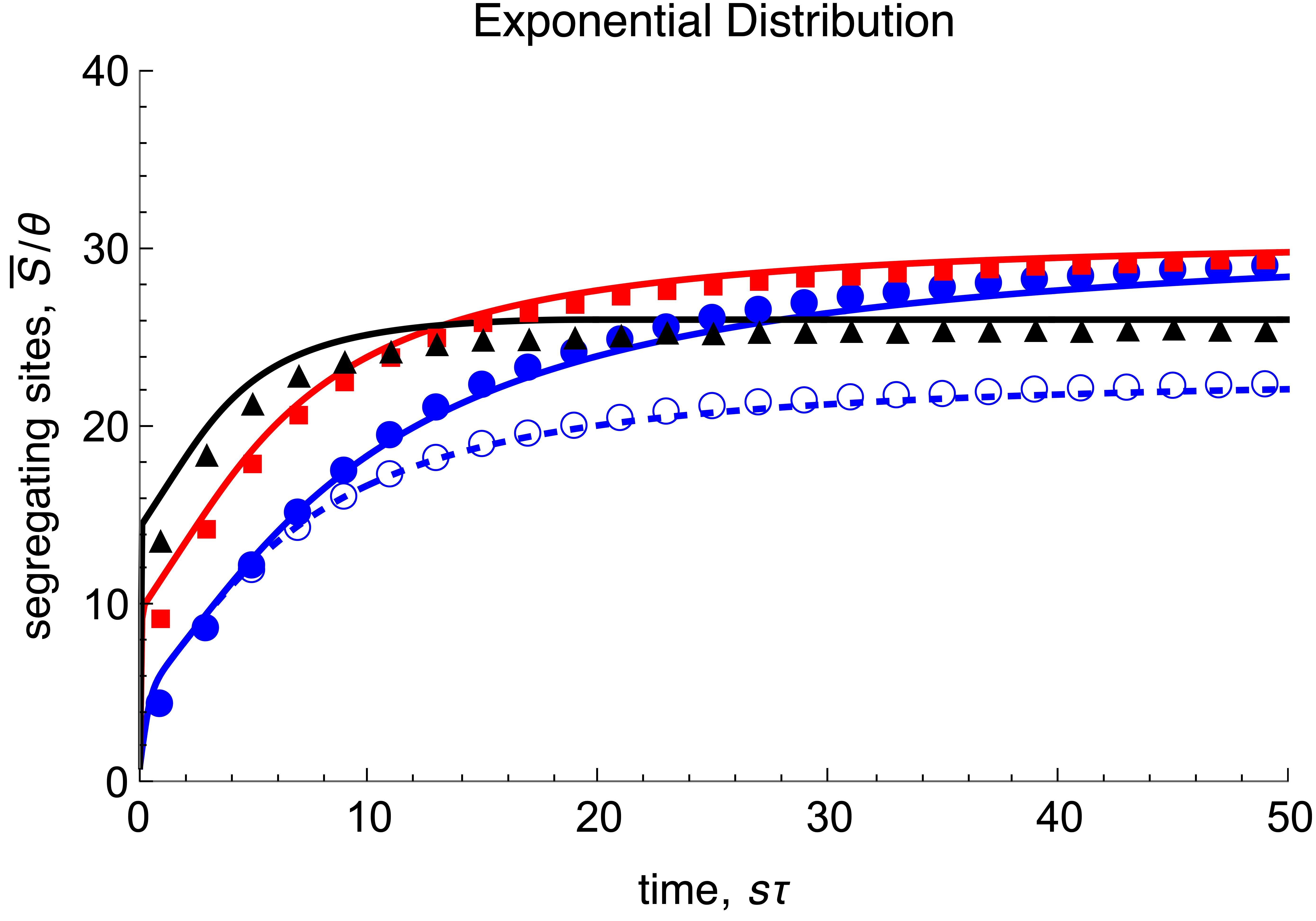} &
\includegraphics[width=0.44\textwidth]{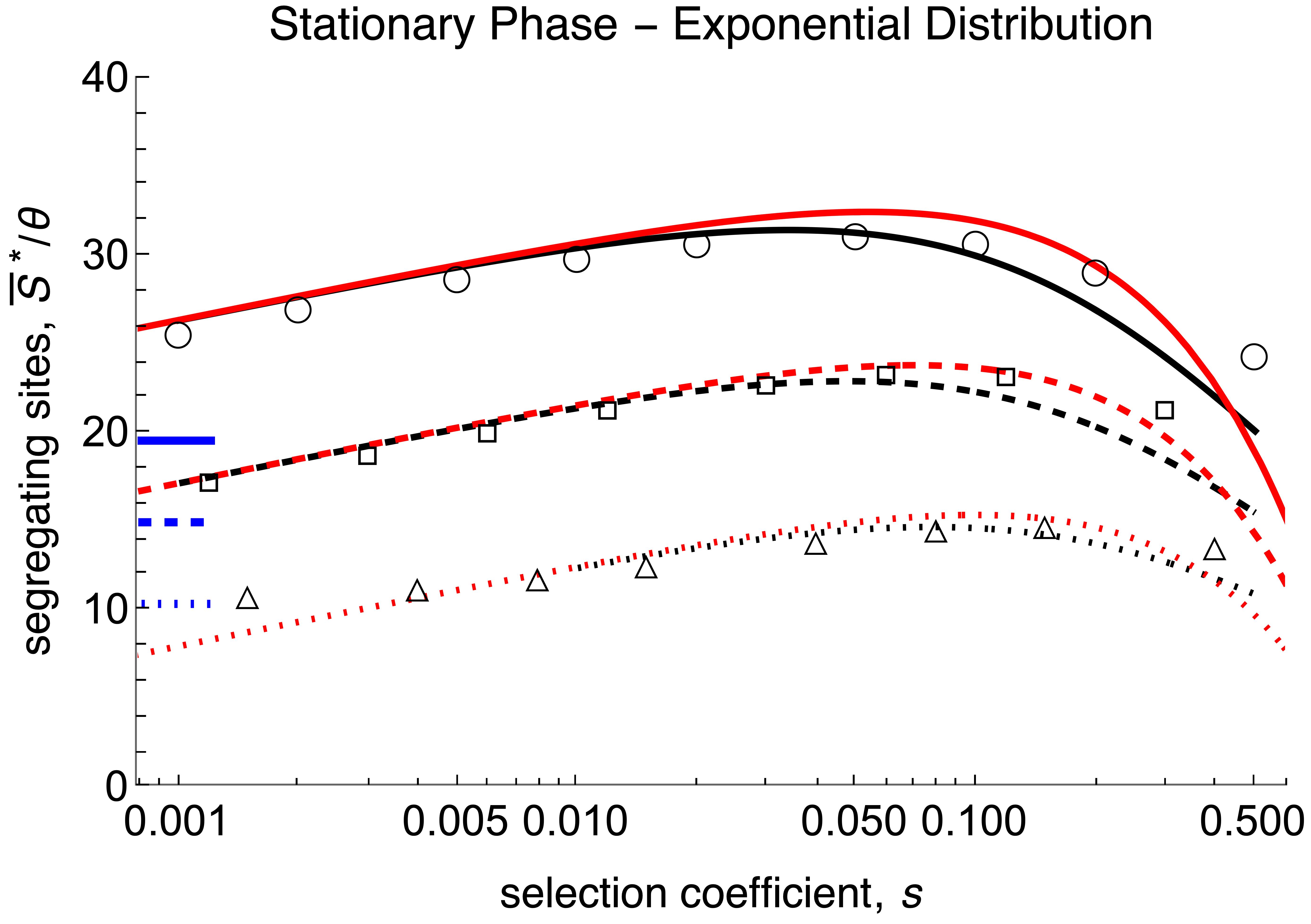}
\end{tabular}
\caption[Segregating sites]{The expected number of segregating sites (rescaled by $\Th$) as a function of $s\tau$ (panels A, C) and at quasi-stationarity as a function of $s$ (panels B, D).  In A and B the mutation effects are equal ($\A=1$), in C and D they are drawn from an exponential distribution with mean $\bar\A =1$. Analytic approximations are computed from \eqref{Seg} and are shown as curves. Wright-Fisher simulation results for $\EV[S]$ (with $\Th = \frac{1}{2}$) are shown as symbols.  In panels A and C, $\bar S/\Th$ is shown for $N=10^4$ and $s=0.1,0.01,0.001$ (solid curves and filled symbols in blue, red, black), and for $N=10^3$ and $s=0.1$ (dashed curve and open symbols in blue).  In panels B and D, $\bar S^*/\Th$ is displayed for $N= 10^4, 10^3,10^2$ (solid, dashed, dotted curves, and corresponding open symbols). Black curves represent \eqref{Seg}, red curves the approximations in \eqref{L*_approx}. Under neutrality, we obtain $\EV[S]\approx 19.6, 15.0, 10.4$ for $N= 10^4, 10^3,10^2$, respectively (indicated by the short horizontal blue lines in B and D). Analogous results for a truncated normal distribution ($\A\ge0$) with mean $\bar\A=1$ are shown in Figure \ref{fig_norSeg}.}
\label{fig_SegSites}
\end{figure}

For the quasi-stationary as well as for the initial phase, we can specify relatively simple explicit approximations by employing diffusion approximtions for $\Pfix$, $\tfix$, and $\tloss$. We recall from \eqref{bartfix} that $\bartfix(s,\bar\A,N)$ is the expected time to fixation of a single mutant with effect $\A$ on the trait drawn from the (exponential) distribution $f$; see also Appendix~\ref{meanfixtime_f}.  In the limit of large $\tau$, i.e., $\tau\gg\bartfix$, we obtain for $\bar S^*(f,s,N,\Th) = \lim_{\tau\to\infty} \bar S(\tau,f,s,N,\Th)$ the following simplifications:
\begin{linenomath}\begin{align} 
	\bar S^* (f,s,N,\Th) &\approx \Th  \int_0^{\infty}  f(\A) \Pfix(\stilde,N) \tfix(\stilde,N) \, d\A  \notag \\
			&\quad+ \Th \int_0^{\infty}  f(\A) (1-\Pfix(\stilde,N)) \tloss(\stilde,N) \, d\A  \notag \\
			&\hspace{-20mm}\approx
			\begin{cases}
				 2\Th  \bigl[(1- s\bar{\A}) \bigl(\ln(N) + \ln (2  N s \bar{\A} )\bigr) + 1+\ga -(\ga + \tfrac12) s \bar{\A} \bigr]
					&\text{if } \A = \bar{\A}  \,,\\ \vspace{-2mm} \\
				 2\Th  \bigl[(1- s\bar{\A}) \bigl(\ln(N) + \ln (2  N s \bar{\A} )\bigr) +1-\tfrac32 s \bar{\A} \bigr] 
					&\hspace{-12mm}\text{if } \A \sim \Exp[1/ \bar{\A}] \,.
			\end{cases}  \label{L*_approx}
\end{align}\end{linenomath}
To evaluate the integrals, we used $\stilde\approx \sA$ from \eqref{stilde}, $\Pfix(s,N)\approx 1-e^{-2s}$ from \eqref{PfixDif}, $\tfix(s,N)\approx\frac{2}{s}(\ln(2Ns)+\ga)$ from \eqref{tfixHP}, and $\tloss(s,N)\approx -2(1+s)\ln(2s)+2(1-\ga)+s(3-2\ga)$ from \eqref{tloss} and \eqref{tloss_diffapp}. Finally, a series expansion for small $s$ was performed. The approximate formulas in \eqref{L*_approx} are accurate if $Ns\bar\A>2$ and $s\bar\A<0.1$; see Figs.~\ref{fig_SegSites}B and D.  

We observe that in both cases of \eqref{L*_approx} the leading-order term is approximately $2\Th\ln(N)$, which is close to the classical neutral result $\EV[S] = \Th \sum_{i=1}^{N-1}\dfrac{1}{i} \approx \Th(\ln(N)+\ga)$, in which the factor 2 is contained in $\Th$ (and $N$ is the sample size) \citep{Ewens1974}. \citet{JainKaushik2022} investigated the site frequency spectrum assuming periodic selection coefficients. In the absence of dominance, their approximation (15b) for strong directional selection (i.e., $N \bar s\gg1$, $\bar s$ the average selection coefficient) becomes independent of $\bar s$ and yields the neutral result, thus essentially our leading order term. We note that for equal mutation effects the stationary value $\bar S^*$ of $\EV[S]$ is slightly higher than that for exponentially distributed effects.

In analogy to $\bartfix(s,\bar\A,N)$, in Appendix~\ref{meanlosstime_f} we introduce the expected time to loss, $\bartloss$, of a single mutant with effect $\A$ on the trait drawn from the (exponential) distribution $f$.
If $\tau$ is small, but not tiny, i.e., $\bartloss < \tau \ll \bartfix$, then
\begin{linenomath}\begin{align} 
	\bar S (\tau,f,s,N,\Th) &\approx \Th (\tau+1) \int_0^{\infty}  f(\A) \Pfix(\stilde,N) \, d\A  \notag \\
			&\quad+ \Th \int_0^{\infty}  f(\A) (1-\Pfix(\stilde,N)) \tloss(\stilde,N) \, d\A  \notag \\
			&\hspace{-20mm}\approx 
			\begin{cases}
				\Th \tau (1-e^{-2s\bar{\A}}) + 2\Th \bigl[-(1-s\bar{\A})(\ln (2s\bar{\A})+\ga)+1+\tfrac12s\bar{\A} \bigr]
					&\text{if } \A = \bar{\A}  \,, \\ \vspace{-2mm} \\
				\Th \tau \dfrac{2s\bar{\A}}{1+2s\bar{\A}} + 2\Th \bigl[-(1-s\bar{\A})\ln (2s\bar{\A}) + 1+ \tfrac32s\bar{\A} \bigr]
					&\hspace{-12mm}\text{if } \A \sim \Exp[1/ \bar{\A}] \,, 
			\end{cases} \label{L*_approx_init}
\end{align}\end{linenomath}
where the same approximations as for \eqref{L*_approx} are used.
If $\tau < \bartloss$, then \eqref{Seg} yields $\bar S (\tau,f,s,N,\Th) \approx \Th (\tau +1)$, which is an overestimate except when $\tau=0$ (cf.\ Fig.~\ref{fig_SegSitesBP}).

As Figure \ref{fig_SegSites} demonstrates, \eqref{Seg} yields a quite accurate approximation for the expected number of segregating sites, both as a function of time (panels A and C), and as function of $s$ at quasi-stationarity (panels B and D). The piecewise linear shape of the curves in A results from the simplifying assumption of taking the minima of $\tau$ and $\tloss$ or $\tfix$, i.e., the kinks occur at $\tloss$ and at $\tfix$. In C, the curves are smooth because effects $\A$ are drawn from the distribution $f$. 
Panels A and C in Figure~\ref{fig_SegSites} and Figure~\ref{fig_SegSitesBP} show that as a function of $s\tau$ the rate of increase of the number of segregating sites is nearly independent of $s$, except for very early generations (e.g.~$s\tau<0.1$). This is also consistent with \eqref{L*_approx_init}.
Therefore, if measured in generations, smaller values of $s$ lead to a slower increase of the number of segregating sites, as is supported by the rough approximations in \eqref{L*_approx_init}. The intuitive reason is that invasion becomes less likely.
Panels B and D (but also A and C) show that the stationary value $\bar S^*$ of the number of segregating sites is maximized at an intermediate selection strength. For small values of $s$, $\bar S^*$ is increasing in $s$ because larger $s$ increases the probability of a sweep. 
For sufficiently large values of $s$, $\bar S^*$ drops again because selected mutants spend less time segregating. 

Based on \eqref{L*_approx}, it is straightforward to show that, as a function of $s$, the maximum value of $\bar S^*$ is achieved close to  
\begin{linenomath}\begin{equation}\label{def_smax}
	s_{\text{max}} = \frac{1}{\bar\A\ProdLog[2e^bN^2]}\approx\frac{1}{2\bar\A\ln(N)}\,,
\end{equation}\end{linenomath}
where $b=\tfrac32+2\ga\approx2.65$ for equal effects and $b=\tfrac52$ for exponentially distributed effects. In each case, the maximum values of $\bar S^*$ are very close to $2\Th(2\ln(N) - \ln\ln(N))$. Thus, the maximum possible value of $\bar S^*$ under directional selection is about twice that under neutrality. 
This shows that once a quasi-stationary response to directional selection has been reached, the number of segregating sites depends only weakly on $N$, essentially logarithmically, and exceeds the neutral value by at most a factor of two, whatever the strength of selection. In summary, by far the most important factor determining the number of segregating sites at quasi-stationarity is $\Th$.

Our branching process model leads to an alternative approximation, $\tilde{S}$, for the expected number of segregating sites. If all mutation effects are equal to $\A$, the approximation
\begin{linenomath}\begin{equation} \label{Seg_BP_equal}
	\tilde{S} (\tau ,\A ,s ,N,\Th) \approx \Th  \sum_{j=0}^{[\tilde{\tau}]} \bigl(1- \Plost (j,e^{\sA}) \bigr) 
\end{equation}\end{linenomath}
is derived in Appendix \ref{app:segsites}. Here, $\tilde{\tau}= \min \{\tfix(\stilde,N),\tau \}$. 
For values $s\tau<2$, this approximation is considerably more accurate than the approximation $\bar S$ given above. This is illustrated by Fig.~\ref{fig_SegSitesBP}.A, which is based on a zoomed-in versions of panel A of Fig.~\ref{fig_SegSites}. A disadvantage of the approximation \eqref{Seg_BP_equal} is that the terms $\Plost (j,e^{\sA})$ have to be computed recursively, whereas the above approximations $\bar S$ are based on analytically explicit expressions.

For mutation effects drawn from a distribution $f$, the following approximation of $\EV[S]$ is derived in Appendix \ref{app:segsites} under the assumption $\tau \ll \bartfix(\stilde,N)$:
\begin{linenomath}\begin{align}  \label{Seg_BP_f_smalltau}
	\tilde{S}(\tau ,f ,s, \Th) \approx   \Th   \sum_{j=0}^{\tau} \biggl(1-\int_0^{\infty}  f(\A) \Plost (j,e^{\sA} )  d\A \biggr) \,. 
\end{align}\end{linenomath}
Figure \ref{fig_SegSitesBP}.B demonstrates that, in contrast to $\bar S$ which is based on diffusion approximations, $\tilde{S}$ is highly accurate if $s\tau<2$. However, it is computationally more expensive than $\bar S$. 

Equations \eqref{Seg_BP_equal} and \eqref{Seg_BP_f_smalltau} show that the expected number of segregating sites is proportional to $\Th$. We already know that for large $\tau$, i.e., in the quasi-stationary phase, the number of segregating sites depends at most logarithmically on $N$. The approximation \eqref{Seg_BP_f_smalltau} of $\tilde S$ is independent of $N$, and $\tilde S$ in \eqref{Seg_BP_equal} depends on $N$ only through $\tilde\tau$, which can be assumed to equal $\tau$ if $\tau\ll\bartfix$. According to the approximation \eqref{bartfix_app3} of $\bartfix$, we have $\bartfix > 2.77/(s\bar\A)$ if $Ns\bar\A\ge2$, because $2\ln 4\approx 2.77$. Therefore, we can expect independence of $\tilde S(\tau)$ from $N$ if $\tau\le2.7/(s\bar \A)$. Comparison of the dashed blue curves with the solid blue curves in panels A and C of Fig.~\ref{fig_SegSites} shows that the expected number of segregating sites is essentially independent of $N$ for a much longer time.  Also the dependence of $\tilde S(\tau)$ on $s$ is much weaker than that on $\Th$.

\FloatBarrier

\subsection{Number of segregating sites as an indicator for sweep-like vs.\ shift-like patterns}\label{sec:indicator-sweep-shift}

\begin{figure}[!th]
	\begin{tabular}{l | c c c}
		& Sweep & & Shift \\			
		& $2\Theta=0.1$ & $2\Theta=1$ & $2\Theta=10$ \\		
		\hline \\	
		$s=0.001$ &
				\multirow{7}{*}{\includegraphics[width=0.25\textwidth]{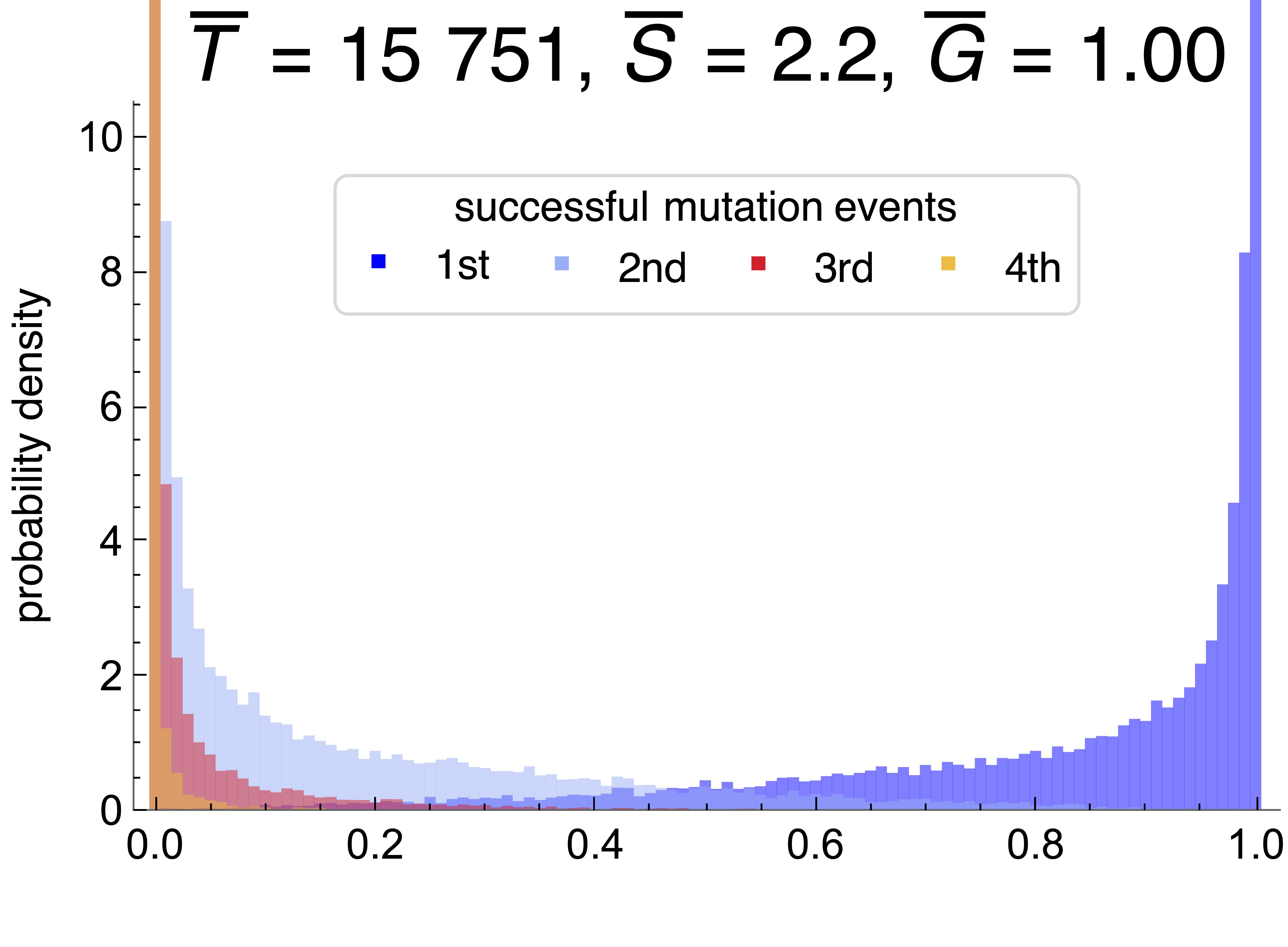}} &
				\multirow{7}{*}{\includegraphics[width=0.25\textwidth]{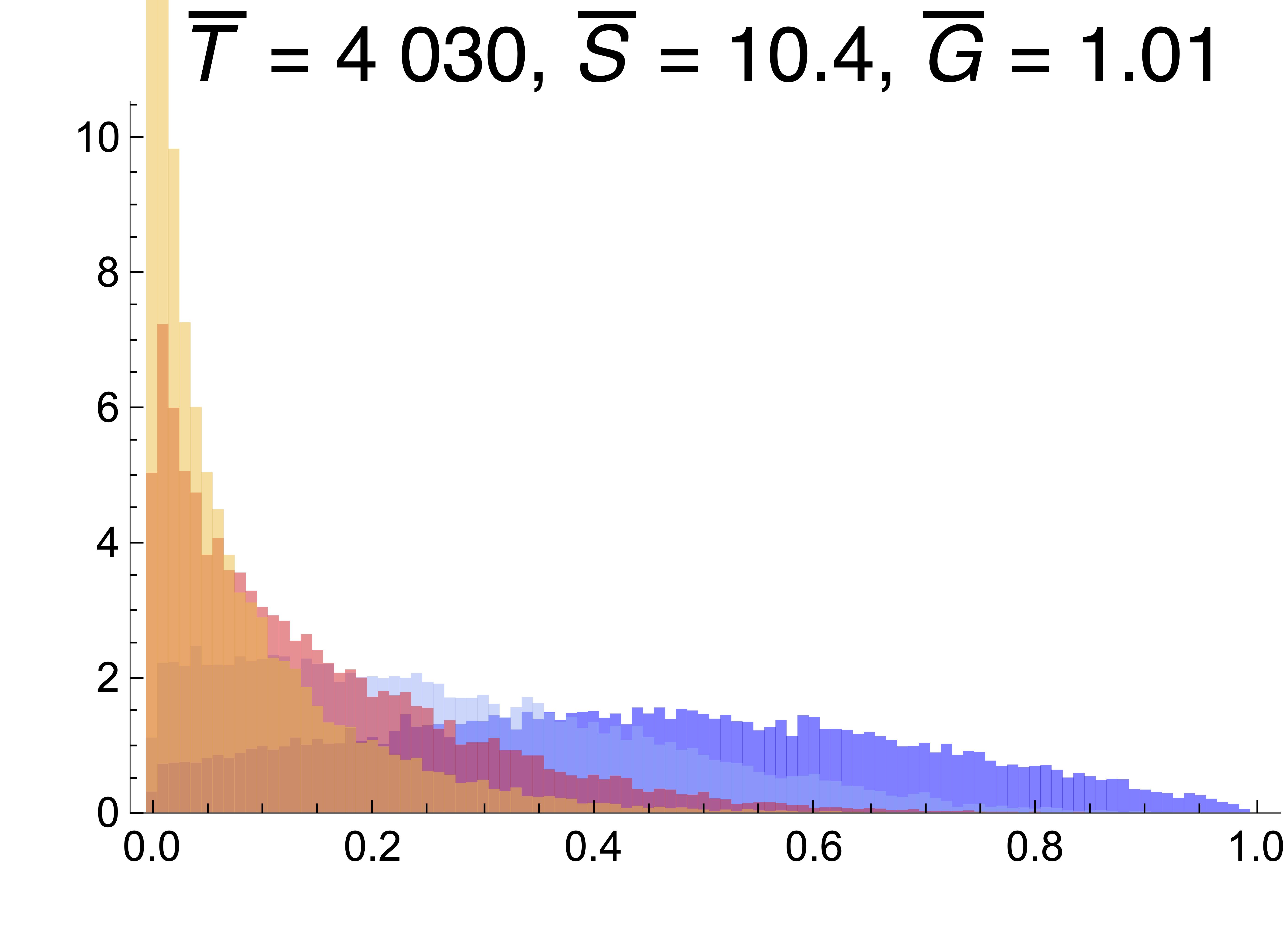}} & 
				\multirow{7}{*}{\includegraphics[width=0.25\textwidth]{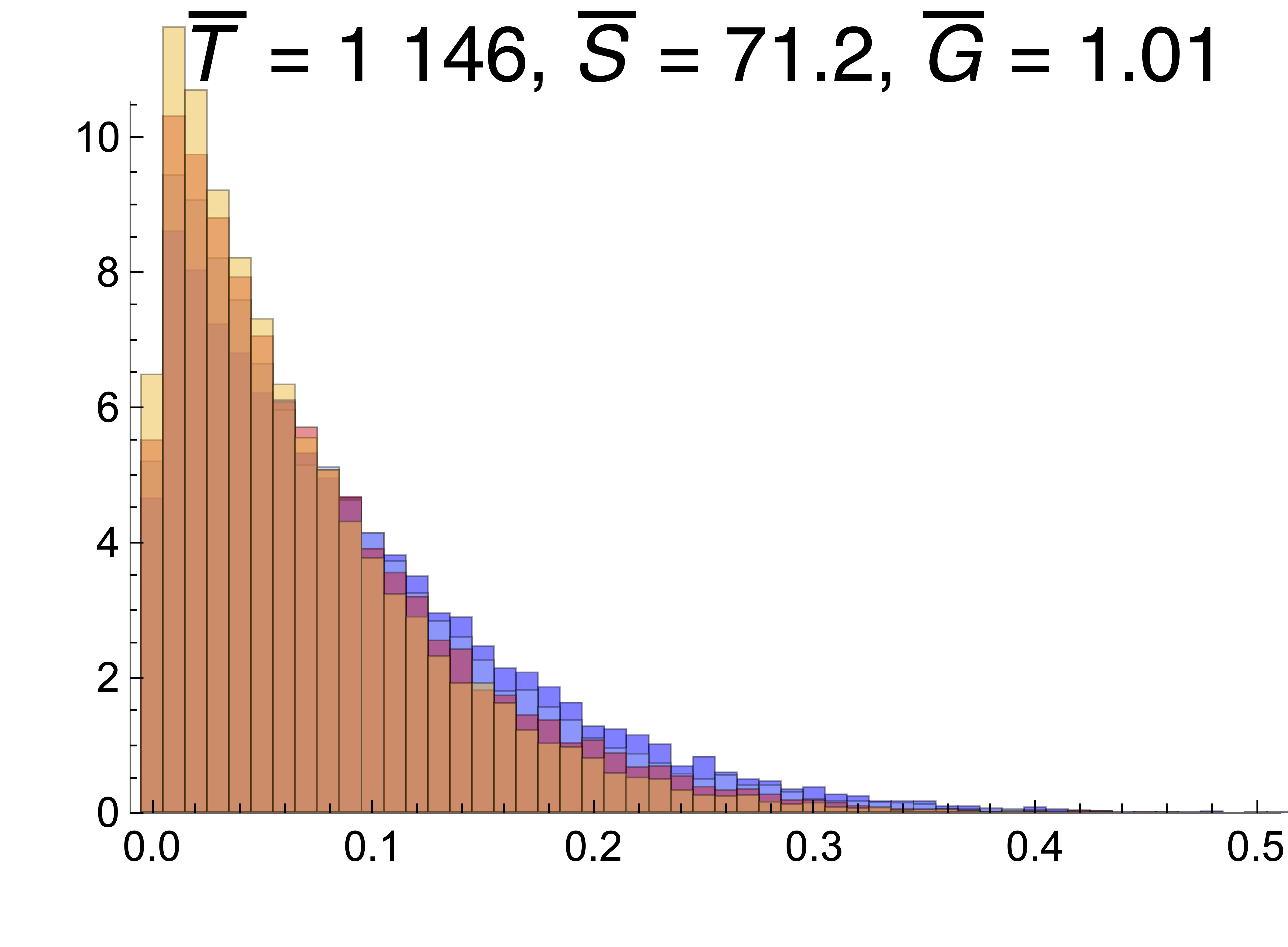}} \\
				$\alpha =1$ \\ \\ \\ \\ \\ \\
		$s=0.1$ &
				\multirow{7}{*}{\includegraphics[width=0.25\textwidth]{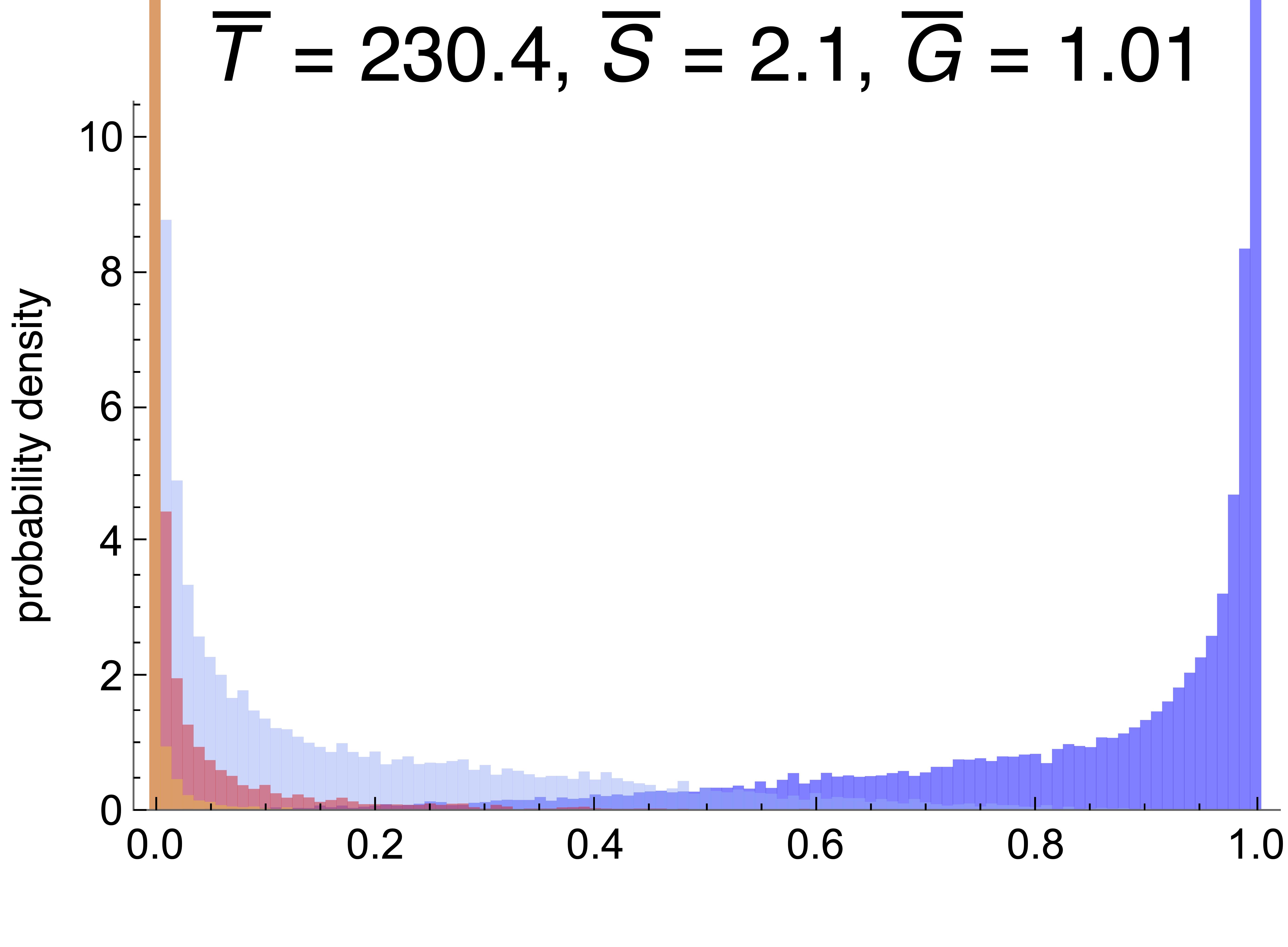}} &
				\multirow{7}{*}{\includegraphics[width=0.25\textwidth]{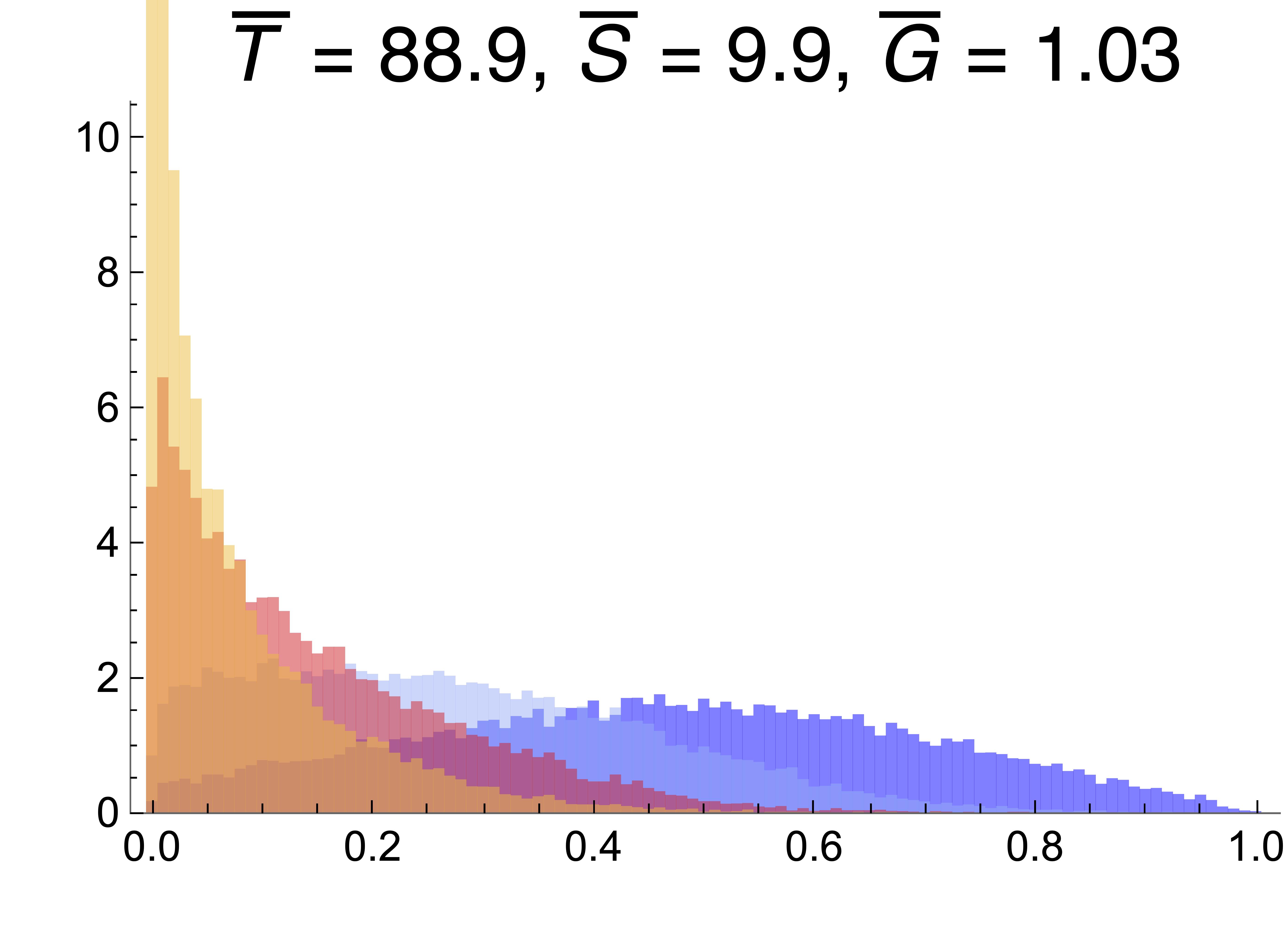}} &
				\multirow{7}{*}{\includegraphics[width=0.25\textwidth]{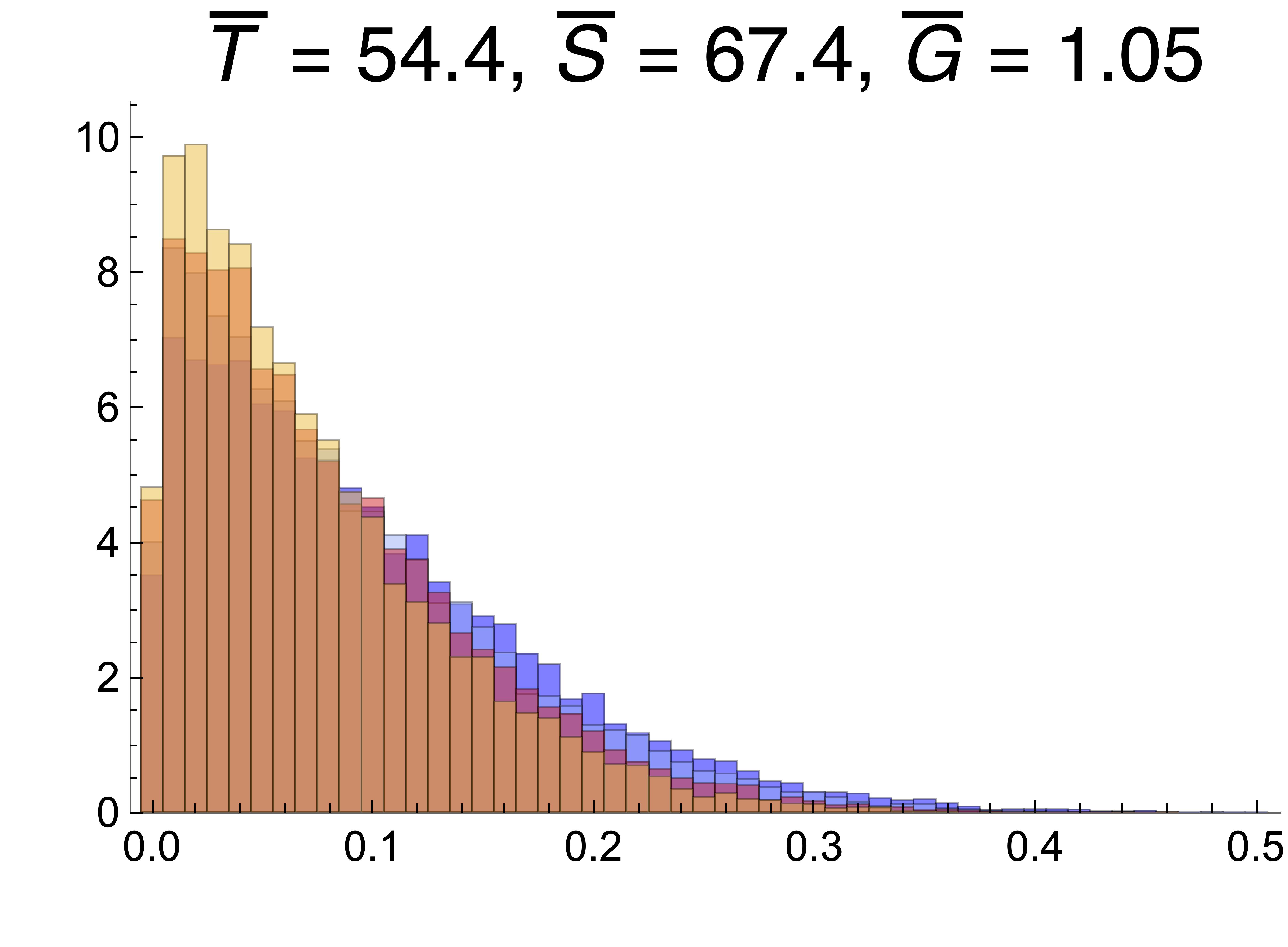}} \\ 	
				$\alpha =1$ \\ \\ \\ \\ \\ \\
		$s=0.1$ &
				\multirow{7}{*}{\includegraphics[width=0.25\textwidth]{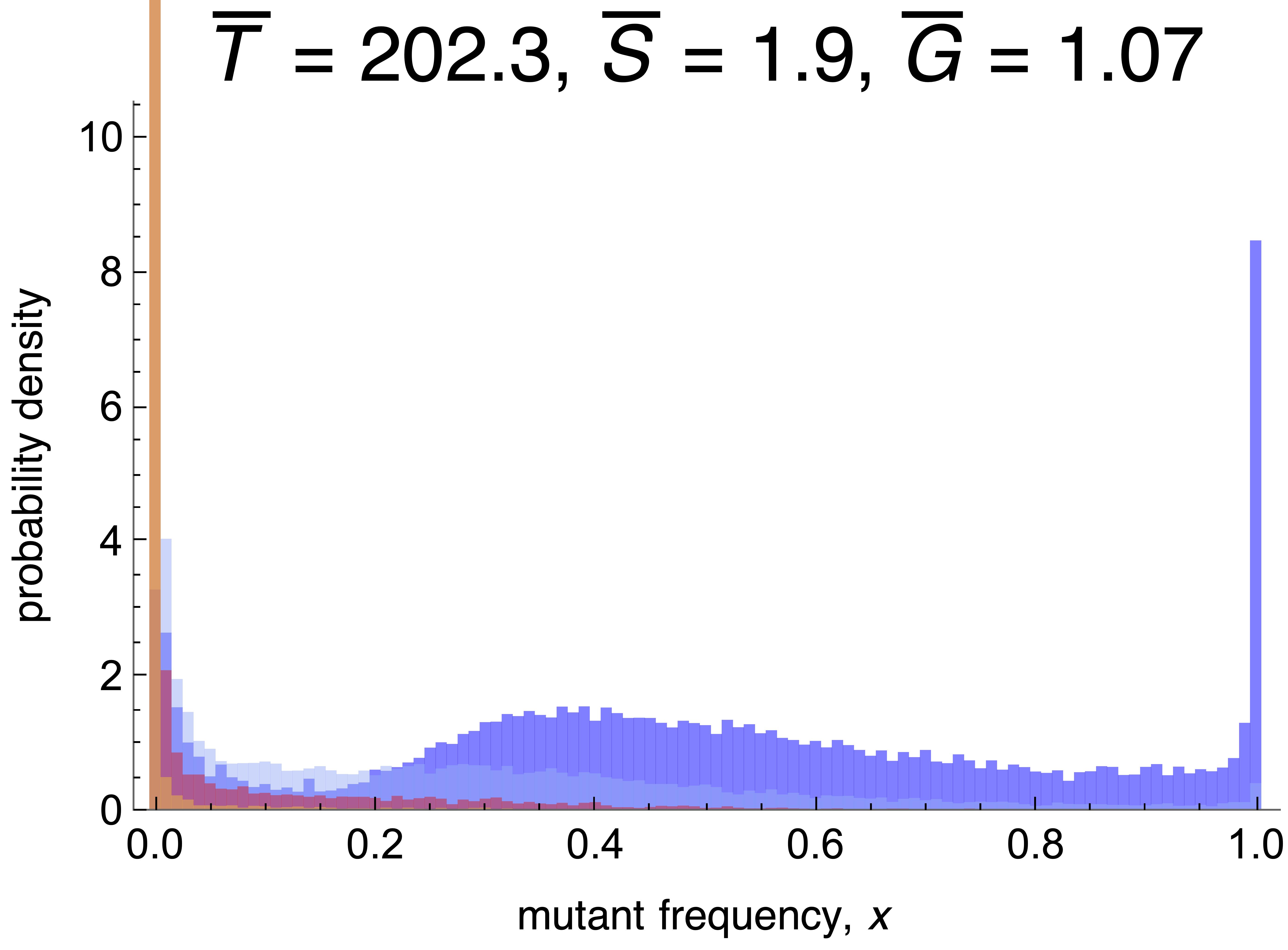}} &
				\multirow{7}{*}{\includegraphics[width=0.25\textwidth]{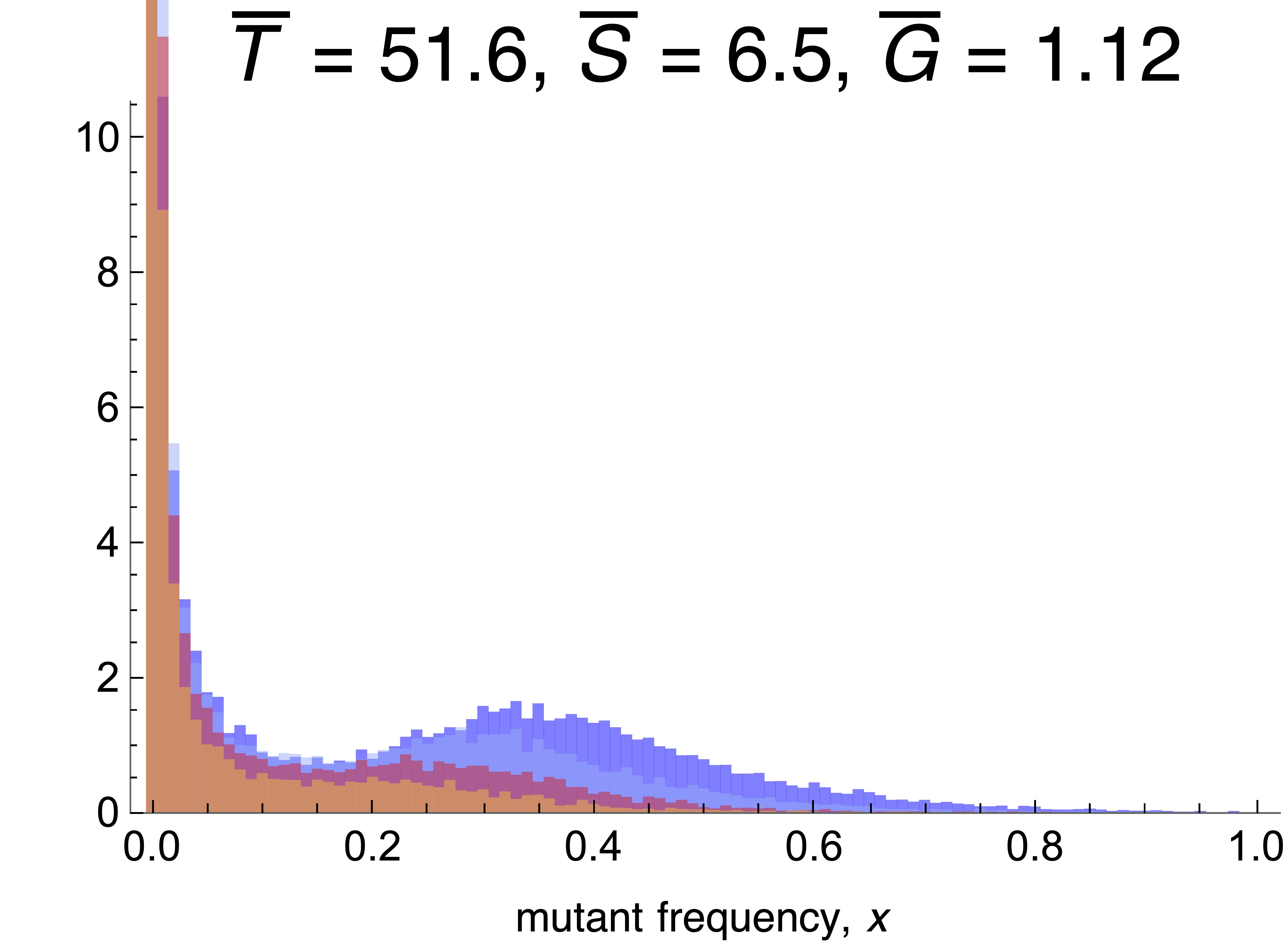}} & 
				\multirow{7}{*}{\includegraphics[width=0.25\textwidth]{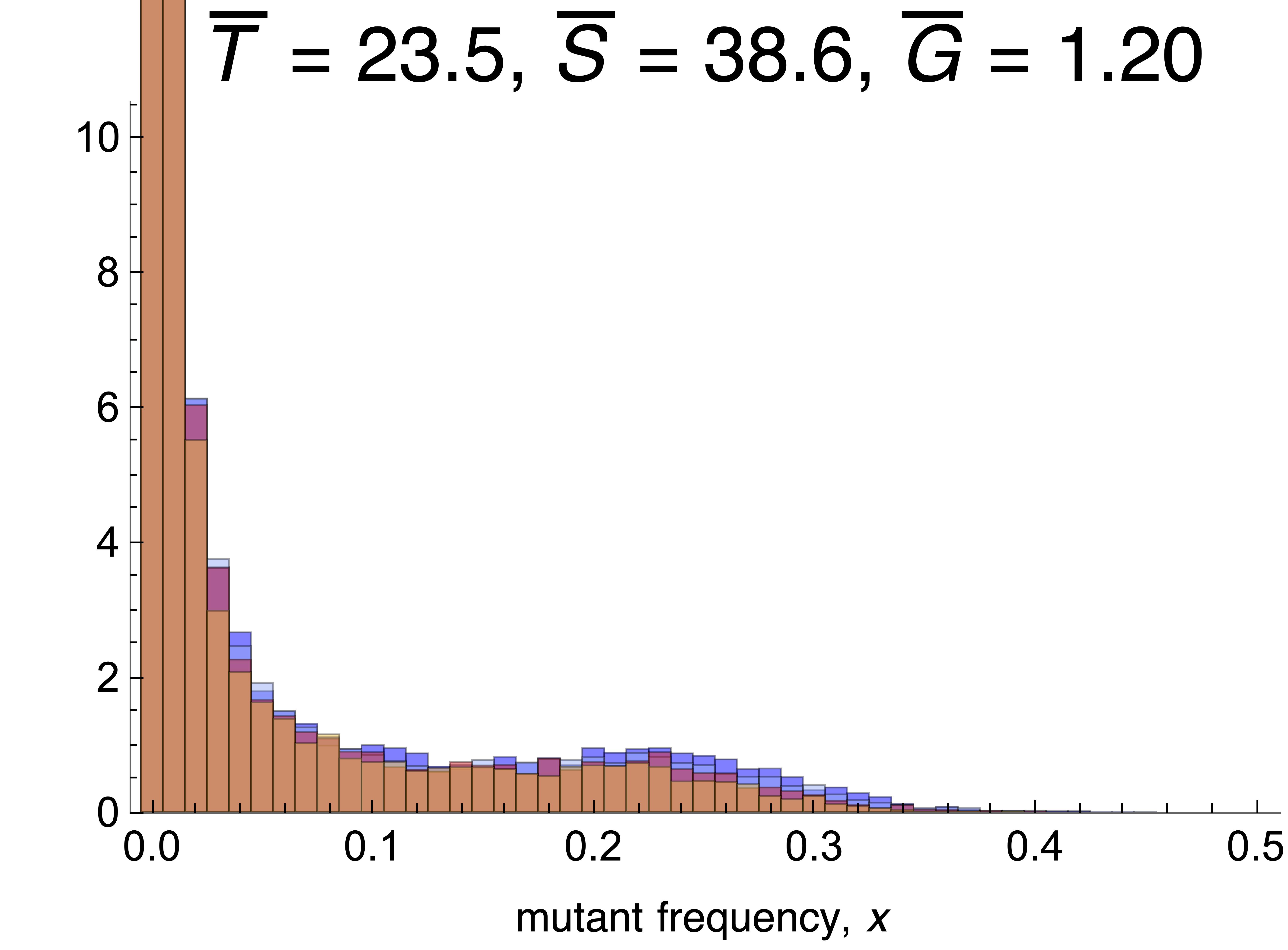}} \\
				$\alpha \sim \text{Exp}[1]$ \\ \\ \\ \\ \\ \\
	\end{tabular}
	\caption[Sweep vs.\ shift]{Histograms of the frequency $x$ of the first four successful mutants at the first generation $T$ in which the mean phenotype $\bar G$ in a single population reaches or exceeds the value $1$. The averages of the stopping time $T$, of the number of segregating sites at $T$, and of the mean phenotypic values at $T$, denoted by $\bar T$, $\bar S (=\bar{S}_T)$, and $\bar G (=\bar{G}_T)$, are given in each panel. Histograms and averages are based on $20\,000$ replicate runs of the Wright-Fisher model.
The dark blue color is for the first successful mutation, light blue for the second, red for the third, and orange for the fourth successful mutation. The selection coefficients $s$ and the mutation effects $\A$ are indicated for each row. Columns show results for different values of $\Th$. The population size is $N = 10^4$ in all cases. One unit on the vertical axis corresponds to $200$ runs. The expected times to fixation for rows 1, 2, and 3 are $\tfix\approx7\,039$, $\tfix\approx156$, and $\bartfix\approx153$, respectively. 
The corresponding scaled stationary values of the number of segregating sites, $S^*/(2\Th)$, are approximately $13.7$, $16.5$, and $16.0$ (calculated from eq.~\eqref{L*_approx}). Thus, in the first column $S$ has reached its stationary value in rows 1 and 2, and $60\%$ of it in row 3.  The distributions of the stopping times $T$ corresponding to the above panels are shown in Fig.~\ref{fig:sweep_shift_tau}. 
}
\label{fig:sweep_shift_new}
\end{figure}

Motivated by the work of \cite{Hoellinger2019}, we explored the polygenic pattern of adaptation at the loci contributing to the trait at the generation $T$ when the mean phenotype $\bar G$ in a population first reaches or exceeds a given value, which for Figure~\ref{fig:sweep_shift_new} we chose to be 1 (in units of mutation effects $\A$). For various parameter combinations and for equal and exponentially distributed mutation effects, Figure~\ref{fig:sweep_shift_new} displays the frequency distributions of the first four \textit{successful} mutants at this generation $T$.
By successful mutant we mean that at the stopping time $T$, we condition on the sites at which mutants are present (segregating or fixed), and we number them from 1 to 4, depending on the order in which these mutations occurred. Thus, the mutation at site 1 is the oldest in the population. The sites segregating the first four successful mutations are only a small subset of sites at which mutations have occurred because their majority has been lost. The histograms are obtained from Wright-Fisher simulations. We computed the average stopping time, $\bar T$, the average number of segregating sites at time $T$, $\bar{S}_T$, and the average of mean phenotypic values in generation $T$, $\bar{G}_T$. These values are given in each panel (with subscripts omitted for visibility reasons). Because time is discrete, $\bar{G}_T \geq 1$. For large selection coefficients (and large mutation effects) the per-generation response tends to be large once several mutations are segregating. Therefore, $\bar{G}_T$ can be noticeably larger than 1. Indeed, the distribution of $\bar G$ (under the stopping condition $\bar G =1$) has a sharp lower bound at 1 and may have substantial skew to the right, especially for exponentially distributed effects (results not shown).

Figure~\ref{fig:sweep_shift_new} shows clear sweep patterns in the panels with $2\Th=0.1$ (where $\bar{S}_T <2.5$) and distinctive small-shift patterns at many sites if $2\Th=10$ (where $\bar{S}_T >35$). The patterns occurring for $2\Th=1$ suggest a series of partially overlapping sweeps. The main effect of varying the selection strength $s$ is on the time $T$ to reach (or exceed) $\bar G=1$. Successive sweeps are indeed expected if $2\Th=0.1$ because the average waiting time between successful mutation events is approximately $1/(2s\Th)$. If $s=0.001$, this waiting time is $10\,000$ and exceeds the expected fixation time of roughly $7\,000$ generations. If $s=0.1$, the waiting time is 100 generations and falls short of the expected fixation time, which is about $156$ generations by \eqref{bartfix_app3}. In addition, we observe that for small $\Th$ the first successful mutant corresponds mostly to the site with the highest mutant frequency at the time of stopping, whereas for large $\Th$ the distributions of the first four successful mutations are hardly distinguishable.

In the panels with $2\Th=0.1$, the expected stopping times $\bar{T}$ exceed their corresponding expected fixation times (given in the caption of Fig.~\ref{fig:sweep_shift_new}) considerably. Hence, most of these populations have already reached the quasi-stationary phase. Somewhat unexpectedly, the observed average numbers of segregating sites (from top to bottom: $2.2$, $2.1$, $1.9$) exceed their respective (approximate) stationary values $\bar S^*$ ($1.37$, $1.65$, $1.60$) considerably. This observation seems to be at variance with the results in panels B and D of Figure~\ref{fig_SegSites}, which show that $\bar S^*$ is a very accurate approximation of the numerically determined $\EV[S]$. The reason for this discrepancy is the different stopping condition, more precisely the fact that the distribution of stopping times $T$ is very wide and strongly skewed to the right if $\Th$ is small (Fig.~\ref{fig:sweep_shift_tau}). A slowly evolving population may carry many more mutations once they reach $\bar G=1$ than the average or a fast evolving population because it has to wait long for the first successful mutant(s). As a result, the stopping condition $\bar G=1$ inflates the number of segregating sites compared to a process that is stopped at a given generation. 

For $2\Th\ge1$ our branching-process approximations for $\tilde S$ in \eqref{Seg_BP_equal} and \eqref{Seg_BP_f_smalltau} predict the observed average $\bar{S}_T$ very well if $\tilde S$ is evaluated at $\bar T$. Indeed, very accurate approximations of the observed $\bar T$ are obtained for equal effects by solving the equation $\bar G(t)=1$ using \eqref{barG_init_equal}. The reason is that the distribution of $T$ becomes increasingly symmetric and less variable as $\Th$ increases. It is nearly perfectly symmetric if $2\Th=10$ (Fig.~\ref{fig:sweep_shift_tau}). For exponentially distributed effects, when the stopping time distribution is more skewed, an accurate approximation of $\bar T$ is obtained from \eqref{barG_init} by solving $\bar G(t)=1$ only if $2\Th\ge10$ (results not shown). Accordingly, for such large values of $\Th$, the Wright-Fisher simulations stopped at the previously determined value $\bar T$ yield distributions very similar to those obtained by stopping when $\bar G=1$ has been reached (results not shown).

Figure~\ref{fig:sweep_shift_new} confirms the above discussed finding that for an exponential distribution of mutation effects, the response of the mean is faster than for equal effects also under the current stopping condition. As a consequence, the number of segregating sites when $\bar G=1$ is reached is smaller, and considerably so if $\bar G=1$ is reached well before the quasi-stationary phase is entered, as is the case of $2\Th\ge1$. Unsurprisingly, the distribution of stopping times $T$ is much wider than for equal effects (Fig.~\ref{fig:sweep_shift_tau}) because mutations of larger (beneficial) effect have a higher fixation probability and lead to faster sweeps, especially for strong selection.  Mutation distributions with a heavier tail than the exponential may intensify this effect. 

\FloatBarrier

\section{Discussion}\label{sec:discussion}

The goal of our study has been to provide essential theory for answering the following questions: Which are the main evolutionary and genetic factors determining the rate and pattern of polygenic adaptation? How do these factors determine the response to selection, both on the phenotypic and on the genetic level? When is the selective response mainly due to sweeps at few loci and when is it due to small frequency shifts at many loci? 

For this purpose, we developed and analysed a simple population-genetic model, starting from a branching process approximation to describe the spreading of a beneficial mutant in a Wright-Fisher population of size $N$. Our basic result is an accurate, analytically tractable, and computationally efficient approximation for the time dependence of the mutant-frequency distribution. Extension to an infinite sites model and the assumption of an additive quantitative trait under exponential directional selection enabled us to derive explicit expressions for the evolution of the expected mean phenotype and the expected phenotypic variance, as well as for the expected number of segregating sites underlying the adaptive response. As demonstrated by comparison with extensive simulations, they provide accurate approximations for the corresponding dynamics resulting from a Wright-Fisher model. Our theory focuses on the response from new mutations and assumes that selection is stronger than random genetic drift ($Ns>1$). In the sequel, and structured according to the main sections (3, 4, and 5) of the paper, we discuss the most relevant findings and their implications for answering the last question posed above. Finally, we provide brief conclusions and discuss the limitations caused by our model assumptions (Sect.~\ref{Disc:Outlook}).
\subsection{Branching process approximation for the time dependence of the mutant frequency distribution}

The spread of a new mutation in a population as a function of time can be described by the transition matrix of an appropriate Wright-Fisher model or the transition density of the approximating diffusion. Although diffusion theory leads to quite simple expressions for many important quantities, explicit and analytically tractable time-dependent results, tracing for instance the distribution of allele frequencies, seem to be out of reach \citep[see][for a semi-explicit approach]{Steinrucken_etal2013}. 

We combined two methods to derive an explicit and accurate time-dependent approximation for the mutant's density in any generation $n$: a branching process approach capturing the stochastic effects and the deterministic logistic growth model. We developed this approach quite generally with the help of a slightly supercritical Galton-Watson process that allows for quite general offspring distributions. 

The approximation $g_a(x)$ in \eqref{g(p)} for the mutant frequency density depends only on the compound parameter $a$, which in turn depends on the generation number $n$, the mean $\sigma=e^s$ of the offspring distribution, the extinction probability $\Plost(n,\sigma)$ until generation $n$ in the branching process, and the population size $N$. Its shape is displayed in Figure~\ref{fig_g3D} and compared with Wright-Fisher simulations in Figures \ref{fig_g(p,a)} and \ref{fig_g(p,a)2}. It is highly accurate in the initial phase, and requires only $Ns\ge1$. If $Ns\ge10$ then it is accurate until about $\frac{1}{s}\ln(2Ns)$, which is approximately one half of the expected fixation time in the Wright-Fisher model. If $Ns\ge100$, then it remains accurate for even longer and even if $x$ is relatively close to 1. Because, in contrast to previous related investigations \citep{Desai2007, Uecker2011, Martin2015}, we condition on non-extinction until generation $n$, our approximation is more accurate than those obtained there, especially in the early initial phase (cf.~Figure~\ref{figpsi}, which shows the density $\psi_n$ in the branching process, from which $g_a$ is obtained by a proper  transformation). However, by relying on a branching process to capture the inherent stochasticity, our model underestimates the effect of genetic drift near fixation, in particular if $Ns$ is not much larger than 1.
Remarkably, the approximations for the mean and variance of this distribution are accurate in a much wider parameter range. This may be explained by the fact that mutants which are rare or near fixation hardly contribute to the variance and to the response of the mean. These approximations clearly display the well known fact that, conditioned on non-extinction (in our case, until the time of observation), the mutant frequency grows faster than predicted by the corresponding deterministic model, particularly when selection is weak (see Fig.~\ref{fig_ContrMeanVar} and the discussion in Sect.~\ref{sec:dyn_allele_freq}).
 
Finally, these results are extended to the infinite sites model, in which the number of new (beneficial) mutants emerging in the population per generation is Poisson distributed with mean $\Th=NU$, and every mutation occurs at a new site. Here, $U$ is the expected number of beneficial mutations per individual per generation affecting the trait. Sites are assumed to be unlinked (and in linkage equilibrium). The monomorphic initial population started to evolve at $\tau=0$. Equation \eqref{hi(p)} provides the probability density $h^{(i)}(x)=h_{\tau,\sigma_i,N,\Th}^{(i)}(x)$ of the $i$th mutation (which occurred at the random time $\tau_i$ at the $i$th site), conditioned on non-extinction until time $\tau$. Mutations of unequal effects are admitted, as signified by $\si_i$. The integration is necessary because the waiting time distribution for the $i$th mutation event (an Erlang distribution) needs to be taken into account. Its cousin, the unconditioned distribution $\tilde X_\tau^{(i)}$ with absolutely continuous part $\tilde h_\tau^{(i)}(x)$ is defined in \eqref{tilde{h}^{(i)}} and \eqref{tildeX^{(i)}}. It forms the basis for the applications to the quantitative genetic model.
\subsection{Evolution of the mean phenotype and the genetic variance}

The approximations \eqref{g(p)} and \eqref{tildeX^{(i)}} for the time dependence of the density $g_{a(\tau)}$ and of the distribution of $\tilde X_\tau^{(i)}$ allow us to obtain highly accurate approximations for the evolution of the expected mean ($\bar G$) and variance ($V_G$) of an additive trait subject to exponential directional selection in a finite population of size $N$. We assume that potentially infinitely many, stochastically independent, loci (sites) can contribute additively to the trait. This assumption of global linkage equilibrium will be a good approximation if sites are physically unlinked because multiplicative fitnesses, as caused by exponential directional selection, do not generate linkage disequilibrium \citep[][p.\ 79]{Buerger2000}. 

The expressions \eqref{barGtau} for the expected mean $\bar G(\tau)$ and \eqref{VGtau} for the expected variance $V_G(\tau)$ of the trait are exact within our model based on the quasi-deterministic branching process approach, i.e., assuming that the density $g_a$ in \eqref{g(p)} is exact. Comparison with the results from Wright-Fisher simulations shows that they are astonishingly accurate for the whole evolutionary process, i.e., from the initial phase until the quasi-stationary response has been achieved (Figure \ref{fig_MeanVarEqu}). These expressions require integration with respect to two parameters: one over the time span until the time $\tau$ of interest, the other with respect to the distribution $f(\A)$ of mutation effects. The former is necessary because the extinction probability is time dependent, at least in the initial phase. The latter integration is unavoidable unless all mutation effects are equal, when accurate approximations involve only the computation of the finite sums given in Remark \ref{rem: mutation_discrete_time}. We note that these approximations hold for very general offspring distributions, so that they can cover cases where the effective population size $N_e$ differs from the actual population size $N$. The offspring distribution enters these equations through the extinction probabilities $\Plost$. 

The general equations \eqref{barGtau} and \eqref{VGtau} do not provide much direct insight into the evolution of $\bar G$ and $V_G$, except that the population-wide mutation rate $\Th$ enters exclusively as a multiplicative factor. Their numerical evaluation is straightforward but not very fast. However, they provide the basis for deriving very simple and highly accurate analytical approximations for the initial phase and for the quasi-stationary phase, when emergence and fixation of new mutations balance and the genetic variance has reached a stationary value.

\subsubsection{The quasi-stationary phase}

Figure \ref{fig_MeanVarEqu} documents that the quasi-stationary phase is reached once time exceeds the expected fixation time $\bartfix$, which is defined in \eqref{bartfix} and has the approximation \eqref{bartfix_app3}. During the quasi-stationary phase the expected per-generation change in the mean, $\De \bar G^*$, and the expected variance, $V_G^*$, are constant. They are given by \eqref{barG*_prop} and \eqref{VG*_prop}, respectively, and invoke the generally unavoidable integration over the distribution of mutation effects. For equal effects, the much simpler expressions \eqref{DebarG*_equal} and \eqref{VG*_equal} are obtained. By recalling $\Th=NU$, equation \eqref{VG*app} reveals the following general and fundamental relation between the change of the mean and the variance:
\begin{linenomath}\begin{equation}\label{DebarG*VG*}
	\De \bar G^* \approx s V_G^* + U\bar\A \,.
\end{equation}\end{linenomath}
If $U\bar\A\approx0$, this is a special case of \citetalias{Lande1976}, which was derived under the assumption of a normal distribution of breeding values. 
An approximation of $\bar G(\tau)$, applicable already during the approach to the quasi-stationary phase, is provided in \eqref{barG_new}, and shown as dashed curves in Figs.~\ref{fig_MeanVarEqu} A,C.

For a Poisson offspring distribution, an exponential distribution $f$ of mutation effects, and weak selection, the response of the mean has the simple approximation $\De\bar G^* \approx 4\Th s\bar\A^2(1-\tfrac52s\bar\A)$. If the mutation effects are equal, the response is about one half of this value (Corollary \ref{cor:equilibrium_simp}). For large population sizes, $N\ge10^4$, the simple approximations in Corollary \ref{cor:equilibrium_simp}, both for $\De\bar G^*$ and $V_G^*$, are highly accurate for selection coefficients $s$ ranging over more than three orders of magnitude (Figure \ref{fig_VarSEqu}).  The relation \eqref{DebarG*VG*} is violated when the population size is so small that $Ns < 2$, as can be observed in Fig.~\ref{fig_VarSEqu} by noting the deviation of the open symbols from the black dashed curves. The reason is that for such small values of $Ns$ random genetic drift becomes an important force and the fixation probability of very slightly beneficial mutations is higher than predicted by the branching process approximation, i.e., in this case the fixation probability under neutrality, $1/N$, exceeds Haldane's approximation, $2s$, for an advantageous mutation.

Formally, the per-generation change of the mean phenotype at stationarity in \eqref{barG*_prop} is equivalent to \citetalias{Hill1982a} approximation for the asymptotic response of a quantitative trait coming from fixation of new mutations. In our terminology and for a haploid population, his general expression [3] becomes 
$\De \bar G = \Th (N_e/N) \int_{-\infty}^\infty \A \Pfix(s\A) f(\A)\, d\A$, where his mutation distribution is not necessarily confined to positive values. Here, $\Pfix$ is the diffusion approximation \eqref{PfixDif} for the fixation probability in the Wright-Fisher model and thus nearly identical to $\Psur$ because we assume $Ns>1$. Thus, we recover classical results, but from a different perspective. It seems somewhat surprising that the branching process approach, which is best suited to study the initial fate of a mutation subject to selection, yields highly accurate approximations for the long-term response of a quantitative trait. For a more general discussion of the long-term response to various forms of directional selection, see \citet[][pp.\ 331--338]{Buerger2000}.

\subsubsection{The transient and the initial phase}

In Proposition~\ref{prop:Dynamics}, we derived the approximations \eqref{barG_init} for $\bar G(\tau)$ and \eqref{VG_init} for $V_G(\tau)$. They can be evaluated much more quickly than \eqref{barGtau} and \eqref{VGtau}, and they are highly accurate for all times, except for a very early phase (e.g., Fig.~\ref{fig_VarInitial_comparison}). They perform particularly well when $\tau > 1/\Th + T_\ep$, where $T_\ep$ is the time needed such that the probability of non-extinction by generation $T_\ep$ has declined to $(1+\ep)\Psur$ (eq.~\eqref{T_ep_gen} and Section \ref{app_frac_lin_compare}). Above this lower bound the (iterative) computation of the time-dependent extinction probabilities $\Plost(t)$ in the general equations \eqref{barGtau} and \eqref{VGtau}, hence the integration over the time, is not needed. Thus, for equal mutation effects, no integration at all is required to evaluate $\bar G(\tau)$ and $V_G(\tau)$, which are then given by \eqref{barG_init_equal} and \eqref{VG_init_equal}, respectively.

For the very early phase when $\tau\lessapprox 1/(4s\bar\A)$, the utterly simple approximations in Proposition \ref{prop:Initial2} are obtained for exponentially distributed mutation effects. The series expansions in \eqref{barG_smalltau_series} and \eqref{VG_smalltau_series} are accurate for a somewhat longer initial phase, even beyond $\tau\approx 1/(s\bar\A)$ provided $s$ is sufficiently small (Figure \ref{fig_MeanVarInitial1}). For mutation effects equal to $\A$, the mean phenotype grows approximately exponentially, i.e., $\bar G(\tau) \approx \frac{\Th}{Ns}(e^{s\A\tau}-1)$ (Remark~\ref{Gbar_smalltau_equal}). This yields an accurate approximation if $\tau\lessapprox 1/(s\bar\A)$ for $Ns=10$ and for a longer time span as $Ns$ increases while $s$ decreases (Figure \ref{fig_MeanVarInitial1}). During this phase, variance and mean are related through $V_G\approx \A\bar G$ and the relation $\De\bar G\approx s V_G$ is violated. For an exponential distribution of mutation effects, the initial increase of the mean and the variance is considerably faster than for equal mutation effects ($=\barA$), indeed we have $\bar G(\tau)\approx\frac{\Th}{N}\bar\A\tau\Bigl(1 + s\barA\tau + (s\barA\tau)^2\Bigr)$ (Proposition \ref{prop:Initial2}). This yields an increase of the mean that is slightly faster than that for equal effects with twice the selection intensity, $2s\barA$. The reason is that with an exponential distribution of effects, mutations of large effect have a much higher probability of becoming established (and eventually fixed) than mutations of small effect. In fact, the average effect of mutations drawn from an exponential distribution with mean $\bar\A$, conditioned on eventual fixation, is approximately $2\barA$.

\subsubsection{Comparison of the response from new mutations with that predicted by the breeder's equation}\label{Disc:breedereq}

We developed our model for an initially monomorphic population and adaptation from \textit{de novo} mutations. Here, we compare our results on the response of the mean phenotype with corresponding results derived from classical quantitative genetic models (thus, essentially statistical models) and provide a rough guide for which parameter regions they conform.

In the classical additive model of quantitative genetics, environmental effects are not ignored, and it is based on a given phenotypic variance. More precisely, the trait value is $P=G+E$, where $G$ is the (additive) genetic contribution and $E$ the environmental contribution. The simplest version assumes that environmental effects have a distribution with mean zero and variance $\si_E^2$ and are independent of the genetic effects, so that $\De\bar P=\De\bar G$ and $\si_P^2=\si_G^2+\si_E^2$. The classical breeder's equation, which reads 
\begin{linenomath}\begin{equation}\label{breeders_equation}
	\De\bar P = h^2 \De_s \bar P = i h^2 \si_P\,,
\end{equation}\end{linenomath}
where $h^2 =\si_G^2/\si_P^2$ is the (narrow sense) heritability of the trait (and $\si_G^2$ the additive genetic variance, which corresponds to our $V_G$), $\De_s \bar P$ is the selection differential, i.e., the difference in mean phenotype before and after selection, and $i=\De_s \bar P/\si_P$ is the so-called selection intensity, or the standardized selection differential \citep[e.g.,][pp.~317-318]{Buerger2000}. Under exponential directional selection one obtains $i=s\si_P$ \citep[][p.~320]{Buerger2000}, so that \eqref{breeders_equation} becomes
\begin{linenomath}\begin{equation}\label{breeders_equation_exponential}
	\De\bar P = h^2 s \si_P^2\,.
\end{equation}\end{linenomath}

\paragraph{Stationary phase}
Our equation \eqref{DebarG*} informs us that, once stationarity is achieved, the response from new mutations is approximately
\begin{linenomath}\begin{equation}\label{DebarG*_simp}
	\De\bar G^* \approx \rho \Th s \A_2(f) = \rho \Th s {\bar\A}^2(1+{\rm CV}_\A^2)\,,
\end{equation}\end{linenomath}
where we have used $\Psur(e^{s\A})\approx \rho s\A$ (Remark \ref{Psur=csalpha}). If the offspring distribution is Poisson, then $\rho=2$ and $N_e\approx N$. Here, ${\rm CV}_\A^2$ is the squared coefficient of variation of the distribution of mutation effects. For equal effects $\rm{CV}_\A^2=0$, for exponentially distributed effects ${\rm CV}_\A^2=1$, for Gamma-distributed effects with density proportional to $\A^{\beta-1}e^{-\A}$, ${\rm CV}_\A^2=1/\be$, and for effects drawn from the truncated normal distribution in Remark~\ref{rem:truncatedGaussian}, ${\rm CV}_\A^2=\tfrac12\pi-1$. As already noted above, an equation equivalent to \eqref{DebarG*_simp} was derived and discussed by \cite{Hill1982a,Hill1982b}. 

Equation \eqref{DebarG*_simp} demonstrates that the response of the mean depends not only on the selection intensity, the mutation rate and the average effect of beneficial mutations, but also on the shape of the mutation distribution, in particular its coefficient of variation. Therefore, distributions with a large coefficient of variation of beneficial effects, such as Gamma-distributions with small $\be$, lead to a faster response. This effect is already clearly visible by comparing panels A and C in Fig.~\ref{fig_MeanVarEqu}, where in A we have ${\rm CV}_\A^2=0$ and in $B$, ${\rm CV}_\A^2=1$; see also panel A in Fig.~\ref{fig_nor}, where ${\rm CV}_\A^2=\tfrac12\pi-1\approx0.507$.

Assuming a Poisson offspring distribution and exponentially distributed mutation effects with mean $\bar\A$, the selection response predicted by \eqref{DebarG*VG*} equals that predicted by \eqref{breeders_equation_exponential} if 
\begin{linenomath}\begin{equation}\label{compare_response1}
	4\Th {\bar\A}^2 = h^2 \si_P^2\,,
\end{equation}\end{linenomath}
where, in the stationary phase and by \eqref{VG*Ns}, $\si_P^2 = V_G^*+\si_E^2 \approx 4\Th\bar\A^2 +\si_E^2$. Inferences from empirical estimates suggest that $\bar\A$ may often be not only of the same order of magnitude as $\si_P$, but actually quite similar to it \citep[][p.~1019]{WalshLynch2018}. In fact, they suggest $\si_\A^2\approx\si_E^2$, where $\si_\A^2$ is the variance of a symmetric distribution of mutation effects with mean zero and $\si_E^2$ is the environmental variance. Under our assumption of exponentially distributed (positive) effects, we obtain $\si_\A^2= 2 \bar\A^2$, which yields $\si_P^2 = 4\Th\bar\A^2+2 \bar\A^2$. Therefore, \eqref{compare_response1} is satisfied if $h^2 = 2\Th/(2\Th+1)$.
Heritabilities are, by definition, between 0 and 1, but are often in the range between $0.1$ and $0.7$ (e.g., Table 10.1 in \citealp{FalconerMackay1996}; Chap.~7 in \citealp{LynchWalsh1998}), for morphological traits more often in the lower half of this range. Thus a value of $h^2\le\tfrac12$ is consistent with $\Th\le\tfrac12$, and a value of $h^2\le\tfrac23$ with $\Th\le1$.

\paragraph{Initial phase}
A comparison of the response of the mean phenotype from standing variation and new mutations during the initial phase of adaptation is delicate. Whereas for the response from standing variation \eqref{breeders_equation_exponential} can still be used, the response from new mutations increases nonlinearly (see Fig.~\ref{fig_MeanVarInitial1}). In particular, equation \eqref{barG_init} as well as its simplification \eqref{barG_init_equal} for equal effects, are too complicated to derive analytical comparisons between the two scenarios. Although numerical solution is simple and computationally fast, there are many scenarios to be compared with of how standing variation is maintained and how it is structured \citep[e.g.,][and references therein]{Chevin2008, deVladar2014, StephanJohn2020-rev, Hayward2021}. Thus, a detailed comparison has to be postponed to a future treatment.
\subsection{Number of segregating sites as an indicator for sweep-like vs.\ shift-like patterns}\label{sect:disc_seg_sites}

Taking up the arguments in Sect.~\ref{sec:sweepshift}, we propose to use the number of segregating sites as a main indicator for the distinction between the scenarios of adaptation being mainly due to selective sweeps or mainly due to small frequency shifts at many loci.  This is supported by our theory in Sect.~\ref{sec:segsites} showing that once a stationary response to directional selection has been achieved, the expected number of segregating sites, $\bar S^*$, is proportional to $\Th$, depends logarithmically on $N$, and exceeds the neutral expectation of $2\Th\ln N$, but never by more than a factor of two if $s$ is varied between 0 and $0.5$ (see eq.~\eqref{L*_approx} and Fig.~\ref{fig_SegSites}). As a function of $s$, the maximum value of $\bar S^*$ is achieved if $\bar\A s\approx 1/(2\ln N)$.

For the initial phase of adaptation, the branching process approximations $\tilde S(\tau)$ in \eqref{Seg_BP_equal} and \eqref{Seg_BP_f_smalltau} are much more accurate than the simple approximations $\bar S(\tau)$ in \eqref{L*_approx_init} (see Fig.~\ref{fig_SegSitesBP}). However, all these approximations show that the expected number of segregating sites is proportional to $\Th$, depends only very weakly on the product of $s$ and the average mutation effect $\bar\A$, and is independent of the population size $N$.

Further support for our proposition derives from Figure~\ref{fig:sweep_shift_new}. Motivated by \citetalias{Hoellinger2019} work on a different model of selection, in this figure we illustrate the polygenic pattern of adaptation at the loci that contribute to the trait at the time $T$ when the mean phenotype $\bar G$ first reaches or exceeds a given value. Although the expected number of segregating sites $\bar{S}_T$ deviates from our approximations in Sect.~\ref{sec:sweepshift} for $2 \Th =0.1$, the main conclusion remains the same. Clear sweep pattern can be observed for $2 \Th =0.1$ (where $\bar{S}_T<2.5$) and shift-like patterns for $2 \Th = 10$ (where $\bar{S}_T>35$).

Because we assume exponential directional selection, beneficial mutations, and weak random genetic drift, all mutants that become `established' will sweep to fixation in the long run. Therefore, we focus on the initial response when characterizing patterns of adaptation as sweep-like vs.\ shift-like. In Section~\ref{Disc:Outlook}, we argue that our results are of relevance for other modes of directional selection. A popular model is one in which a population is initially in equilibrium under mutation-stabilizing selection-drift balance and then starts to evolve due to a sudden or a continuous change in the environment, modeled by a sudden shift in the optimum phenotype or by a continuously moving optimum \citep[e.g.][]{Buerger1995,GomulkiewiczHolt1995,Matuszewski_etal2015}. For models like this, our findings may be of relevance either for the initial response (in case of a sudden shift) or even in the long term (in case of a moving optimum). Below, we briefly review recent treatments focusing on the polygenic dynamics underlying trait evolution.

If a small sudden shift in the optimum phenotype occurs, mutations of relatively small effect may be most conducive to selective sweeps because then large-effect mutations become deleterious when the population closely approaches the optimum. If a sudden shift in the optimum of appreciable magnitude occurs and the initial population has been in mutation-stabilizing selection balance before, then a kind of threshold behavior has been identified for deterministic models \citep{deVladar2014, Jain2017}. Under mutation-selection balance the initial variance will be high if allelic effects are small because such alleles are maintained at intermediate frequencies. By contrast, the initial variance will be low if allelic effects are very large because such alleles will be nearly fixed or absent. Alleles of very small effect will respond slowly after the onset of selection because they are nearly neutral. If their number is large, as assumed by the infinite sites model, then the total response of the mean phenotype can still be large (proportional to the variance, as in Lande's equation discussed above). Alleles of very large effect will initially also respond very slowly because they are so rare, and this will lead to a slow response of the mean phenotype because there is little variation on which selection can act. As a result, alleles with effects close to an intermediate threshold value will be the first to respond quickly and start sweeping. On a longer time scale and for a large shift in the optimum, the response from a few loci with large effects is comparable with that from many loci with small effects if the sum of effect sizes is the same. If most loci have large effects, then the response will clearly be much faster than the response from the same number of small-effect loci; the larger the effect the faster the sweeps \citep{deVladar2014, Jain2017, JainStephan2017review}. 

In finite populations the conclusions are even more subtle and may deviate from those in the deterministic case. Mutations of small effect may contribute much less to the response than in large populations because their evolutionary fate is dominated by genetic drift and their initial distribution under mutation-stabilizing selection-drift balance will be U-shaped. Again, mutations of moderate effects may contribute most to the selective response after a sudden shift in the optimum \citep{JohnStephan2020, DeviJain2023, Hayward2021}. The latter authors also conclude that mutations of large effect are unlikely to sweep, but this is a consequence of their assumptions that the trait is highly polygenic (i.e., $\sqrt{\Th}\gg1$), the population is initially in mutation-stabilizing selection-drift balance (whence alleles of large effect are extremely rare), and there is only a small shift in the optimum of a few phenotypic standard deviations. The latter assumption implies that the equilibration phase is reached quickly, and then fine-tuning by alleles of small effect is essential for adaptation. If alleles of large effect contribute substantially to the initial variance and the shift is large, then they will sweep \citep{Chevin2008, Thornton2019, JohnStephan2020, StephanJohn2020-rev}. 

For a gradually moving optimum, after a relatively short initial phase, a quasi-stationary phase is entered during which, except for stochastic fluctuations, a constant average variance is attained and the mean phenotype lags behind the optimum by an, on average, constant amount. Of course, a prolonged quasi-stationary response will depend crucially on a constant input of new mutations.
During this phase, the trait experiences a mixture of directional and stabilizing selection, and the variance settles to a value that is lower than the variance under mutation-drift balance (which is nearly the same as that achieved under exponential selection), but higher than that under mutation-stabilizing selection-drift balance \citep{LynchLande1993, Buerger1995, JonesArnoldBuerger2004}. \cite{Matuszewski_etal2015} showed that compared to adaptation from \emph{de novo} mutations, adaptation from standing variation proceeds by the fixation of more alleles of small effect. If the optimum is stopped after some time or new mutations are ignored, then the phenotypic dynamics becomes more complex and may depend on details of the underlying genetics \citep{deVladar2014}. 

In our finite-population model, in which initial mutant frequencies are $1/N$, all mutants are beneficial, and no epistatic interactions occur, the response of the mean is faster the larger the allelic effects. In addition, mutations that sweep, sweep faster than predicted by deterministic growth. Nevertheless, the response of the mean phenotype depends linearly on $N$ through $\Th=NU$ (Corollary \ref{cor:equilibrium_simp} and Proposition \ref{prop:Dynamics}), except in the very early phase when it depends mainly on $U\bar\A$ but is nearly independent of $N$ (Proposition \ref{prop:Initial2}). 
\subsection{Conclusions, limitations, and outlook}\label{Disc:Outlook}

We conclude from our analyses that $\Th$, the expected number of beneficial mutations occurring in the population per generation (and contributing to the trait) is the main determinant of the pattern of adaptation at the genetic level. Selective sweeps at a few loci are the dominant pattern if $\Th\le0.1$, and small allele frequency shifts at many loci are observed if $\Th\ge10$. Given the same mean (beneficial) mutation effect, $\bar\A$, higher second and third moments of the mutation distribution favour sweeps. The expected number of segregating sites is an excellent indicator for the pattern of adaptation. However, it may be challenging to estimate the number of segregating sites contributing to a trait.
The selection coefficient $s$ and the average effect $\bar\A$ primarily determine the rate of adaptation via their product $s\bar\A$. Except for $\Th$, the population size has a weak effect on the rate of adaptation and on the population variance; often it enters expressions approximately logarithmically. We emphasize that our analysis is based on the assumption $Ns\bar\A\gg1$, i.e., that selection is considerably stronger than random genetic drift. Most of our approximations, however, perform well when $Ns\bar\A>2$. 

Our conclusions concerning the influence of the parameters on the pattern of adaptation are similar to those of \cite{Hoellinger2019}, although their model and approach deviate from ours in important respects. Most significantly, their genetic architecture differs drastically from ours. They study a polygenic trait that assumes only two or very few values, such as resistance and non-resistance. The contributing loci are thus highly redundant, and a mutation at one or at very few loci is sufficient to complete adaptation. This is a very strong form of epistasis. In addition, mutation effects are the same at all loci. Selection on the discrete trait acts similarly as in our model: they assume a linear fitness function in continuous time, which corresponds to an exponential in discrete time. In contrast to our model, their initial populations are polymorphic and (mostly) assumed to be in mutation-stabilizing selection-drift balance. Nevertheless, they find that their population-scaled background mutation rate $\Th_{\rm{bg}}$, which corresponds closely to our $2\Th$, explains the main differences in the patterns of adaptation, i.e., few sweeps versus many slight shifts. This main conclusion is confirmed by \cite{Hoellinger2023} for an additive trait subject to a sudden shift in the optimum phenotype, again for equal mutation effects.


The validity of our conclusions may be limited by a number of simplifying assumptions. \textblue{In the introduction we provided empirical support for the assumptions that selection is stronger than random drift and that the initial selection response results from new mutations, i.e., that we ignore the initial response from standing variation. To relax the latter assumption, our model could be combined with the breeder's equation to predict the joint response of the mean phenotype from new mutations and standing variation. However, as discussed in Sect.~\ref{Disc:breedereq}, the (initially constant) response from standing variation depends on the initial genetic variance of the trait. To infer the selection response from standing variation at individual loci, our approach could be augmented by that developed in \cite{JohnStephan2020} and \cite{Stephan2021}, who derived diffusion approximations for the allele-frequency distributions under mutation-selection-drift balance (cf.\ Sect.~\ref{sect:disc_seg_sites}). 
}

\textblue{Our results are derived for haploid populations, but will apply as well to diploids without dominance by substituting $2N$ for $N$.
The assumption of unlinked sites will be biologically reasonable unless the number of (simultaneously) segregating sites exceeds the number of chromosomes by far or most sites occur in regions of reduced recombination. Although exponential directional selection by itself does not induce linkage disequilibrium, deviations from this form of selection as well as random genetic drift will do so. Therefore, it would be desirable to extend our results to loosely linked loci and perform a kind of quasi-linkage equilibrium analysis. Below we summarize theoretical support for the proposition that the neglect of epistasis and of linkage disequilibrium poses only mild limitations to our analyses and conclusions.
}  

Quite extensive simulation results for models of artificial directional selection (such as truncation selection) by \cite{KeightleyHill1983, KeightleyHill1987} suggest that linkage reduces the asymptotic variance, hence the response of the mean, compared to unlinked loci. The reduction will be small unless linkage is very tight. A similar observation was reported in \cite{Buerger1993} for exponential directional selection. However, these results included only population sizes of a few hundred individuals. For a moving optimum model and population sizes up to about $10^4$, a qualitatively similar finding was obtained in \citet[][pp.\ 335--338]{Buerger2000}. In all these investigations, mutation distributions with positive and negative effects were used. Then, mutations of both signs appear on the same chromosome, whence low recombination will reduce fixation probabilities, especially of weakly beneficial mutants, and hamper adaptation \citep{Barton1995}. In concordance with these results, a simulation study by \cite{Thornton2019} found that weak or moderate linkage has almost no effect on the response of the mean and on the variance of the trait, and also little effect on the fixation probabilities of advantageous mutations, but the resulting interference increases fixation times and affects haplotype diversity. 

Analytical work on the effects of linkage is scarce, based on specific assumptions, and shows that linkage can increase or decrease the variance of a trait under directional selection, depending on the model assumptions and the magnitude of the parameters \citep[e.g.][]{Robertson1977, HospitalChevalet1996, ZhangHill2005}. The most comprehensive analysis may be that by \cite{TurelliBarton1994}. They showed that even under strong epistatic selection, such as truncation selection, the trait distribution is significantly affected by linkage disequilibrium only if loci are tightly linked. This provides additional support for our expectation that our results derived under the assumptions of linkage disequilibrium and absence of epistasis will provide decent approximations if recombination among the contributing loci is moderate or strong. However, the explicit determination of the effects of linkage on the likelihood of complete sweeps will pose a formidable challenge \citep[e.g.][]{KimStephan2003}. 

Moderate or strong linkage can lead to selective interference. An extreme scenario is provided by a single locus with recurrent mutation. For this case our theory remains applicable if $2N\mu \leq 1$, where $\mu$ is the per-locus mutation rate. Then it is sufficient to consider the first \emph{successful} mutation event. If $2N\mu > 1$, competition of mutants from different mutation events can no longer be ignored and complicates the evolutionary dynamics (unpublished results by HG).

We expect that our assumption of exponential directional selection provides a good approximation for the early phases of adaptation under other forms of directional selection. A property of exponential selection that is crucial for our analysis is that the strength of selection remains constant over time. It shares this property with truncation selection (if the truncation probability remains constant), and with selection imposed by a steadily moving optimum during the phase when the population mean follows the optimum at the same rate and with a constant average lag. If a sudden shift in the optimum phenotype occurs then selection will remain approximately constant initially, but eventually gets increasingly weaker as the mean proceeds towards the optimum. Some of the above discussed results provide support for the expectation that our results may be applicable to these forms of selection. Exponential directional selection will not be suitable as an approximation if the selection strength is fluctuating in time, if there are pleiotropic fitness effects, or if selection is epistatic, strong, and recombination is weak.

Our results concerning the time dependence of the distribution of sweeping \emph{de novo} mutations will be applicable to very rare allelic variants already present in the population. This will be the case if their effect is large and the population has been under stabilizing selection before directional selection started. If more than one copy is initially present,  the quasi-deterministic phase could be started from a gamma-distributed effective initial population size \citep{Martin2015} instead of an exponential. The extent to which our results on the pattern of adaptation from \emph{de novo} mutations remain applicable in the presence of standing variation would be an interesting and challenging topic for future research.

\newgeometry{left=3cm,right=3cm}
\begin{scriptsize}
\setlength{\LTcapwidth}{\textwidth}
\begin{longtable}[c]{l|p{0.02\textwidth}p{0.02\textwidth}|ll}
\caption{Glossary of symbols. For both the Roman and Greek alphabets, uppercase letters precede lowercase ones. For each uppercase or lowercase letter, listing is in order of appearance of the definition in the text. The second column indicates whether the symbol is used in the one-locus model or in the infinite sites (and/or quantitative-genetic) model. The references are to the equation closest to the definition of each symbol. Thus, (2.1), (2.1)+, (2.1)- refers to (2.1), the text below (2.1), the text above (2.1), respectively. Sometimes the respective subsection is given. Symbols that occur only in the Discussion (mainly Sect.~\ref{Disc:breedereq}) or only in the Appendix are not listed.} \label{table_notation} \\

Symbol & \multicolumn{2}{c|}{1L/ISM} & Reference & Definition \\
\hline &&&&\\ \endfirsthead

\caption{(continued)} \\

Symbol & \multicolumn{2}{c|}{1L/ISM} & Reference & Definition \\
\hline &&&&\\ \endhead

$a$ & & & \eqref{a(t)}, \eqref{a(t)_continuous} & important parameter for $g_a$ \\

$f$ & & x & Sect.~\ref{sec:QGmodel} & density of the distr. of mutation effects $\A_i$ \\

$G$ & & x & Sect.~\ref{sec:QGmodel} & genotype (or phenotype) \\
$\bar{G}_i$ & & x & \eqref{Gi} & contributions of locus $i$ to $\bar{G}$ \\
$\bar{G}$ & & x & \eqref{def_barGtau} & expected phenotypic mean \\
$\bar{G}^*$ & & x & \eqref{nu}- & expected phenotypic mean in the quasi-stationary phase \\
$\Delta \bar{G}^*$ & & x & \eqref{barG*_prop} & expected per-generation response in the quasi-stationary phase \\
$\bar{G}_T$ & & x & Sect.~\ref{sec:indicator-sweep-shift} & average of mean phenotypic value in gen. $T$ \\
$g_a$ & x & & \eqref{g(p)} & density of $X_n$ \\

$h^{(i)}$ & & x & \eqref{hi(p)} & density of r.v. measuring the freq. of the $i$-th mut. cond. on non-ext. by gen. $\tau$ \\
$\tilde{h}^{(i)}$ & & x & \eqref{tilde{h}^{(i)}} & describes the positive part of $\tilde{X}_{\tau}^{(i)}$ \\

$k$ & & x & \eqref{Poi(Thetatau)}- & total number of mutation events until time $\tau$ \\

$M_{i,\Th}$ & & x & \eqref{Mi} & normalization constant \\
$m_{i,\Th}$ & & x & \eqref{Erlang} & density of Erlang distr. (approx. waiting time $\tau_i$) \\

$N$ & & & Sect.~\ref{sec:Model} & population size \\
$N_e$ & & & \eqref{PfixDif_Ne}- & effective population size \\
$n$ & x & & Sect.~\ref{sec:1Lmodel} & generation \\

$\Poi_{\Th\tau}$ & & x & \eqref{Poi(Thetatau)} & Poisson distr. \\
$\Pexinf$ & x & & \eqref{defW}-, \eqref{plost*} & probability of ultimate extinction \\
$\Psur$ & x & & \eqref{Plost*app}+ & (long-term) survival probability (Galton-Watson process) \\
$\Pfix$ & x & & \eqref{PfixDif} & probability of ultimate fixation (diffusion theory) \\
$\Plost(n)$ & x & & \eqref{plost(tau)} & probability of extinction before gen. $n$ \\
$\Plost^{(i)}(\tau)$ & & x & \eqref{plosti(tau)} & probability that the mutation at locus $i$ has been lost by gen. $\tau$ \\
$p(\cdot)$ & x & & \eqref{eqdet} & solution of the discrete, deterministic selection equation \\

$S$ & & x & \eqref{Seg}- & number of segregating sites \\
$\bar{S}$ & & x & \eqref{Seg} & expected number of seg. sites \\ 
$\bar{S}^*$ & & x & \eqref{L*_approx}- & expected number of seg. sites in the stat. phase \\ 
$\tilde{S}$ & & x & \eqref{Seg_BP_f}, \eqref{Seg_BP_equal}, \eqref{Seg_BP_f_smalltau} & expected number of seg. sites under branching process model \\ 
$\bar{S}_T$ & & x & Sect.~\ref{sec:indicator-sweep-shift} & average number of seg. sites at time $T$ \\
$s$ & & & Sect.~\ref{sec:QGmodel} & selection coefficient \\
$\tilde{s}$ & x & & \eqref{stilde} & selection coefficient used in relation to diffusion theory \\
$s_{\rm{max}}$ & & x & \eqref{def_smax} & sel. coeff. for which $\bar{S}^*$ is maximized \\

$T_\ep$ & x & & \eqref{Tep_FL_general} & time needed such that $1-\Plost(T_\ep) = (1+\ep) (1-\Pexinf)$ for $\ep>0$\\
$T$ & & x & Sect.~\ref{sec:indicator-sweep-shift} & stopping time \\
$\bar{T}$ & & x & Sect.~\ref{sec:indicator-sweep-shift} & average stopping time \\
$t_M$ & x & & \eqref{gammas}+ & time at which the variance $\ga$ is maximized \\
$\tfix$ & x & & \eqref{tfix} & expected time to fixation \\
$\bartfix$ & x & & \eqref{bartfix} & expected mean time to fixation \\
$\tloss$ & x & & \eqref{tloss} & expected time to loss \\
$\bartloss$ & x & & \eqref{app_bartloss} & expected mean time to loss \\

$U$ & & x & Sect.~\ref{sec:QGmodel} & expected number of beneficial mutations per individual per generation \\

$V_{G_i}$ & & x & \eqref{VGi} & contributions of locus $i$ to $V_G$ \\
$V_G$ & & x & \eqref{def_VGtau} & expected phenotypic variance \\
$V_G^*$ & & x & \eqref{nu}- & expected phenotypic variance in the quasi-stationary phase \\
$v$ & x & & \eqref{def_momentsxi} & variance of the offspring distribution $\xi_j$ \\

$W_n$ & x & & \eqref{defW}- & rescaled (discrete) random variable obtained from $Z_n$ \\
$W$ & x & & \eqref{defW} & limiting random variable of $W_n$ \\
$W^+$ & x & & \eqref{defW+} & absolutely continuous part of $W$ \\
$w^+$ & x & & \eqref{defW+}+ & density of $W^+$ \\
$W_n^+$ & x & & \eqref{defWn+} & positive part of $W_n$ \\
$\bar{w}(\cdot)$ & x & & \eqref{eqdet}+ & mean fitness needed for the discrete, det. sel. equation \\

$X_n$ & x & & \eqref{def_X_n} & random variable measuring the (positive) mut.-frequency in gen. $n$ \\
$\tilde{X}_{\tau}^{(i)}$ & & x & \eqref{tildeX^{(i)}} & (unconditioned) r.v. for the frequency of the $i$-th mutant \\
$X_{\tau}^{(i)}$ & & x & \eqref{Gi}- & $X_n$ for locus $i$ (with $\tau=n$, $\si_i=\si$) \\
$\hat{x}$ & x & & \eqref{g(p)}+ & mut. freq. for which $g_a$ has unique maximum for $a<2$ \\

$Y_n$ & x & & \eqref{Psin_exp}+ & random variable with distr. $\Psi_n$ (approx. for positive part of $Z_n$) \\

$Z_n$ & x & & \eqref{def_momentsxi}- & Galton-Watson process \\

&&&&\\ \hline &&&&\\

$\A_i$ & & x & Sect.~\ref{sec:QGmodel} & mutation effect on phenotype of locus $i$ \\
$\bar{\A}$ & & x & Sect.~\ref{sec:QGmodel} & mean of the mutation effects \\
$\A_n (f)$ & & x & \eqref{DebarG*}- & $n$th moment about zero of the mutation distribution $f$ \\

$\ga_1$ & x & & \eqref{mean_g} & first moment of $g_a$ \\
$\ga_2$ & x & & \eqref{moment2_g} & second moment of $g_a$ \\
$\ga$ & x & & \eqref{var_g} & within-population variance of the mutant’s allele frequency \\

$\Th$ & & x & Sect.~\ref{sec:QGmodel} & expected total number of mutations occurring in the population per generation \\

$\lambda_n$ & x & & \eqref{Psin_exp} & parameter of exponential distribution $\Psi_n$ \\

$\sN$ & & x & \eqref{nu} & notational simplification \\

$\xi_j$ & x & & \eqref{def_momentsxi}- & (random) number of offspring of individual $j$ \\

$\rho$ & x & & Rem.~\ref{Psur=csalpha} & constant, such that $\Psur\approx \rho s\A$ \\

$\si_i$ & & x & Sect.~\ref{sec:QGmodel} & fitness effect of a mutation at locus $i$ \\
$\si$ & x & & Sect.~\ref{sec:1Lmodel} & fitness of the derived state (mean of $\xi_j$, mean offspring number) \\

$\tau$ & & x & Sect.~\ref{sec:ISmodel} & (discrete) generation since the initial population started to evolve \\
$\tau_i$ & & x & \eqref{Erlang}- & random time at which $i$th mutation event occurred \\
$\tau_c$ & & x & \eqref{tau_c} & expected time by which the first mutation becomes fixed \\
$\tau_1$ & & x & \eqref{barG_init}- & bound for time \\
$\tilde{\tau}$ & & x & \eqref{Seg_BP_equal}+ & bound for time \\

$\Phi$ & x & & \eqref{Phi_varphi}- & Laplace transform of $W$ \\
$\varphi$ & x & & \eqref{def_momentsxi}+ & offspring generating function of $\xi_j$ (of $Z_1$) \\

$\Psi_n$ & x & & \eqref{Psin_exp} & exponential distribution \\
$\psi_n$ & x & & \eqref{Psin_exp}+ & density of distr. $\Psi_n$ \\

&&&&\\ \hline \hline 
\multicolumn{3}{l}{}&&\\

\multicolumn{3}{l}{$\EV$} & & expectation \\
\multicolumn{3}{l}{$\Var$} & & variance \\
\multicolumn{3}{l}{$\Prob$} & & probability \\
\multicolumn{3}{l}{$\ProdLog$} & \eqref{plost*} & product logarithm (Lambert function) \\
\multicolumn{3}{l}{$\Ga$} & \eqref{Mi}+ & incomplete Gamma function \\
\multicolumn{3}{l}{$[\cdot]$} & \eqref{a(t)_continuous}- & nearest integer to $\cdot$ \\
\multicolumn{3}{l}{$E_1$} & \eqref{gammas}+ & exponential integral \\
\multicolumn{3}{l}{$\Exp$} & & exponential distribution \\
\multicolumn{3}{l}{$\ga$} & & Euler–Mascheroni constant \\

\end{longtable}

\end{scriptsize}
\restoregeometry

\section*{Acknowledgments}
We thank Joachim Hermisson, Emmanuel Schertzer, Ben W\"olfl, and Ilse H\"ollinger for helpful discussions.
We gratefully acknowledge extraordinarily detailed and helpful comments by four reviewers and the editor, John Wakeley. 
Financial support by the Austrian Science Fund (FWF) through the Vienna Graduate School of Population Genetics (GrantDK W1225-B20) to HG is gratefully acknowledged. For open access purposes, the authors have applied a CC BY public copyright license to any author accepted manuscript version arising from this submission.

\section*{Additional materials}
We provide a comprehensive Wolfram \emph{Mathematica} notebook, containing additional visualizations of the analytical predictions and efficient numerical procedures. Upon publication, the notebook and the simulation code (C++) will be made available via github: \url{https://github.com/Ha-nn-ah/EvolutionOfQuantitativeTraits}.

\appendix\appendixpage
\renewcommand{\thesection}{\Alph{section}}
\section{Fractional linear offspring distributions and associated Galton-Watson processes}\label{app_frac_linear}

\subsection{Basics}\label{app_fl_basics}

The fractional linear, or modified geometric, distribution is given by
\begin{equation}\label{frac_lin}
	\pfl_0 = r \;\text{ and }\; \pfl_k = (1-r)(1-p)p^{k-1} \; \text{ if }\; k\ge1 \,,
\end{equation}
where $0<r<1$ and $0<p<1$; e.g.~\citet[pp.\ 6-7]{Athreya1972} or \citet[pp.\ 16]{Haccou2005}. The name fractional linear derives from the fact that its generating function is
\begin{equation}\label{frac_lin_gen}
	\varphi(x) = \frac{r + x(1-p-r)}{1 - x p}\,,
\end{equation}
hence fractional linear. With $r=1-p$, the geometric distribution is recovered. It is straightforward to show that every fractional linear generating function generates a modified geometric distribution. 

Mean and variance of $\{\pfl_k\}$ are
\begin{equation}
	m = \frac{1-r}{1-p} \text{ and } v = \frac{(1-r)(p+r)}{(1-p)^2}\,.
\end{equation}
In terms of $m$ and $v$, we obtain
\begin{equation}
	p = 1-\frac{2m}{m+m^2+v} \text{ and } r = 1- \frac{2m^2}{m+m^2+v}\,.
\end{equation}
Therefore $m>1$ if and only if $0<r<p<1$. In this case the associated Galton-Watson process $\{Z_n\}$ is supercritical. We note that $m>v$ if and only if $2p<1-r$, which implies $p<\tfrac12$.  

If $m\neq1$, the $n$-times iterated generating function $\varphi_{(n)} = \varphi_n$ is (rearranged from the parameterization in \citet[p.~7]{Athreya1972})
\begin{equation}
	\varphi_n(x) = \frac{r(1-x) - m^{-n}(r-px)}{p (1-x)- m^{-n}(r-px)}\,.
\end{equation}
It is fractional linear with parameters
\begin{equation}\label{pnrn}
	p_n = \frac{p(1-m^{-n})}{p-rm^{-n}} \text{ and } r_n = \frac{r(1-m^{-n})}{p-rm^{-n}}\,. 
\end{equation}

From now on we assume $m>1$, i.e., $r<p$. The probability of extinction by generation $n$, $\Pexnfl$, is given by
\begin{equation}\label{Pexnfl=r_n}
	\Pexnfl = \varphi_n(0) = \frac{r(1-m^{-n})}{p-rm^{-n}} = r_n\,,
\end{equation}
and the (ultimate) extinction probability is 
\begin{equation}\label{Pexinftyfl}
	\Pexinftyfl = \frac{r}{p}\,.
\end{equation}
From \eqref{Pexnfl=r_n} and \eqref{Pexinftyfl} we obtain the relation
\begin{linenomath}\begin{subequations}
\begin{equation}
	\frac{1-\Pexinftyfl}{1-\Pexnfl} = 1 - m^{-n} \Pexinftyfl  \label{Psurinfty/Psurvn_FL} 
\end{equation}
and its equivalent
\begin{equation}
	\frac{1-\Pexnfl}{1-\Pexinftyfl} = 1 + \frac{\Pexinftyfl}{m^n-\Pexinftyfl} \,. \label{Psurvn/Psurinfty_FL_app}
\end{equation}
\end{subequations}\end{linenomath}
This allows to compute the time needed for the probability of non-extinction by generation $n$, $1-\Pexnfl$, to decline to $(1+\ep)(1-\Pexinftyfl)$.  For $\ep>0$ (not necessarily small) we define $T_\ep$ as the (real) solution of
\begin{linenomath}
\begin{equation}\label{Tep_definition}
	1-\Plost(T_\ep) = (1+\ep)(1-\Pexinftyfl)\,.
\end{equation}
\end{linenomath}
By \eqref{Psurvn/Psurinfty_FL_app}, this time can be calculated explicitly:
\begin{linenomath}
\begin{equation}\label{Tep_FL_general}
	T_\ep = \frac{\ln\Bigl(\Pexinftyfl \Bigl(1+\frac{1}{\ep} \Bigr)\Bigr)}{\ln m}  \,.
\end{equation}
\end{linenomath}
Therefore, $\dfrac{1}{\ln m}\ln\Bigl(1+\frac{1}{\ep} \Bigr)$ is always an upper bound to $T_\ep$.

\subsection{Distributions of $W_n$ and $W$}\label{app:fraclin_W_n_W}

The cumulative distribution function of $W_n=Z_n/m^n$ can be calculated explicitly (e.g., from $\varphi_n$):
\begin{equation}
	P(W_n\le x) = 1-\frac{(p-r)p_n^{\lfloor m^n  x \rfloor}}{p-rm^{-n}}\,,
\end{equation}
where $p_n$ is given by \eqref{pnrn} and $\lfloor z \rfloor$ denotes the largest integer smaller than $z$. From the equation
\begin{equation}
	\Prob[W_n\le x] = \Pexnfl + (1-\Pexnfl)\Prob[W_n^+\le x]\,,
\end{equation}
we infer
\begin{equation}
	\Prob[W_n^+\le x] = 1 - p_n^{\lfloor m^n  x \rfloor}\,.
\end{equation}
This shows that the cumulative distribution of $W_n^+$ is imbedded into (hence approximated by) the exponential distribution with parameter 
\begin{equation}
	\la_n = -m^n \ln p_n = -m^n \ln \frac{p(1-m^{-n})}{p-rm^{-n}}\,. 
\end{equation}
We note that $\lim_{n\to\infty} p_n^{m^n x} = e^{-(1-r/p)x}$.
As a consistency check, it can be confirmed directly from the CDF that $\EV[W_n^+]=(1-\Pexnfl)^{-1}$. 

If we take the exponential distribution with parameter $\la_n^*=1-\Pexnfl$, this provides an approximation to the (discrete) distribution of $W_n^+$ which has the property that the integral of the difference of the two distribution functions vanishes, i.e., such that 
\begin{equation}
	\int_0^\infty (1 - p_n^{\lfloor m^n  x \rfloor})- (1 - e^{\la_n^*x})\,dx = 0 \,.
\end{equation}
Conditional on this requirement, the exponential with parameter $\la_n^*=1-\Pexnfl$ provides the best possible exponential approximation to $W_n^+$.

We obtain that $W^+=\lim_{n\to\infty}W_n^+$, which exists quite generally \citep[e.g.,]{Athreya1972}, is exponentially distributed with parameter
\begin{equation}\label{la_W^+}
	\la = \lim_{n\to\infty} \la_n^* = 1- \frac{r}{p} = 1 - \Pexinftyfl\,.
\end{equation}
This follows as well from Poincar\'e's functional equation \eqref{Phi_varphi}, i.e., $\Phi(u) = \varphi(\Phi(u/m))$, which implies that the Laplace transforms of $W$ is
\begin{equation}
	\Phi(u) = \EV[e^{-u W}] = \frac{p-r + ru}{p-r+pu} = 1-\la + \la\,\frac{1}{1+u/\la}\,,
\end{equation}
with $\la$ given by \eqref{la_W^+}.

It is easy to show that if $W^+$ is exponential, then the generating function of the offspring distribution is fractional linear. This follows because the Laplace transform $\Phi$ is strictly monotone decreasing so that $\varphi$ is uniquely determined by eq.~\eqref{Phi_varphi}.
Therefore, as pointed out by a reviewer, $W^+$ is exponential if and only if the offspring distribution is fractional linear.

\subsection{Comparison with the Galton-Watson process originating from a Poisson offspring distribution} \label{app_frac_lin_compare}

We assume a fractional linear distribution with $v=m$, as for a Poisson distribution with mean $m$.
A simple calculation shows that $v=m$ if and only if $p=\frac{m}{2+m}$ and $r = \frac{2-m}{2+m}$, where we assume $m<2$. Then we obtain
\begin{equation}\label{Pexnfl[m,n]}
	\Pexinftyfl = \frac{2}{m} - 1 \text{ and } \Pexnfl = \frac{(2-m)(1-m^{-n})}{m - (2-m)m^{-n}} \,.
\end{equation}

If $m=e^{\sA}$, as we assume in the main text, then $\Pexinftyfl=2e^{-\sA}-1$ and 
\begin{equation}\label{Psur_fracLin}
	\Psur = 1-\Pexinftyfl = 2\bigl(1-e^{-\sA}\bigr) = 2\sA - (\sA)^2 + O((\sA)^3)\,.
\end{equation}
We note that $\Psur<1$ if and only if $\sA<\ln 2\approx 0.69$ (otherwise, $\Psur=1$). For comparison, the series expansion of $\Psur$ for the corresponding Poisson offspring distribution is $2\sA - \tfrac{5}{3}(\sA)^2 + O((\sA)^3)$, and the survival probability is always $<1$. In particular, the survival probability for the Poisson distribution is always smaller than that for the corresponding linear fractional. In the limit of $\sA\to0$, they are asymptotically equivalent.

If we assume in the fractional linear case that $v$ and $m$ are proportional as $m=e^{s\A}$ is varied, i.e., $v=2m/\rho$ (cf.~Remark~\ref{Psur=csalpha}), equation \eqref{Tep_FL_general} simplifies to
\begin{linenomath}\begin{subequations}\label{Tep_FL_mrho_sA}
\begin{align}
	T_\ep(\sA) &= \frac{1}{s\A}\,\ln\Bigl(\Pexinftyfl\Bigl(1+\frac{1}{\ep} \Bigr)\Bigr)  \label{Tep_FL_mrho_sA_a} \\
		& = \frac{1}{s\A}\ln\Bigl(1+\frac{1}{\ep}\Bigr) - \rho + O(s\A)\,,  \label{Tep_FL_mrho_sA_b}
\end{align}
\end{subequations}\end{linenomath}
where the asymptotic estimate in the second line holds if $s\A\to0$ because $\Pexinftyfl\approx 1-\rho s\A$.
Comprehensive numerics show that \eqref{Tep_FL_mrho_sA_a} apparently always (i.e., if $\Pexinftyfl\Bigl(1+\frac{1}{\ep} \Bigr)\ge1$) provides an upper bound for any Poisson offspring distribution with mean $e^{s\A}$ if the corresponding $\Pexinftyfl$ is used. This bound is a highly accurate approximation with relative errors of order $s$ or smaller. Finally, we note that $\Pexinftyfl=0$ if $m\ge 1+2/\rho$.
\section{Mean times to fixation or loss of a mutant}\label{meanfixtime}

Approximations for the expected times to absorption conditioned on either fixation or loss of a new mutation, for simplicity called mean fixation time or mean loss time, have been an important tool in theoretical population genetics ever since its inception in the 1920s. Even in large populations, the initial fate of a mutant, as well as its fate in the final stages before fixation, is subject to strong stochastic influences, and the derivation of simple and accurate approximations is still a topic of research \citep[e.g.,][and references therein]{Charlesworth2020}. In this appendix, we add simple estimates to this literature that are particularly useful to compute the mean fixation or loss times of beneficial mutants drawn from a distribution of effects.

For the computation of mean times to fixation or loss in a finite population, we use diffusion approximations because the branching process model is not directly applicable. Although the approximation $g_a(p)$ in \eqref{g(p)} for the distribution of the number of mutations is quite accurate for late stopping times, the derivation of useful approximations for the mean fixation and loss times requires more delicate transformations than that used for Result~\ref{thm:g(p)} and is work in progress.

\subsection{Diffusion approximations}\label{sec:tfix_diff}

Here, we collect results on diffusion approximations of the fixation probability and of the mean times to fixation or loss of favorable mutants with a fixed selection coefficient $s>0$. In a haploid population, the well-known diffusion approximation for the fixation probability of a single beneficial mutant is
\begin{linenomath}\begin{equation}\label{PfixDif}
	\Pfix(s, N) = \frac{1-e^{-2s}}{1-e^{-2Ns}}\,.
\end{equation}\end{linenomath}
In the applications to our model, we replace $s$ by $\stilde$ defined in \eqref{stilde}.
The diffusion approximation $\Pfix(s, N)$ is always an upper bound to the true fixation probability in the Wright-Fisher model, and is highly accurate for every $N\ge8$ if $0<s\le0.1$ \citep{BuergerEwens1995}. 

The mean times to fixation or loss can be expressed only in terms of integrals of the corresponding sojourn time densities \citep[][Sects.~4.6 and 5.3]{Ewens2004}.
For the mean fixation time, the simple approximation 
\begin{linenomath}\begin{equation}\label{tfixHP}
	  \tfix^{\text{(HP)}}(s, N) = \frac{2}{s} \left(\ln (2Ns) + \gamma -\frac{1}{2Ns} \right) 
\end{equation}\end{linenomath}
was provided by \citet{Hermisson2005}, where $\gamma$ is the Euler gamma. It is very accurate if $2Ns\ge3$. 

\begin{figure}[t!]
\centering
\begin{tabular}{ll}
A & B \\
\includegraphics[width=0.47\textwidth]{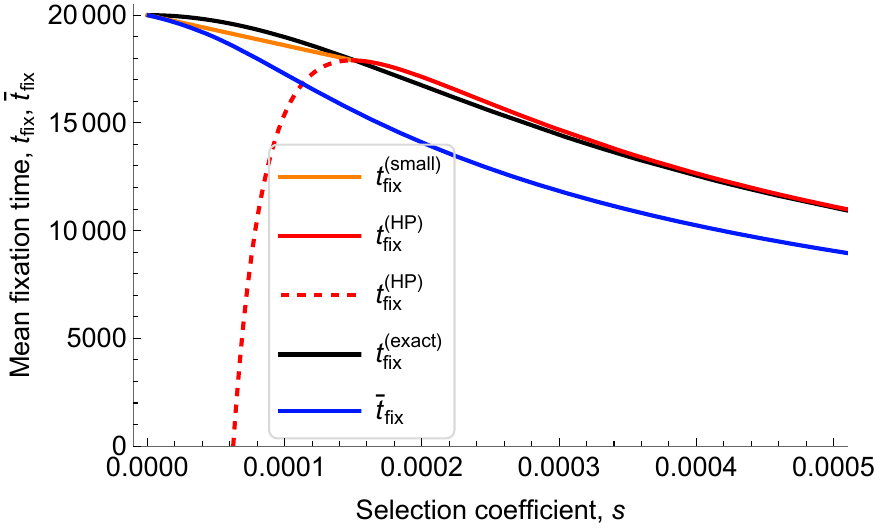} &
\includegraphics[width=0.47\textwidth]{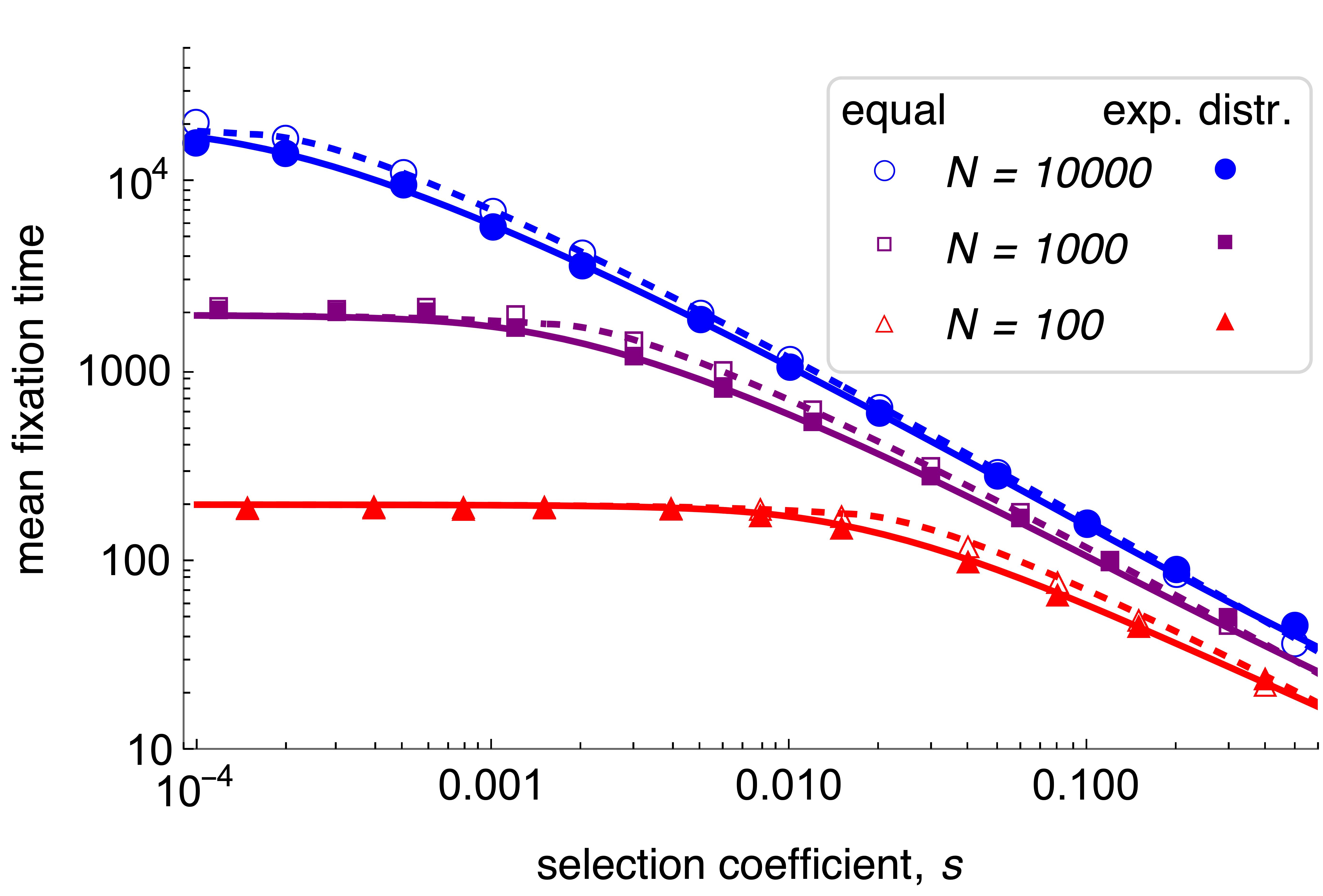} \\
C & D \\
\includegraphics[width=0.47\textwidth]{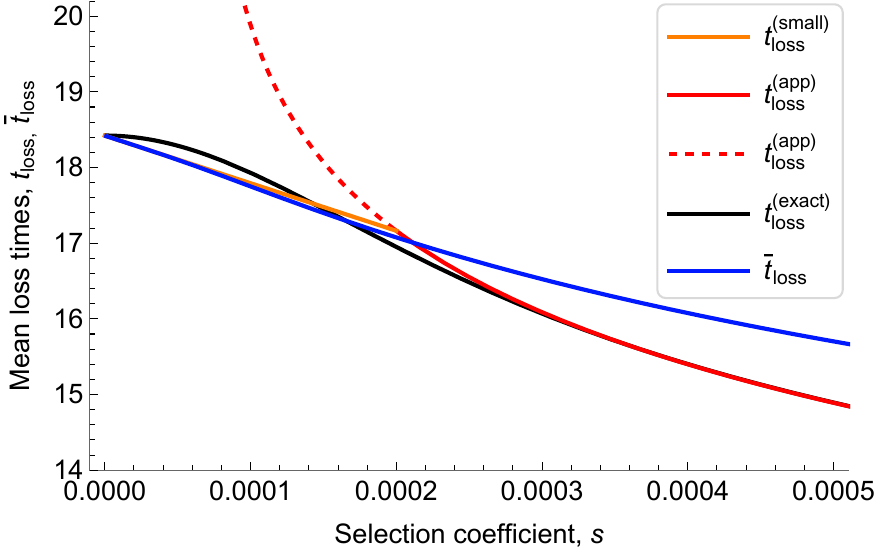} &
\includegraphics[width=0.47\textwidth]{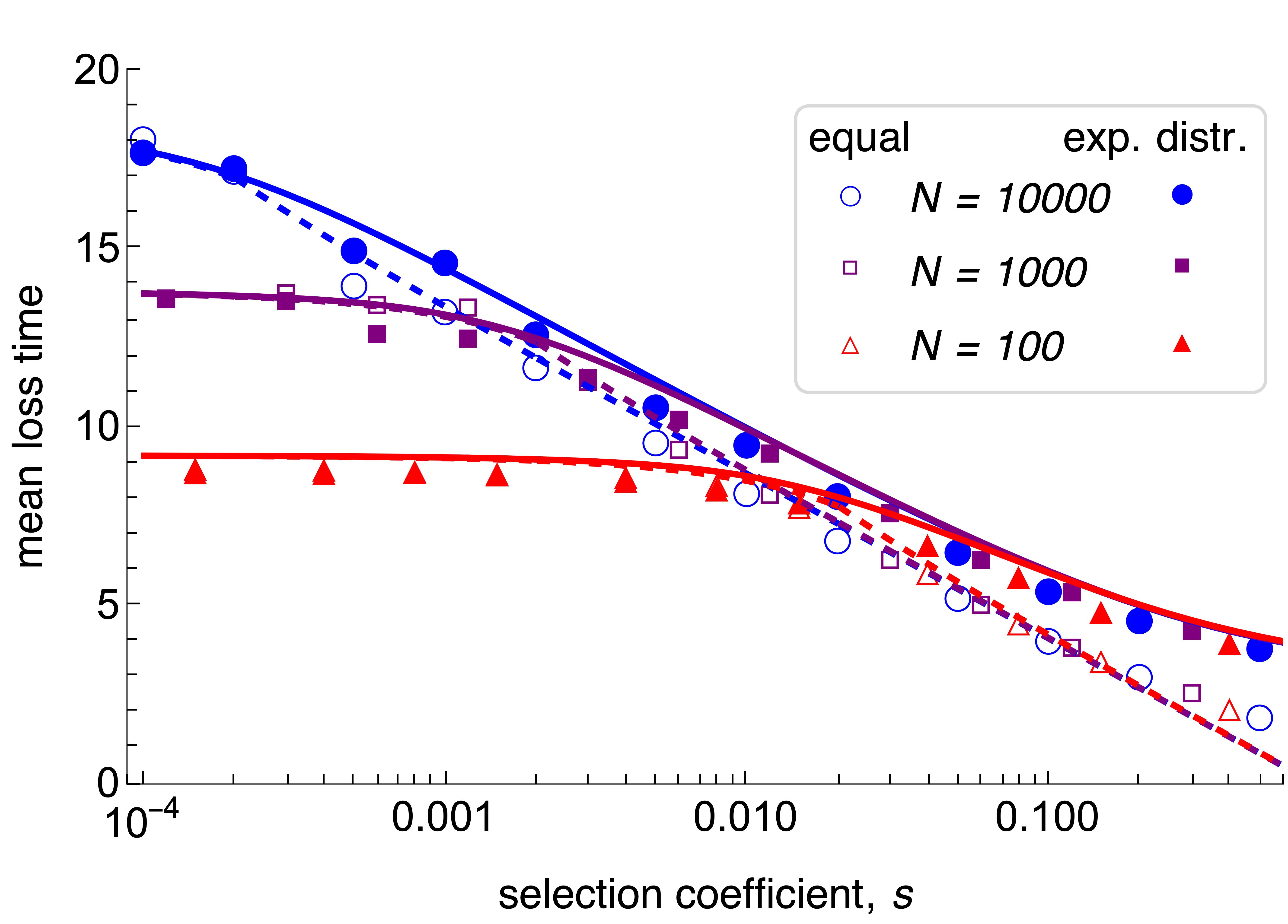} 
\end{tabular}
\caption[Mean fixation and loss times]
{Panels A and C demonstrate the accuracy of our method of approximating the true diffusion approximation for the mean time to fixation ($\tfix$) or loss ($\tloss$) by concatenation of a linear approximation for small $s$ and a simplified approximation, i.e., of eqs.~\eqref{tfix} and \eqref{tloss}. In panel A, the mean fixation times $\tfix^{\text{(small)}}(s,N)$ for $s<3/(2N)$ (orange, \eqref{consttfix_small}) and $\tfix^{\text{(HP)}}(s,N)$ for $s\ge3/(2N)$ (red, \eqref{tfixHP}) are compared with the exact diffusion approximation (black) for $N=10^4$. Panel C is  analogous to A and displays the mean loss times $\tloss^{\text{(small)}}(s,N)$ for $s<2/N$ (orange, \eqref{consttloss_small}) and $\tloss^{\text{(app)}}(s,N)$ for $s\ge2/N$ (red, \eqref{tloss_diffapp}).
In both panels, the exact diffusion approximations are computed by numerical integration of the corresponding sojourn time densities \citep[][Sects.~4.6 and 5.3]{Ewens2004},
and the dashed red curves show $\tfix^{\text{(HP)}}$ or $\tloss^{\text{(app)}}$, where they are inaccurate and substituted by $\tfix^{\text{(small)}}$ or $\tloss^{\text{(small)}}$, respectively. In addition, for mutation effects drawn from an exponential distribution with mean 1, the expected mean times to fixation ($\bartfix$ in \eqref{app_bartfix}) or loss ($\bartloss$ in \eqref{app_bartloss}) are displayed in blue. 
Panel B shows $\tfix(\stilde,N)$ from \eqref{tfix} (dashed curves) and $\bartfix(s,1,N)$ from \eqref{app_bartfix} (solid curves) as functions of $s\ge10^{-4}$ for different population sizes (blue for $N=10^4$, purple for $N=10^3$, red for $N=10^2$).  Simulation results from the Wright-Fisher model for equal mutation effects $\A=1$ and exponentially distributed effects with mean $\bar{\A}=1$ are shown as open and filled symbols, respectively.
Panel D is analogous and depicts $\tloss(\stilde,N)$ from \eqref{tloss} (dashed curves) and $\bartloss(s,1,N)$ from \eqref{app_bartloss} (solid curves).  In accordance with \eqref{tloss_diffapp}, panel D illustrates that $\tloss$ and $\bartloss$ depend only very weakly on $N$ if $s>2/N$.}
\label{fig_meanfixloss}
\end{figure}

If $2Ns<3$, we use 
\begin{linenomath}\begin{equation}\label{consttfix_small}
	 \tfix^{\text{(small)}}(s, N) = 2N\left(1 - 2N s\, \frac{11 - 6\ga - 6 \ln 3}{27} \right)\,.
\end{equation}\end{linenomath}
This is the linear function in $s$ which connects the mean fixation time $2N$ at $s=0$ with the value of $\tfix^\text{(HP)}(s, N)$ at $s = 3/(2N)$. As a function of $s$, $\tfix^{\text{(HP)}}$ assumes its maximum very close to this value and becomes negative for very small $s$. As illustrated by Fig.~\ref{fig_meanfixloss}, the function 
\begin{linenomath}\begin{equation}\label{tfix}
	\tfix(s,N) \approx \begin{cases}  
				\tfix^{\text{(HP)}}(s, N) & \text{if } s \ge 3/(2N) \,, \vspace{2mm} \\
				\tfix^{\text{(small)}}(s, N)  & \text{if } s < 3/(2N) \,,
					\end{cases}
\end{equation}\end{linenomath}
obtained by concatenation of $\tfix^{\text{(HP)}}$  and $\tfix^{\text{(small)}}$ yields a highly accurate approximation of the true diffusion approximation.

For the mean loss time of a single advantageous mutant, we obtain (see below)
\begin{linenomath}\begin{subequations}\begin{align}\label{tloss_diffapp}
	\tloss^{\text{(app)}}(s,N) &\approx -2(1+s)\ln(2s) + 2(1 -\ga) + s(3-2\ga) + \frac{1}{Ns} \\
				&\approx -2\ln(2s) + 2(1 -\ga) + \frac{1}{Ns}\,,
\end{align}\end{subequations}\end{linenomath}
which is very accurate if $Ns\ge2$ but breaks down if $Ns\le1$.

If $Ns<2$, we use
\begin{linenomath}\begin{equation}\label{consttloss_small}
	 \tloss^{\text{(small)}}(s, N) = 2\ln(N) - sN\left(\ln 4 + \gamma - \tfrac54  \right)\,.
\end{equation}\end{linenomath} 
This is the linear function in $s$ which connects the (approximate) mean loss time $2\ln(N)$ at $s=0$ \citep{Kimura&Ohta1969} with the value of $\tloss^{\text{(app)}}(s, N)$ at $s = 2/N$. As illustrated by Fig.~\ref{fig_meanfixloss}, the function 
\begin{linenomath}\begin{equation}\label{tloss}
	\tloss(s,N) \approx \begin{cases}  
				\tloss^{\text{(app)}}(s, N) & \text{if } s \ge 2/N \,, \vspace{2mm} \\
				\tloss^{\text{(small)}}(s, N)  & \text{if } s < 2/N 
					\end{cases}
\end{equation}\end{linenomath}
yields a highly accurate approximation of the true diffusion approximation.

\begin{proof}[Proof of \eqref{tloss_diffapp}]
To derive this approximation, we start with the well-known diffusion approximations for the conditional sojourn time densities  \citep[eqs.~(5.54) and (5.55) in Sect.~5.4 of][]{Ewens2004}.
To account for haploidy, we set $\A=2Ns$ and $p=1/N$ in these expressions ($p$ the initial mutant frequency). By integration with respect to $x$, we obtain with the help of \citetalias{Mathematica} and after simple algebraic rearrangement
\begin{linenomath}\begin{align}
		\tloss^{(1)}(s, N) &= \frac{1}{(e^{2Ns}-1)s}\biggl[ e^{2Ns}\Bigl(-\Ei(-2s) + \ga+\ln(2Ns)-\ln(N-1) \biggr) - \Ei(2Ns) + \Ei(2(N-1)s) \notag \\
		&\quad- \Ei(2s) + \ga +\ln(2Ns)-\ln(N-1) + e^{2Ns}\bigl(\Ei(-2(N-1)s) -\Ei(-2Ns)\Bigr) 	\biggr] \label{tex1}
\end{align}\end{linenomath}
for the expected time the mutant frequency spends in $(0,1/N]$, and
\begin{linenomath}\begin{align}
		\tloss^{(2)}(s, N) &= \frac{e^{2s}-1}{(e^{2Ns}-1)(e^{2Ns}-e^{2s})s} \biggl[ -e^{4Ns}\Ei(-2s) + e^{2Ns}\Ei(2(N-1)s) \notag \\
		&\quad+ e^{4Ns}\Ei(-2Ns)+\Ei(2Ns)+2e^{2Ns}\bigl(\ga+\ln(2Ns)-\ln(N-1)\bigr)  \notag\\
		&\quad- \Ei(2s)+e^{2Ns}\Ei(-2(N-1)s) \biggr] 		\label{tex2}
\end{align}\end{linenomath}
for the time spent in $(1/N,1)$. We note that the times $\tloss^{(1)}(s, N)$ and $\tloss^{(2)}(s, N)$ have been rescaled to generations, i.e., the diffusion time has been multiplied by $N$.
Omitting terms of order $e^{-2Ns}$ or smaller in \eqref{tex1} and using the approximation $\Ei(x)\approx e^x/(x-1)$ if $|x| > 4$, we obtain
\begin{linenomath}\begin{align*}
		\tloss^{(1)}(s, N) &= \frac{1}{(e^{2Ns}-1)s}\biggl[ e^{2Ns}\Bigl(-\Ei(-2s) + \ga+\ln(2Ns)-\ln(N-1) \biggr) - \Ei(2Ns) + \Ei(2(N-1)s) \biggr] \notag\\
				&\approx\frac{1}{s}\biggl[-\Ei(-2s) + \ga+\ln(2s)+\ln(N)-\ln(N-1) - \frac{1}{e^{2Ns}}\Bigl(\Ei(2Ns) - \Ei(2(N-1)s)\Bigr) \biggr] \notag \\
				&\approx 2 - s + 1/N\,.
\end{align*}\end{linenomath}
The latter approximation is obtained by performing a series expansion up to order $s^2$ and by keeping $Ns$ constant (as in the diffusion approximation).
For $\tloss^{(2)}$ we obtain in an analogous way, by omitting terms of order $e^{-2Ns}$ or smaller,
\begin{linenomath}\begin{align*}
		\tloss^{(2)}(s, N) &\approx \frac{e^{2s}-1}{(e^{2Ns}-1)(e^{2Ns}-e^{2s})s} \biggl[- e^{4Ns}\Ei(-2s) + e^{2Ns}\Ei(2(N-1)s) \biggr] \notag \\
		&\approx \frac{e^{2s}-1}{s} \Bigl[-\Ei(-2s) + e^{-2Ns}\Ei(2(N-1)s) \Bigr] \notag\\
		&\approx -2\ga-2(1+s)\ln(2s) + 2(2-\ga)s +\frac{1-s}{Ns}\,.
\end{align*}\end{linenomath}
By summing up, i.e., $\tloss= \tloss^{(1)}+\tloss^{(2)}$, and omitting terms of order $1/N$, we arrive at \eqref{tloss_diffapp}.
\end{proof}

\subsection{Mean fixation time averaged over the mutation distribution $f$}\label{meanfixtime_f}

Now we derive an approximation for the mean fixation time of a single mutant if its effect $\A$ on the selected trait is drawn from an exponential distribution $f$ with mean $\bar\A$.
For given $N$ and $s$, we defined in \eqref{bartfix} the corresponding expected mean fixation time by  
\begin{linenomath}\begin{equation}\label{app_bartfix}
	\bartfix(s,f, N) = \frac{\int_0^{\infty}\tfix(\stilde, N) \Pfix(\stilde, N) f(\A)\, d\A}{\int_0^{\infty}\Pfix(\stilde, N) f(\A)\, d\A}\,.
\end{equation}\end{linenomath}
We recall from \eqref{stilde} that $\stilde=e^{s\A}-1$ is the selection coefficient of a mutant of effect $\A$. For $\Pfix(\stilde, N)$ we use the approximation \eqref{PfixDif}, and for $\tfix(\stilde, N)$ we use \eqref{tfix}.
With this simplified diffusion approximation for $\tfix(\stilde, N)$, the integrals in \eqref{app_bartfix} are readily computed numerically. Some values are given in the legend of Fig.~\ref{fig_MeanVarEqu}.

An analytically explicit, still quite accurate, approximation of $\bartfix$ can be derived as follows. 
We use the approximations $\Pfix(s,N) \approx 1-e^{-2s}$ and $\tfix^{\text{(HP)}}(s, N) \approx \frac{2}{s} \left(\ln \bigl(2N s \bigr) + \gamma \right)$, which are valid for sufficiently large $Ns$. In addition, we assume that the selective advantage of the mutant is $s\A$ instead of $\stilde$. If $f$ is exponentially distributed with mean $\bar\A$, we obtain
\begin{linenomath}\begin{equation}
	 \int_0^{\infty}\Pfix(s\A, N) f(\A)\, d\A \approx \frac{2 s\bar\A}{1+2s\bar\A}
\end{equation}\end{linenomath}
and
\begin{linenomath}\begin{equation}
	 \int_0^{\infty}\tfix^{\text{(HP)}}(s\A, N) \Pfix(s\A, N) f(\A)\, d\A 
			\approx \frac{\ln(1+2s\bar\A)}{s\bar\A} \bigl[2\ln(2N s\bar\A) - \ln(1+2s\bar\A) \bigr]\,.
\end{equation}\end{linenomath}
Therefore,
\begin{linenomath}\begin{equation}\label{bartfix_app1}
	 \bartfix(s,\bar\A,N) \approx 
		\frac{(1+2s\bar\A)\ln(1+2s\bar\A)}{2(s\bar\A)^2} \bigl[2\ln(2N s\bar\A) - \ln(1+2s\bar\A) \bigr]\,.
\end{equation}\end{linenomath}
The high accuracy of this approximation is demonstrated in the supplementary \citetalias{Mathematica} notebook.
Assuming $s\bar\A\ll1$ and performing a series expansion in $s\bar\A$, we arrive at 
\begin{linenomath}\begin{equation}
 	 \bartfix(s,\bar\A,N) \approx 2 \ln(2Ns\bar\A) \left(\frac{1}{s\bar\A} +1 - \frac{2}{3}s\bar\A \right) - 2 \,. \label{bartfix_app2}
\end{equation}\end{linenomath}

If the selective coefficient of the mutants is $\stilde$, instead of $s\A$, then some of the above integrations cannot be performed analytically. However, numerics shows that the simple approximation \eqref{bartfix_app3} 
still works very well, especially if $2/N\le s\bar\A \le 0.2$ (see \citetalias{Mathematica} notebook).

\subsection{Mean loss time averaged over the mutation distribution $f$}\label{meanlosstime_f}

In analogy to \eqref{bartfix} and \eqref{app_bartfix}, we define the expected mean loss time of a single mutant, with effect $\A$ on the selected trait drawn from an exponential distribution $f$ with mean $\bar\A$, by  
\begin{linenomath}\begin{equation}\label{app_bartloss}
	\bartloss(s,f, N) = \frac{\int_0^{\infty}\tloss(\stilde, N) (1-\Pfix(\stilde, N)) f(\A)\, d\A}{\int_0^{\infty}(1-\Pfix(\stilde, N)) f(\A)\, d\A}\,.
\end{equation}\end{linenomath}
Here, $\tloss(\stilde, N)$ is taken from \eqref{tloss}.
With this simplified diffusion approximation for $\tloss(\stilde, N)$, the integrals in \eqref{app_bartloss} are readily computed numerically (see \citetalias{Mathematica} notebook). For graphs of $\tloss$ and $\bartloss$, see Fig.~\ref{fig_meanfixloss}. It shows that, unless $s$ is very small, $\bartloss>\tloss$. The reasons are that mutations of larger effect are lost with lower probability than those of smaller effect, and small effect mutations stay longer in the population.
The following is a simple and accurate approximation if (approximately) $Ns\bar\A>1$ and $s\bar\A<0.2$ (see \citetalias{Mathematica} notebook):
\begin{linenomath}\begin{equation}\label{app_bartloss_app}
	\bartloss(s,\bar\A, N) \approx 2(-\ln (2s\bar\A)+1+2s\bar\A)-\frac{1.42-\ln(Ns\bar\A)}{Ns\bar\A}\,.
\end{equation}\end{linenomath}
Usually, these times to loss are very small. For instance, if the mean of the exponential distribution is $\bar\A=1$, $N=10^4$, we obtain $\bartloss\approx5.5$, $8.8$, and $14.5$ for $s=0.1$, $0.01$, and $0.001$, respectively. For larger values of $N$ almost the same numbers are obtained. 

\FloatBarrier
\section{Mutations occur at their expected time - approximations for the phenotypic mean and variance}\label{app:mutation_discrete}

We outline a simpler approach than using the Erlang distribution for the waiting time to new mutations. We assume that the $i$th mutation occurs at its mean waiting time $i/\Th$ and the number of mutation events until time $\tau$ is (approximately) $[\Th\tau]$. This is most efficient if the mutation effects $\A_i$ are given, in particular if they are equal across sites. Then we can approximate \eqref{Ek} by 
\begin{linenomath}\begin{equation}\label{int_p^k_tilde_hi}
	\EV \left[  (\tilde X^{(i)}_\tau)^k \right]  \approx \bigl( 1-\Plost([ \tau - i/\Th]) \bigr) \int_0^1 x^k g_{a(\tau - i/\Th)}(x)\, dx \,.
\end{equation}\end{linenomath}
Then, instead of \eqref{barGtau}, we obtain the approximation
\begin{linenomath}\begin{align}  
	\bar{G}(\tau) &\approx \sum_{j=1}^{[\Th\tau]} \bigl(1- \Plost([\tau-i/\Th],e^{s\A_i})\bigr) \,\A_i \ga_1(a(\tau-i/\Th,e^{s\A_i})) \notag\\
		& = \sum_{j=0}^{[\Th\tau]-1} \bigl(1- \Plost([j/\Th],e^{s\A_j})\bigr) \, \A_j \ga_1(a(j/\Th,e^{s\A_j}))\,, \label{Efinal_app}
\end{align}\end{linenomath}
where we have reversed the summation order (i.e., $j=[\Th\tau]-i$) to obtain the second equation. 
This yields \eqref{Efinal} by using \eqref{Gi}. 
Analogously, we obtain instead of \eqref{VGtau} the approximation \eqref{Varfinal}.

For equal mutation effects, \eqref{Efinal} and \eqref{Varfinal} provide highly accurate approximations of \eqref{barGtau} and \eqref{VGtau}, respectively (results not shown, but expected because the right-hand sides of \eqref{Efinal} and \eqref{Varfinal} can be interpreted as approximating Riemann sums). 

\begin{figure}[th]
\centering
\begin{tabular}{ll}
A & B \\
\includegraphics[width=0.4\textwidth]{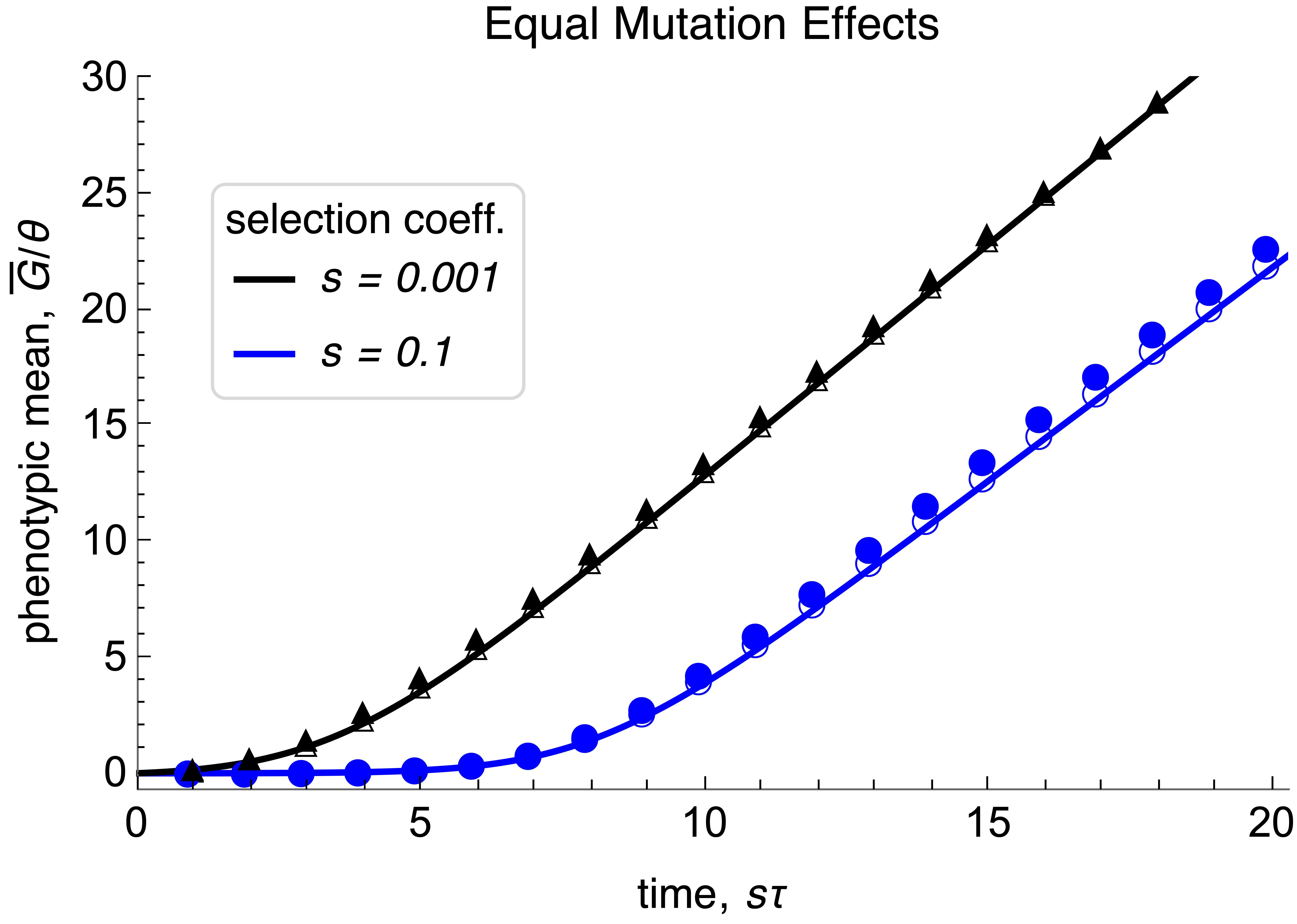} &
\includegraphics[width=0.4\textwidth]{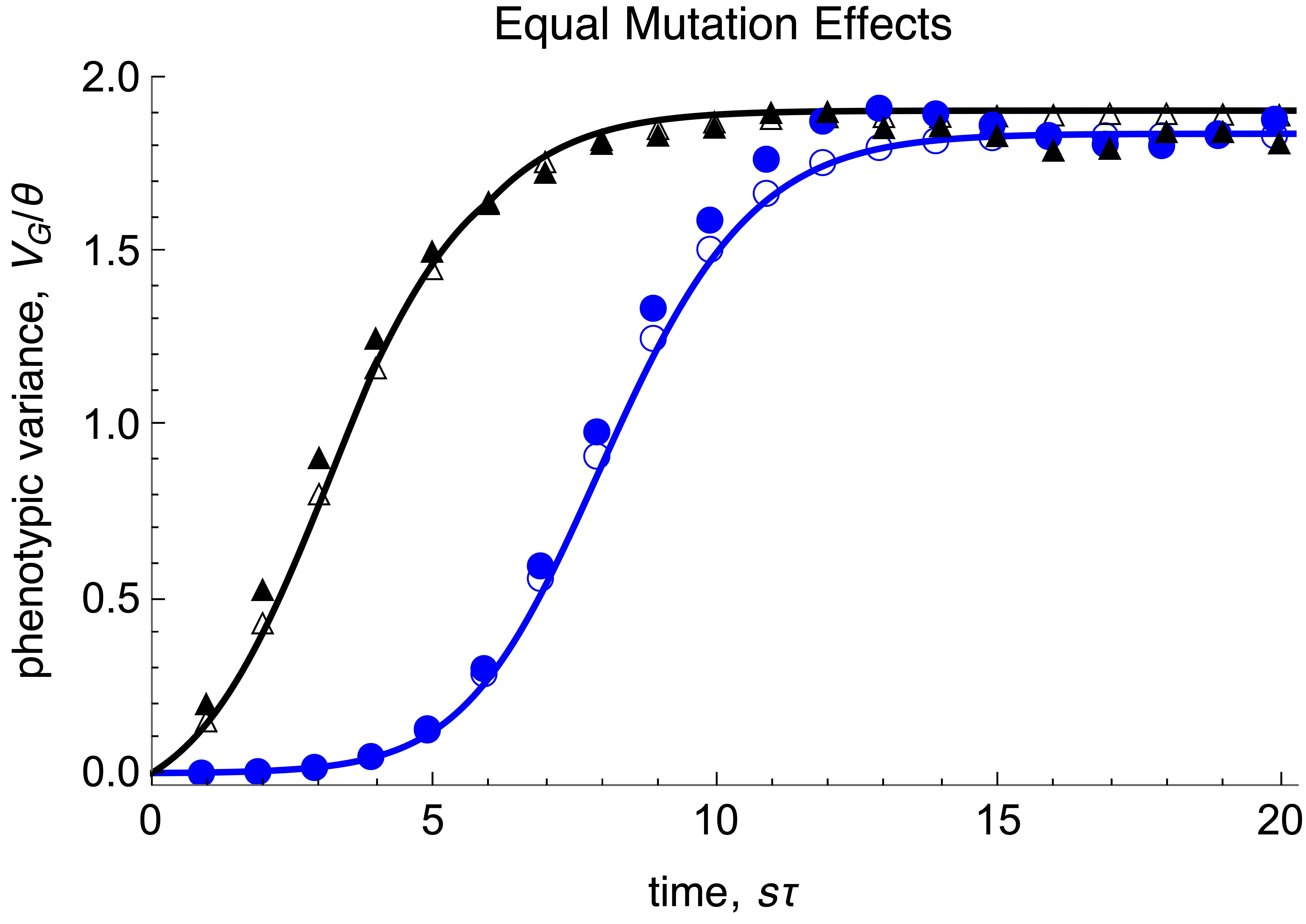} \\
\multicolumn{1}{c}{C}  \\
\multicolumn{2}{c}{
\includegraphics[width=0.45\textwidth]{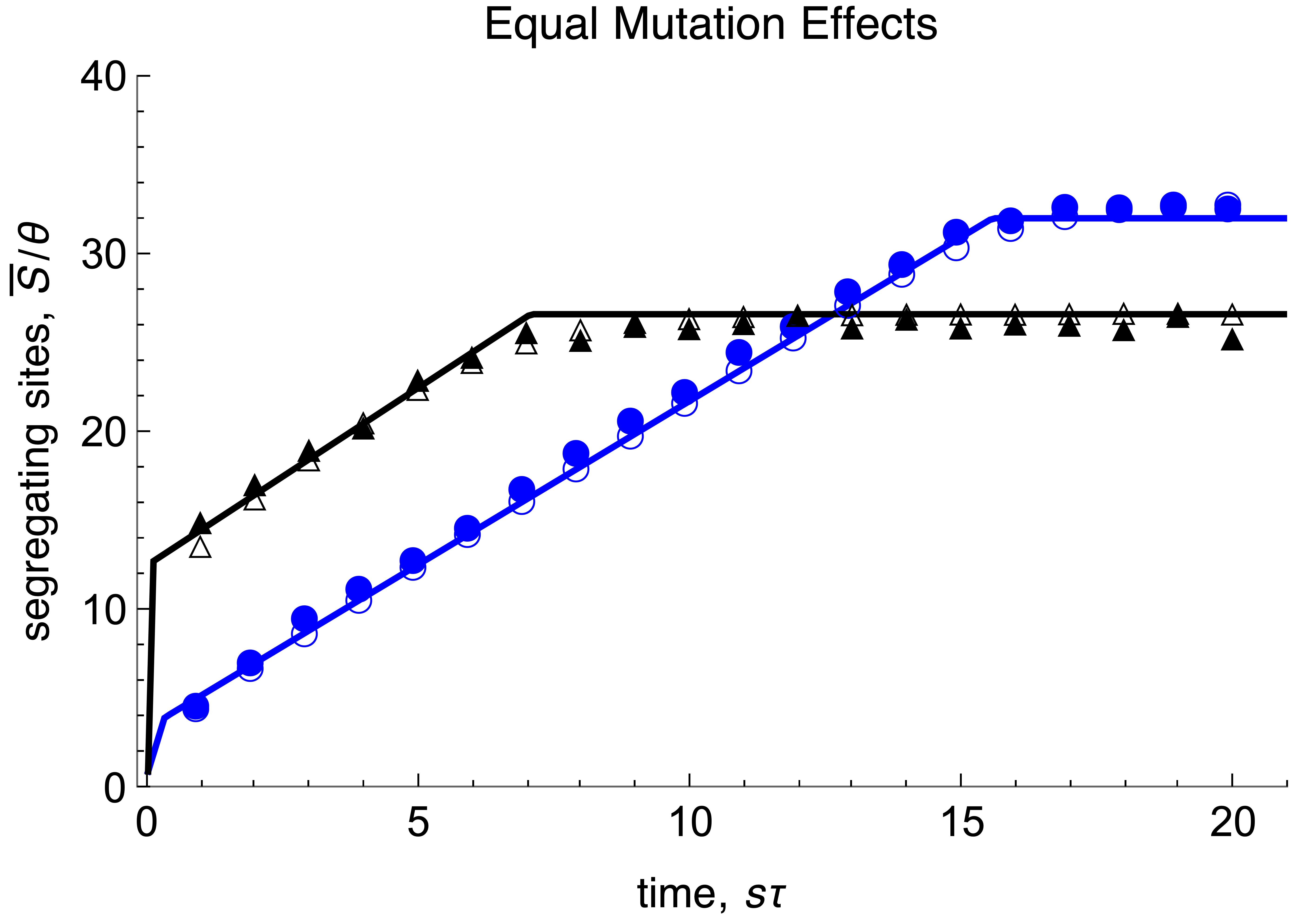}} 
\end{tabular}
\caption[$\bar{G}$ and $V_G$]{The scaled expected phenotypic mean $\bar G(s\tau)/\Th$ (panel A), the scaled expected phenotypic variance $V_G(s\tau)/\Th$ (panel B), and the scaled expected number of segregating sites $\bar S(s\tau)/\Th$ (panel C) are shown as functions of $s\tau$ for the selection coefficients $s=0.1$ and $s=0.001$ (blue and black, respectively). Equal mutation effects are assumed, i.e., $\A_i=1$. Analytic approximations, with $\bar G(s\tau)$ computed from \eqref{Efinal}, $V_G(s\tau)$ from \eqref{Varfinal}, and $\bar S(s\tau)$ from \eqref{Seg}, are shown as solid curves. Wright-Fisher-simulation results with $2\Th=10$ and $2\Th=0.01$ (open and filled symbols, respectively) are shown as symbols. Whereas $\Th$ is a multiplicative factor in the analytic approximations, these simulation results show fluctuations around the predicted curve if $2\Th=0.01$.  Presumably, the reason is that with such small $\Th$, the waiting time for successful mutations is very long, about $1/(2s\Th)$. On the time scale of the figure, this yields $1/(2\Th)=100$ units. Such rare successful mutations will entail substantial stochastic effects. We note that very slight fluctuations are also visible in Figures \ref{fig_MeanVarEqu} and \ref{fig_SegSites}, where $2\Th=1$.}
\label{fig_MeanVar}
\end{figure}

For values of $\Th$ much smaller than 1, few new mutations are expected until generation $\tau$ (unless $\tau$ is sufficiently large) and the rounded, expected number of mutation events, $[\Th \tau]$, can be zero for many generations. Moreover, the Poisson distribution is highly asymmetric for small $\Th \tau$ and gets more symmetric for larger $\Th \tau$.  This also implies that the phenotypic mean and variance may be only poorly described by their first moments when one sweep after another occurs (as with very small $\Th$). 
By Proposition~\ref{prop:general_barG_VG(tau)}, the mutation rate $\Th$ enters the expected phenotypic mean and variance only as a multiplicative factor (see also Fig.~\ref{fig_MeanVar}). Hence, we use $\bar{G}(\tau,\Th) \approx \Th\cdot\bar{G}(\tau,1)$ and $V_G(\tau,\Th) \approx \Th \cdot V_G(\tau,1)$ to obtain rough approximations for small values of $\Th$ from equations \eqref{Efinal} and \eqref{Varfinal}. 
\section{Proofs of the results in Section \ref{sec:MeanVar}}\label{Proofs_MeanVar}

Here, we give the proofs of the main results of Section \ref{sec:MeanVar}. We start by collecting some formulas needed in the proofs. Integrals and series expansions were mostly calculated with the help of \citetalias{Mathematica}. However, integrals involving $E_1$ can be obtained as well by partial integration using $\int E_1(x)\,dx = -e^{-x} + x E_1(x)$.

\subsection{Some integral formulas}

We recall the abbreviation $\sN = N\Psur(e^{\sA})$ from \eqref{nu}. In analogy to \eqref{a(t)_continuous} we define
\begin{equation}\label{ainfty}
	a^\infty(t,e^{\sA}) = \sN e^{-\sA t},
\end{equation}
where we omit the dependence on $N$. Furthermore, we introduce the (simplified) notation
\begin{linenomath}\begin{subequations}\label{gamma_infty}
\begin{equation}
	\ga_1(t,e^{\sA}) = \ga_1(a(t,e^{\sA})) \;\text{and}\; \ga_1^\infty(t,e^{\sA}) = \ga_1(a^\infty(t,e^{\sA})) 
\end{equation}
as well as 
\begin{equation}
	\ga(t,e^{\sA}) = \ga(a(t,e^{\sA})) \;\text{and}\; \ga^\infty(t,e^{\sA}) = \ga(a^\infty(t,e^{\sA})) \,.
\end{equation}\end{subequations}
\end{linenomath}
We note that $a(t,e^{\sA})>a^\infty(t,e^{\sA})$ and, because $\ga_1$ is monotone decreasing in $a$, $\ga_1^\infty(t,e^{\sA})>\ga_1(t,e^{\sA})$.

We will need the following integrals:
\begin{linenomath}\begin{subequations}\label{approx_ga_tau}
\begin{equation}
	\int_0^\tau \ga^\infty(t,e^{\sA}) \, dt
			=\frac{1}{\sA}\left[e^\sN \sN E_1(\sN)  - \exp(\sN e^{-s\A\tau}) \sN e^{-s\A\tau} E_1(\sN e^{-s\A\tau})\right]\,,
\label{approx_ga_tau1}
\end{equation}
\begin{equation}
	\lim_{\tau\to \infty}\int_0^\tau \ga^\infty(t,e^{\sA}) \, dt
			=\frac{\sN}{\sA}\, e^\sN E_1(\sN) \,.
\label{approx_ga_infty1}
\end{equation}\end{subequations}\end{linenomath}
In addition, we obtain for $T_1>0$:
\begin{linenomath}\begin{equation}
	\int_{T_1}^{\tau} \ga_1^\infty(t,e^{\sA}) \, dt
			=\frac{1}{\sA} \left[ \exp(\sN e^{-\sA\tau })E_1(\sN e^{-\sA\tau})  - \exp(\sN e^{-\sA T_1})E_1(\sN e^{-\sA T_1})  \right]
\label{approx_ga1a}
\end{equation}\end{linenomath}
and
\begin{linenomath}\begin{equation}
	\int_0^{T_1} \ga_1^\infty(t,e^{\sA}) \, dt
			=\frac{1}{\sA} \left[\exp(\sN e^{-\sA T_1})E_1(\sN e^{-\sA T_1}) - e^\sN E_1(\sN)  \right]\,.
\label{approx_ga1b}
\end{equation}\end{linenomath}
By setting $T_1=\tau-1$ in \eqref{approx_ga1a} and using $\lim_{t\to\infty}\frac{1}{s\A t}\exp(\sN e^{-\sA t})E_1(\sN e^{-\sA t}) = 1$, we arrive at
\begin{linenomath}\begin{equation}
	\lim_{\tau\to\infty} \int_{\tau-1}^{\tau} \ga_1^\infty(t,e^{\sA}) \, dt = 1\,.
\label{approx_ga1_int01}
\end{equation}\end{linenomath}

From the asymptotic properties of the exponential integral \citep[][Chap.~5.1]{Abramowitz1964}, in particular
\begin{linenomath}\begin{equation}\label{e^x E_1(x)}
	e^x E_1(x) = \sum_{k=0}^{\infty} (-1)^k \frac{k!}{x^{k+1}} = \frac{1}{x}-\frac{1}{x^2} + \frac{2}{x^3} - \frac{6}{x^4} +\ldots \;(x\to\infty)\,,
\end{equation}\end{linenomath}
we derive for arbitrary $T_1>0$
\begin{linenomath}\begin{equation*}
	\lim_{T\to\infty}
			\frac{1}{\sA T} \left[\exp(\sN e^{-\sA (T_1+T)})E_1(\sN e^{-\sA (T_1+T)}) -  \exp(\sN e^{-\sA T_1})E_1(\sN e^{-\sA T_1}) \right] = 1\,.
\end{equation*}\end{linenomath}
Combining this with \eqref{approx_ga1a}, we obtain
\begin{linenomath}\begin{equation}\label{int_tau-bartfix}
	\lim_{\tau\to\infty} \frac{1}{\tau-T_1} \int_{T_1}^{\tau} \ga_1^\infty(t,e^{\sA}) \, dt = 1 \,.
\end{equation}\end{linenomath}

\subsection{A relation between the Poisson and Erlang distribution}\label{app:Poisson_Erlang}

We prove
\begin{linenomath}\begin{equation}
	\sum_{n=1}^{\infty} \Poi_{\Th \tau}(n) \left(\sum_{i=1}^{n} \frac{ m_{i,\Th}(t)}{M_{i,\Th}(\tau)}\right) 
		=\Th \,.  \label{ErlangPoissonSum}
\end{equation}\end{linenomath}
Indeed, we have
\begin{linenomath}\begin{align}
	\sum_{n=1}^{\infty} \Poi_{\Th \tau}(n) \left(\sum_{i=1}^{n} \frac{ m_{i,\Th}(t)}{M_{i,\Th}(\tau)}\right) 
		&= \sum_{n=1}^{\infty} e^{-\Th \tau}\frac{(\Th\tau)^n}{n!}\sum_{i=1}^{n} e^{-\Th t}  \frac{\Th (\Th t)^{i-1}}{\Ga(i)M_{i,\Th}(\tau)} \notag \\
		&=\Th e^{-\Th t} \sum_{i=1}^{\infty} \frac{(\Th t)^{i-1}}{\Ga(i)M_{i,\Th}(\tau)} \sum_{n=i}^\infty e^{-\Th \tau}\frac{(\Th\tau)^n}{n!} \notag \\
		&= \Th e^{-\Th t} \sum_{i=1}^{\infty} \frac{(\Th t)^{i-1}}{\Ga(i)M_{i,\Th}(\tau)} M_{i,\Th}(\tau) 
		=\Th \,. 
\end{align}\end{linenomath}
The first equality is simply based on the definitions of $\Poi_{\Th \tau}(n)$ in \eqref{Poi(Thetatau)} and $m_{i,\Th}(t)$ in \eqref{Erlang}. The second is obtained by reordering the sums, the third from a well-known relation between the Poisson distribution and the incomplete Gamma function \citep[e.g.][6.5.13]{Abramowitz1964} and the definition of $M_{i,\Th}(\tau)$ in \eqref{Mi}, and the last from the exponential series.

\subsection{Proof of Proposition \ref{prop:general_barG_VG(tau)}}

We will need the following integrals. Let $\phi(x)$ be a (continuous) function. Then we obtain directly from the definition of $\tilde{h}^{(i)}$ by a change in the order of integration:
\begin{linenomath}\begin{equation}
	\int_0^1 \phi(x) \tilde{h}^{(i)}_{\tau,\si_i,\Th}(x)\, dx 
		= \int_0^\tau \frac{ m_{i,\Th}(t)}{M_{i,\Th}(\tau)} \bigl(1-\Plost([\tau-t],\si_i) \bigr)  \int_0^1 \phi(x) g_{a(\tau-t,\si_i)}(x)\, dx \, dt \,. \label{int_phi_htilde}
\end{equation}\end{linenomath}
With $\phi(x)=x$, we get from \eqref{mean_g} and \eqref{int_phi_htilde}
\begin{linenomath}\begin{equation} \label{xhtilde}
	\int_0^1 x \tilde{h}^{(i)}_{\tau,\si_i,\Th}(x)\, dx = 
			\int_0^\tau \frac{ m_{i,\Th}(t)}{M_{i,\Th}(\tau)} \bigl(1-\Plost([\tau-t],\si_i) \bigr) \ga_1(a(\tau-t, \si_i))\, dt \,.
\end{equation}\end{linenomath}
With $\phi(x)=x(1-x)$, we get from \eqref{var_g} and \eqref{int_phi_htilde}
\begin{linenomath}\begin{equation}\label{x(1-x)htilde}
		\int_0^1 x(1-x) \tilde{h}^{(i)}_{\tau,\si_i,\Th}(x)\, dx = 
			\int_0^\tau \frac{ m_{i,\Th}(t)}{M_{i,\Th}(\tau)} \bigl(1-\Plost([\tau-t],\si_i) \bigr) \ga(a(\tau-t, \si_i))\, dt \,.
\end{equation}\end{linenomath}

\begin{proof}[Proof of \eqref{barGtau}]
From the definition \eqref{def_barGtau} of $\bar{G}(\tau)$ we deduce:
\begin{subequations}
\begin{linenomath}\begin{align}  
	\bar{G}(\tau) &= \sum_{n=1}^{\infty} \Poi_{\Th \tau}(n) \left( \sum_{i=1}^{n} \int_0^\infty \A \EV \left[ \tilde X^{(i)}_{\tau,e^{\sA},\Th} \right] f(\A)\, d\A \right) \label{Etheor0w1} \\
	&= \sum_{n=1}^{\infty} \Poi_{\Th \tau}(n)\int_0^\infty \A f(\A) \left(\sum_{i=1}^{n} \int_0^1 x \tilde{h}^{(i)}_{\tau,e^{\sA},\Th}(x)\, dx \right) d\A \label{Etheor0w2} \\
	&= \sum_{n=1}^{\infty} \Poi_{\Th \tau}(n) \int_0^\infty \A f(\A) \notag \\
	&\qquad \times \int_0^\tau \left(\sum_{i=1}^{n} \frac{ m_{i,\Th}(t)}{M_{i,\Th}(\tau)}\right) \bigl(1-\Plost([\tau-t],e^{\sA}) \bigr) \ga_1(a(\tau-t, e^{\sA}))\, dt \, d\A \label{Etheor0w3} \\
	&= \Th \int_0^\infty \A f(\A) \int_0^\tau \bigl(1-\Plost([\tau-t],e^{\sA}) \bigr) \ga_1(a(\tau-t, e^{\sA}))\, dt \, d\A \,. \label{barGtau_app}
\end{align}\end{linenomath}\label{barGtau_deriv}
\end{subequations}
We obtained \eqref{Etheor0w1} by using that the mutation effects $\A_i$ are drawn independently from the same distribution $f$; \eqref{Etheor0w2} by using \eqref{Ek}; and \eqref{Etheor0w3} by employing \eqref{xhtilde} and changing the order of integration and summation. Then \eqref{Etheor0w3} yields \eqref{barGtau_app} by applying \eqref{ErlangPoissonSum}, and \eqref{barGtau_app} yields \eqref{barGtau} by applying the time transformation $\tau-t\to T$, and returning to $t$ (instead of $T$). 
Note also that we suppressed the dependence on $N$ in $\tilde X^{(i)}$, $\tilde h^{(i)}$, and $a$.
\end{proof}

\begin{proof}[Proof of \eqref{VGtau}]
Analogously, by using \eqref{x(1-x)htilde} instead of \eqref{xhtilde}, we obtain for the variance from its definition \eqref{def_VGtau},
\begin{subequations}
\begin{linenomath}\begin{align}  
	V_G(\tau) & =  
		\sum_{n=1}^{\infty} \Poi_{\Th \tau}(n) \left( \sum_{i=1}^{n} \int_0^\infty \A^2 \EV \left[ \tilde X^{(i)}_{\tau,e^{\sA},\Th}(1-\tilde X^{(i)}_{\tau,e^{\sA},\Th} ) \right] f(\A)\, d\A \right) \notag \\
		&=\sum_{n=1}^{\infty} \Poi_{\Th \tau}(n) \int_0^\infty \A^2 f(\A) \left( \sum_{i=1}^{n} \int_0^1 x(1-x) \tilde{h}^{(i)}_{\tau,e^{\sA},\Th}(x)\, dx \right) d\A \notag\\
		&=\sum_{n=1}^{\infty} \Poi_{\Th \tau}(n) \int_0^\infty \A^2 f(\A) \notag\\
		&\qquad \times\int_0^\tau \left( \sum_{i=1}^{n}\frac{ m_{i,\Th}(t)}{M_{i,\Th}(\tau)}\right) \bigl(1-\Plost([\tau-t],\si_i) \bigr) \ga(a(\tau-t, \si_i))\, dt   \, d\A \label{Vartheor0w3} \\
		&=\Th \int_0^\infty \A^2 f(\A) \int_0^\tau \bigl(1-\Plost([\tau-t],e^{\sA}) \bigr) \ga(a(\tau-t, e^{\sA}))\, dt \, d\A\,, \label{VGtau_app}
\end{align}\end{linenomath}
\end{subequations}
where \eqref{Vartheor0w3} yields \eqref{VGtau_app} with the help of \eqref{ErlangPoissonSum}. Finally, \eqref{VGtau} is obtained by the time transformation $\tau-t\to T$ (and returning from $T$ to $t$).
Note that for the derivation of the variance, the independence of the random variables $\tilde X^{(i)}$ is crucial.
\end{proof}

\subsection{Proof of Proposition \ref{prop:equilibrium}}

\begin{proof}[Proof of \eqref{barG*_prop}]
From \eqref{barGtau} we obtain
\begin{linenomath}\begin{equation}\label{G(tau+1)-G(tau)}
	\bar{G}(\tau+1) -\bar{G}(\tau) = \Th \int_0^\infty \A f(\A) H(e^{\sA},\tau)\, d\A \,,
\end{equation}\end{linenomath}
where
\begin{linenomath}\begin{align}
	H(e^{\sA},\tau) &=  \int_0^{\tau+1} \bigl(1-\Plost([t],e^{\sA}) \bigr) \ga_1(t, e^{\sA})\, dt \notag \\
			&\quad - \int_0^{\tau} \bigl(1-\Plost([t],e^{\sA}) \bigr) \ga_1(t, e^{\sA})\, dt \notag \\
			&=\int_\tau^{\tau+1} \bigl(1-\Plost([t],e^{\sA}) \bigr) \ga_1(t, e^{\sA})\, dt \,. 	\label{G(a,tau)}
\end{align}\end{linenomath}
Now we use $\lim\limits_{\tau\to\infty} \bigl(1-\Plost([\tau],e^{\sA})\bigr) = \Psur(e^{\sA})$, $\lim\limits_{\tau\to\infty} \dfrac{\ga_1^\infty(\tau, e^{\sA})}{\ga_1(\tau, e^{\sA})} =1$ (by continuity of $\ga_1$ and the fact that $\lim\limits_{\tau\to\infty} \dfrac{a^\infty(\tau, e^{\sA})}{a(\tau, e^{\sA})} =1$), and 
\begin{linenomath}\begin{equation}
	\lim_{\tau\to\infty} \int_\tau^{\tau+1} \ga_1(t, e^{\sA})\, dt = \lim_{\tau\to\infty} \int_\tau^{\tau+1} \ga_1^\infty(t, e^{\sA})\, dt = 1\,, 
\end{equation}\end{linenomath}
where we used \eqref{approx_ga1_int01}. 
Therefore, $\lim\limits_{\tau\to\infty} H(e^{\sA},\tau) = \Psur(e^{\sA})$ and \eqref{G(tau+1)-G(tau)} yields
\begin{linenomath}\begin{equation}
	\De\bar G^* = \lim_{\tau\to\infty} \Bigl(\bar{G}(\tau+1) -\bar{G}(\tau)\Bigr) = \Th \int_0^\infty \A f(\A) \Psur(e^{\sA})\, d\A \,,
\end{equation}\end{linenomath}
which is \eqref{barG*_prop}.
\end{proof}

\begin{proof}[Proof of \eqref{VG*_prop}]
We start from
\begin{linenomath}\begin{equation}
	V_G(\tau)  = \Th \int_0^\infty \A^2 f(\A) \int_0^\tau \bigl(1-\Plost([t],e^{\sA}) \bigr) \ga(t, e^{\sA})\, dt \, d\A\,.	\tag{\ref{VGtau}}
\end{equation}\end{linenomath}
In order to derive \eqref{VG*_prop}, we need to estimate
\begin{linenomath}\begin{align}\label{D_V}
D_V(\tau) &=  \int_0^\infty \A^2 f(\A) \int_0^\tau \bigl(1-\Plost([t],e^{\sA}) \bigr) \ga(t, e^{\sA})\, dt \, d\A \ \notag\\
	&\quad  -\int_0^\infty \A^2 f(\A) \int_0^\tau \bigl(1-\Pexinf(e^{\sA}) \bigr) \ga^\infty(t, e^{\sA})\, dt \, d\A \notag\\
	&= \int_0^\infty \A^2 f(\A) \int_0^\tau I_V(t,e^{\sA})\, dt \, d\A
\end{align}\end{linenomath}
as $\tau\to\infty$, where
\begin{linenomath}\begin{equation}
	I_V(t,e^{\sA}) = \bigl(1-\Plost([t],e^{\sA}) \bigr) \ga(t, e^{\sA}) - \bigl(1-\Pexinf(e^{\sA}) \bigr) \ga^\infty(t, e^{\sA})\,. \label{I_V}
\end{equation}\end{linenomath}
With $D_V(\infty)=\lim\limits_{\tau\to\infty}D_V(\tau)$ we obtain
\begin{linenomath}\begin{align}
	V_G^*/\Th = \lim_{\tau\to\infty }V_G(\tau)/\Th 
		&= \int_0^\infty \A^2 f(\A) \bigl(1-\Pexinf(e^{\sA}) \bigr) \int_0^\infty \ga^\infty(t, e^{\sA})\, dt \, d\A + D_V(\infty)\notag\\
		&= \int_0^\infty \frac{\A}{s} f(\A) \Psur(e^{\sA}) \sN e^\sN E_1(\sN) \, d\A  + D_V(\infty) \,, 
\end{align}\end{linenomath}
where we have used \eqref{approx_ga_infty1} in the last step. Because $V_G^*/\Th$ is of order 1 ($x e^x E_1(x)<1$ and $\Psur\le\rho\sA$), proving that $|D_V(\infty)|$ becomes arbitrarily small as $N\to\infty$ (in a sense to be specified), is sufficient to establish the desired result.

In view of the approximations \eqref{VG*Nsconst} and \eqref{VG*Ns}, it is desirable to have an error of order $o(s)+o(1/(Ns))$ as $Ns\to\infty$ and $s\to0$. For this reason, we will need the scaling Assumption \ref{A1}, i.e., we start with $Ns^K=C^K$ as $N\to\infty$, where $C>0$ and $K>1$ are constants, and $K$ will be determined at the end of the proof.

(1) In this first main step we fix the mutational effect $\A$. Then we deal with the complications arising from drawing $\A$ from a distribution $f$ with mean $\barA$ in a second main step.

Let $\ep>0$ be sufficiently small, where we will quantify this at the end of step (1). 
\textblue{Motivated by \eqref{Tep_FL_general}, we define 
\begin{equation}
	T_\ep = \frac{\ln\Bigl(\Pexinftyfl \Bigl(1+\frac{1}{\ep} \Bigr)\Bigr)}{\ln\si}   \label{T_ep_gen}
\end{equation}
for offspring distributions with mean $\si$ that satisfy the inequality \eqref{Psurvn/Psurinfty}. 
By simple algebra this is equivalent to $\ep = \frac{\Pexinf}{\si^{T_\ep}-\Pexinf}$.
Therefore \eqref{Psurvn/Psurinfty} implies
\begin{linenomath}\begin{equation}\label{Tep_general_ineq}
	1-\Pexinf < 1-\Plost([t]) \le (1+\ep) \bigl(1-\Pexinf \bigr) \;\text{ for every } t\ge T_\ep\,,
\end{equation}\end{linenomath}
where here and below we suppress the dependence of $\Plost([t])$, $\Pexinf$, and $T_\ep$ on $\si=e^{s\A}$. For later use, we note that 
\begin{linenomath}\begin{equation}
	e^{s\A T_\ep} = \Pexinftyfl\frac{1+\ep}{\ep} \,. \label{e^saTep} 
\end{equation}\end{linenomath}
}

In the limit $\tau\to\infty$, we split the integral with respect to time in \eqref{D_V} as follows
\begin{linenomath}\begin{equation}
	\int_0^\infty I_V(t,e^{\sA})\, dt  = \int_0^{T_\ep} I_V(t,e^{\sA})\, dt +\int_{T_\ep}^\infty I_V(t,e^{\sA}) \, dt\,. \label{int_T1_infty}
\end{equation}\end{linenomath}

(i) First, we treat $\int_{T_\ep}^\infty I_V(t,e^{\sA}) \, dt$ by assuming $t\ge T_\ep$. We recall the notation from \eqref{ainfty} and \eqref{gamma_infty}. Then \eqref{Tep_general_ineq} implies $a^\infty(t,e^{\sA}) < a(t,e^{\sA}) \le (1+\ep)a^\infty(t,e^{\sA})$. Using
\begin{linenomath}\begin{equation}
	a(t) = (1+u(t)\ep)a^\infty(t)\,,
\end{equation}\end{linenomath}
where $0<u(t)\le1$ if $t\ge T_\ep$ (indeed, $u(t)\to0$ as $t\to\infty$, and for a fractional linear offspring distribution $u(t)$ can be calculated explicitly from \eqref{Psurvn/Psurinfty_FL_app}), 
we obtain
\begin{linenomath}\begin{equation}
	\bigl(1-\Plost([t]) \bigr) \ga(a(t)) = (1+u(t)\ep)(1-\Pexinf)\,\ga^\infty\bigl((1+u(t)\ep)a^\infty(t)\bigr)\,.
\end{equation}\end{linenomath}
This yields
\begin{linenomath}\begin{align}\label{I_V(t)_ep}
	I_V(t) &=  (1+u(t)\ep)(1-\Pexinf)\,\ga^\infty\bigl((1+u(t)\ep)a^\infty(t)\bigr) - (1-\Pexinf) \ga^\infty(a^\infty(t)) \notag \\
		&= \ep u(t) (1-\Pexinf) \ga_D(a^\infty(t))+ \ep^2u(t)^2(1-\Pexinf)R_1(a^\infty(t)) \,,
\end{align}\end{linenomath}
where
\begin{linenomath}\begin{equation}\label{gamma_D}
	\ga_D(x) = x\Bigl[\Bigl(2+4x+x^2\Bigr) e^x E_1(x) - 3 - x \Bigr] 
\end{equation}\end{linenomath}
and $R_1$ are obtained by series expansion for $\ep\to0$. The function $\gamma_D(x)$ is positive and bounded (by $\approx0.217$) and can be integrated. If we write $T_\ep= \ln(Z)/(s\A)$ (with $Z=\Pexinf (1+\ep)/\ep$) and recall $a^\infty(t)=\nu e^{-s\A t}$ (with $\sN=N(1-\Pexinf)$), we obtain
\begin{linenomath}\begin{align}\label{int_gamma_D}
	\int_{T_\ep}^\infty \ga_D(a^\infty(t)) \,dt &= -\frac{\sN/Z}{s\A}\Bigl(1-e^{\sN/Z}(2+\sN/Z)E_1(\sN/Z)  \Bigr) \notag \\
			& = \frac{1}{\sA}\Bigl(1- 2(\sN/Z)^{-2} + O\Bigl((\sN/Z)^{-3}\bigr)\Bigr)
\end{align}\end{linenomath}
as $\sN/Z\to\infty$. In essentially the same way
\begin{linenomath}\begin{align}\label{int_gamma_R1}
	\int_{T_\ep}^\infty R_1(a^\infty(t))\,dt \le\frac{4}{s\A}(\sN/Z)^{-2}
\end{align}\end{linenomath}
follows.

In order to satisfy $\sN/Z\to\infty$, we need $Ns\ep\to\infty$ as $N\to\infty$. In particular, it follows from the above results that $I_V(t)>0$ if $t\ge T_\ep$. Therefore, we obtain
\begin{linenomath}\begin{align}\label{int_Tep_infty_IV}
	\int_{T_\ep}^\tau I_V(t) \,dt & \le \int_{T_\ep}^\infty I_V(t) \,dt \notag \\
		& \le \ep (1-\Pexinf) \int_{T_\ep}^\infty \ga_D(a^\infty(t)) \,dt + \ep^2(1-\Pexinf)\int_{T_\ep}^\infty R_1(a^\infty(t))\,dt \notag \\
		& \le \ep \rho \Bigl(1 + O\bigl((N\ep)^{-2}\bigr)\Bigr) + \ep^2 O\Bigl((N\ep)^{-2} \Bigr)  \le 2 \ep \rho \,.
\end{align}\end{linenomath}
Here we used \eqref{I_V(t)_ep} together with $0<u(t)\le1$ in the second estimate, and \eqref{int_gamma_D} and \eqref{int_gamma_R1} together with $1-\Pexinf\le\rho s\A$ (Remark~\ref{Psur=csalpha}) in the third. 

Thus, to achieve our desired result, we will need the conditions $\ep=o(s)$, $\ep=o((Ns)^{-1})$, and $(N\ep)^{-1} = o(1)$.

(ii) Second, we derive an estimate for $\int_0^{T_\ep} I_V(t,e^{\sA})\, dt$, where we recall $I_V(t,e^{\sA})$ from \eqref{I_V}. For small $t$, $a(t)$ is large because it is decreasing from $N$ at $t=0$ to $a(T_\ep)\approx \rho \A N s\ep$ (if $s\A$ is small), where we recall $a(t)=N(1-\Plost([t],e^{\sA}))e^{-\sA t}$. We will choose $\ep$ and the scaling factor $K$ such that $N s\ep\to\infty$ as $N\to\infty$.  Then we use the approximation $\ga(a(t))=\frac{1}{a(t)}-\frac{4}{a(t)^2} \textblue{+\frac{18}{a(t)^3} +O(a(t)^{-4})}$ which is readily derived from \textblue{\eqref{var_g}} and \eqref{e^x E_1(x)}. Therefore, we obtain by using the expansion of $\ga$:
\begin{linenomath}\begin{align}
	I_V(t,e^{\sA}) &= \bigl(1-\Plost([t]) \bigr) \ga(t) - \bigl(1-\Pexinf \bigr) \ga^\infty(t) \notag \\
	& = \Bigl(\frac{e^{\sA t}}{N} - \frac{4e^{2\sA t}}{N^2(1-\Plost([t]))} \Bigr) - \Bigl(\frac{e^{\sA t}}{N} - \frac{4e^{2\sA t}}{N^2(1-\Pexinf)} \Bigr) + O(a(t)^{-3}) \notag \\
	& = \frac{4e^{2\sA t} }{N^2} \Bigl(\frac{1}{1-\Pexinf} -  \frac{1}{1-\Plost([t])} \Bigr) + O(a(t)^{-3})\,, \label{I_V_1}
\end{align}\end{linenomath}
where again we have suppressed the dependence on $e^{\sA}$ in the terms $a$ and $\Plost$. 

From \eqref{Psurvn/Psurinfty} we obtain by a brief calculation
\begin{equation}
	  \frac{1}{1-\Pexinf} - \frac{1}{1-\Plost([t])} \le \frac{\Pexinf}{1-\Pexinf}\, e^{-\sA t} \,.
\end{equation}
Now we integrate $I_V(t,e^{\sA})$ with respect to $t$:
\begin{linenomath}\begin{align}\label{int_0_Tep_IV}
	\int_0^{T_\ep}I_V(t,e^{\sA})\,dt &\le \frac{4}{N^2}\, \frac{\Pexinf}{1-\Pexinf} \int_0^{T_\ep} e^{\sA t}\,dt + R\notag \\
		& = \frac{4}{N^2}\,\frac{\Pexinf}{1-\Pexinf}\, \frac{e^{s\A T_\ep}-1}{s\A} + R \notag \\
		& = \frac{4}{N^2}\,\frac{\Pexinf}{1-\Pexinf}\, \frac{\Pexinf(1+\ep)-\ep}{s\A\ep} + R \notag \\
		&\le \frac{4}{N^2s\A\ep(1-\Pexinf)} + R \notag\\
		&\le \frac{8}{\rho\A^2}\frac{1}{(Ns)^2\ep} + \frac{1}{\textblue{\rho^2\A^3}}\,O\Bigl(\frac{1}{(Ns)^3\textblue{\ep^2}}\Bigr) \,, 
\end{align}\end{linenomath}
where $R$ arises from integration of terms of order $a(t)^{-3}$ or smaller \textblue{(see below)}, hence is of smaller order than the first term, and \textblue{\eqref{e^saTep} was used}. In the final step we also used $(1-\Pexinf)\ge\tfrac12\rho s\A$ for small $s\A$ (Remark~\ref{Psur=csalpha}). To keep the error in \eqref{int_0_Tep_IV} sufficiently small, we need that $((Ns)^2\ep)^{-1} = o((Ns)^{-1})$ (and $((Ns)^2\ep)^{-1} =o(s)$, which is weaker).

\textblue{\begin{proof}[Derivation of the estimate of $R$] We assume that $\A$ is fixed. The $O(a(t)^{-3})$ term in \eqref{I_V_1} is
\begin{linenomath}\begin{equation}
	\frac{18e^{3\sA t} }{N^3} \biggl[\frac{1}{(1-\Plost([t]))^2} - \frac{1}{(1-\Pexinf)^2} \biggr]\,,
\end{equation}\end{linenomath}
which is negative.
From \eqref{Psurvn/Psurinfty}, we obtain 
\begin{linenomath}\begin{equation*}
	\frac{(1-\Plost([t]))^2}{(1-\Pexinf)^2} \le \biggl(1+ \frac{\Pexinf}{\si^t-\Pexinf} \biggr)^2
\end{equation*}\end{linenomath}
which, by inversion and further rearrangement, becomes
\begin{linenomath}\begin{align}
	\frac{1}{(1-\Pexinf)^2} - \frac{1}{(1-\Plost([t]))^2}  &\le 2\si^{-t} \frac{\Pexinf}{(1-\Pexinf)^2} - \si^{-2t} \frac{(\Pexinf)^2}{(1-\Pexinf)^2} \notag \\ 
	&< 2\si^{-t} \frac{\Pexinf}{(1-\Pexinf)^2} \,. \label{R_estimate_1}
\end{align}\end{linenomath}
Therefore, we obtain
\begin{linenomath}\begin{subequations}\begin{align}
	|R| &= \frac{18}{N^3}\int_0^{T_\ep} e^{3\sA t} \biggl[\frac{1}{(1-\Pexinf)^2} -\frac{1}{(1-\Plost([t]))^2} \biggr] \, dt \label{|R|a} \\
		&\le \frac{18}{N^3} \int_0^{T_\ep} 2e^{2\sA t} \frac{\Pexinf}{(1-\Pexinf)^2} \, dt \notag\\
		&= \frac{36}{N^3} \frac{\Pexinf}{(1-\Pexinf)^2} \frac{e^{2s\A T_\ep}-1}{2s\A}  \notag \\
		&\le \frac{18}{N^3} \frac{\Pexinf}{(1-\Pexinf)^2} \frac{(\Pexinf)^2(1+\ep)^2-\ep^2}{s\A\ep^2} \label{|R|b}\\
		&\le \frac{36}{N^3} \frac{1}{(1-\Pexinf)^2} \frac{1}{s\A\ep^2} \label{|R|c}\\
		&\le \frac{144}{\rho^2\A^3}\frac{1}{(Ns)^3\ep^2} \,. \label{|R|} 
\end{align}\end{subequations}\end{linenomath}
Here, we used \eqref{e^saTep} to arrive at \eqref{|R|b}, and $(1-\Pexinf)\ge\tfrac12\rho s\A$ to arrive at \eqref{|R|}.
\end{proof}}

In addition to $Ns^K=C^K$, choose $\ep=s^{K_1}$, where $K_1>0$. The lower bound $K>1$ arises from the fact that in our model selection is stronger than random genetic drift. With this choice, we obtain $s=O(N^{-1/K})$, $\ep=O(N^{-K_1/K})$, $1/(Ns)=O(N^{-1+1/K})$, $1/((Ns)^2\ep) = O(N^{-2+(2+K_1)/K})$, and $Ns\ep=O(N^{1-(1+K_1)/K})$.
To keep all error terms sufficiently small, we need $\ep=o(s)$, $1/((Ns)^2\ep)=o((Ns)^{-1})$, $N\ep\to\infty$ and $Ns\ep\to\infty$ as $N\to\infty$. These asymptotic properties are satisfied if and only if $K_1>1$ and $K>K_1+1$. 
With this choice, we get error terms of smaller order than $s$ and $a(T_\ep)=\rho\A Ns\ep\to\infty$ (as $N\to\infty$), which is required in (i) and (ii) above. 
In summary, we obtain
\begin{linenomath}\begin{subequations}\label{order_of_integrals_of I_V}
\begin{align}
	\int_0^{T_\ep}I_V(t,e^{\sA})\,dt &=  O\Bigl(N^{-2+(2+K_1)/K} \Bigr) \,, \label{int_IV_0Tep_error}\\
	\int_{T_\ep}^\infty I_V(t,e^{\sA})\,dt &=  O\Bigl(N^{-K_1/K} \Bigr)\,, \label{int_IV_Tepinfty_error}
\end{align}
\end{subequations}\end{linenomath}
where $K_1/K<2-(2+K_1)/K$ because $K_1>1$ and $K>K_1+1$.
A simple choice for $K_1$ and $K$ is $K_1=\tfrac32$ and $K=3$. Then $s=O(N^{-1/3})$, $\ep=O(N^{-1/2})$, $1/(Ns)=O(N^{-2/3})$, $1/((Ns)^2\ep) =1/( (Ns^3)(Ns^{1/2}))= O(N^{-5/6})$, and $Ns\ep=O(N^{1/6})$. Then the error term in \eqref{int_IV_0Tep_error} is $O(N^{-5/6})$, and that in \eqref{int_IV_Tepinfty_error} is $O(N^{-1/2})$.

(2) In this second main step we assume that \textblue{the distribution of mutation effects $\A$ has density $f$, mean $\barA$, and all higher moments are finite.} We adopt the scaling assumption from step (1) with $K_1>1$ and $K>K_1+1$.

(i) First we treat the case $t\le T_\ep$. 
We assume $Ns>1$ and set \textblue{$z= K_2\ln(Ns)$, where $K_2\ge1$ is a constant such that $z>1$, and} we choose $s$ small enough such that $\Psur\ge\tfrac12\rho s\A$ holds for every $\A\le z\barA$. This is possible by Remark \ref{Psur=csalpha} and our scaling assumption \textblue{\eqref{scaling_N_s}}. Furthermore, we choose 
\textblue{
\begin{linenomath}\begin{equation}
	\ep = \frac{r(N)}{\min\{\A,z\barA\}} \text{ and } r(N)=N^{-K_1/K} \,.  \label{def_ep_r(N)}
\end{equation}\end{linenomath}
}
Assuming $\A\le z\barA$ and using \textblue{\eqref{e^saTep}}, we obtain
\begin{linenomath}\begin{align}
	a(T_\ep) &= N (1+\ep)\Psur e^{-s\A T_\ep} = N (1+\ep) \frac{\Psur}{\Pexinf}\frac{\ep}{1+\ep} \notag \\
	&\ge  \tfrac12\rho Ns\, r(N) \frac{\A}{\min\{\A,z\barA\}} = \tfrac12\rho \, Ns r(N) \,,
\end{align}\end{linenomath}
and this inequality holds for every $t\le T_\ep$. 
Therefore, the bound (and approximation) $\ga(a(t))\le\frac{1}{a(t)}$ will be accurate in this parameter region since $a(t) \to \infty$ as $N \to \infty$. 
 
Therefore, we can use \eqref{int_0_Tep_IV} and obtain
\begin{linenomath}\begin{align}\label{int_IV_part1}
	\int_0^{z\barA}\A^2 f(\A)\int_0^{T_\ep}I_V(t,e^{\sA})\,dt\,d\A 
		& \le  \frac{2\tilde c}{(Ns)^2 r(N)} \int_0^{z\barA}\A^2 f(\A) \frac{\A}{\A^2} \, d\A \,,  \notag \\
		& \le \frac{2\tilde c\barA }{(Ns)^2 r(N)} \notag \\
		& = O\Bigl( N^{-2+(2+K_1)/K} \Bigr)\,,
\end{align}\end{linenomath}
where we have used $\sA T_\ep \le \ln\Bigl(1+\frac{\min\{\A,z\barA\}}{r(N)} \Bigr) = \ln\Bigl(1+\frac{\A}{r(N)} \Bigr)$ if $\A\le z\barA$, $\tilde c>1$ is a constant which also accounts for the error term $R$, \textblue{and our scaling assumption \eqref{scaling_N_s}}. Thus, the order of this error term is the same as in \eqref{int_IV_0Tep_error}.

\textblue{
If $\A>z\barA$, we use $1-\Pexinf\ge\tfrac12\rho sz\barA$. This inequality indeed holds because $sz\to0$ as $N\to\infty$ and $\ep = r(N)/(z \barA)$. Then, we obtain from the second-but-last line in \eqref{int_0_Tep_IV} and from \eqref{|R|c}:
\begin{linenomath}\begin{align*}
	\int_0^{T_\ep}I_V(t,e^{\sA})\,dt &\le \frac{4}{N^2s\A\ep(1-\Pexinf)} + R \notag\\
		&\le \frac{8}{\rho\A}\frac{1}{(Ns)^2 r(N)} + \frac{144}{\rho^2\A} \frac{1}{(Ns)^3 r(N)^2} \notag \\
		&\le \frac{C}{\A}\frac{1}{(Ns)^2 r(N)}\,,
\end{align*}\end{linenomath}
where $C$ is an appropriate constant (independent of $\A$, $N$ and $s$).
As a consequence we derive
\begin{linenomath}\begin{align}\label{int_IV_part2}
	&\int_{z\barA}^\infty \A^2 f(\A) \int_0^{T_\ep} I_V(t)\, dt \, d\A 
	 \le \int_{z\barA}^\infty \A^2 f(\A) \frac{C}{\A}\frac{1}{(Ns)^2 r(N)} \, d\A \notag \\
	&\qquad=  \frac{C}{(Ns)^2 r(N)} \int_{z\barA}^\infty \A f(\A)\, d\A  \notag \\
	&\qquad\le \frac{C}{(Ns)^2 r(N)}\frac{\A_2(f)}{z\barA} 	\notag \\
	&\qquad= \frac{\tilde C}{(Ns)^2 r(N) \ln(Ns)} =  o\biggl( \frac{1}{(Ns)^2 r(N)} \biggr) \text{ as } N\to\infty\,,  
\end{align}\end{linenomath}
where $\A_2(f)$ denotes the second moment of the mutation distribution. Importantly, the term in the last line of \eqref{int_IV_part2} is of smaller order than the term $\frac{1}{(Ns)^2 r(N)}$ obtained in \eqref{int_IV_part1}. Thus,
\begin{linenomath}\begin{equation}\label{int_IV_part1+2}
	\int_0^\infty\A^2 f(\A)\int_0^{T_\ep}I_V(t,e^{\sA})\,dt\,d\A = O\Bigl( N^{-2+(2+K_1)/K} \Bigr) \text{ as } N\to\infty\,,
\end{equation}\end{linenomath}
which is analogous to \eqref{int_IV_0Tep_error}.
}

(ii) Now we assume $t>T_\ep$. Then we obtain from \eqref{int_Tep_infty_IV}
\begin{linenomath}\begin{align}\label{int_IV_part3}
	\int_0^\infty \A^2 f(\A) \int_{T_\ep}^\tau I_V(t) \,dt \,d\A 
	& \le  2 \rho \int_0^\infty \A^2 f(\A) \frac{r(N)}{\min\{\A,z\bar{\A}\}} \,d\A \notag \\
	& =  2 \rho r(N) \int_0^{z\bar{\A}} \A f(\A) \,d\A + 
	\frac{2\rho}{z\bar{\A}} r(N) \int_{z\bar{\A}}^\infty \A^2 f(\A) \,d\A \notag \\
	& <  2 \rho r(N) \bar{\A} + 2\rho\frac{\A_2(f)}{\bar{\A}} r(N) 	= O\Bigl(N^{-K_1/K}\Bigr) 
\end{align}\end{linenomath}
(because $z\ge 1$). With our standard choice of $K_1=\tfrac32$ and $K=3$, this is $O(N^{-1/2})$.

\textblue{In summary, we obtain by the same reasoning as below \eqref{order_of_integrals_of I_V} that $0<D_V(\infty)\le O(N^{-K_1/K})$ as $N\to\infty$. This finishes the proof of eq.~\eqref{VG*_prop}.}
\end{proof}

Because for fractional linear offspring distributions with $v=2m/\rho$, we have $\Pexinf=0$ if $m\ge1+2/\rho$ (Appendix \ref{app_frac_lin_compare}), in some of the integrals above the upper bound $\infty$ should be reduced accordingly because then $I_V(t)=0$ for large $\A$.

\begin{rem} \label{Rem:barG*}
We provide an approximation for the time dependence of $\bar{G}(\tau)$ for sufficiently large $\tau$.
We assume $\tau\gg\bartfix$ (in particular, $\tau>\tau_c$) and suppress the dependence of $a$ and $\nu$ on $N$.
Then we obtain from \eqref{barGtau}
\begin{linenomath}\begin{subequations}\begin{align}
	\bar{G}(\tau) &= \Th \int_0^\infty \A f(\A)  \int_0^\tau \bigl(1-\Plost([t],e^{\sA}) \bigr) \ga_1(a(t, e^{\sA}))\, dt \, d\A  \notag \\
		&\approx \Th \int_0^\infty \A f(\A) \bigl(1-\Pexinf(e^{\sA})\bigr) \int_0^\tau \ga_1(a(t, e^{\sA})) \,dt\,  d\A  \label{barG_new2}\\
		&=\Th \int_0^\infty \A f(\A) \Psur(e^{s\A}) \biggl(\int_{\bartfix}^{\tau} \ga_1(a(t))\,dt
			+ \int_0^{\bartfix} \ga_1(a(t)) \,dt \biggr)  d\A \label{barG_new3}\\
		&\approx \Th(\tau-\bartfix) \int_0^\infty \A f(\A) \Psur(e^{s\A}) \, d\A \notag \\
		&\quad + \Th  \int_0^\infty f(\A) \frac{1}{s}\Psur(e^{s\A}) \left[\exp(\sN e^{-\sA\bartfix})E_1(\sN e^{-\sA\bartfix}) - e^\sN E_1(\sN)  \right]\,  d\A \,. \label{barG_new}
\end{align}\end{subequations}\end{linenomath}
Here, we used $\Plost([\tau-t],e^{\sA})\approx \Pexinf(e^{\sA})$ for sufficiently large $\tau$ to obtain \eqref{barG_new2}, a simple integral splitting (and $\Psur=1-\Pexinf$) for \eqref{barG_new3}, and finally \eqref{int_tau-bartfix} and \eqref{approx_ga1b} to arrive at \eqref{barG_new}.

The second term in \eqref{barG_new} accounts for the contribution of the segregating mutations to the phenotypic mean. On average, these are the mutations that appeared less than $\bartfix$ generations in the past and have not yet had enough time to become fixed. A permanent response of the mean is achieved by mutations that have gone to fixation. These are accounted for by the first term in \eqref{barG_new}, which is consistent with \eqref{barG*_prop}.
\end{rem}

\subsection{Proof of Corollary \ref{cor:equilibrium_simp}}\label{proof_of_Corollary}

(1) The approximation \eqref{DebarG*} follows directly from \eqref{barG*_prop} by using \eqref{Plost*app}.

(2) For equal effects, \eqref{DebarG*} immediately simplifies to \eqref{DebarG*equal} by using \eqref{Plost*app}.

(3) follows from \eqref{DebarG*} by the properties of the exponential distribution.

(4) Using $\sN e^\sN E_1(\sN)\approx 1-1/\sN$ (because $\sN = N\Psur(e^{\sA})\gg1$) in \eqref{VG*_prop} we obtain
\begin{linenomath}\begin{subequations}\begin{align}
		V_G^*(f, s, N, \Th) 
    &\approx \Th \int_0^{\infty}  \frac{\A }{s} \Psur(e^{\sA}) \Bigl(1- \frac{1}{\sN}\Bigr) f(\A)\, d\A \label{VG*_DeltaGbar_a}\\
	&= \frac{\Th}{s} \int_0^{\infty} \A \Psur(e^{\sA}) f(\A)\, d\A - \frac{\Th}{sN} \int_0^{\infty} \A f(\A)\, d\A\,,	\label{VG*_DeltaGbar_b}
\end{align}\label{VG*_DeltaGbar}
\end{subequations}\end{linenomath}
which yields \eqref{VG*app}. 

We prove the exponential case (6) before (5).

(6) We require Assumption \ref{A1}. In \eqref{VG*_prop}, we substitute $2Ns\A$ for $\sN$ and $2s\A-\tfrac53(s\A)^2$ for $\Psur(e^{\sA})$ to obtain
\begin{linenomath}\begin{subequations}
\begin{align} 
	&V_G^*(f, s, N, \Th) 
  \approx 4\Th Ns \int_0^{\infty}  \A^3 \Bigl(1-\frac{5}{6}s\A \Bigr) e^{2Ns\A} E_1(2Ns\A) f(\A)\, d\A  \label{VG_proofCorsimple_a}\\ 
  &\quad=\frac{4\Th Ns{\barA}^3}{(2Ns\barA-1)^4}\left[(2Ns\barA-1)\bigl(8(Ns\barA)^2-14Ns\barA + 11\bigr) - 6\ln(2Ns\barA)  \right] \notag \\
	&\qquad -\frac{20\Th Ns^2{\barA}^4}{3(2Ns\barA-1)^5}\left[(2Ns\barA-1)\bigl(24(Ns\barA)^3 - 52(Ns\barA)^2 + 46Ns\barA -25\bigr)  + 12\ln(2Ns\barA) \right]\,. \label{VG_proofCorsimple_b}
\end{align}
\end{subequations}\end{linenomath}
By letting $N\to\infty$, we obtain to order $1/N$:
\begin{linenomath}\begin{equation}
	V_G^*(f, s, N, \Th)\approx 4\Th{\barA}^2\left(1-\frac{5}{2}s\barA - \frac{1}{4Ns\barA} +\frac{5}{12N}\right)\,,
\end{equation}\end{linenomath}
where Assumption \ref{A1} yields $s=O(N^{-1/K})$ and $1/(Ns) = O(N^{-1+1/K})$, where $K>2$.
This yields \eqref{VG*Ns} by omitting the term of order $N^{-1}$.

(5) For equal effects, \eqref{VG_proofCorsimple_a} simplifies to  $4\Th Ns\A^3 \Bigl(1-\frac{5}{6}s\A \Bigr) e^{2Ns\A} E_1(2Ns\A)$. Using $e^{x} E_1(x) \approx x^{-1} - x^{-2}$ if $x\to\infty$ (with $x=2Ns\A$ and the asymptotic assumption in (6)), we obtain
\begin{linenomath}\begin{equation}
	V_G^*(\A, s, N, \Th) \approx 2\Th{\A}^2\left(1-\frac{5}{6}s\A - \frac{1}{2Ns\A} +\frac{5}{12N}\right)\,,
\end{equation}\end{linenomath}
which yields \eqref{VG*Nsconst}, in which the term of smallest order, $\frac{5}{12N}$, has been omitted.

\subsection{Proof of Proposition \ref{prop:Dynamics}}\label{App:proof of prop:Dynamics}

The proof of the approximation \eqref{VG_init} for $V_G(\tau)$ is based on that of the approximation \eqref{VG*_prop} for the stationary variance $V_G^*$. 

We recall the definition of $D_V(\tau)$ from \eqref{D_V} and that of $I_V(t)$ from \eqref{I_V}. Then we obtain
\begin{linenomath}\begin{align}
V_G(\tau) &=  \Th \int_0^\infty \A^2 f(\A) (1-\Pexinf) \int_0^\tau \ga^\infty(t, e^{\sA})\, dt \, d\A + \Th D_V(\tau) \notag\\
		&= \Th\, \int_0^{\infty} \frac{\A}{s} \Psur \Bigl( \sN e^\sN E_1(\sN) - \sN e^{-\sA\tau} \exp (\sN e^{-\sA\tau}) E_1(\sN e^{-\sA\tau}) \Bigr) f(\A)\, d\A\ \notag \\
		&\qquad + \Th D_V(\tau)\,,
\end{align}\end{linenomath}
where the second equality follows from \eqref{approx_ga_tau1}.

As in step (2) of the proof of \eqref{VG_init}, we choose $T_\ep=\frac{1}{\sA}\ln\Bigl(\Pexinf\frac{1+\ep}{\ep}\Bigr)$, with  $\ep = r(N)/\min\{\A,z\barA\}$, $z= K_2\ln(Ns)$ and, $r(N)=N^{-K_1/K}$.

If $\tau\le T_\ep$, we obtain from \eqref{int_IV_part1} and \eqref{int_IV_part2}
\begin{linenomath}\begin{equation}
	D_V(\tau) \le D_V(T_\ep) = O\Bigl( N^{-2+(2+K_1)/K} \Bigr)
\end{equation}\end{linenomath}
because the integrands $I_V(t)$ and their bounds are nonnegative.  
If $\tau > T_\ep$, we obtain from \eqref{int_IV_part3}
\begin{linenomath}\begin{equation}
	D_V(\tau) \le D_V(\infty) = O\Bigl(N^{-K_1/K} \Bigr) \,.
\end{equation}\end{linenomath}

We leave the proof of \eqref{barG_init} for $\bar G(\tau)$ to the interested reader. It is based on analogous estimates.
The main difference is that instead of $\gamma_D$ in \eqref{gamma_D} the function 
\begin{equation}
	\tilde\gamma_D(x) = 1+x-(2+x)xE_1(x)
\end{equation}
 occurs, which has a diverging integral. 
However, instead of $u(t)\le1$ we can use the estimate
\begin{equation}
	u(t) \le \frac{\Pexinf  e^{-s\A t}}{\ep (1-\Pexinf e^{-s\A t})} \le \frac{\Pexinf  e^{-s\A t}}{\ep(1-\Pexinf)}\,,
\end{equation}
which follows from \eqref{Psurvn/Psurinfty}. Then
\begin{equation}
	\int_{T_\ep}^\infty e^{-s\A t}\tilde\gamma(a^\infty(t))\,dt = \frac{1}{s\A\sN} + \text{terms of lower order}
\end{equation}
replaces \eqref{int_gamma_D}. However, the leading-order term is now $1/(s\A\sN)$, from which the stronger requirement $K>3$ is deduced.

\subsection{Proof of Proposition \ref{prop:Initial2}}\label{app:proof_prop:Initial2}

\begin{proof}[Proof of \eqref{barG_smalltau_a}]
We recall Assumption \ref{A1} and let $N\to\infty$. As a consequence $Ns\to\infty$ and $s\to0$. The mean mutational effect $\barA$ is given and fixed. From the asymptotic expansion \eqref{e^x E_1(x)} and the definition of $\ga_1$ in \eqref{mean_g}, we obtain the asymptotic equivalence 
\begin{linenomath}\begin{equation}\label{ga1_asymptotic}
	\ga_1(a) = 1- a e^a E_1(a)\sim 1/a - 2/a^2 \;\text{as}\; a\to\infty\,.
\end{equation}\end{linenomath}
In addition, $\max\{0,1/a-2/a^2\}<\ga_1(a)<\min\{1/a,1\}$ holds for every $a>0$ and $\ga_1(a(t))$ is monotonically increasing from 0 to 1 as $t$ increases from 0 to $\infty$; see text below \eqref{var_g}. 

The basic idea of the proof is to use the approximation
\begin{linenomath}\begin{equation}\label{approx_gamma1_smallt}
	\ga_1(a(t,e^{s\A},N)) \approx \frac{1}{a(t,e^{\sA},N)} = \frac{e^{\sA t}}{N\bigl(1-\Plost([t],e^{\sA})\bigr)}
\end{equation}\end{linenomath}
for small $t$ because $a(t,e^{\sA},N)$, defined in \eqref{a(t)}, is monotonically decreasing in $t$ from $a(0,e^{\sA},N)=N$ to 0 as $t$ increases to $\infty$. Then one could proceed as follows by first substituting \eqref{approx_gamma1_smallt} into \eqref{barGtau}:
\begin{linenomath}\begin{subequations}\label{approx_barGtau_smallt}
\begin{align}
	\bar G(\tau) &= \Th \int_0^\infty \A f(\A) \int_0^\tau \bigl(1-\Plost([t],e^{\sA}) \bigr) \ga_1(a(t, e^{\sA},N))\, dt \, d\A \label{approx_barGtau_smallt_a1} \\
	&\approx \frac{\Th}{N} \int_0^\infty \A f(\A) \int_0^\tau e^{\sA t}\, dt\, d\A \label{approx_barGtau_smallt_a2}\\
	&= \frac{\Th}{Ns} \int_0^\infty f(\A) (e^{\sA\tau}-1)\, d\A \,. \label{approx_gamma1_smallt_a}
\end{align}\end{subequations}\end{linenomath}
The order of the error of the approximation in \eqref{approx_barGtau_smallt_a2} can be obtain by integrating $2/a(t,e^{\sA},N)^2$ accordingly.
However, on the way several technical problems occur. In particular, if mutation effects are drawn from a distribution, in our case the exponential distribution with mean $\barA$, then for given $\tau$, $s$, and $N$, the approximation \eqref{approx_barGtau_smallt} will be accurate if $\A$ is sufficiently small, but will fail if $\A$ is extremely large because then $a(\tau,e^{\sA},N)$ can drop from $N$ to close to 0 in the first generation. 

Therefore, we will proceed as in \eqref{approx_barGtau_smallt} only if $\A\le z\barA$, where we choose $z=\tilde K\ln(Ns\barA)$ and determine $\tilde K(\ge1)$ below. Because we assume an exponential mutation distribution $f$ with mean $\barA$, we have $\Prob[\A > z\barA]= e^{-z}$. If $\A > z\barA$, we use $0<\ga_1(a)<1$ (see above). Straightforward integration shows that 
\begin{linenomath}\begin{align}\label{gbar_init_help1}
	0 < &\int_{z\barA}^\infty \A f(\A) \int_0^\tau \bigl(1-\Plost([t],e^{\sA}) \bigr) \ga_1(a(t, e^{\sA},N))\, dt \, d\A \notag \\
	& <\int_{z\barA}^\infty \A f(\A) \int_0^\tau 1 \, dt \, d\A \notag \\
	& = \tau\barA e^{-z}(1+z) \notag \\
	&	= \tau\barA\,\frac{1+\tilde K\ln(Ns\barA)}{(Ns\barA)^{\tilde K}} \notag \\
	&	= \tau \,O\biggl(\frac{\ln(Ns)}{(Ns)^{\tilde K}}\biggr) \;\text{ as } N\to\infty \,,
\end{align}\end{linenomath}
and this holds for every $\tau>0$. 

For $\A \le z\barA$, we start from \eqref{barGtau} and use \eqref{ga1_asymptotic} and \eqref{approx_gamma1_smallt} to obtain 
\begin{linenomath}\begin{align}
	&\int_0^{z\barA} \A f(\A) \int_0^\tau \bigl(1-\Plost([t],e^{\sA}) \bigr) \ga_1(a(t, e^{\sA},N))\, dt \, d\A  \notag\\
	&\quad= \int_0^{z\barA} \A f(\A) \int_0^\tau \frac{1}{N} e^{\sA t}\, dt\, d\A  + R(\tau)\notag\\
	&\quad= \frac{1}{Ns} \int_0^{z\barA} f(\A) (e^{\sA\tau}-1)\, d\A  + R(\tau) \notag\\
	&\quad= \frac{1}{Ns}\biggl(\frac{\barA s \tau}{1-\barA s \tau} - R_1(\tau) \biggr) + R(\tau)\,. \label{gbar_init_help2}
\end{align}\end{linenomath}
Here, 
\begin{linenomath}\begin{equation}\label{R1-term}
	R_1(\tau) = e^{-z}\biggl( \frac{e^{z\barA s\tau}}{1-\barA s \tau} -1 \biggr)
\end{equation}\end{linenomath}
and $R(\tau)$ is, to leading order in $1/a^2$, the error term arising from the approximation \eqref{approx_gamma1_smallt}, i.e.,
\begin{linenomath}\begin{equation}\label{R-term}
	R(\tau) = -2 \int_0^{z\barA} \A f(\A) \int_0^\tau \frac{1-\Plost([t],e^{\sA})}{a(t,e^{\sA},N)^2}\, dt\, d\A \,.
\end{equation}\end{linenomath}
To obtain bounds on $R_1(\tau)$ and $R(\tau)$, we assume $s\barA\tau\le \tfrac14$ (this is not best possible; any upper bound $<\tfrac12$ would be sufficient to arrive at \eqref{help_Rtau1}). 
For $R_1(\tau)$ we obtain by observing its convexity and bounding it by the linear function assuming 0 at $\tau=0$ and $e^{-z}(\tfrac43e^{z/4}-1)$ at $\tau=1/(4\barA s)$:
\begin{linenomath}\begin{align}\label{R1-term_bd}
	R_1(\tau) &\le 4\barA s\tau e^{-z}\biggl(\frac{4}{3} e^{z/4} -1 \biggr) \le \frac{16}{3}\barA s\tau e^{-3z/4} \notag\\
		& = \frac{16}{3}\barA s\tau (Ns\barA)^{-3\tilde K/4}\,,
\end{align}\end{linenomath}
which is of smaller order in magnitude than the leading order term $\frac{\barA s \tau}{1-\barA s \tau}$ in \eqref{gbar_init_help2} (this holds for every $\tilde K\ge1$).

Now we derive a bound for $R(\tau)$. 
From the monotonic decay in $t$ of the probabilities of non-extinction ($1-\Plost([t],e^{\sA})$) and the approximation $1-\Pexinf(e^{\sA})=\Psur(e^{\sA})\approx\rho\sA$ in Remark~\ref{Psur=csalpha}, which requires that $\sA$ is sufficiently small, we infer
\begin{linenomath}\begin{equation}\label{lowerbound_csalpha}
	1-\Plost([t],e^{\sA}) > 1-\Pexinf(e^{\sA}) > c\sA
\end{equation}\end{linenomath}
if $c<\rho$. We can choose, for instance, $c=\rho/2$. Henceforth, we assume that $s$ is small enough such that \eqref{lowerbound_csalpha} holds for all $\A\le z\barA$. (This is possible because with $Ns^K=C^K$ we obtain
$sz = s\tilde K\ln(Ns\barA) = CN^{-1/K}\tilde K\ln(N^{(K-1)/K}C\barA) \to0$ as $N\to\infty$. Hence, $czs\barA\to0$ as $N\to\infty$, which guaranties that \eqref{lowerbound_csalpha} holds for every $\A\le z\barA$ if  $N$ is large enough).
Then, we obtain for every $t$ and $\A\le z\barA$:
\begin{equation}\label{help_bound_Rterm}
	\frac{2\bigl(1-\Plost([t],e^{\sA})\bigr)}{a(t,e^{\sA},N)^2} = \frac{2e^{2 \sA t}}{N^2\bigl(1-\Plost([t],e^{\sA}) \bigr)} < \frac{2e^{2\sA t}}{N^2\bigl(1-\Pexinf(e^{\sA}) \bigr)} < \frac{2e^{2\sA t}}{N^2 c\sA}\,.
\end{equation}
Therefore, the integral with respect to $t$ in \eqref{R-term}, is bounded by
\begin{equation}\label{error_bound_a^2_equal}
	 \int_0^\tau \frac{2e^{2 \sA t}}{N^2 c\sA} = \frac{1}{N^2s^2c}\frac{e^{2\sA\tau}-1}{\A^2}\,.
\end{equation}
Substituting this into \eqref{R-term} and integrating with respect to $\A$ we obtain
\begin{linenomath}\begin{align}
	|R(\tau)|&\le  \frac{1}{N^2s^2c}\int_0^{z\barA} f(\A)\frac{e^{2\sA\tau}-1}{\A}\, d\A \notag\\
		&=\frac{1}{N^2s^2c\barA}\Bigl(-\ln(1-2s\barA\tau) - R_2(\tau) \Bigr) \notag \\
	 &< \frac{1}{N^2s^2c\barA}\Bigl(3s\barA\tau - R_2(\tau)\Bigr)\,, \label{help_Rtau1}
\end{align}\end{linenomath}
where $R_2(\tau) = E_1(z(1-2s\barA\tau))-E_1(z)$ and the linear bound on the logarithmic term applies if $s\barA\tau<\tfrac14$. Upon two-fold differentiation, we observe that $R_2(\tau)$ is convex and increases from 0 at $\tau=0$ to $E_1(z/2))-E_1(z)$ at $\tau=1/(4s\barA)$. Therefore, we bound $R_2(\tau)$ by the corresponding linear function in $\tau$ and obtain
\begin{linenomath}\begin{equation} \label{help_R2tau}
	R_2(\tau) \le  4s\barA\tau \Bigl( E_1(z/2)-E_1(z)\Bigr) = s\tau \,O\biggl( \frac{1}{(Ns)^{\tilde K/2}\ln (Ns)} \biggr)
\end{equation}\end{linenomath}
by series expansion for $Ns\to\infty$ and using $z=\tilde K\ln(Ns\barA)$. In summary, we obtain from \eqref{help_Rtau1} and \eqref{help_R2tau}, and then choosing $\tilde K=4$:
\begin{linenomath}\begin{equation}\label{bound_Rtau}
	|R(\tau)|\le \frac{3\tau}{N^2sc} \biggl(1 + O\biggl( \frac{s^{2-\tilde K/2}}{N^{\tilde K/2}\ln(Ns)} \biggr) \biggr) = \tau\,O\biggl( \frac{1}{N^2s} \biggr) \,.
\end{equation}\end{linenomath}

We note that estimates similar to those in \eqref{help_bound_Rterm} show that
\begin{equation}
	a(\tau,e^{sz\barA},N) \ge {\rm const}\, (Ns)^{3/4}\ln(Ns\barA)\,,
\end{equation}
where we have used $s\barA\tau\le\tfrac14$. This shows that the approximation \eqref{ga1_asymptotic} is indeed justified for every $\A\le z\barA$.

By combining \eqref{gbar_init_help2} and \eqref{gbar_init_help1}, and substituting the bounds \eqref{R1-term_bd} for $R_1(\tau)$ and \eqref{bound_Rtau} for $R(\tau)$, we obtain
\begin{linenomath}\begin{align}
	\bar G(\tau) &= \Th \int_0^\infty \A f(\A) \int_0^\tau \bigl(1-\Plost([t],e^{\sA}) \bigr) \ga_1(a(t, e^{\sA},N))\, dt \, d\A  \notag \\
		&= \frac{\Th}{Ns}\biggl(\frac{\barA s \tau}{1-\barA s \tau} - R_1(\tau) \biggr) + \Th R(\tau) + \Th\tau \,O\Bigl(\frac{\ln(Ns)}{(Ns)^{\tilde K}}\Bigr) \notag\\
		&= \frac{\Th\tau}{Ns}\biggl(\frac{\barA s}{1-\barA s \tau} + O\Bigl(\frac{s}{(Ns)^{3\tilde K/4}}\Bigr)\biggr) +  \Th\tau\, O\Bigl(\frac{1}{N^2s}\Bigr) + \Th\tau\, O\biggl(\frac{\ln(Ns)}{(Ns)^{\tilde K}}\biggr) \notag\\
		&= \frac{\Th\tau}{N}\biggl[\frac{\barA}{1-\barA s \tau} + O\Bigl((Ns)^{-3\tilde K/4}\Bigr) + O\Bigl((Ns)^{-1}\Bigr) + O\biggl(\frac{\ln(Ns)}{N^{\tilde K-1}s^{\tilde K}}\biggr) \biggr] \notag\\
		&=  \frac{\Th\tau}{N}\biggl[\frac{\barA}{1-\barA s \tau} + O\Bigl(N^{-1+1/K}\Bigr)\biggr]\,,
\end{align}\end{linenomath}
where in the last step we chose $\tilde K=4$ so that the first error term is $O\Bigl((Ns)^{-3}\Bigr)$ and the last error term is 
\begin{equation}
	O\biggl(\dfrac{\ln(Ns)}{Ns}\dfrac{1}{N^2s^3}\biggr) = O\biggl(\dfrac{\ln(Ns)}{Ns} N^{-2+3/K}\biggr) = o\Bigl( N^{-1+1/K}\Bigr) = o\Bigl((Ns)^{-1}\Bigr)
\end{equation}
by using once again the scaling assumption \ref{A1}. This finishes the proof of \eqref{barG_smalltau_a}.
\end{proof}

\begin{proof}[Proof of \eqref{VG_smalltau_a}]
From \eqref{var_g} and \eqref{e^x E_1(x)} we get $\ga(a)\sim 1/a-4/a^2$ as $a\to\infty$. Therefore, the proof is analogous to that of \eqref{barG_smalltau_a}, i.e., it is based on the idea of substituting this approximation into \eqref{VGtau}, which yields
\begin{linenomath}\begin{align}  
	V_G(\tau) &=  \Th \int_0^\infty \A^2 f(\A) \int_0^\tau \bigl(1-\Plost([t],e^{\sA}) \bigr) \ga(a(t, e^{\sA}))\, dt \, d\A \notag \\
	&\approx \frac{\Th}{N} \int_0^\infty \A^2 f(\A) \int_0^\tau e^{s\A(t)} \, dt  \, d\A \notag \\
	&= \frac{\Th}{Ns} \int_0^\infty \A f(\A) (e^{\sA\tau}-1)\, d\A\,,	
\end{align}\end{linenomath}
which yields \eqref{VG_smalltau_a} if $f$ is exponential with mean $\bar\A$.

We omit several of the details of the proof and present only the most relevant expression. 
Instead of \eqref{gbar_init_help1} we obtain
\begin{linenomath}\begin{align}\label{VG_init_help1}
	0 < &\int_{z\barA}^\infty \A^2 f(\A) \int_0^\tau \bigl(1-\Plost([t],e^{\sA}) \bigr) \ga_1(a(t, e^{\sA},N))\, dt \, d\A \notag \\
	& < \tau \,O\biggl(\frac{\ln(Ns)^2}{(Ns)^{\tilde K}}\biggr)\,.
\end{align}\end{linenomath}

Instead of \eqref{gbar_init_help2} we obtain
\begin{linenomath}\begin{align}
	&\int_0^{z\barA} \A^2 f(\A) \int_0^\tau \bigl(1-\Plost([t],e^{\sA}) \bigr) \ga_1(a(t, e^{\sA},N))\, dt \, d\A  \notag\\
	&\quad= \frac{\barA}{Ns}\biggl(\frac{1}{(1-s\barA\tau)^2} - 1 - R_1^V(\tau) \biggr) + R^V(\tau)\,, \label{help_Vsmall_lead}
\end{align}\end{linenomath}
where $R_1^V(\tau)$ and $R^V(\tau)$ are defined analogously to $R_1(\tau)$ and $R(\tau)$ above. For $R_1^V(\tau)$ one obtains
\begin{linenomath}\begin{align}
	R_1^V(\tau) &= \barA e^{-z}\biggl(\frac{e^{sz\barA\tau}(1+z(1-s\barA\tau)}{(1-s\barA\tau)^2} -(1+z) \biggr) \notag\\
		&\le \tau\, O\biggl( \frac{s\ln(Ns)}{(Ns)^{3\tilde K/4}} \biggr)\,,
\end{align}\end{linenomath}
where the expression in $z$ is monotone increasing and convex in $\tau$, and we bound it by the linear function assuming $0$ at $\tau=0$ and the corresponding value at $\tau=1/(4s\barA)$. Finally, we obtain after similar calculations and series expansions 
\begin{linenomath}\begin{align}
	|R^V(\tau)|&\le  \frac{2}{N^2s^2c}\int_0^{z\barA} f(\A)(e^{2\sA\tau}-1)\, d\A \notag\\
		&= \frac{2}{N^2s^2c\barA}\Bigl(\frac{2s\barA\tau}{1-2s\barA\tau} + R_2(\tau)\Bigr) = \tau O\Bigl(\frac{1}{N^2s} \Bigr) \,. 
\end{align}\end{linenomath}
Putting all this together, writing the leading order term in \eqref{help_Vsmall_lead} as 
\begin{equation}
	\frac{1}{(1-s\barA\tau)^2} - 1 = 2s\barA\tau\,\frac{1-s\barA\tau/2}{(1-s\barA\tau)^2}
\end{equation}
and choosing $\tilde K=4$, we obtain
\begin{linenomath}\begin{align}
	V_G(\tau) &= \Th \int_0^\infty \A^2 f(\A) \int_0^\tau \bigl(1-\Plost([t],e^{\sA}) \bigr) \ga_1(a(t, e^{\sA},N))\, dt \, d\A  \notag \\
			&=  \frac{2\Th\barA^2\tau}{N}\biggl[\frac{1-s\barA\tau/2}{(1-s\barA\tau)^2} + O\Bigl(N^{-1+1/K}\Bigr)\biggr]\,.
\end{align}\end{linenomath}
\end{proof}

The approximations \eqref{barG_smalltau_series} and \eqref{VG_smalltau_series} are then obtained by a series expansion of the leading order terms. 

\begin{proof}[Proof of \eqref{VG_smalltau_const}]
For equal mutation effects the proof is much simpler because we do not have to integrate with respect to $\A$. Indeed, we observe directly from \eqref{approx_gamma1_smallt} that, to leading order in $Ns$, $\bar G(\tau)= \frac{\Th}{Ns} (e^{\sA\tau}-1)$. The estimate of the error term, analogous to $R(\tau)$, is also much simpler and straightforwardly yields (cf.~\eqref{error_bound_a^2_equal})
\begin{equation}
	 |R(\tau)| \le \int_0^\tau \frac{2e^{2 \sA t}}{N^2 c\sA} = \frac{1}{N^2s^2c}\frac{e^{2\sA\tau}-1}{\A^2}\,,
\end{equation}
which is of smaller order, by a factor of $1/(Ns)$, than the leading order term. Of course, this requires $e^{2\sA\tau}=O(1)$, but it does not require the stronger constraint $\sA\tau<\tfrac14$. The approximation for $V_G(\tau)$ is derived analogously.
\end{proof}

\FloatBarrier
\section{Segregating sites: A branching process approximation}\label{app:segsites}

Here we derive an alternative approximation for the number of segregating sites by relying on our branching process approach, similar as in Sect.~\ref{sec:inf_many_sites}, and amending it by a diffusion approximation for the mean fixation time. 
At a given generation $\tau$, $\Th \tau$ mutations have occurred on average.  A mutant with fitness $\si$ is still present in the population with probability $1-\Plost(\tau-t,\si)$, where $t$ is the number of generations since its occurrence and $\Plost(n,\si)$ is the probability of loss until generation $n$; see \eqref{plost(tau)}. On average, $t= i/\Th$ for the $i$th mutation event. Some of those mutations may have already reached fixation and are no longer segregating in the population. Fixation occurs on average after $\bartfix$ generations (see eqs.~\ref{bartfix} and \ref{bartfix_app3}).

With this in mind and following the definitions the expectations of the phenotypic mean and variance in \eqref{def_barGtau} and \eqref{def_VGtau}, respectively, we define random variables $S^{(i)}_{\tau,e^{\sA},\Th}$, which are $1$ when locus $i$ is segregating in the population and $0$ otherwise. Using the expected fixation time $\tfix$ in \eqref{PfixDif} from the diffusion approximation, we set
\begin{linenomath}\begin{align} 
	\tilde{\tau}= \min \{\tfix(\stilde,N),\tau \}  \, ,
\end{align}\end{linenomath}
which is the expected time a mutant present in generation $\tau$ has already been segregating.
Then 
\begin{linenomath}\begin{align} \label{EV_Si}
	\EV[S^{(i)}_{\tau,e^{\sA},\Th}] \approx \int_0^{\tilde{\tau}} \frac{m_{i,\Th}(t)}{M_{i,\Th}(\tau)} (1- \Plost([\tilde{\tau}-t],e^{\sA})) dt \,.
\end{align}\end{linenomath}
We define the expected number of segregating sites in our model, where mutation effects are drawn from a distribution $f$ and the offspring distribution is Poisson, by
\begin{linenomath}\begin{equation}
	\EV[S](\tau ,f ,s ,N,\Th) = \sum_{n=1}^{\infty} \Poi_{\Th \tau}(n) \left( \sum_{i=1}^{n} \int_0^\infty \EV \left[ S^{(i)}_{\tau,e^{\sA},\Th} \right] f(\A)\, d\A \right)\,, \label{EV_S_def}
\end{equation}\end{linenomath}
By assuming equality in \eqref{EV_Si}, we obtain the following approximation for $\EV[S]$:
\begin{linenomath}\begin{align}  \label{Seg_BP_f}
	\tilde{S}(\tau ,f ,s ,N,\Th) &= \sum_{n=1}^{\infty} \Poi_{\Th \tau}(n) \left( \sum_{i=1}^{n} \int_0^\infty \int_0^{\tilde{\tau}} \frac{m_{i,\Th}(t)}{M_{i,\Th}(\tau)} \bigl(1- \Plost([\tilde{\tau}-t],e^{\sA}) \bigr)\, dt\,  f(\A)\, d\A \right) \nonumber \\
	& = \int_0^\infty \int_0^{\tilde{\tau}} \sum_{n=1}^{\infty} \Poi_{\Th \tau}(n) \left( \sum_{i=1}^{n}  \frac{m_{i,\Th}(t)}{M_{i,\Th}(\tau)} \right) \bigl(1- \Plost([\tilde{\tau}-t],e^{\sA})\bigr)\, dt\,   f(\A)\, d\A \nonumber \\
	& = \Th \int_0^{\infty}  f(\A)  \int_0^{\tilde{\tau}} \bigl(1-\Plost ([\tilde{\tau} -t ],e^{\sA} ) \bigr) \, dt \, d\A \nonumber \\
	&\approx \Th \int_0^{\infty}  f(\A)  \sum_{i=0}^{[\tilde{\tau}]} \bigl(1-\Plost ([\tilde{\tau}]-i,e^{\sA} )\bigr) \, d\A\nonumber \\
	& =  \Th \int_0^{\infty}  f(\A)  \sum_{j=0}^{[\tilde{\tau}]} \bigl(1-\Plost (j,e^{\sA} )\bigr)  d\A  \,. 
\end{align}\end{linenomath}
The above calculations are analogous to those in the derivation of \eqref{barGtau} for $\bar G(\tau)$ in \eqref{barGtau_deriv} upon substitution of $\A \EV[\tilde{X}^{(i)}_{\tau,e^{\sA},\Th}]$ by $\EV[S^{(i)}_{\tau,e^{\sA},\Th}]$. 

\begin{figure}[t!]
\centering
\begin{tabular}{ll}
A & B \\
\includegraphics[width=0.45\textwidth]{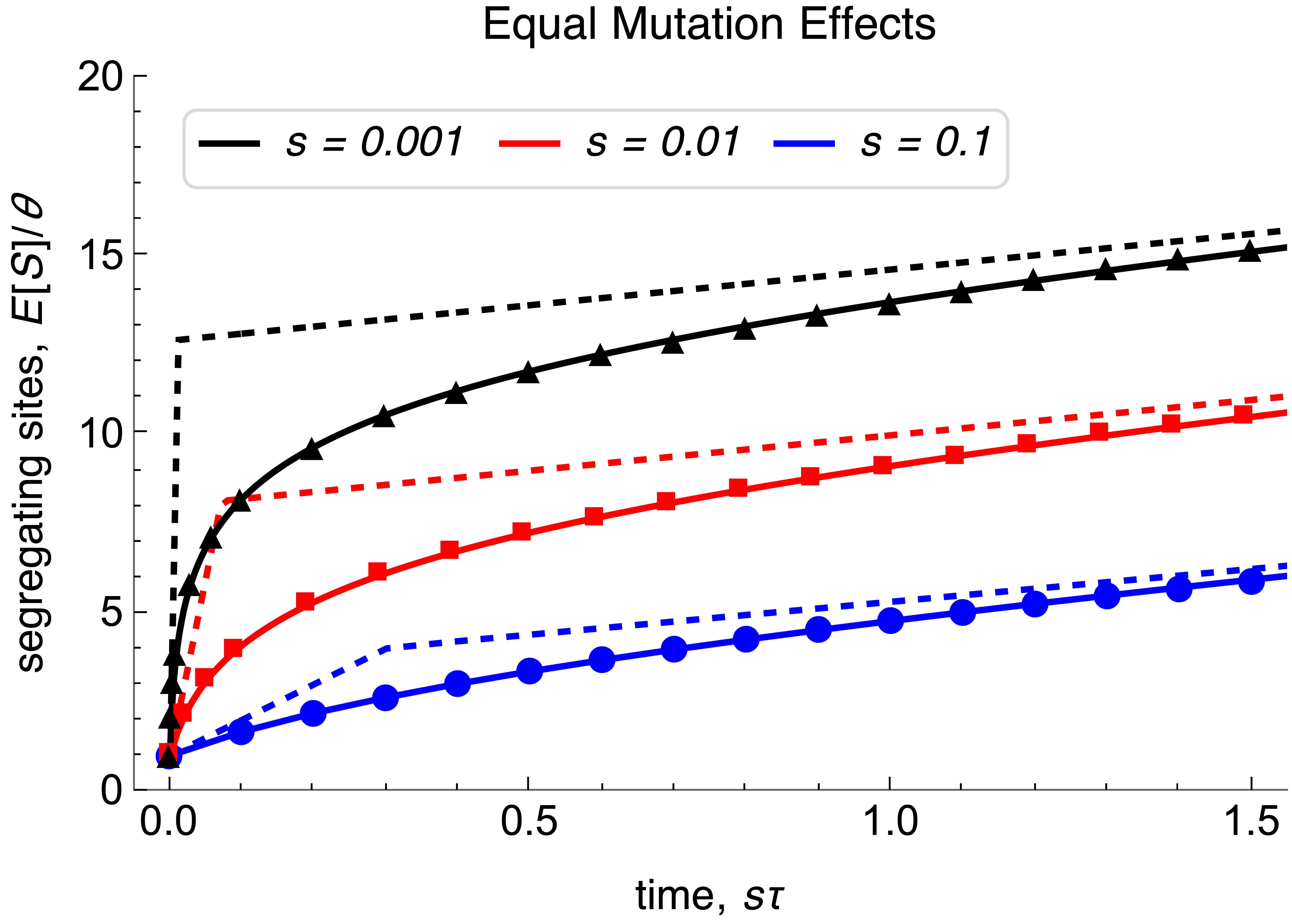} &
\includegraphics[width=0.45\textwidth]{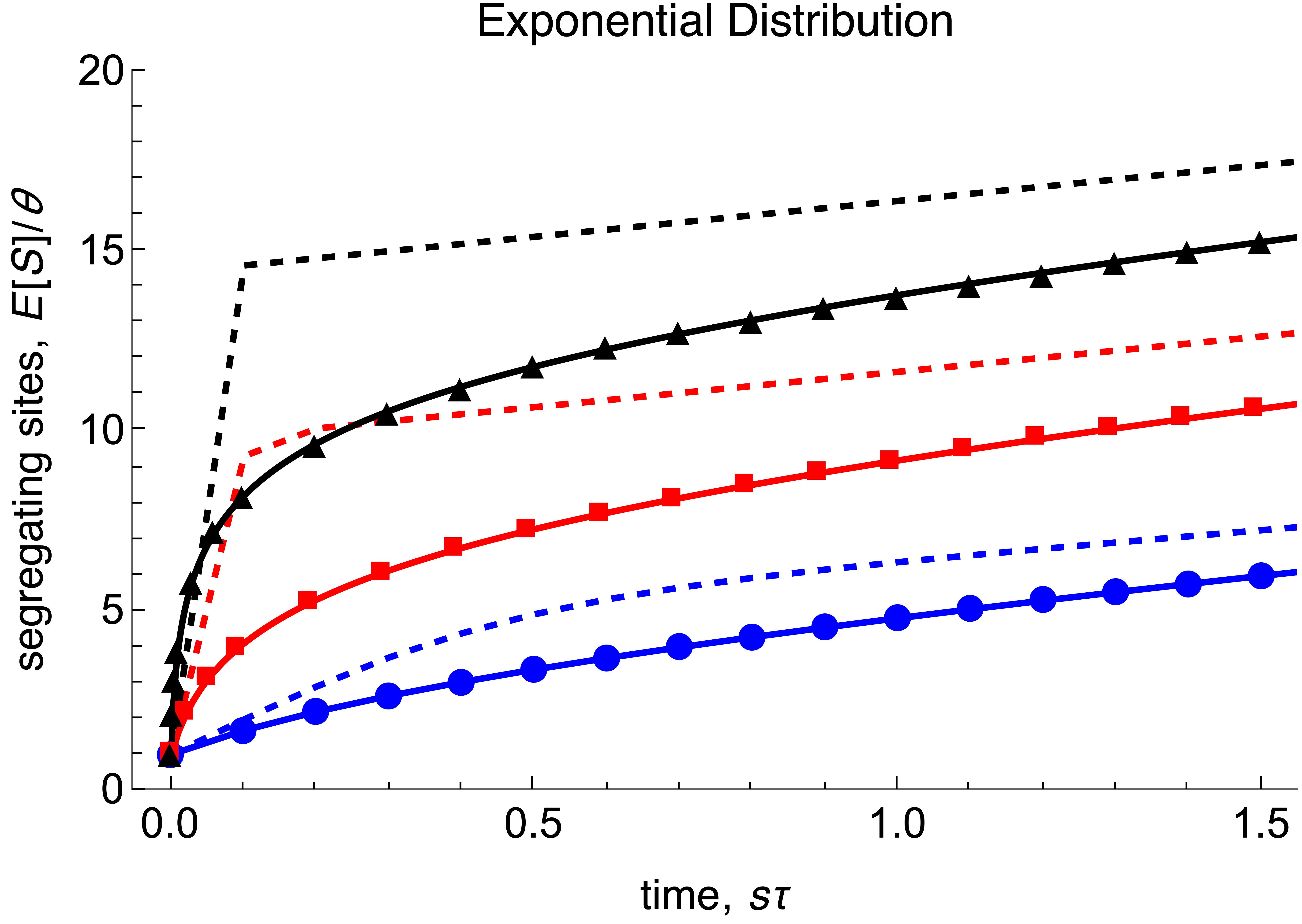} 
\end{tabular}
\caption[Segregating sites - branching process vs. diffusion]{Scaled expected number of segregating sites, $E[S]/\Th$ as a function of $s\tau$ in the early phase of adaptation ($s\tau\le2$) for the selection intensities $s=0.1,0.01,0.001$ (blue, red, black). In A the mutation effects are equal ($\A = 1$), in B they are drawn from an exponential distribution with mean $\bar \A = 1$. The branching process approximations $\tilde{S}$ in \eqref{Seg_BP_equal} and \eqref{Seg_BP_f_smalltau} are  shown by solid curves, the diffusion approximation $\bar S$ in \eqref{Seg} by dashed curves. The kinks occur at $s\tloss$. Results from Wright-Fisher simulations (with $\Th = \frac{1}{2}$) are shown as symbols. The mutation effects are $\A=1$ and the population size is $N=10^4$. }
\label{fig_SegSitesBP}
\end{figure}

If all effects are equal to $\A$, then $\tilde\tau=\tilde\tau(\A)$ is constant and the integration in \eqref{Seg_BP_f} collapses to the computation of the sum, i.e., to \eqref{Seg_BP_equal} in the main text.
Unless the distribution $f(\A)$ is a delta function, as for equal mutation effects, no further simplification of \eqref{Seg_BP_f} seems possible, except when $\tau \ll \bartfix(\stilde,N)$. Then, for given small $\tau$, we have $\tilde\tau=\tau$ except for the (very rare) very large values of $\A$ in the tail of $f$ for which $\tfix(\stilde,N)<\tau$. As a consequence, we obtain from \eqref{Seg_BP_f} the approximation \eqref{Seg_BP_f_smalltau} in the main text.

\FloatBarrier
\setcounter{section}{18}
\newpage
\section{Supplementary figures}

\FloatBarrier

\vspace{5mm}
\begin{figure}[ht]
\centering
\begin{tabular}{c}
\includegraphics[width=0.7\textwidth]{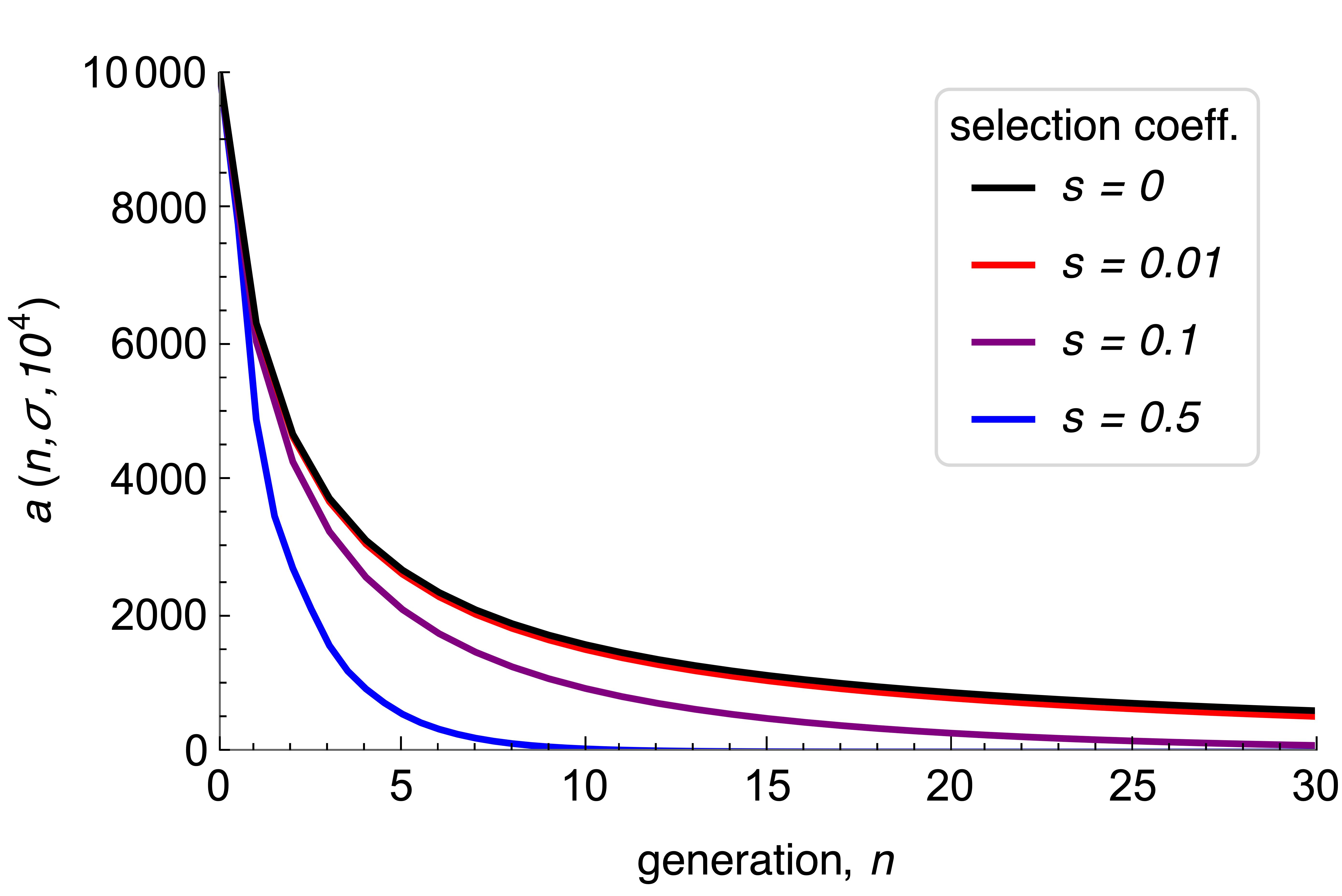}
\end{tabular}
\caption[$a(n, \si)$]{Shape of $a(n, \si, N)$ in eq.~\eqref{a(t)} as a function of the number of generations $n$ for different selection coefficients $s$. The population size is $N = 10^4$ and the offspring distribution is Poisson with mean $\si=e^s$. }
\label{fig_a}
\end{figure}

\newpage

\newpage

\begin{figure}[ht]
\centering
\begin{tabular}{c|ccc}
& $a = 100$ & $a = 2$ & $a = 0.5$\\
\hline
$s=0.1$ \\
& \includegraphics[width=0.25\textwidth]{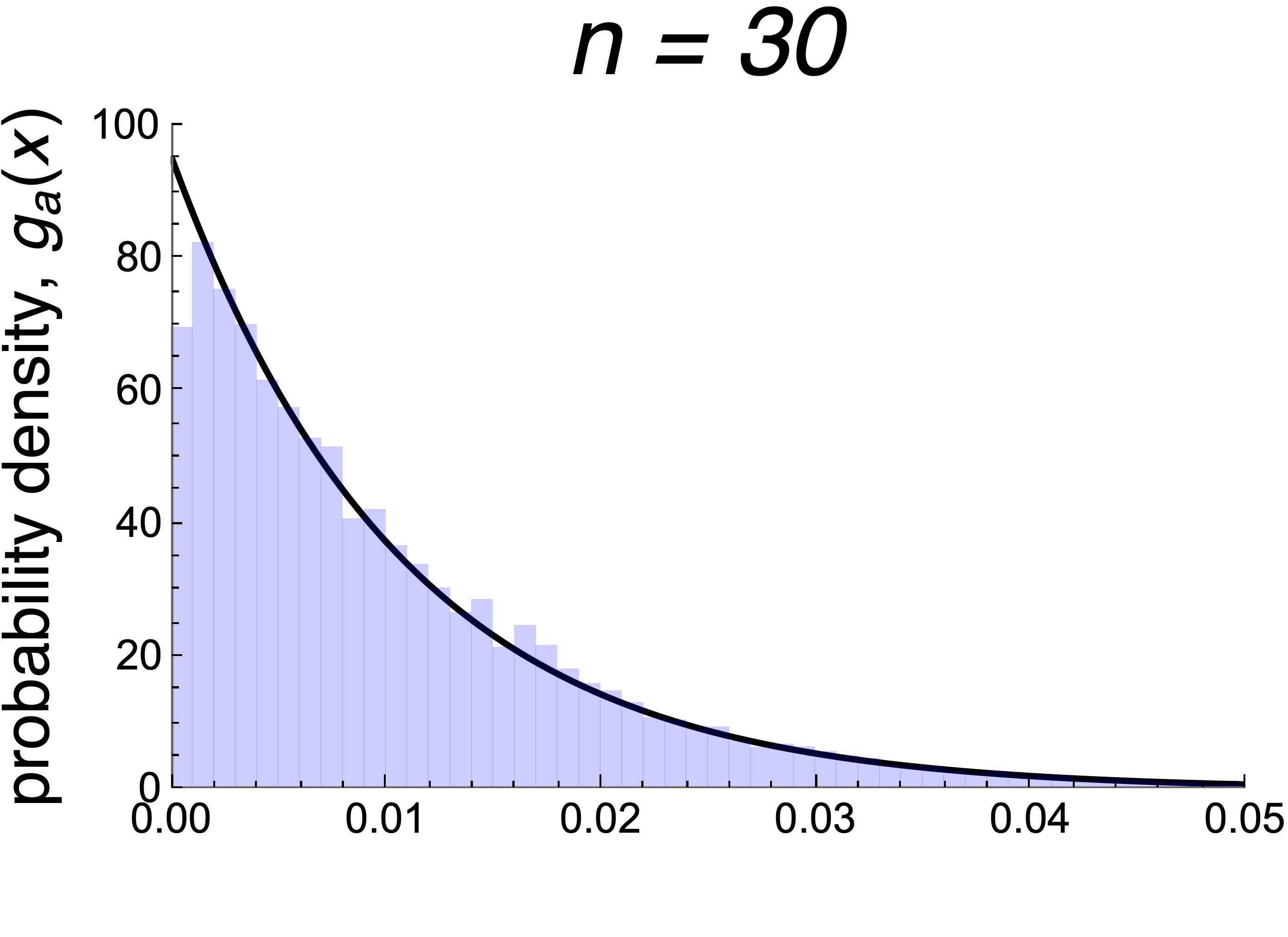}
& \includegraphics[width=0.25\textwidth]{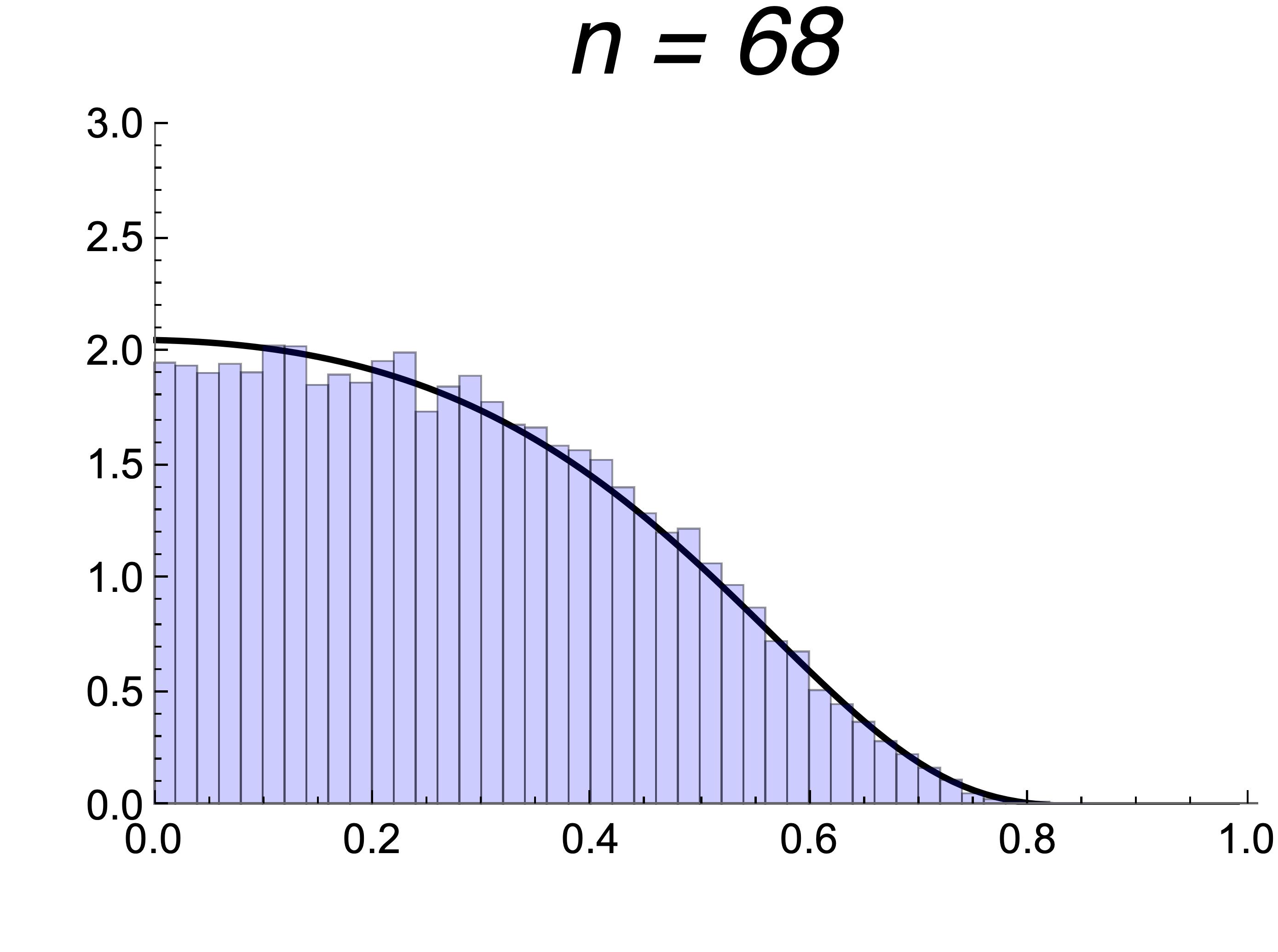}
& \includegraphics[width=0.25\textwidth]{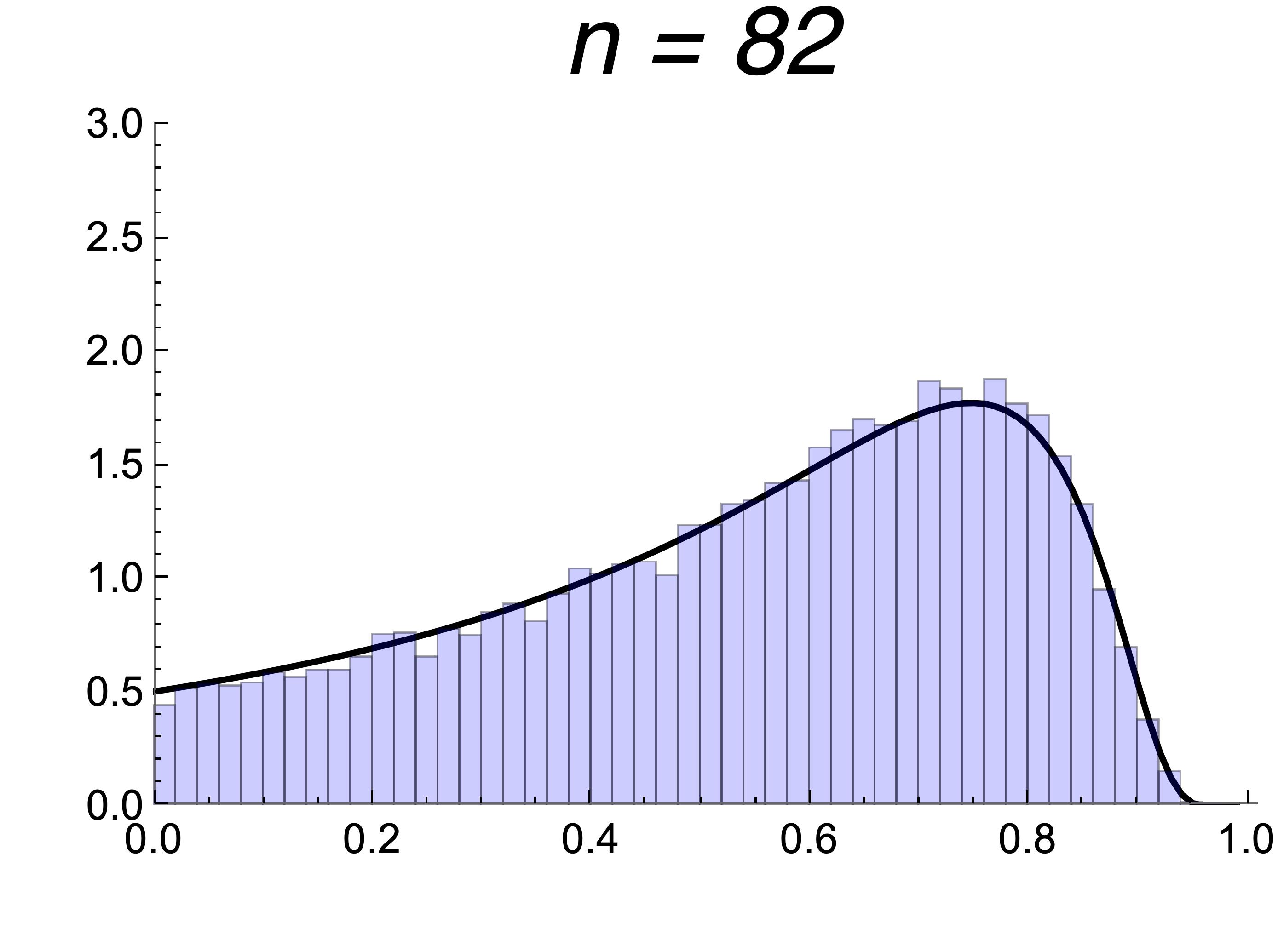} 
\\ $s=0.01$ \\
& \includegraphics[width=0.25\textwidth]{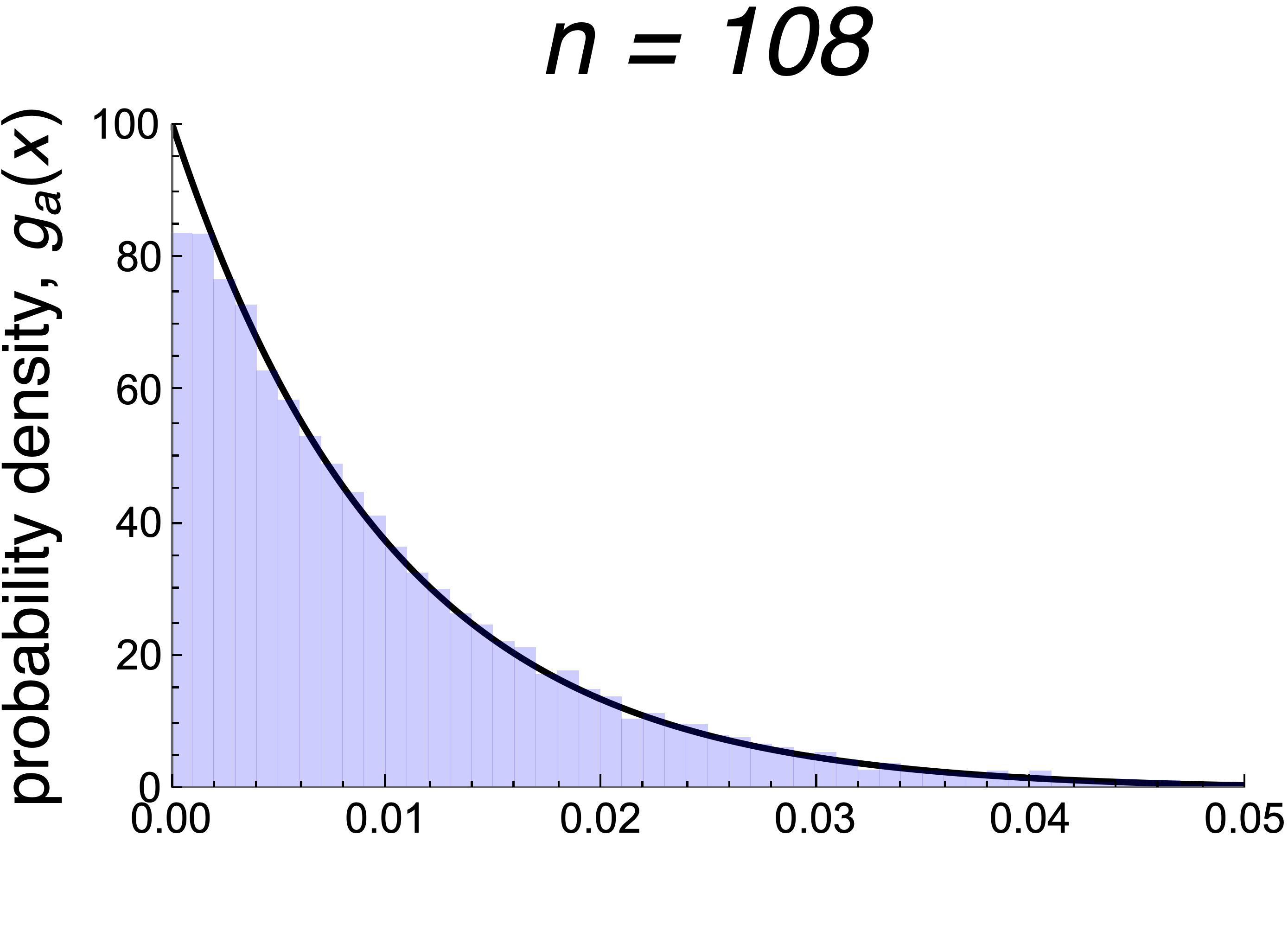}
& \includegraphics[width=0.25\textwidth]{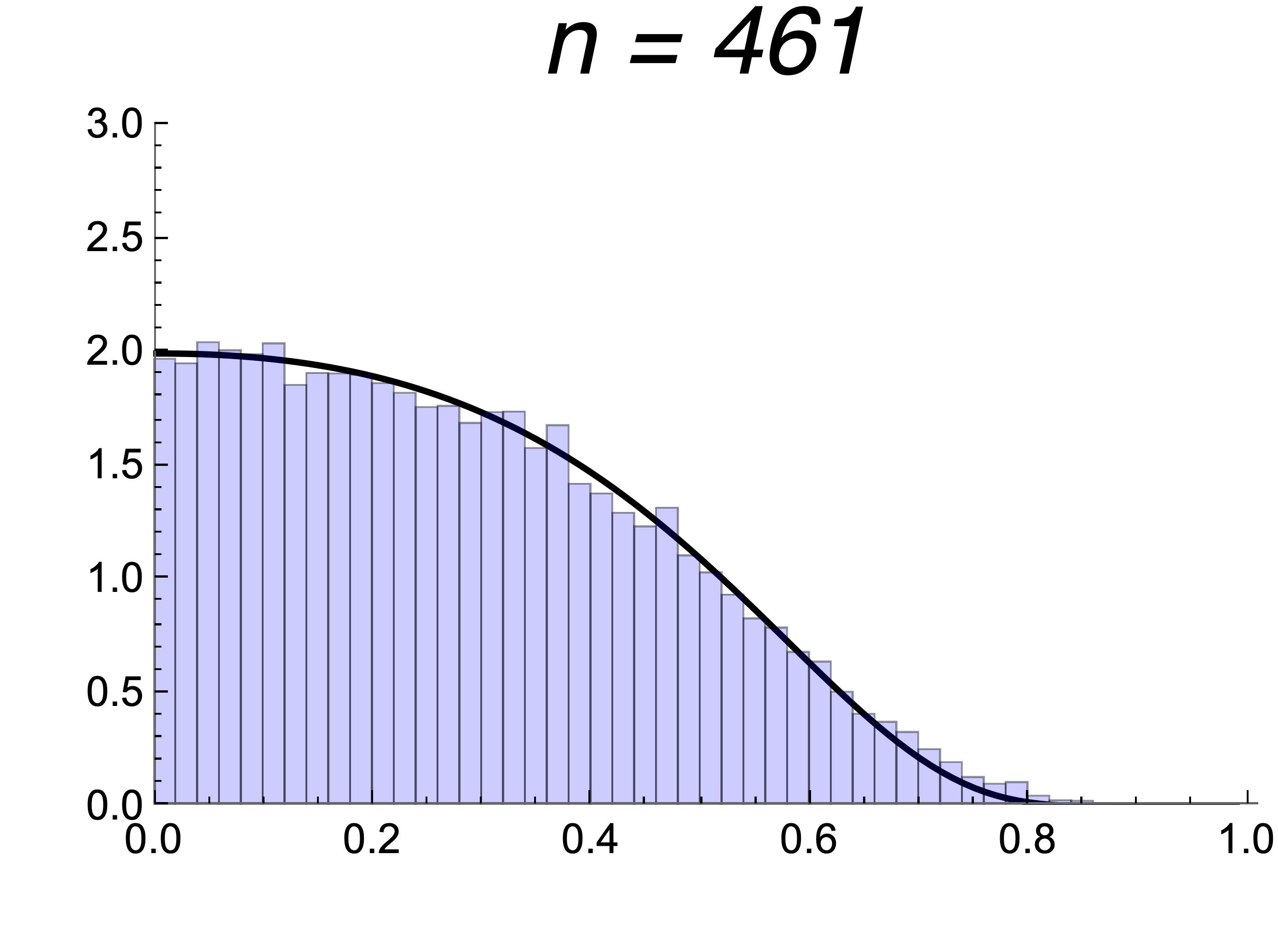}
& \includegraphics[width=0.25\textwidth]{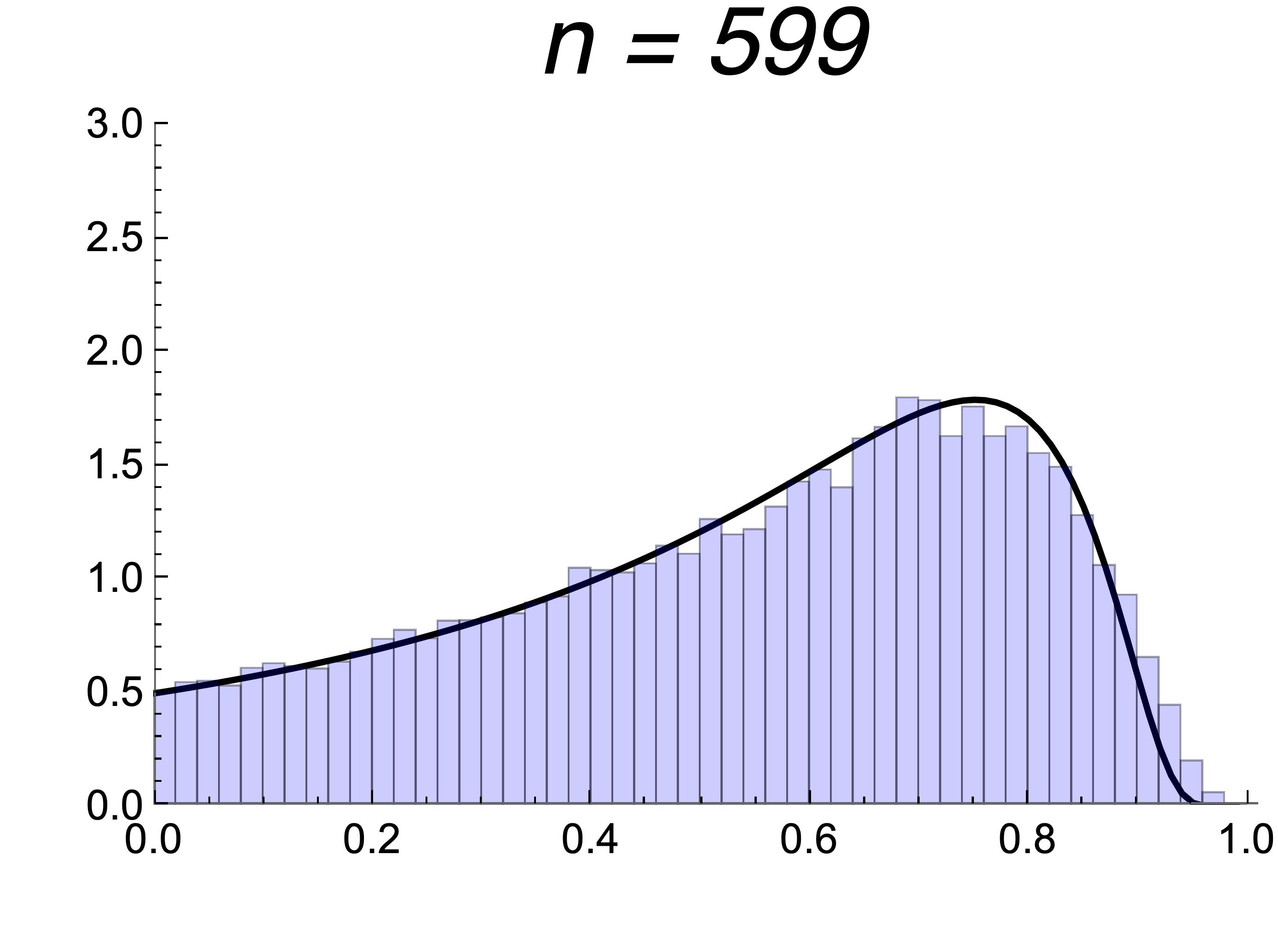} 
\\ $s=0.001$ \\
& \includegraphics[width=0.25\textwidth]{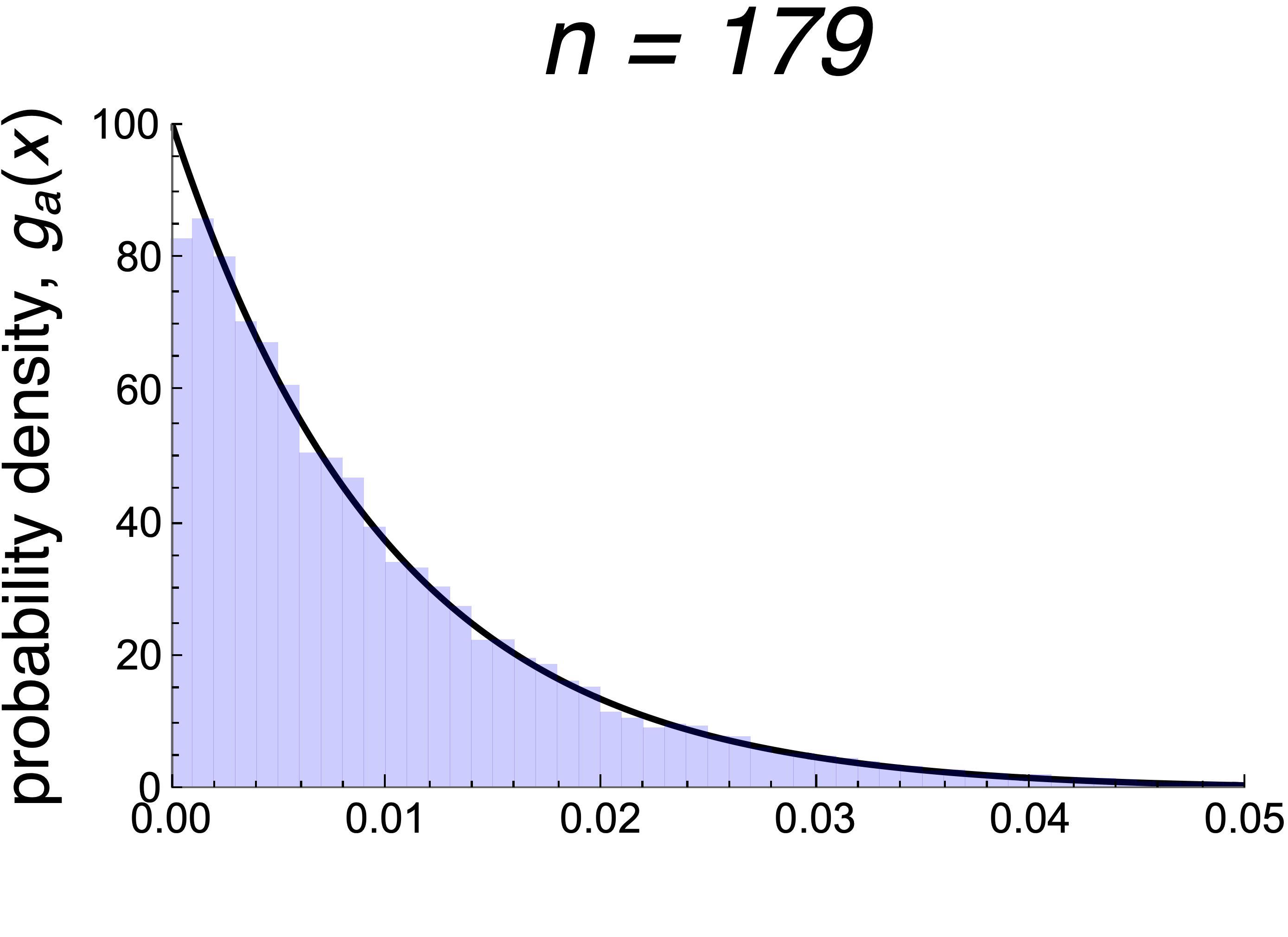}
& \includegraphics[width=0.25\textwidth]{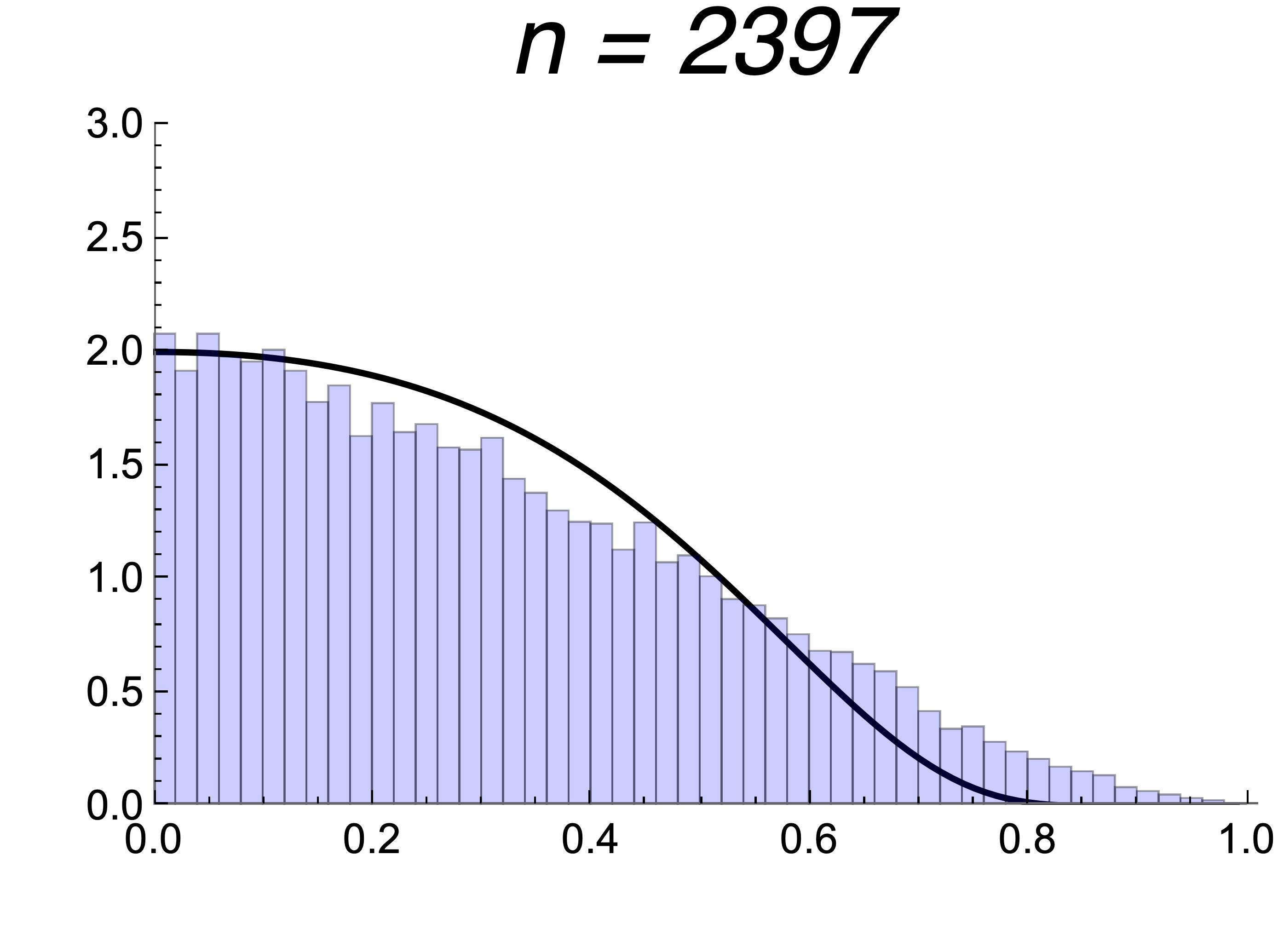}
& \includegraphics[width=0.25\textwidth]{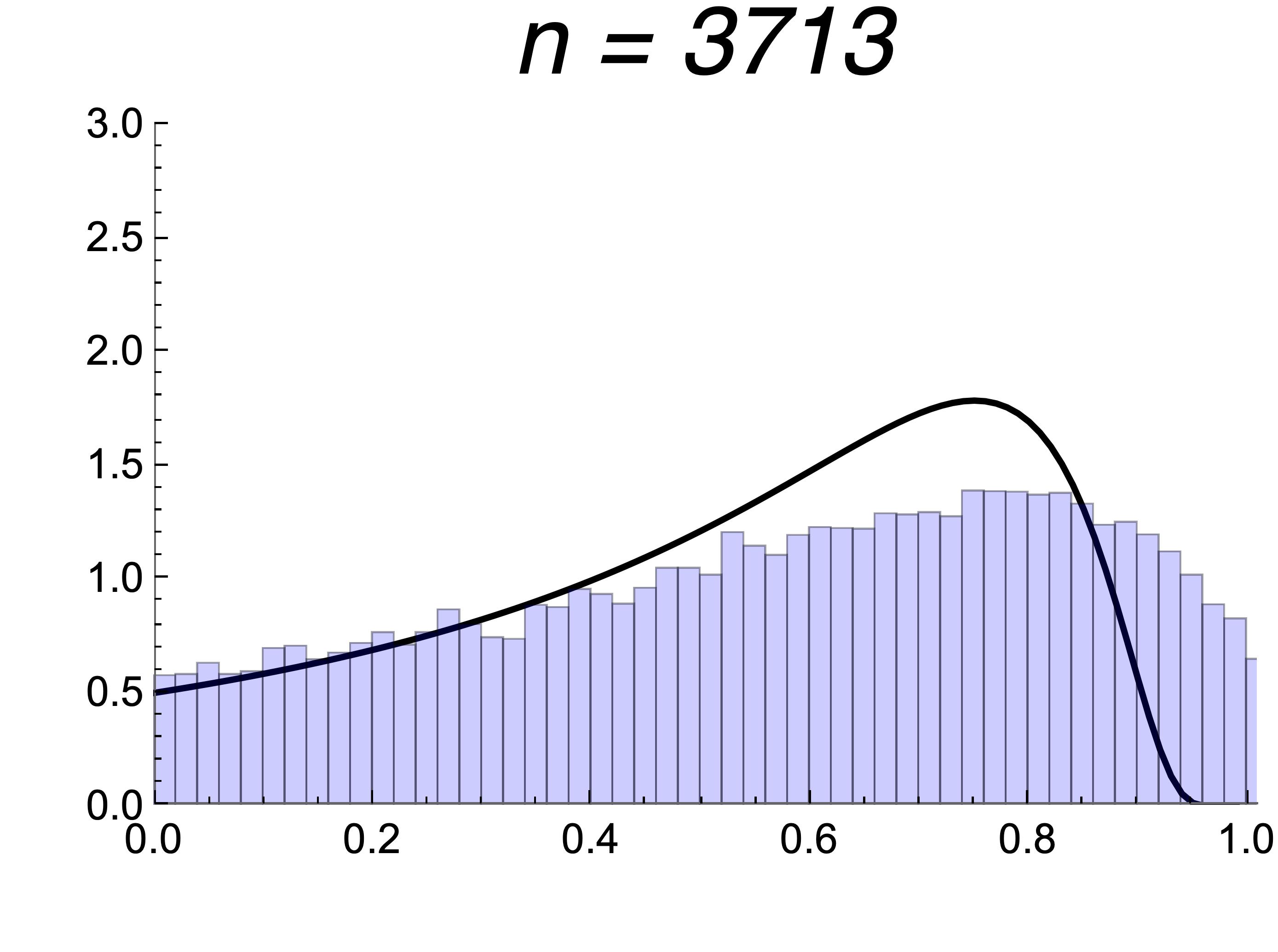} 
\\ $s=0.0001$ \\
& \includegraphics[width=0.25\textwidth]{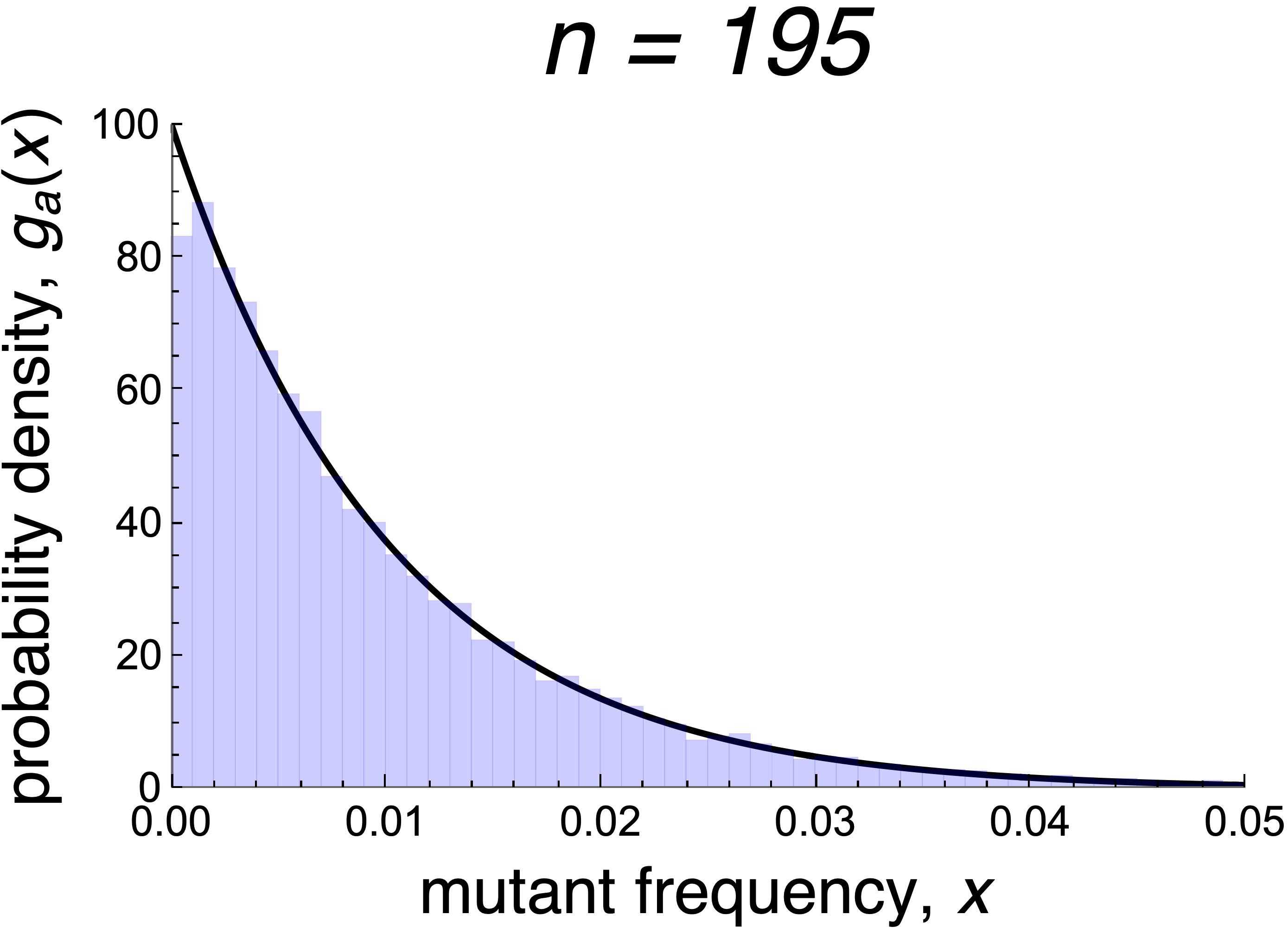}
& \includegraphics[width=0.25\textwidth]{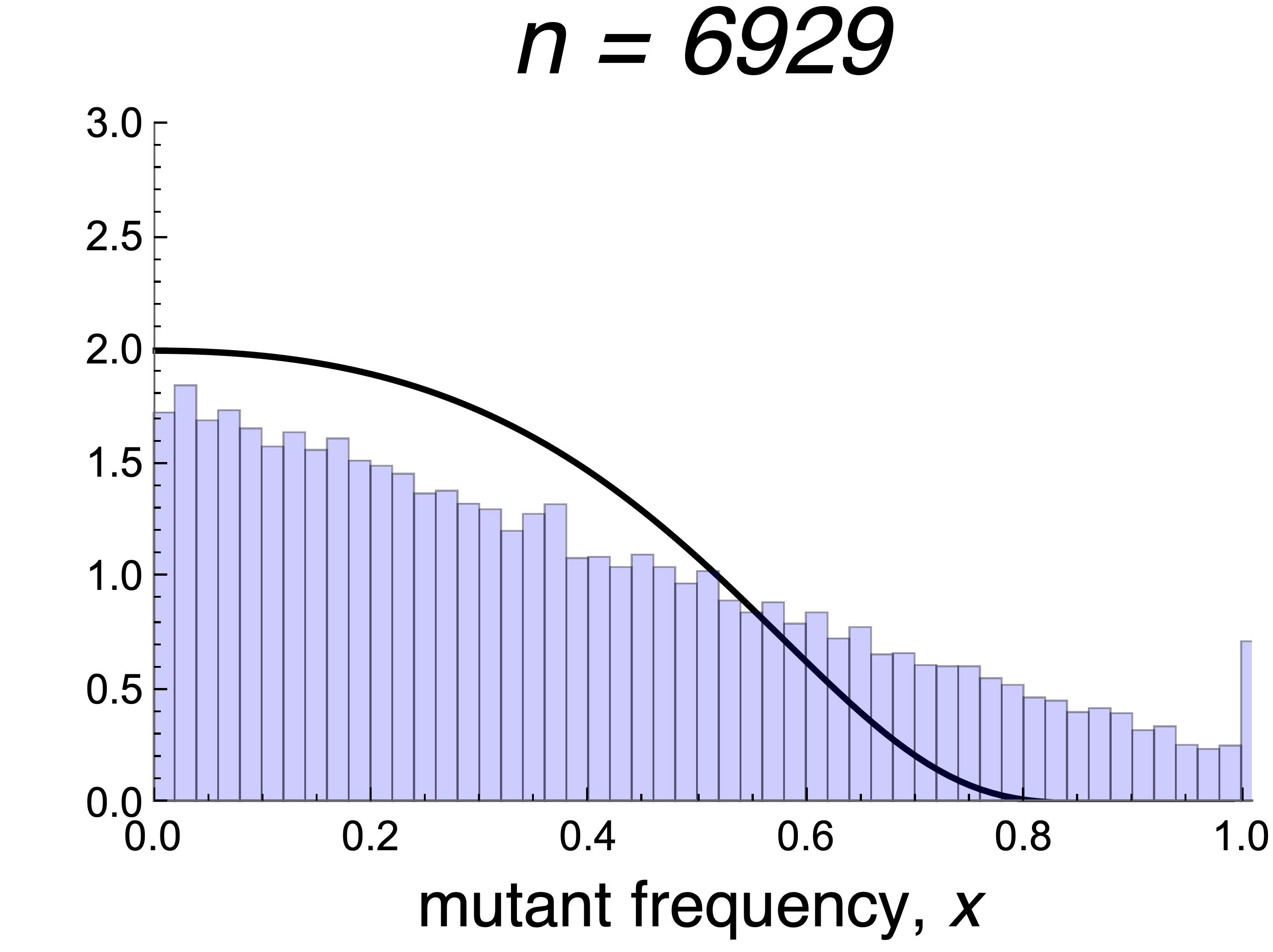}
& \includegraphics[width=0.25\textwidth]{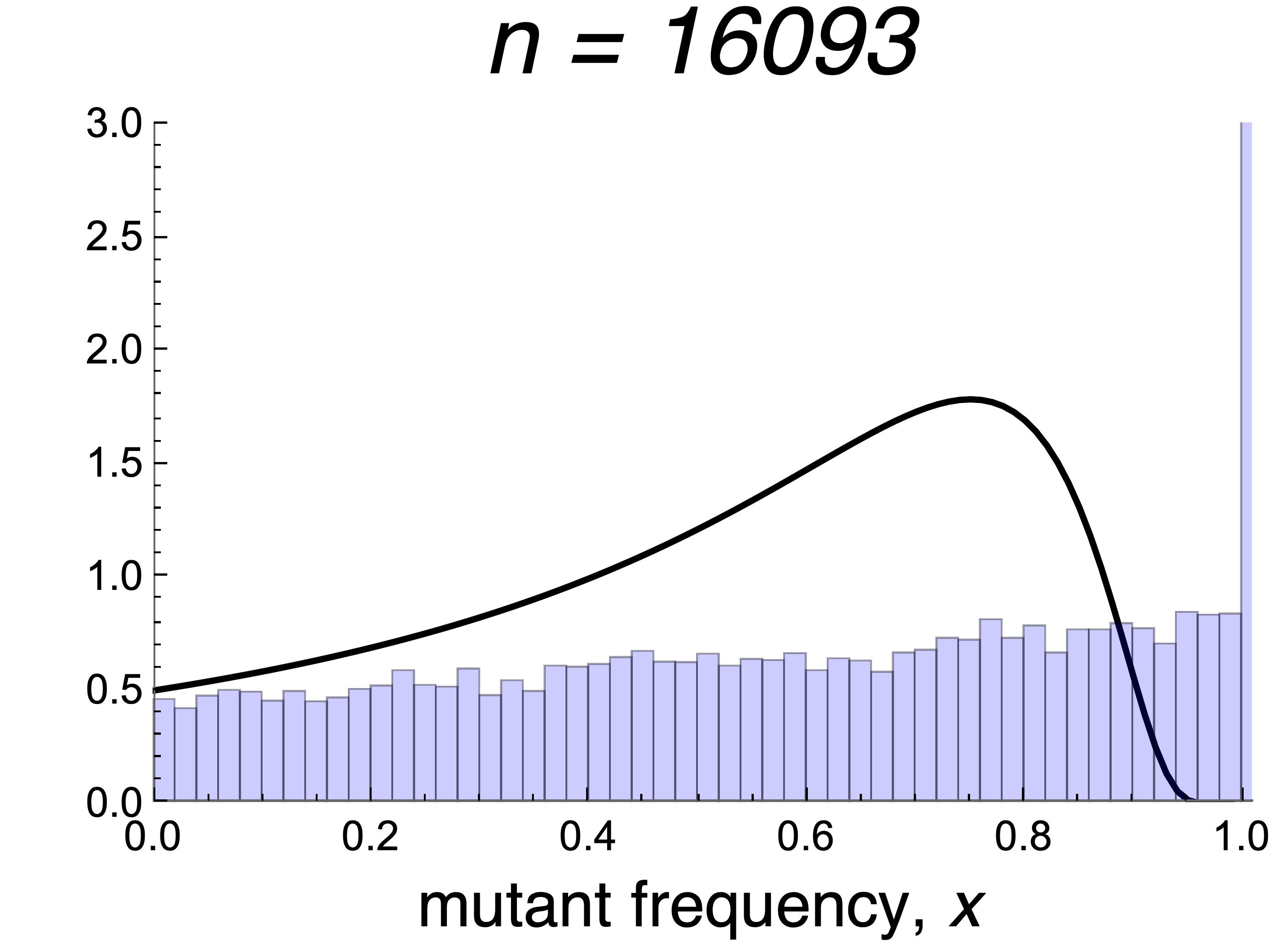} 
\end{tabular}
\caption[$g_a (x)$]{The black curves show the probability density $g_a(x)$ in \eqref{g(p)} of the mutant frequency in different generations $n$ for various combinations of $a = a(n,\si,N)$ and $\si=e^s$ (conditioned on non-extinction until generation $n$). The number of discrete generations $n$ (indicated above each graph) is chosen such that $a$ (see eq.~\eqref{g(p)}) is closest to $100$, $2$ and $0.5$, respectively. The population size is $N=10^4$. The blue histograms are obtained from simulations of the corresponding Wright-Fisher model. The analytic approximation $g_a (x)$ underestimates the effect of random genetic drift near fixation ($x$ large and $n >\tfrac12\tfix\approx \tfrac1s\ln(2Ns)$), unless selection is very strong relative to random drift (top row, where $Ns=10^3$). For $s=0.1$, $0.01$, $0.001$, and $0.0001$, we have $\tfix\approx 156$, $1060$, $5991$, and $13\,863$, respectively. In general, and according to intuition and assumptions, the accuracy of the approximation by $g_a(x)$ decreases with larger $x$ and $n$, and smaller $Ns$.  Nevertheless, for very early stopping times $n$ relative to $\tfix$, the approximation remains useful even if $Ns=1$ (left bottom). Its mean, given by \eqref{mean_g}, stays accurate in a much wide parameter range than $g_a(x)$ (Fig.~\ref{fig_ContrMeanVar}.A).} 
\label{fig_g(p,a)2}
\end{figure}

\newpage

\begin{figure}[ht]
\centering
\begin{tabular}{c}
\includegraphics[width=0.7\textwidth]{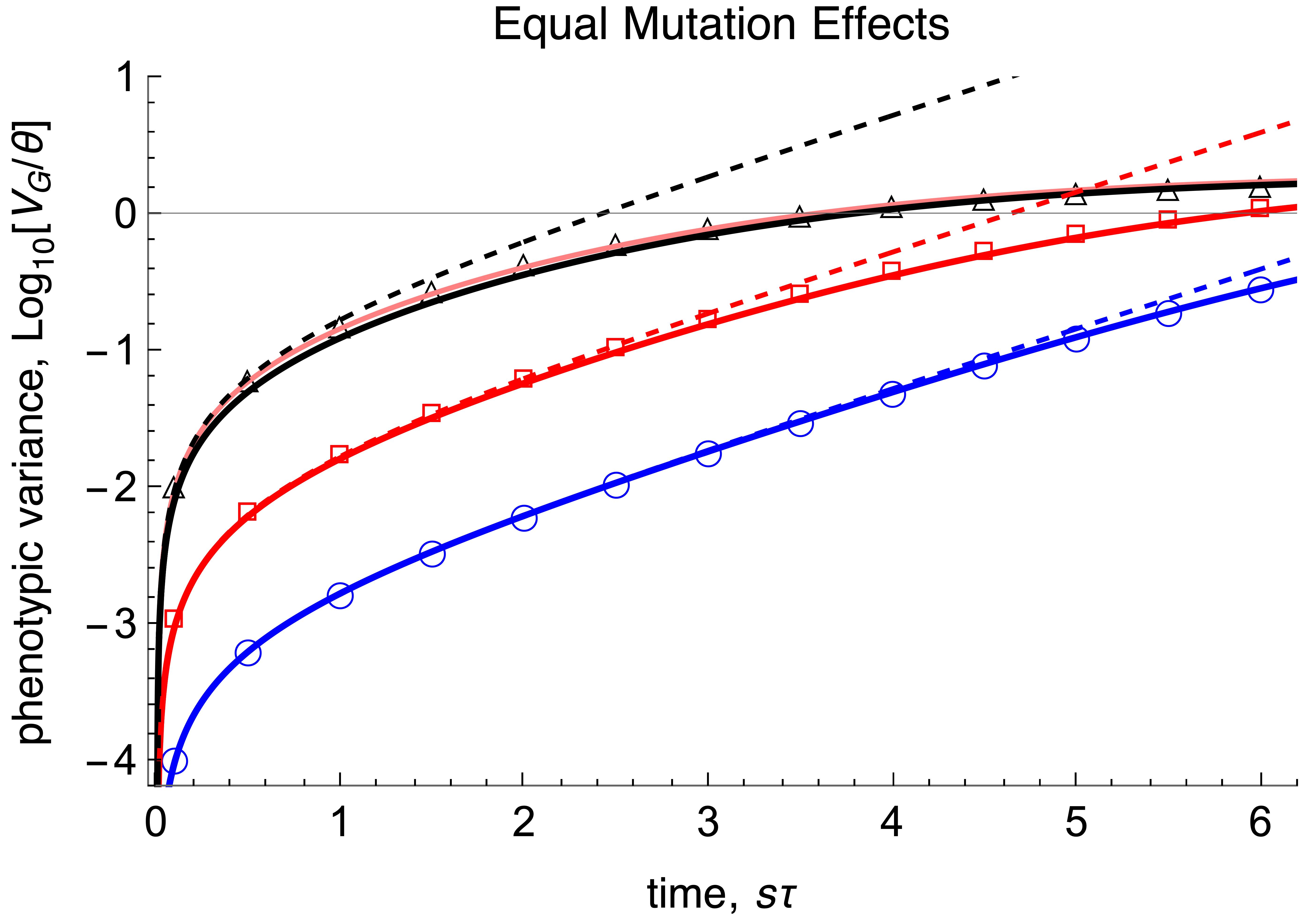}
\end{tabular}
\caption[$V_{G}$ - Prop.~\ref{prop:general_barG_VG(tau)}]{This is analogous to panel B of Figure \ref{fig_MeanVarInitial1}, the scaled phenotypic variance $\log_{10}(V_G(s\tau)/\Th)$ in the initial phase for equal mutation effects $\A=1$ and selection coefficients $s=0.001,0.01,0.1$ (represented by black, red and blue color, respectively) is shown. The solid curves show the approximation \eqref{VG_init_equal}. The dashed curves show the simplified approximation in \eqref{VG_smalltau_const}. For $s=0.001$, the pink solid curve shows the most accurate approximation \eqref{VGtau}. For all other parameter combinations in Figure \ref{fig_MeanVarInitial1} (also for the phenotypic mean) no deviation from the solid curves to the approximations from Proposition \ref{prop:general_barG_VG(tau)} is visible. Symbols present results from Wright-Fisher simulations with $\Th=\tfrac12$, the population size is $N=10^4$, and the offspring distribution is Poisson.}
\label{fig_VarInitial_comparison}
\end{figure}

\newpage

\begin{figure}[ht]
\centering
\begin{tabular}{ll}
A & B \\
\includegraphics[width=0.45\textwidth]{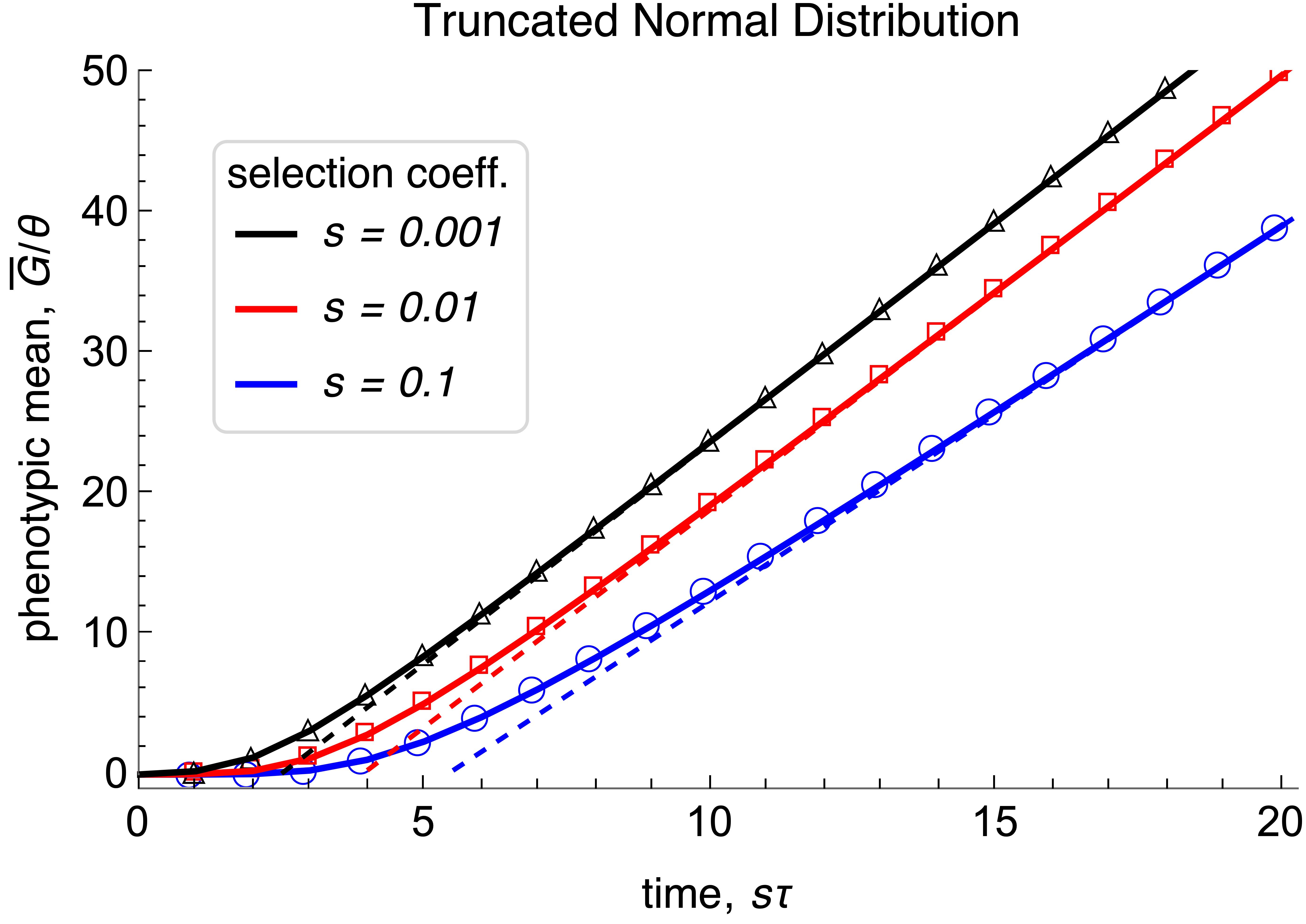} &
\includegraphics[width=0.45\textwidth]{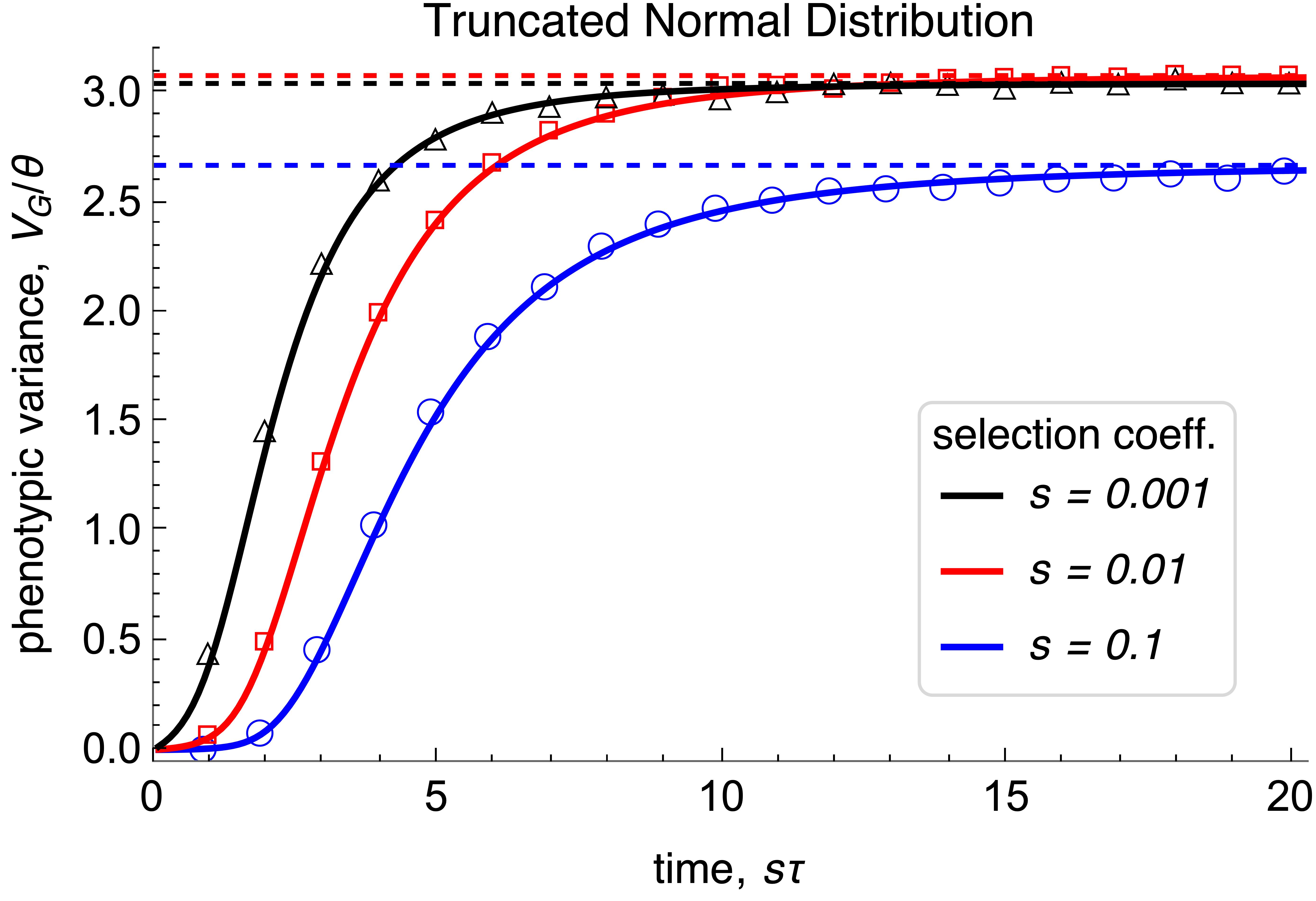} \\
\multicolumn{1}{c}{C}  \\
\multicolumn{2}{c}{
\includegraphics[width=0.45\textwidth]{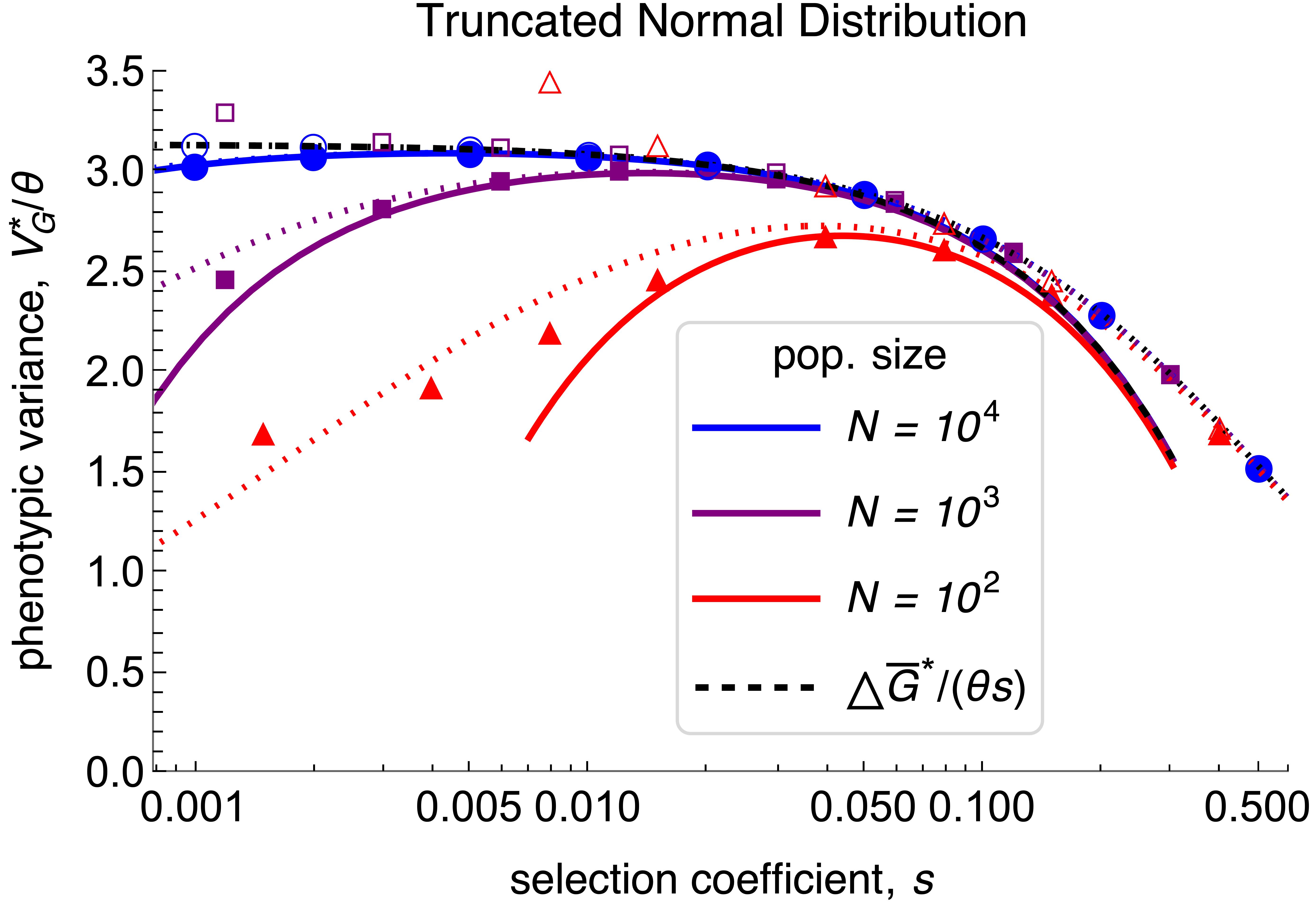}} 
\end{tabular}
\caption[Truncated normal distribution - $\bar{G}$ and $V_G$]{
The scaled expected phenotypic mean and variance, $\bar G(s\tau)/\Th$ and $V_G(s\tau)/\Th$ as functions of scaled time (panels A and B), and the scaled expected phenotypic variance at stationarity, $V^*_G(s\tau)/\Th$ (panel C) as a function of $s$ for a truncated normal distribution with $\A\ge0$ and $\bar\A=1$ (cf.~Remark \ref{rem:truncatedGaussian}). Panels A and B are analogous to Fig.~\ref{fig_MeanVarEqu} and panel C is analogous to Fig.~\ref{fig_VarSEqu}. Analytic approximations are shown as curves, Wright-Fisher simulation results (for $\Th=\frac{1}{2}$) as symbols. \\
\hglue5mm
In panels A and B, we have $N=10^4$ and selection coefficients and color code are shown in the legends. The solid curves show $\bar{G}(s\tau)/\Th$ and $V_G(s\tau)/\Th$ as computed from \eqref{barGtau} and \eqref{VGtau}. The dashed lines show the equilibrium values given by \eqref{barG*_prop} and \eqref{VG*_prop}. The values of the mean fixation times multiplied by $s= 0.1, 0.01, 0.001$ are $s\bartfix \approx 16.1,11.0,6.2$, respectively (computed from \eqref{bartfix}).\\
\hglue5mm
In panel C, different colors correspond to different population sizes, as shown in the legend.  The most accurate, available approximation is \eqref{VG*_prop}; it is shown as dotted curves. In addition, the scaled variance $V_G^*/ \Th$ is compared with the scaled expected change in the phenotypic mean $\Delta \bar{G}^*/ (\Th s)$, as calculated from \eqref{DebarG*tn} (black dashed curve). For the Wright-Fisher simulation results (circles for $N = 10^4$, squares for $N = 10^3$, triangles for $N = 10^2$), filled symbols are used for the scaled phenotypic variance and open symbols for the change in the scaled phenotypic mean. 
}
\label{fig_nor}
\end{figure}

\newpage

\begin{figure}[ht!]
\centering
\begin{tabular}{ll}
A & B \\
\includegraphics[width=0.45\textwidth]{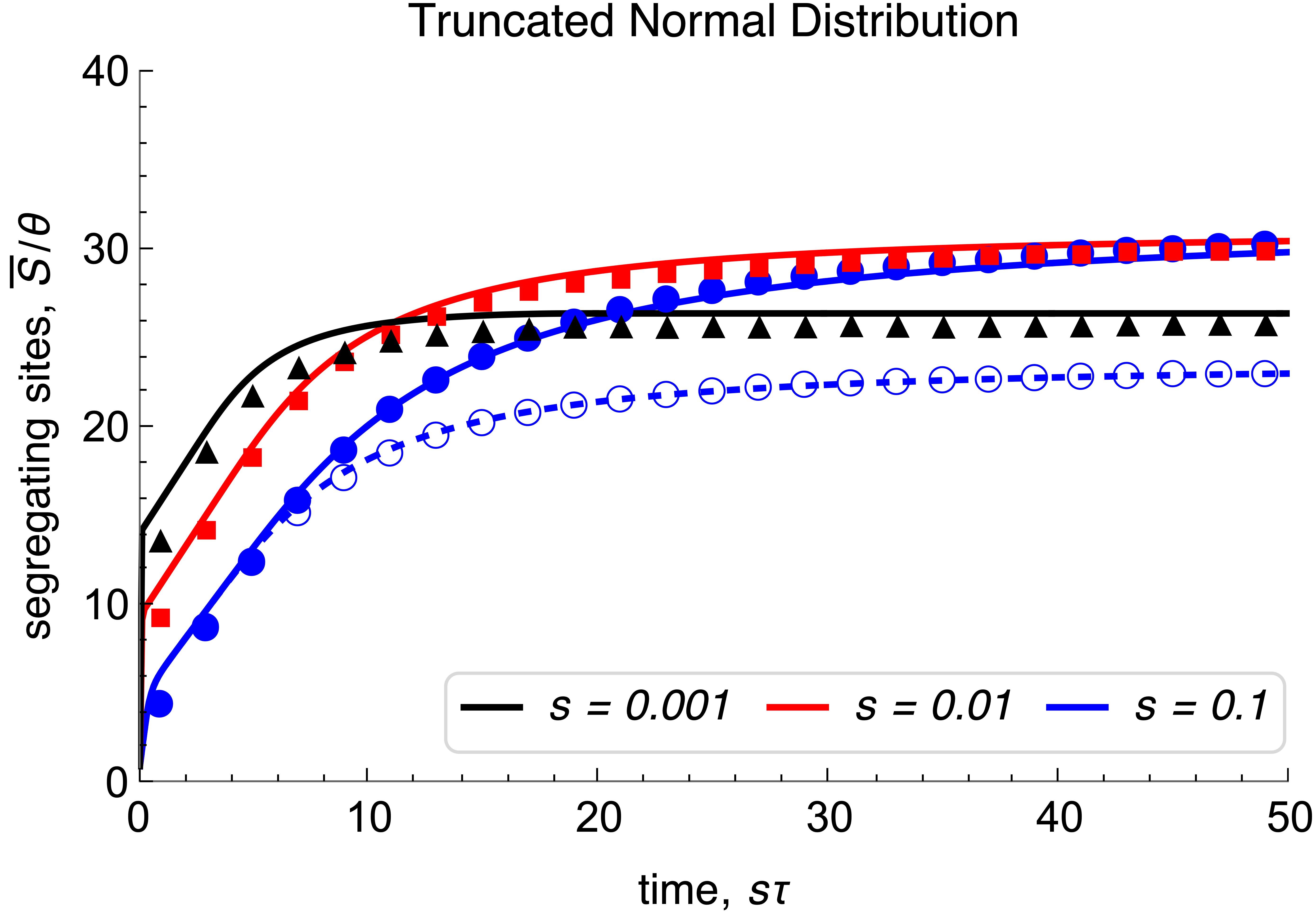} &
\includegraphics[width=0.45\textwidth]{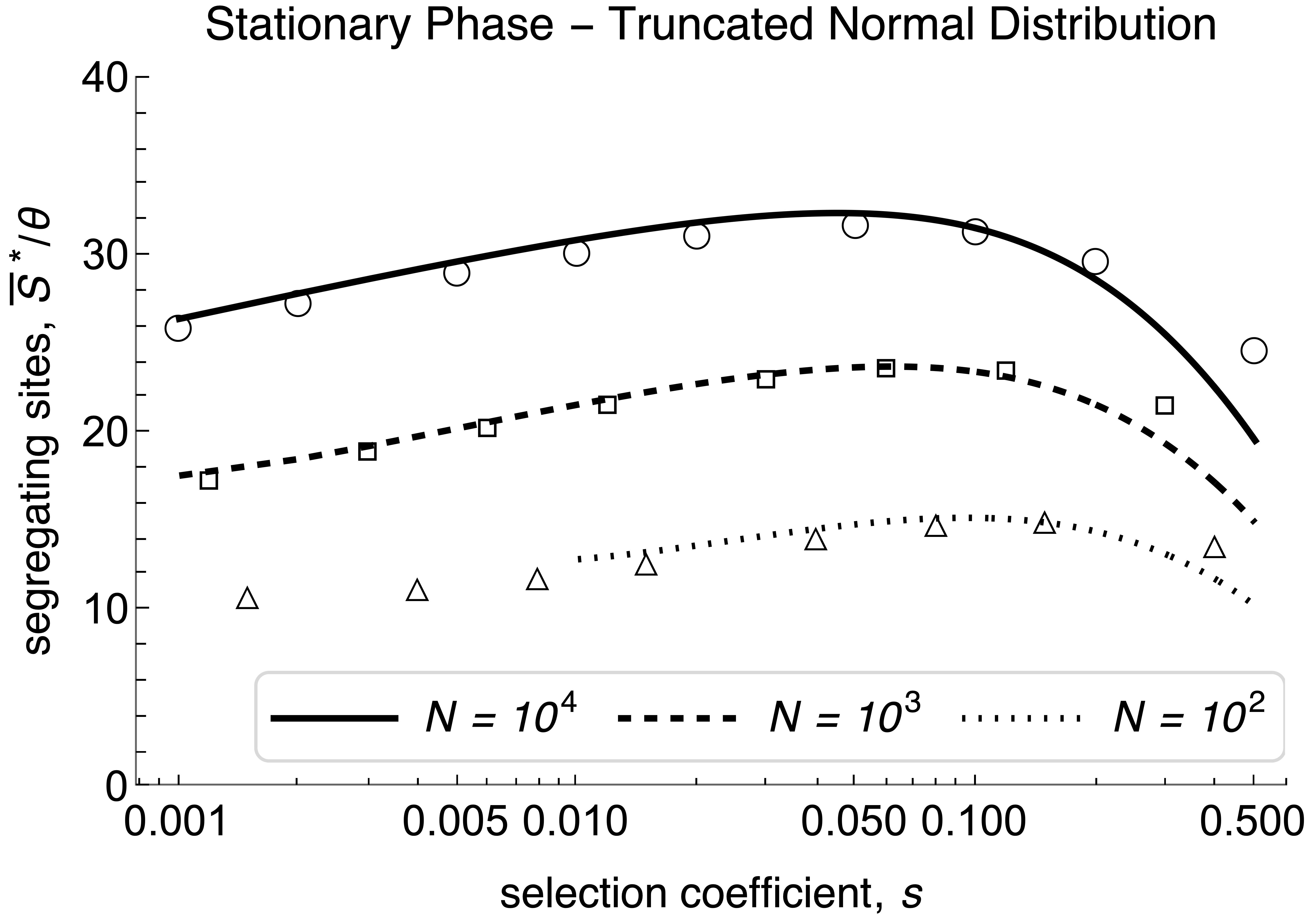}
\end{tabular}
\caption[Truncated normal distribution - segregating sites]{
The expected number of segregating sites (rescaled by $\Th$) as a function of $s\tau$ (panel A) and at stationarity as a function of $s$ (panel B). This figure is analogous to the rows in Fig.~\ref{fig_SegSites} but mutation effects are drawn from a normal distribution truncated at 0 with mean $\bar{\A} = 1$ (cf.~Remark \ref{rem:truncatedGaussian}). Analytic approximations are computed from \eqref{Seg} and are shown as curves. Wright-Fisher simulation results for $\EV[S]$ (with $\Th = \frac{1}{2}$) are shown as symbols.  In A, $\bar S/\Th$ is shown for $N=10^4$ and $s=0.1,0.01,0.001$ (solid curves and filled symbols in blue, red, black), and for $N=10^3$ and $s=0.1$ (dashed curve and open symbols in blue).  In B, $\bar S^*/\Th$ is displayed for $N= 10^4, 10^3, 10^2$ (solid, dashed, dotted curves, and corresponding open symbols).}
\label{fig_norSeg}
\end{figure}

\newpage

\begin{figure}[!htb]
	\begin{tabular}{l | c c c}
		& Sweep & & Shift \\			
		& $2\Theta=0.1$ & $2\Theta=1$ & $2\Theta=10$ \\		
		\hline \\	
		$s=0.001$ &
				\multirow{7}{*}{\includegraphics[width=0.25\textwidth]{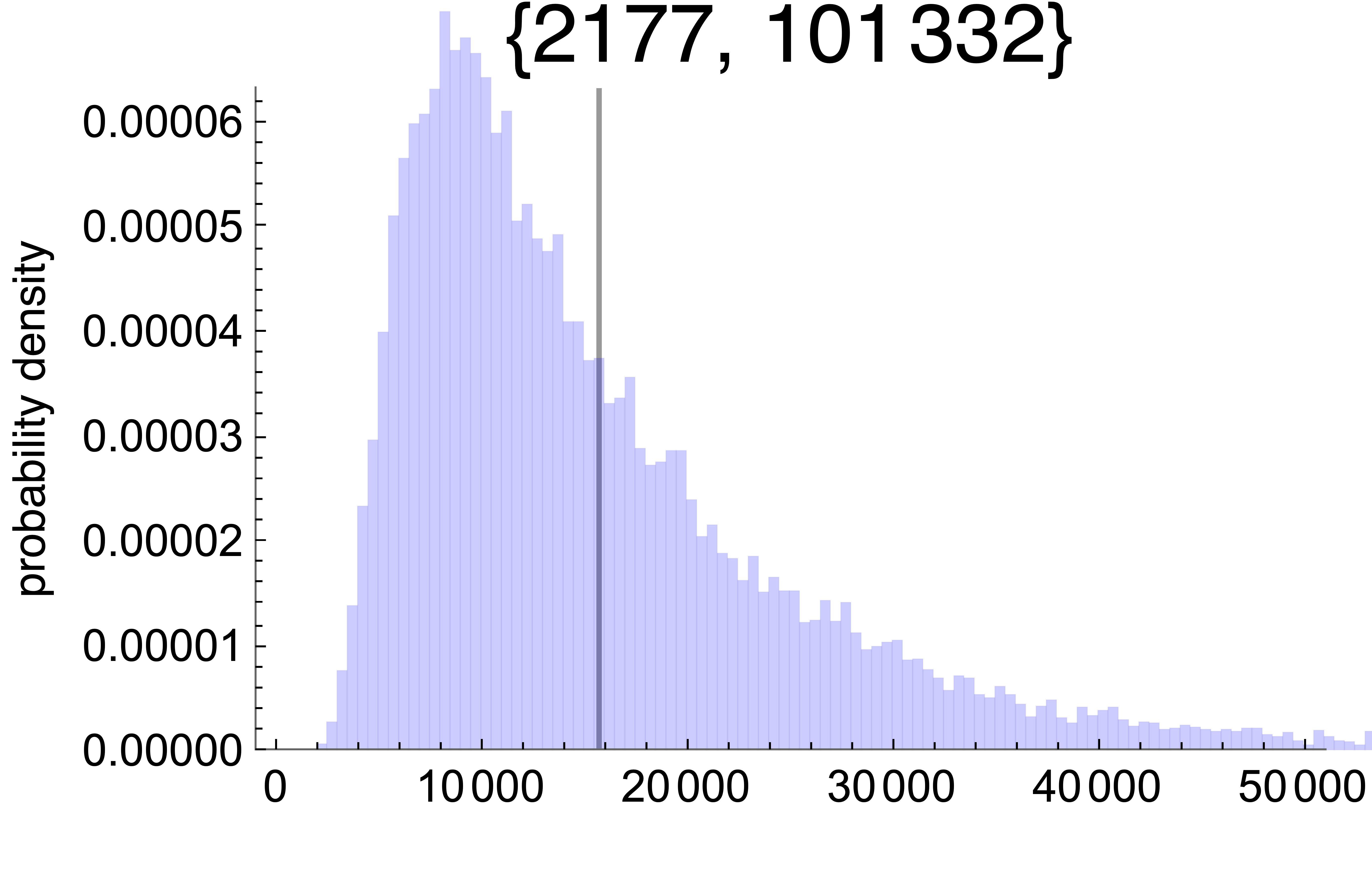}} &
				\multirow{7}{*}{\includegraphics[width=0.25\textwidth]{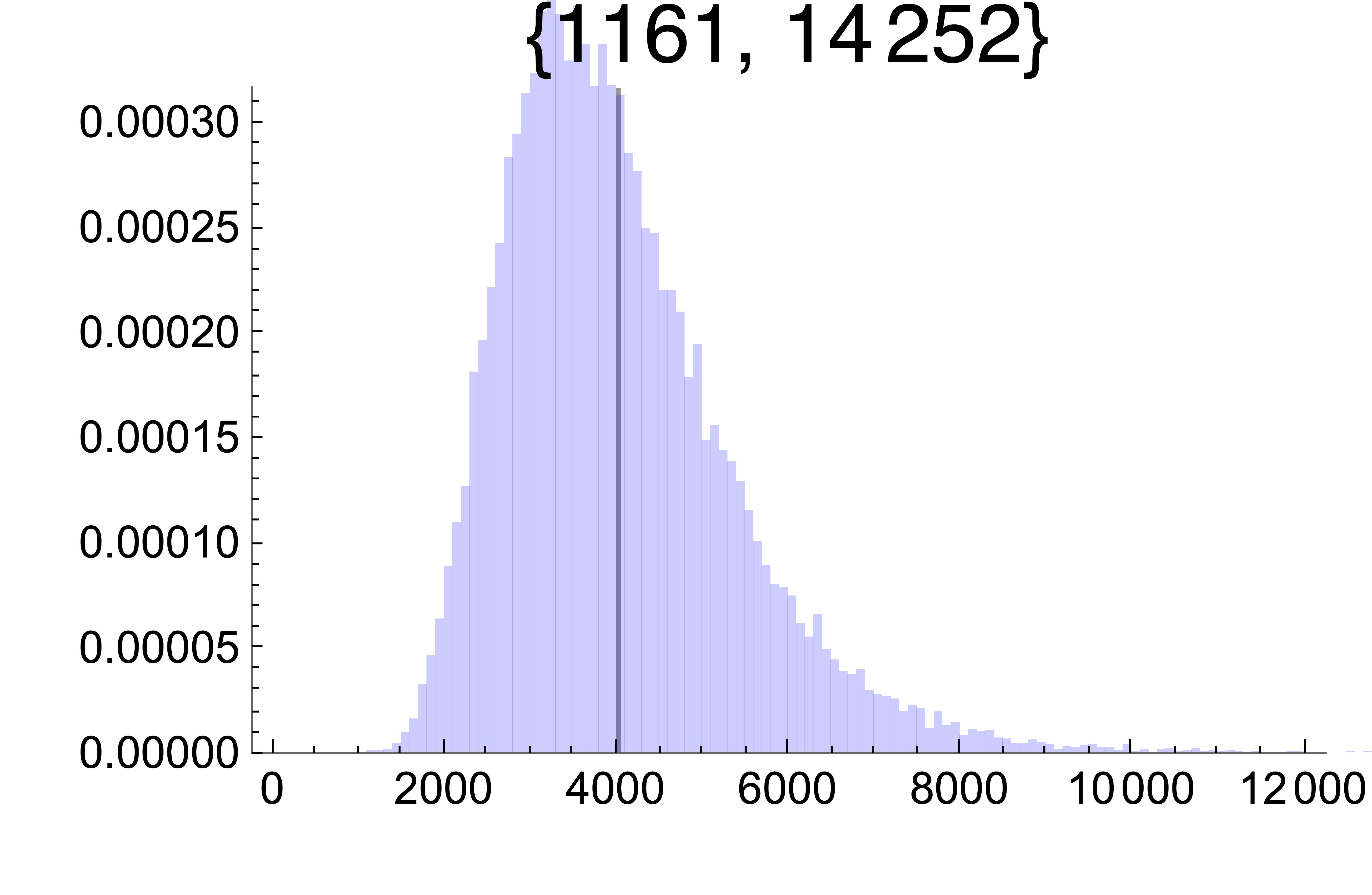}} & 
				\multirow{7}{*}{\includegraphics[width=0.25\textwidth]{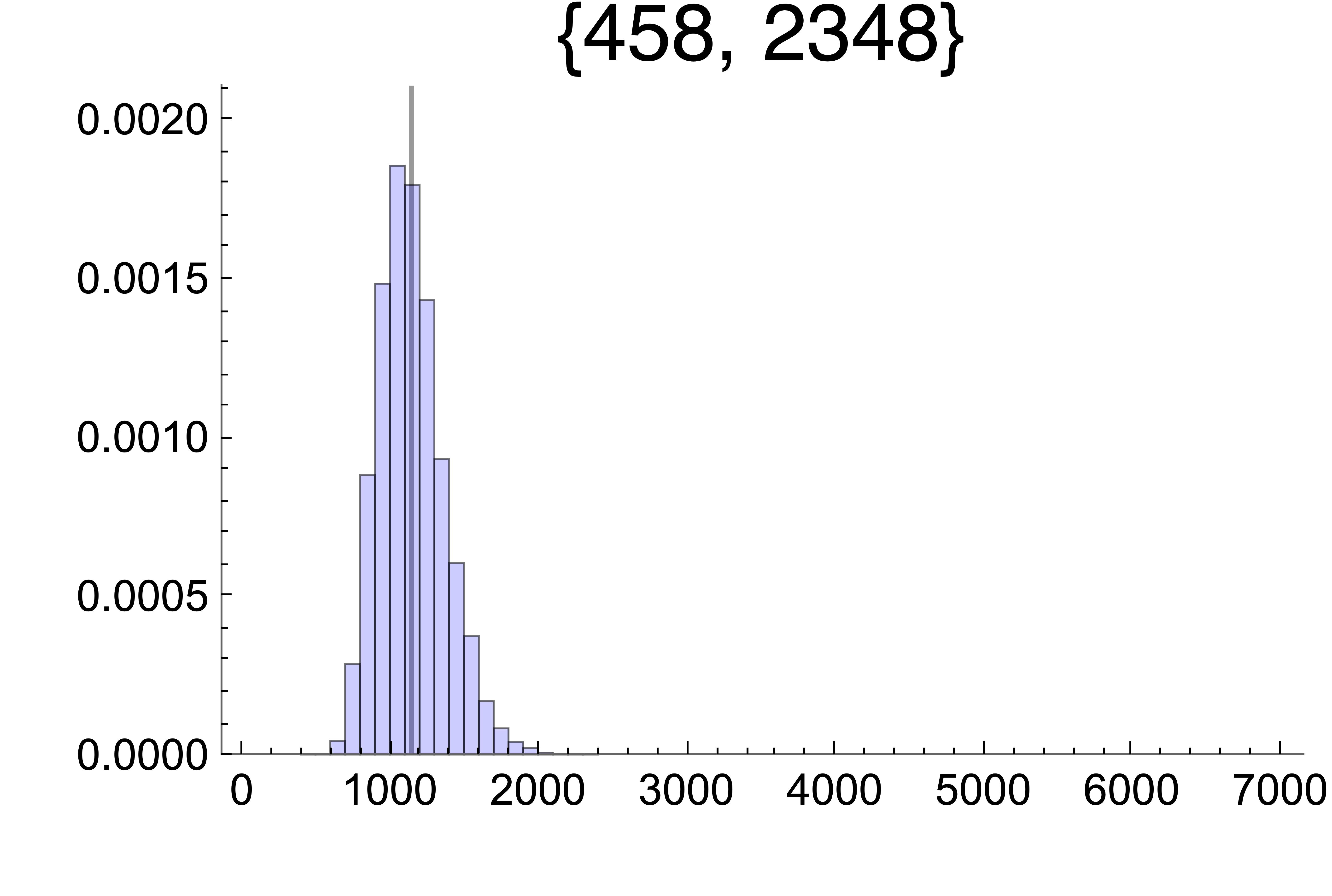}} \\
				$\alpha =1$ \\ \\ \\ \\ \\ \\
		$s=0.1$ &
				\multirow{7}{*}{\includegraphics[width=0.25\textwidth]{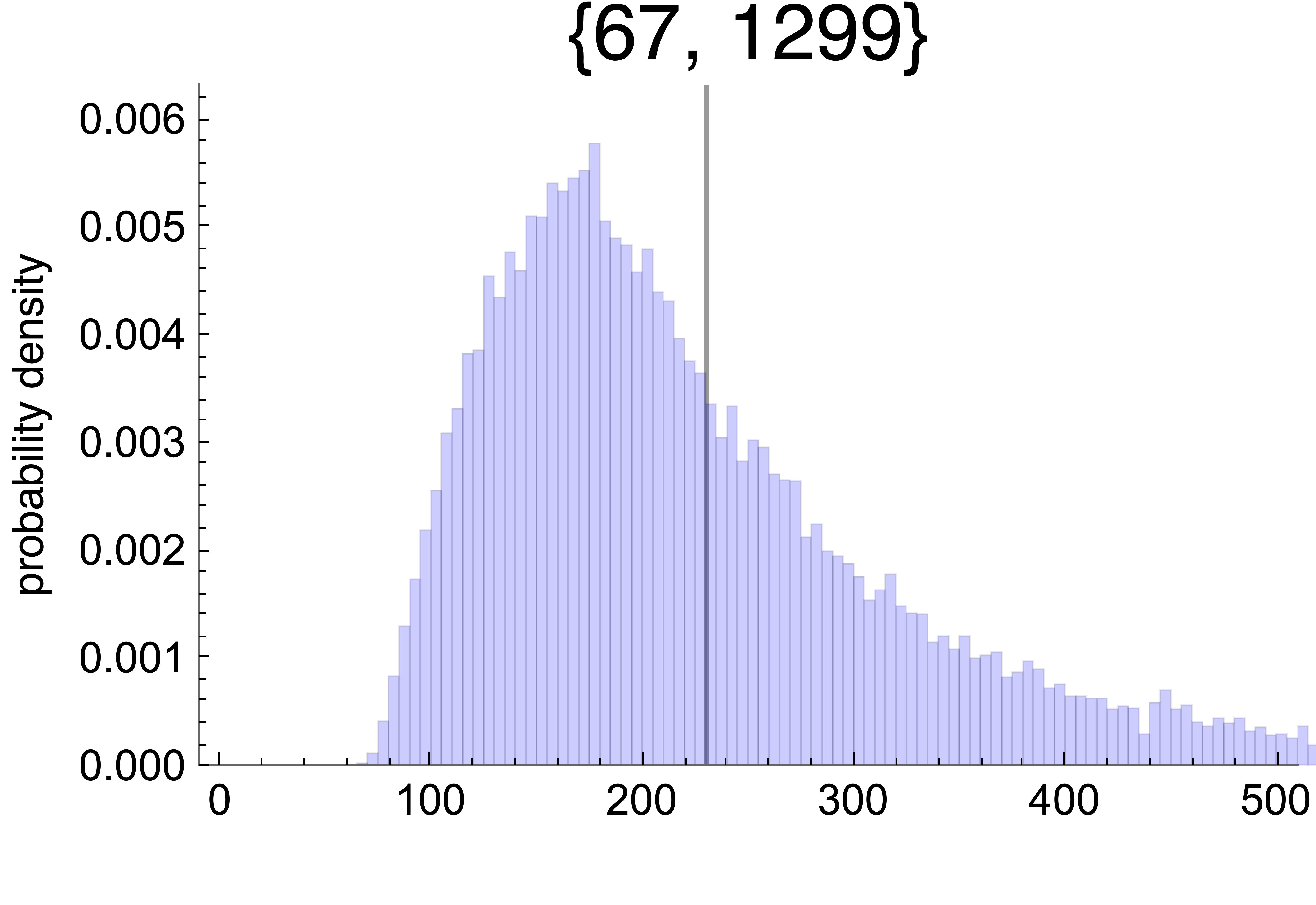}} &
				\multirow{7}{*}{\includegraphics[width=0.25\textwidth]{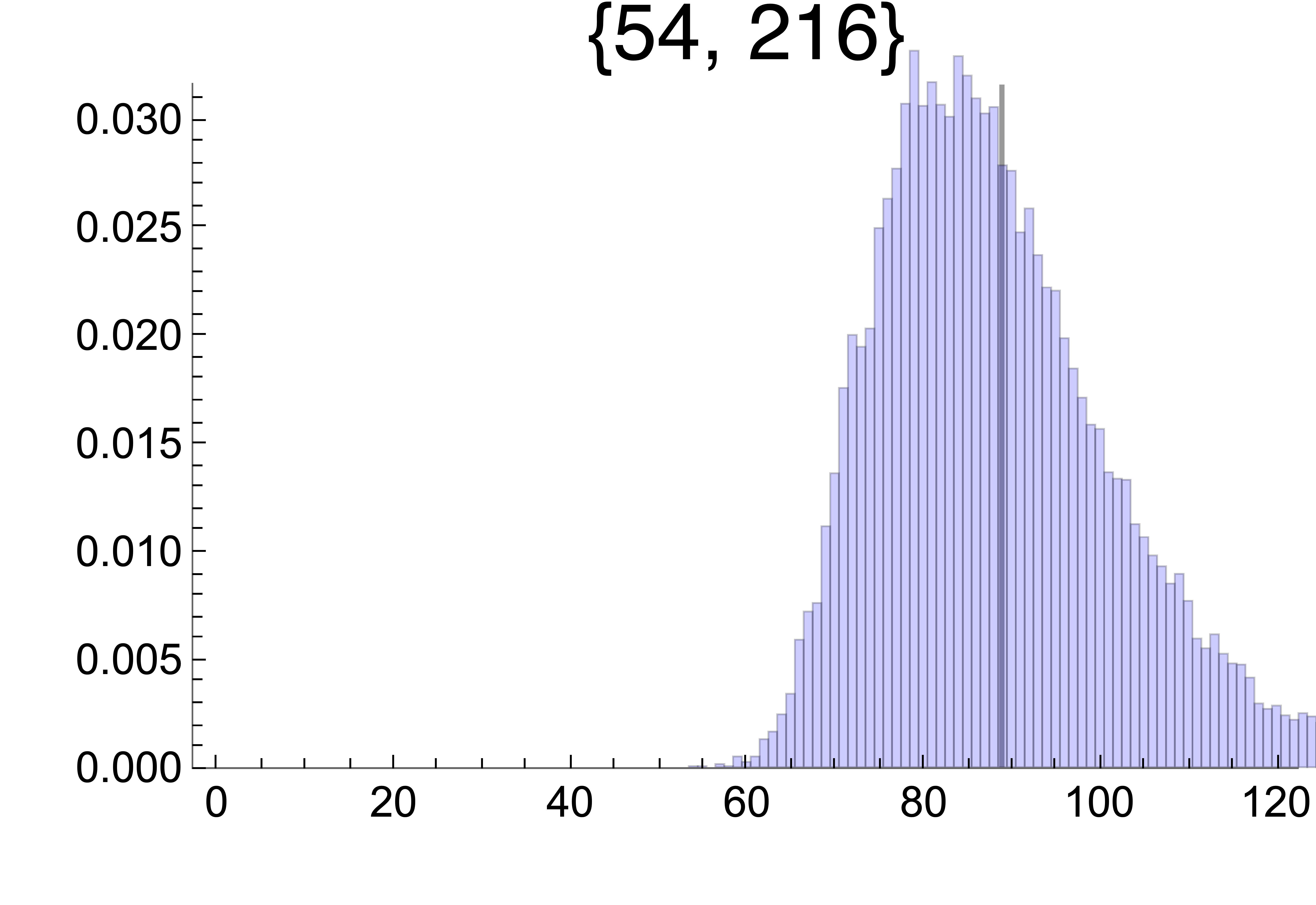}} &
				\multirow{7}{*}{\includegraphics[width=0.25\textwidth]{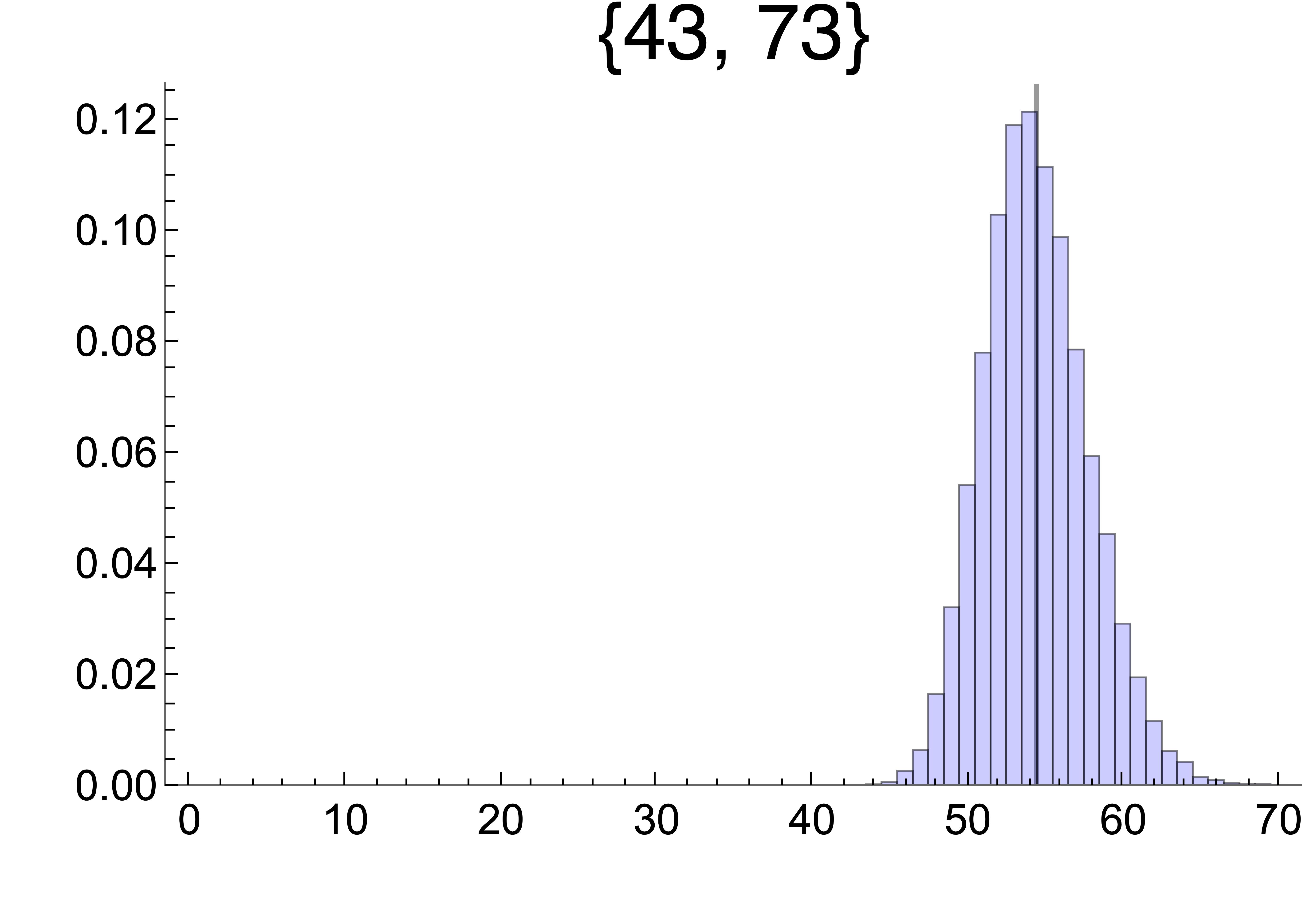}} \\ 	
				$\alpha =1$ \\ \\ \\ \\ \\ \\
		$s=0.1$ &
				\multirow{7}{*}{\includegraphics[width=0.25\textwidth]{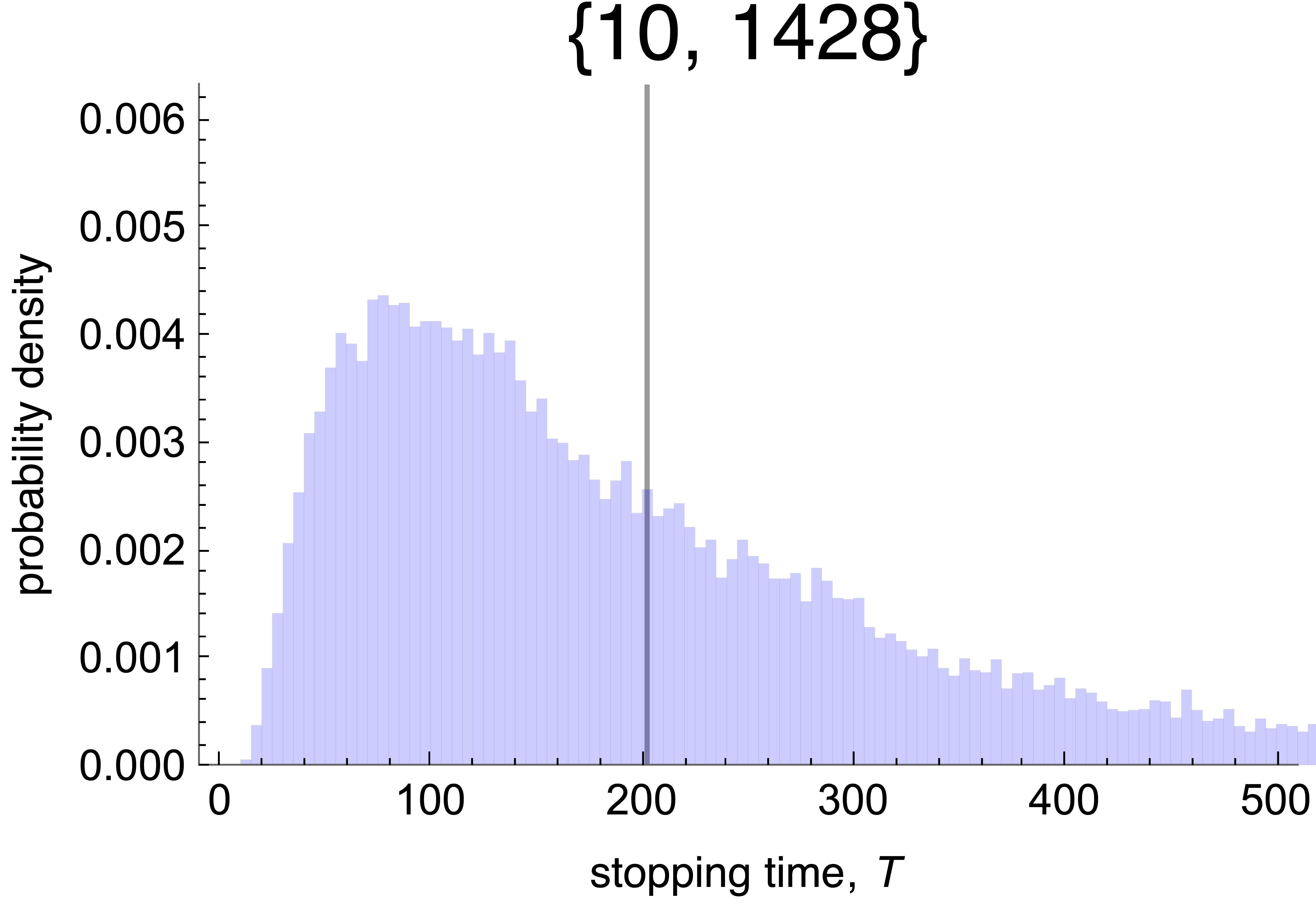}} &
				\multirow{7}{*}{\includegraphics[width=0.25\textwidth]{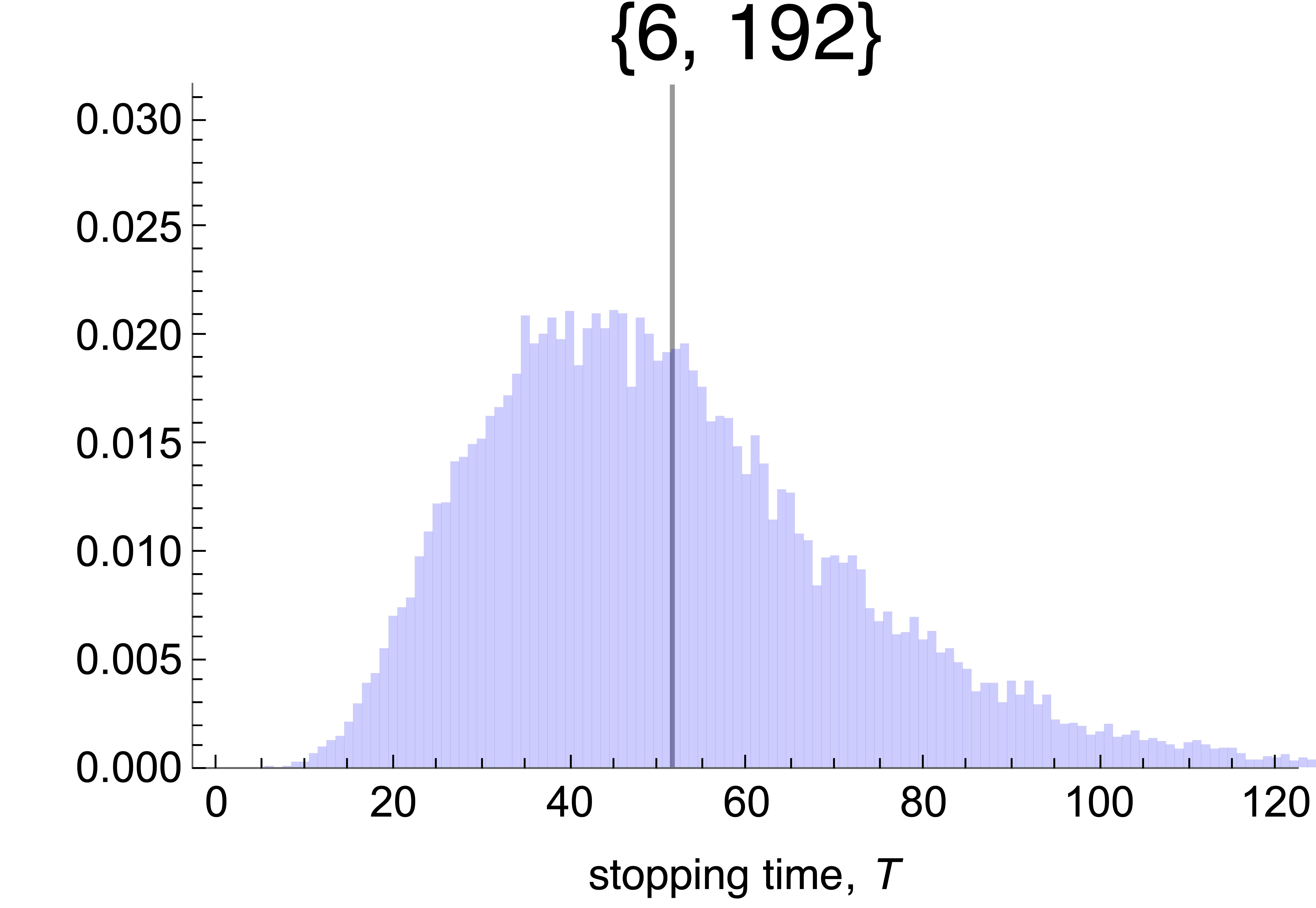}} & 
				\multirow{7}{*}{\includegraphics[width=0.25\textwidth]{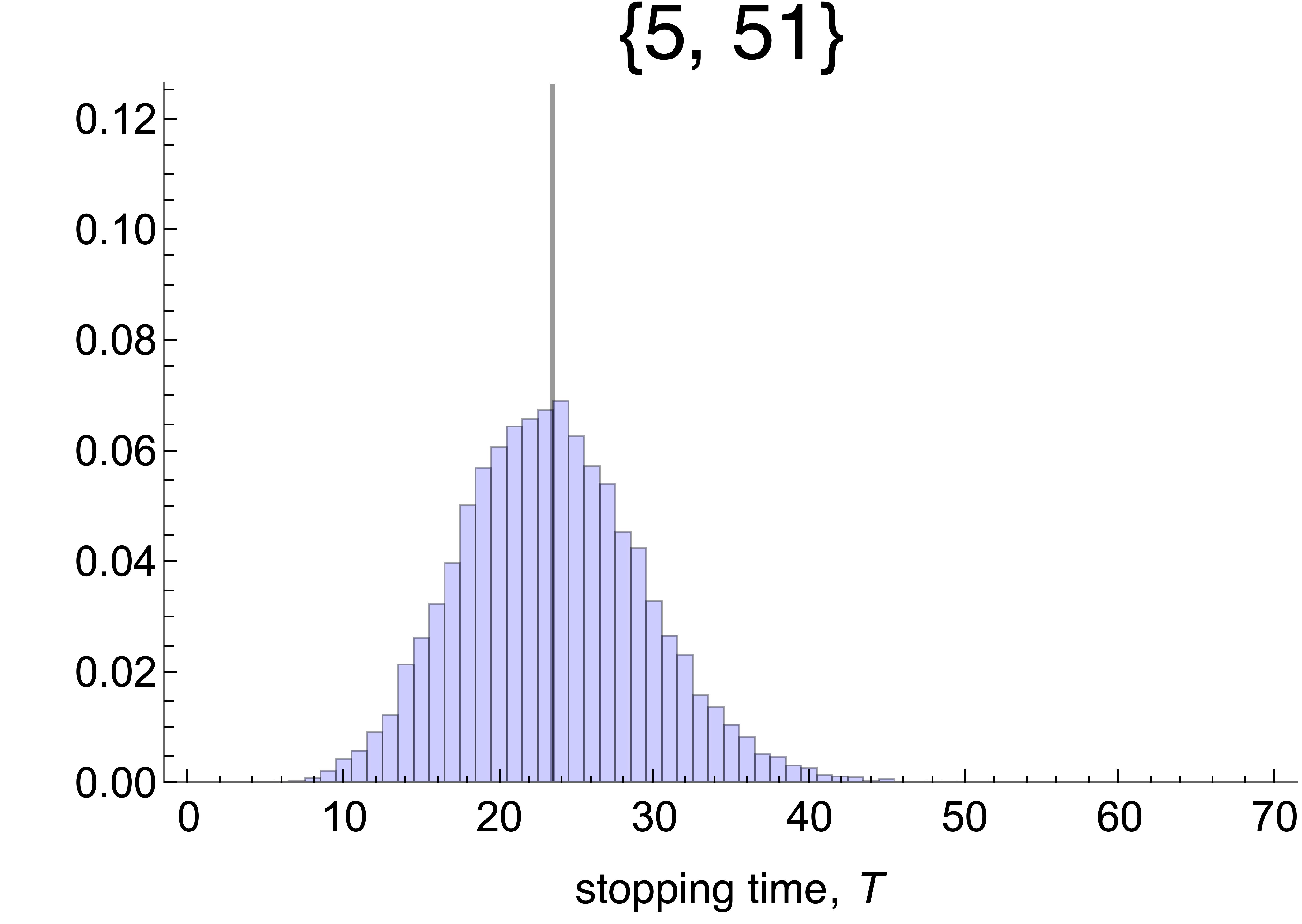}} \\
				$\alpha \sim \text{Exp}[1]$ \\ \\ \\ \\ \\ \\
	\end{tabular}
	\caption[Sweep vs.\ shift - $\tau$]{Histograms of the stopping times $T$ at which the mean phenotype $\bar G$ in a single population first reaches or exceeds the value $1$ (this figure complements Fig.~\ref{fig:sweep_shift_new}).  Histograms are based on $20\,000$ replicate runs of the Wright-Fisher model. The vertical black lines represent the corresponding means $\bar{T}$, which are given in Fig.~\ref{fig:sweep_shift_new}. The corresponding minimum and maximum values of $T$ are given in curly braces. The selection coefficients $s$ and the mutation effects $\A$ are indicated for each row. Columns show results for different values of $\Th$. The population size is $N = 10^4$ in all cases.}
\label{fig:sweep_shift_tau}
\end{figure}

\FloatBarrier

\singlespacing 
\bibliographystyle{elsarticle-harv}
\bibliography{./content/references}

\end{document}